\documentclass[floatfix,
aps,
10pt,
twocolumn,
superscriptaddress,
amsmath,amssymb,
pra,
]{revtex4-2}

\usepackage{amsmath}
\usepackage{amssymb}
\usepackage[]{graphicx}
\usepackage[colorlinks=true,citecolor=blue,linktocpage=true,linkcolor=blue,filecolor=blue,urlcolor=blue]{hyperref}
\usepackage{physics}
\usepackage{color}
\usepackage{textcase}
\usepackage{bm}

\begin{document}
\title{Parametrically Driven iSWAP Gate Using a Capacitively Shunted \\ Double-Transmon Coupler at the Zero-Flux Sweet Spot}

\author{Shinichi Inoue}
\email{inoue@qipe.t.u-tokyo.ac.jp}
\affiliation{RIKEN Center for Quantum Computing (RQC), Wako, Saitama 351--0198, Japan}
\affiliation{Department of Applied Physics, Graduate School of Engineering, The University of Tokyo, Bunkyo-ku, Tokyo 113-8656, Japan}
\author{Rui Li}
\affiliation{RIKEN Center for Quantum Computing (RQC), Wako, Saitama 351--0198, Japan}
\author{Kentaro Kubo}
\affiliation{Corporate Laboratory, Toshiba Corporation, Kawasaki, Kanagawa 212-8582, Japan}
\author{Yinghao Ho}
\affiliation{Corporate Laboratory, Toshiba Corporation, Kawasaki, Kanagawa 212-8582, Japan}
\author{Yasunobu Nakamura}
\email{yasunobu@ap.t.u-tokyo.ac.jp}
\affiliation{RIKEN Center for Quantum Computing (RQC), Wako, Saitama 351--0198, Japan}
\affiliation{Department of Applied Physics, Graduate School of Engineering, The University of Tokyo, Bunkyo-ku, Tokyo 113-8656, Japan}
\author{Hayato Goto}
\email{hayato.goto.d36@mail.toshiba}
\affiliation{Corporate Laboratory, Toshiba Corporation, Kawasaki, Kanagawa 212-8582, Japan}
\affiliation{RIKEN Center for Quantum Computing (RQC), Wako, Saitama 351--0198, Japan}

\date{\today}

\begin{abstract}
    A double-transmon coupler (DTC) enables a fast, high-fidelity CZ gate between two highly detuned, fixed-frequency transmon qubits. Moreover, a recently proposed capacitively shunted DTC~(CSDTC) realizes a small residual $ZZ$ interaction over a wide flux-bias range around zero flux, eliminating the necessity of static flux biasing while maintaining high CZ-gate fidelity. However, CZ gates with the DTC and CSDTC require baseband flux pulses with large amplitudes, which are vulnerable to pulse distortion and decoherence due to large qubit--coupler hybridization. To address these issues, we experimentally demonstrate a parametrically driven iSWAP gate operated at zero flux bias between highly detuned, fixed-frequency transmon qubits coupled through a CSDTC. Using a simple flux-drive waveform without predistortion, we realize an average gate fidelity of 99.92(2)\% at a total gate time of $112~\mathrm{ns}$. The observed high-fidelity performance is consistent with small qubit--coupler hybridization and small effective $ZZ$ interaction during the gate. Our numerical simulations reproduce the experimentally observed iSWAP interaction rate and effective $ZZ$ interaction, demonstrating the applicability of the theoretical model not only to spectral information but also to time-domain dynamics such as gate operations. These results boost further progress in the research of superconducting quantum computers.
\end{abstract}

\maketitle
\section{Introduction}
    Reliable quantum computation hinges on high-fidelity two-qubit entangling gates. The utility of current noisy intermediate-scale quantum (NISQ) devices based on superconducting circuits is largely limited by noisy two-qubit gates~\cite{kim2023evidence, google2025observation}. High-fidelity two-qubit gates are also key to fault-tolerant quantum computation. Recent work on superconducting processors has been advancing from small-distance quantum error correction~\cite{zhao2022realization,andersen2020repeated,krinner2022realizing} toward logical error suppression beyond break-even~\cite{google2023suppressing,google2024quantum,he2025experimental,google2025demonstrating,lacroix2025scaling}, where overall system performance, including not only two-qubit gates but also other operations, such as measurement, reset, and idling, is crucial~\cite{google2025demonstrating,he2025experimental}. Nevertheless, errors in two-qubit gates remain a major bottleneck of the system performance.

    From the perspective of scalable architecture, it is preferable to design data qubits as fixed-frequency transmon qubits because they are robust against low-frequency flux noise and exhibit relatively long coherence times. While direct interactions between fixed-frequency transmon qubits via capacitive coupling~\cite{sheldon2016procedure, wei2022hamiltonian} are conceptually simple, this approach faces a trade-off between gate efficiency and residual $ZZ$ interaction, which is optimal only in a narrow straddling regime~\cite{tripathi2019operation, malekakhlagh2020first}, severely restricting the designable frequency range and scalability in large processors~\cite{hertzberg2021laser, morvan2022optimizing, zhang2022high, zhang2025efficient}. The trade-off can be overcome by mediating interactions through a dedicated coupler applicable to highly detuned qubits.

    Recent proposals have engineered coupler structures to further reduce unwanted $ZZ$ interactions between data qubits inside~\cite{kandala2021demonstration, mundada2019suppression, kumph2024demonstration} and outside~\cite{shirai2023all, shirai2025high} the straddling regime. The double-transmon coupler (DTC) exploits two internal modes of the coupler and suppresses the $ZZ$ interaction over a broad range of data-qubit detuning at a device-specific optimal flux-bias condition~\cite{goto2022double,kubo2023fast,kubo2024high, li2024realization, li2025capacitively, campbell2023modular, cai2025multiplexed}. Furthermore, its capacitively shunted variant (CSDTC)~\cite{li2025capacitively} enables this cancellation over a wide flux-bias range around zero flux bias.

    Flux-activated two-qubit gates can be broadly classified according to two aspects: (i)~whether they use a baseband flux pulse or an ac flux drive, and (ii)~whether they are activated around the zero-flux sweet spot (a first-order flux-insensitive point) or away from it. A representative class is a baseband flux-controlled gate dynamically tuning the qubit and/or coupler frequencies to approach an avoided level crossing adiabatically~\cite{chen2014qubit,yan2018tunable, li2019realisation, marxer2023long, campbell2023modular, glaser2024sensitivity, sung2021realization, xu2020high, xu2021realization, rol2019fast, collodo2020implementation,stehlik2021tunable} or non-adiabatically~\cite{negirneac2021high, barends2019diabatic, foxen2020demonstrating, xu2021realization, li2019realisation, scarato2025realizing}. These gates exploit strong exchange interactions and therefore can achieve fast two-qubit operations. Some baseband flux-controlled gates are implemented with so-called net-zero flux pulses around the zero-flux sweet spot, thereby reducing their sensitivity to flux noise~\cite{andersen2020repeated, negirneac2021high, li2025capacitively}. However, they require large flux excursions, which enhance incoherent errors during gate execution. Furthermore, the use of baseband pulses necessitates time-consuming calibration of flux-line distortions~\cite{rol2020time, barends2014superconducting, li2024realization,li2025capacitively, hellings2025calibrating}, a process that scales poorly with system size. 
    Another class is flux-driven parametric gates activated by first-order parametric modulation~\cite{mckay2016universal, roth2017analysis, mundada2019suppression, ganzhorn2020benchmarking, kubo2023fast, han2020error, caldwell2018parametrically, abrams2020implementation, sung2021realization, reagor2018demonstration}, offering an attractive alternative by avoiding the above calibration issue and enabling frequency-selective activation of a desired transition. In particular, the iSWAP-type interaction can confine the dynamics within the computational subspace. However, the first-order parametric drive typically requires a flux bias away from the zero-flux sweet spot of the tunable element, reintroducing the trade-off between gate speed and dephasing.
    Finally, a less explored class comprises flux-driven parametric gates activated by the second-order modulation of the energy levels around the zero-flux sweet spot, where the flux-drive frequency is converted to the doubled modulation frequency of the coupling strength, analogous to second-harmonic generation. This class of gates offers a route to the use of an ac flux-drive pulse while retaining a flux-insensitive idle point~\cite{caldwell2018parametrically, abrams2020implementation, hong2020demonstration, sete2024error, sete2021parametric}. By activating the interaction without introducing a static offset from the zero-flux sweet spot, this approach is naturally compatible with tunable-coupler architectures designed for zero-flux-bias idling, such as the CSDTC, where the qubits are first-order insensitive to 1/$f$ flux noise. This feature makes second-harmonic activation especially attractive for the CSDTC.

    Here, we experimentally demonstrate the parametrically driven iSWAP gate at the zero-flux sweet spot between highly detuned, fixed-frequency transmons coupled through a CSDTC. The gate is operated by a flux drive centered at the coupler's zero-flux bias point, where the exchange interaction is generated through the second-harmonic response of the coupler. Through randomized benchmarking, we demonstrate an iSWAP gate with a fidelity of $99.92(2)\%$ for a total gate time of $112~\mathrm{ns}$. We attribute the high gate performance to two distinct mechanisms: smaller flux excursions compared with the baseband-controlled CZ gate realized on the same device~\cite{li2025capacitively}, which leads to a smaller contribution of the coupler to decoherence, and smaller coherent $ZZ$ errors during the gate. The latter effect is due to the wide-range small static $ZZ$ interaction of the CSDTC as well as an additional cancellation of it by the dynamical $ZZ$ interaction induced by the flux drive.

\section{Device} \label{sec:device}
    \begin{figure*}[t]
        \centering
        \includegraphics[]{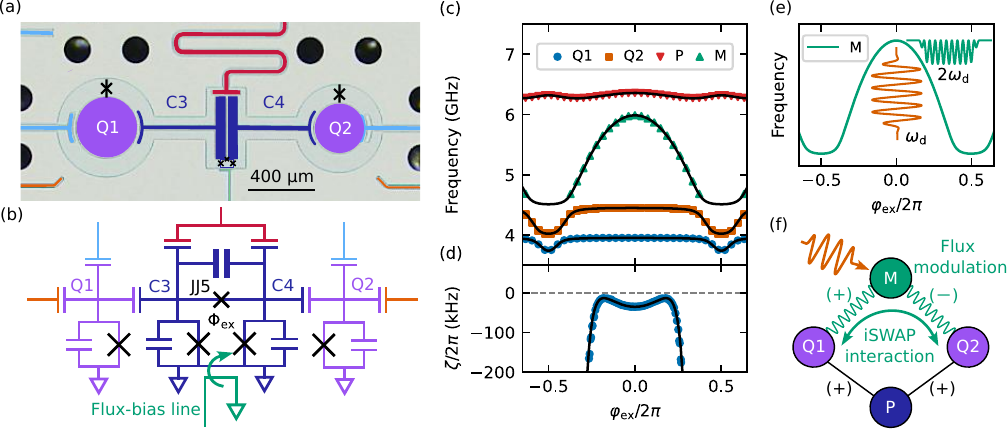}
        \caption{
            Device and operating principle of the parametrically driven iSWAP gate at the zero-flux sweet spot.
            (a)~False-colored optical micrograph. In the present work, we use the same device as in Ref.~\citenum{li2025capacitively}.
            (b)~Circuit diagram of the device. Black crosses represent Josephson junctions. The device has two fixed-frequency transmon qubits (Q1 and Q2) coupled via a CSDTC. The CSDTC consists of two transmons~(C3 and C4), which hybridize into two internal modes~(the P~mode and the M~mode) through the interaction mediated by a Josephson junction (JJ5) and a shunt capacitor. A local magnetic flux $\Phi_{\mathrm{ex}}$ generated by a dedicated bias line~(green) threads the coupler loop. Microwave flux drive is applied through the same line.
            (c)~Energy-level spectrum of the device. Symbols indicate measured spectroscopy data, and solid black lines depict numerical fits using the model Hamiltonian given by Eqs.~\eqref{eq-H}--\eqref{eq-Hm}.
            (d)~Static $ZZ$ interaction $\zeta$ between Q1 and Q2 as a function of the reduced external flux $\varphi_{\mathrm{ex}}$.
            (e)~Parametric drive at the zero flux bias. As the coupler M-mode frequency $\omega_{\mathrm{m}}(\varphi_{\mathrm{ex}})$ is an even function of the external flux $\varphi_{\mathrm{ex}}$, a sinusoidal flux drive at frequency $\omega_{\mathrm{d}}$ modulates the energy level $\omega_{\mathrm{m}}(\varphi_{\mathrm{ex}})$ at $2\omega_{\mathrm{d}}$, resulting in modulation of the interaction between Q1 and Q2 at $2\omega_{\mathrm{d}}$ (see Appendix~\ref{sec:two_photon_iswap_toy_model}).
            (f)~Schematic picture of the exchange interaction driven by the coupler-flux modulation. Choosing $2\omega_{\mathrm{d}}\simeq \Delta_{21}=\omega_2 - \omega_1$ resonantly drives the $\ket{10}$$\Leftrightarrow$$\ket{01}$ exchange interaction.
        }
        \label{fig:device_and_spectrum}
    \end{figure*}
    Here we use the CSDTC device developed in our previous work~\cite{li2025capacitively}. A micrograph of the device and the circuit diagram are shown in Figs.~\ref{fig:device_and_spectrum}(a) and (b), respectively. Details of the CSDTC device design, fabrication, and characterization can be found in Ref.~\citenum{li2025capacitively}. The device parameters remeasured in this work are summarized in Tables~\ref{tab:csdtc_circuit_parameters} and~\ref{tab:device_parameters}.
    
    The device consists of two fixed-frequency transmon qubits (Q1 and Q2) coupled via a CSDTC with two coupler transmons (C3 and C4). An external flux bias $\Phi_{\mathrm{ex}}$ threads the coupler flux loop between the coupler transmons. Defining the reduced external flux as ${\varphi_{\mathrm{ex}} = 2\pi\Phi_{\mathrm{ex}}/\Phi_0}$, where $\Phi_0 = h/2e$ is the magnetic flux quantum, we write the system Hamiltonian depending on $\varphi_{\mathrm{ex}}$ in terms of the phase-difference operators and Cooper-pair number operators as~\cite{goto2022double}
    \begin{align}
        \hat{H}(\varphi_{\mathrm{ex}})
        &= \hat{H}_0 + \hat{V}(\varphi_{\mathrm{ex}}), \label{eq-H} \\
        \hat{H}_{0}
        &= 4\hbar \hat{\mathrm{\mathbf{n}}}^T W \hat{\mathrm{\mathbf{n}}} - \sum_{j=1}^4 E_{\mathrm{J}j} \cos \hat{\varphi}_j - E_{\mathrm{J}5} \cos (\hat{\varphi}_3 - \hat{\varphi}_4), \label{eq-H0}
        \\
        \hat{V}(\varphi_{\mathrm{ex}})
        &=- E_{\mathrm{J}5} \qty[\cos \qty(\hat{\varphi}_3 - \hat{\varphi}_4 + \varphi_{\mathrm{ex}}) - \cos\qty(\hat{\varphi}_3 - \hat{\varphi}_4)],
        \label{eq-Hm}
    \end{align}
    where $\hat{H}_0$ is the Hamiltonian at the zero-flux idle point~[$\hat{H}_0=\hat{H}(0)$], $\hat{V}(\varphi_{\mathrm{ex}})$ describes the flux-dependent part used for the parametrically driven iSWAP gate~[$\hat{V}(0)=0$], $\hbar$ is the reduced Planck constant, and $\hat{\mathrm{\mathbf{n}}}$ is a column vector consisting of the Cooper-pair number operators $\hat{n}_j$ ($j=1, 2, 3, 4$), which are conjugate to the phase-difference operator $\hat{\varphi}_j$ for the $j$th Josephson junction. The phase difference across the coupler junction (JJ5) follows $\hat{\varphi}_5=\hat{\varphi}_3 - \hat{\varphi}_4 + \varphi_{\mathrm{ex}}$ due to the flux quantization, $\hbar W=e^2 \mathsf{C}^{-1}/2$ is a $4\times 4$ charging-energy matrix~[$e$ is the elementary charge and $\mathsf{C}$ is the capacitance matrix defined by $\mathsf{C}_{jj}=\sum_{l=1}^4 C_{jl}$ and $\mathsf{C}_{jl}=-C_{jl}$~($j\neq l$)~\cite{kubo2023fast}], and $E_{\mathrm{J}j}=\Phi_0 I_{\mathrm{c}j}/(2\pi)$~($j=1,\ldots,5$) is the Josephson energy of the $j$th Josephson junction~($I_{\mathrm{c}j}$ is the corresponding critical current).

    The two coupler transmons, C3 and C4, strongly hybridize into two internal modes, the symmetric P~mode and the antisymmetric M~mode, which are observed in the spectroscopy data in Fig.~\ref{fig:device_and_spectrum}(c). The phase-difference operators of the hybridized P~mode and M~mode are expressed as~\cite{li2024realization}
    \begin{align}
        \hat\varphi_{\mathrm{p}}=\frac{\hat\varphi_3+\hat\varphi_4}{2},\qquad
        \hat\varphi_{\mathrm{m}}=\frac{\hat\varphi_3-\hat\varphi_4}{2}.
    \end{align}

    The M~mode is largely flux-tunable, while the P~mode is only weakly flux-dependent, as shown in Fig.~\ref{fig:device_and_spectrum}(c).

    We diagonalize the Hamiltonian $\hat{H}(\varphi_{\mathrm{ex}})$ and denote its $j$th eigenvalue and eigenvector as $\hbar \widetilde{\omega}_j(\varphi_{\mathrm{ex}})$ and $\ket*{\widetilde{\Psi}_{j}(\varphi_{\mathrm{ex}})}$, respectively, with $j$ in ascending order of $\widetilde{\omega}_j(\varphi_{\mathrm{ex}})$ (see Sec.~\ref{sec-operator-matrix} for the matrix representations of the operators). The eigenvectors can be regarded as hybridized states of the two data qubits and two coupler modes (the P~mode and the M~mode) and are denoted by~\cite{li2024realization,li2025capacitively}
    \begin{align}
    |\widetilde{\Psi}_0 \rangle= |\widetilde{0000} \rangle, \quad |\widetilde{\Psi}_1 \rangle= |\widetilde{1000} \rangle, \nonumber \\
    |\widetilde{\Psi}_2 \rangle= |\widetilde{0100} \rangle, \quad |\widetilde{\Psi}_3 \rangle= |\widetilde{0001} \rangle, \nonumber \\
    |\widetilde{\Psi}_4 \rangle= |\widetilde{0010} \rangle, \quad |\widetilde{\Psi}_5 \rangle= |\widetilde{2000} \rangle, \nonumber \\
    |\widetilde{\Psi}_6 \rangle= |\widetilde{1100} \rangle, \quad |\widetilde{\Psi}_7 \rangle= |\widetilde{0200} \rangle, \label{eq-eigenstates}
    \end{align}
    where the ket with four numbers and a tilde is the hybridized state dominated by the state with the excitation numbers of Q1, Q2, the P~mode, and the M~mode, respectively. For convenience, we introduce the following simplified and mutually compatible notations for the single-excitation eigenfrequencies:
    \begin{align}
        \omega_{1} &= \omega_{\widetilde{1000}} = \widetilde{\omega}_1, \quad \omega_{2} = \omega_{\widetilde{0100}} = \widetilde{\omega}_2, \nonumber \\
        \omega_{\mathrm{p}} &= \omega_{\widetilde{0010}} = \widetilde{\omega}_4, \quad
        \omega_{\mathrm{m}} = \omega_{\widetilde{0001}} = \widetilde{\omega}_3,
    \end{align}
    and for the two-qubit ground state with no coupler excitation:
    \begin{align}
        |00\rangle &= |\widetilde{0000} \rangle = |\widetilde{\Psi}_0 \rangle, \\
        \omega_{00} &= \omega_{\widetilde{0000}} = \widetilde{\omega}_0, \label{eq-two_qubit_ground_state_frequency}
    \end{align}
    and similarly for other two-qubit states. Unless otherwise noted, eigenstates and frequencies in Eqs.~\eqref{eq-eigenstates}--\eqref{eq-two_qubit_ground_state_frequency} are evaluated at the zero-flux bias point.

    Figure~\ref{fig:device_and_spectrum}(c) shows the measured energy-level spectrum of the device as a function of the external flux $\varphi_{\mathrm{ex}}$, namely, $\{\widetilde{\omega}_j(\varphi_{\mathrm{ex}})\}$, together with the fits to the circuit Hamiltonian in Eqs.~\eqref{eq-H}--\eqref{eq-Hm}. (The fitting parameters and their results are shown in Table~\ref{tab:csdtc_circuit_parameters}.) The qubit\nobreakdash{--}qubit detuning at the zero-flux idle point in our device is
    \begin{align}
        \Delta_{21}/2\pi
        = (\omega_2 - \omega_{1})/2\pi = 498~\mathrm{MHz},
    \end{align}
    which is larger than the qubit anharmonicities~($-176~\mathrm{MHz}$ and $-213~\mathrm{MHz}$, respectively). Therefore, the two qubits are highly detuned, namely, outside the straddling regime.

    The measured $ZZ$ interaction, together with the theoretical curve obtained from the above fitting, is shown in Fig.~\ref{fig:device_and_spectrum}(d), where $\zeta$ is defined as 
    \begin{align}
       \zeta(\varphi_{\mathrm{ex}}) = \omega_{11}(\varphi_{\mathrm{ex}}) - \omega_{10}(\varphi_{\mathrm{ex}}) - \omega_{01}(\varphi_{\mathrm{ex}}) + \omega_{00}(\varphi_{\mathrm{ex}}).
    \end{align}

    The $ZZ$ interaction is $\zeta/2\pi = -35.3~\mathrm{kHz}$ at zero flux bias and remains small over a wide flux-bias range, reaching the minimum absolute value with $\zeta/2\pi = -13.8~\mathrm{kHz}$ at $\varphi_{\mathrm{ex}}/2\pi = \pm 0.18$. This broad flux-bias region for small $ZZ$ interaction is a key advantage of the CSDTC architecture, which relaxes the constraint on the flux bias and is also beneficial for suppressing coherent $ZZ$ errors during the iSWAP gate operation, as we will discuss in Sec.~\ref{sec:error_analysis}.

\section{\NoCaseChange{iSWAP}-gate implementation}
    \begin{figure}[t]
        \centering
        \includegraphics[]{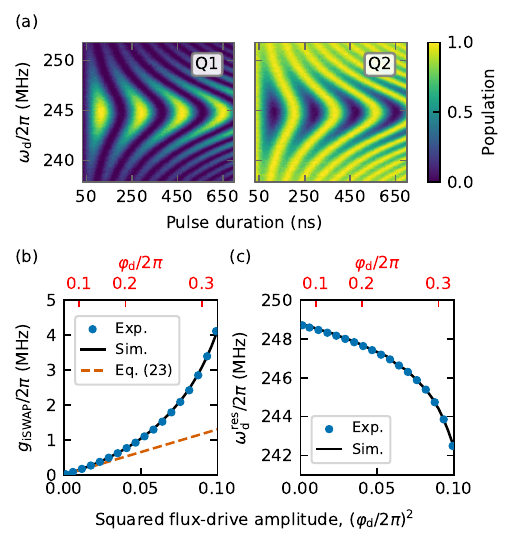}
        \caption{Characterization of parametrically driven iSWAP interaction. (a)~Chevron patterns of the excited-state populations of Q1 and Q2, plotted as a function of the frequency $\omega_{\mathrm{d}}$ and pulse duration of the flux drive with the amplitude $\varphi_\mathrm{d}/2\pi = 0.28$. The initial state is $|01\rangle$. (b)~Experimentally measured iSWAP interaction strength $g_{\mathrm{iSWAP}}$ and (c)~resonant drive frequency $\omega_{\mathrm{d}}^{\mathrm{res}}$ as functions of the squared flux-drive amplitude $\varphi_{\mathrm{d}}^2$, together with numerical simulation results (black solid curves) and analytic prediction, $g_{\mathrm{iSWAP}}$ (orange dashed line) given by Eq.~\eqref{eq:semi1}.}
        \label{fig:iswapscaling}
    \end{figure}
    In this section, we demonstrate a parametrically driven iSWAP gate between the two fixed-frequency transmon qubits (Q1 and Q2) using the CSDTC. We first explain the operating principle in Sec.~\ref{subsec:gate_principle}, showing how ac flux drive at half the qubit--qubit detuning activates an exchange interaction between data qubits (see also Appendix~\ref{sec:two_photon_iswap_toy_model} for an alternative explanation). We then define the gate waveform and summarize calibration procedures in Sec.~\ref{sec:waveform_and_tuneup}. Finally, we present randomized benchmarking (RB) results of the optimized gate in Sec.~\ref{sec:gate_characterization}.

    \subsection{Gate principle} \label{subsec:gate_principle}
        To activate the parametrically driven iSWAP interaction, we apply a sinusoidal flux drive to the coupler flux loop, $\varphi_{\mathrm{ex}}(t)$, around zero flux bias as
        \begin{align}
            \varphi_{\mathrm{ex}}(t) = \varphi_{\mathrm{d}} \cos(\omega_{\mathrm{d}} t + \phi_{\mathrm{d}}),
            \label{eq:flux_drive}
        \end{align}
        where $\varphi_{\mathrm{d}}$, $\omega_{\mathrm{d}}$, and $\phi_{\mathrm{d}}$ are the flux-drive amplitude, frequency, and phase, respectively.

        Moving to the rotating frame with the unitary operator $e^{i\hat{H}_0 t / \hbar}$, we obtain the following Hamiltonian:
        \begin{align}
            \hat{H}_{\mathrm{I}}(t)
            &= e^{i\hat{H}_0 t / \hbar} [\hat{H}(\varphi_{\mathrm{ex}}(t)) - i\hbar \partial_t] e^{-i\hat{H}_0 t / \hbar} \\
            &=
            \sum_{j, k}
            e^{i \Delta_{jk} t} \hbar g_{jk}(t) |\widetilde{\Psi}_j \rangle \langle \widetilde{\Psi}_k |, \label{eq_Heff}
        \end{align}
        where ${\Delta_{jk}=\widetilde{\omega}_j - \widetilde{\omega}_k}$ and 
        \begin{align}
            g_{jk}(t) = \bra*{\widetilde{\Psi}_j} \hat{V}(\varphi_{\mathrm{ex}}(t)) \ket*{\widetilde{\Psi}_k}/\hbar
        \end{align}
        are the detuning and the time-dependent coupling rate between the states $\ket*{\widetilde{\Psi}_j}$ and $\ket*{\widetilde{\Psi}_k}$, respectively. Note that we have dropped a negligible term proportional to the time derivative of $\varphi_{\mathrm{ex}}(t)$ from Eq.~\eqref{eq_Heff}~[see Eq.~(\ref{eq:H_sim}) in Appendix~\ref{sec:cooper_pair_numerical} for the term]. Here, the full interaction-picture Hamiltonian still contains couplings between many different pairs of states, including non-computational states and excited states of the coupler modes.

        We focus on the direct iSWAP interaction between $|10\rangle$ and $|01\rangle$ and consider the following terms in the summation in Eq.~\eqref{eq_Heff}:
        \begin{align}
            \hat{H}_{\mathrm{I}}^{\mathrm{dir}}(t)
            =
            e^{i \Delta_{21} t} \hbar g_{21} (t) |01 \rangle \langle 10 | + \mathrm{h.c.}.
            \label{eq_HiSWAP} 
        \end{align}
        By expanding $\hat{V}(\varphi_{\mathrm{ex}}(t))$ to second order in $\varphi_{\mathrm{ex}}$, 
        \begin{equation}
            \hat{V} (\varphi_{\mathrm{ex}}(t)) \simeq
            E_{\mathrm{J5}} \qty[\frac{\varphi_{\mathrm{ex}}(t)^2}{2} \cos(2\hat{\varphi}_{\mathrm{m}}) + \varphi_{\mathrm{ex}}(t) \sin (2\hat{\varphi}_{\mathrm{m}})], \label{eq:V_expansion}
        \end{equation}
        $g_{21}(t)$ approximately becomes 
        \begin{align}
            g_{21}(t)
            \simeq
            \frac{E_{\mathrm{J}5} \varphi_{\mathrm{ex}}(t)^2}{2\hbar} \langle 01| \cos (2\hat{\varphi}_{\mathrm{m}}) |10\rangle,
            \label{eq-g1001}
        \end{align}
        where the second term in Eq.~\eqref{eq:V_expansion} does not contribute to it because of the parity of the sine function (see Appendix~\ref{sec:hybridization_factor}). 

        Equations~\eqref{eq_HiSWAP} and \eqref{eq-g1001} suggest that the iSWAP interaction is parametrically activated by setting the flux-drive frequency $\omega_{\mathrm{d}}$ at half the detuning as
        \begin{align}
            \omega_{\mathrm{d}} \simeq \frac{\Delta_{21}}{2}. \label{eq:drive-freq}
        \end{align}
        We thus obtain a time-independent iSWAP Hamiltonian in the rotating-wave approximation:
        \begin{align}
            \hat{H}_{\mathrm{iSWAP}}^{\mathrm{dir}}
            &\simeq
            \hbar g_{\mathrm{iSWAP}}^{\mathrm{dir}} e^{-2i \phi_{\mathrm{d}}} |01 \rangle \langle 10 | + \mathrm{h.c.},
            \\
            g_{\mathrm{iSWAP}}^{\mathrm{dir}}
            &=
            \frac{E_{\mathrm{J}5}\varphi_{\mathrm{d}}^2}{8\hbar} \langle 01| \cos (2\hat{\varphi}_{\mathrm{m}}) |10\rangle. \label{eq:semi1}
        \end{align}

        On the other hand, the second term in Eq.~\eqref{eq:V_expansion} results in the indirect iSWAP interaction due to the following terms in the interaction Hamiltonian in Eq.~\eqref{eq_Heff}:
        \begin{equation}
            \hat{H}_\mathrm{I}^\mathrm{indir} = \sum_{j \neq 1,2} \sum_{k = 1,2} e^{i \Delta_{jk} t} \hbar g_{jk}(t) |\widetilde{\Psi}_j \rangle \langle \widetilde{\Psi}_k | + \mathrm{h.c.}, \label{eq:H_virt}
        \end{equation}
        which leads to another iSWAP interaction rate (see Appendix~\ref{sec-giSWAPindirect} for the derivation),
        \begin{align}
            &g_{\mathrm{iSWAP}}^{\mathrm{indir}} \nonumber \\
            &= -\frac{E_{\mathrm{J}5}^2 \varphi_{\mathrm{d}}^2}{4\hbar^2} 
            \sum_{j\neq 1,2} \frac{\bra*{\widetilde{\Psi}_2}  \! \sin (2\hat{\varphi}_{\mathrm{m}}) \! \ket*{\widetilde{\Psi}_j} \bra*{\widetilde{\Psi}_j}  \! \sin (2\hat{\varphi}_{\mathrm{m}}) \! \ket*{\widetilde{\Psi}_1}}{\tilde{\omega}_j-(\widetilde{\omega}_1+\widetilde{\omega}_2)/2}.
        \label{eq:giswap-virtual}
        \end{align}

        Thus in total, the iSWAP interaction rate is given by
        \begin{align}
            g_{\mathrm{iSWAP}}
            =g_{\mathrm{iSWAP}}^{\mathrm{dir}} + g_{\mathrm{iSWAP}}^{\mathrm{indir}}.
            \label{eq:semi1}
        \end{align}
        An alternative description of the iSWAP interaction using bosonic operators is given in Appendix~\ref{sec:two_photon_iswap_toy_model}.
        
        Figure~\ref{fig:iswapscaling}(a) shows the experimentally measured patterns for the $|10\rangle$ and $|01\rangle$ populations, as functions of the flux-drive frequency and pulse duration. The chevron patterns demonstrate coherent population exchange between $|10\rangle$ and $|01\rangle$, namely, the iSWAP interaction. We note that the center frequencies of the chevron patterns are slightly lower than $(\Delta_{21}/2)/(2\pi)=249$~MHz. We attribute this shift to flux-drive-induced ac Zeeman shifts of the qubit frequencies.

        From patterns measured at different flux-drive amplitudes, we extract the iSWAP interaction rate $g_{\mathrm{iSWAP}}$ and resonant drive frequency $\omega_\mathrm{d}^\mathrm{res}$ for each amplitude and summarize the results in Figs.~\ref{fig:iswapscaling}(b) and (c), where we have determined a conversion factor between the generated drive signal amplitude at room temperature and the flux-drive amplitude at the device such that the experimental results for $g_{\mathrm{iSWAP}}$ are in agreement with our full-Hamiltonian numerical simulation results, as shown in Fig.~\ref{fig:iswapscaling}(b) (see Appendix~\ref{sec:cooper_pair_numerical} for the details of the simulation). It is notable that we obtain quantitative agreement between the experimental data and the simulation results over the entire range of flux-drive amplitude by using the theoretical model with the parameter values predetermined by the energy-level fitting in Table~\ref{tab:csdtc_circuit_parameters}. The present theoretical model accurately reproduces experimental results even in time-dependent cases. 

        In Fig.~\ref{fig:iswapscaling}(b), we also show the analytic expression in Eq.~\eqref{eq:semi1} as the dashed line. The analytic result is in good agreement with the experimental result around ${\varphi_{\mathrm{d}}=0}$. The discrepancy at large $\varphi_{\mathrm{d}}$ indicates the limitation of the present analytic approach based on the idle-state basis and the lowest-order approximation in $\varphi_{\mathrm{d}}$.

    \subsection{Waveform and tune-up} \label{sec:waveform_and_tuneup}
        We implement the parametrically driven iSWAP gate by applying a smoothly ramped envelope to the sinusoidal flux drive,
        \begin{align}
            \varphi_{\mathrm{ex}}(t)
            = \varphi_{\mathrm{d}}\, f_{\mathrm{env}}(t)\cos(\omega_{\mathrm{d}} t + \phi_{\mathrm{d}}),
            \label{eq:flux_waveform_def}
        \end{align}
        where $f_{\mathrm{env}}(t)$ is the envelope of the waveform. To suppress leakage while keeping the gate time short, we use a tanh-shaped envelope~\cite{kubo2024high},
        \newcommand{\condshift}{\hspace*{-11.5em}} 
        \begin{equation}
            f_{\mathrm{env}}(t)=
            \left\{
            \begin{array}{@{}l@{\qquad}l@{}}
                \tanh\!\left[\beta \left(t-t_{\mathrm{idle}}\right)\right]
                \tanh\!\left[\beta \left(t_{\mathrm{d}}+t_{\mathrm{idle}}-t\right)\right], & \\[2pt]
                & \condshift (t_{\mathrm{idle}} \le t \le t_{\mathrm{d}}+t_{\mathrm{idle}}), \\
                0, & \condshift (\text{otherwise}).
            \end{array}
            \right. \label{eq:25}
        \end{equation}
        where $t_{\mathrm{d}}$ is the flux-drive duration and the ramp parameter is fixed to $\beta/2\pi = 290/2\pi~\mathrm{MHz}$. Here, we additionally insert $t_{\mathrm{idle}} = 6~\mathrm{ns}$ of zero-amplitude idling before and after the flux pulse to reduce unintentional overlap with other gate pulses. Note that we do not apply any predistortion to compensate for possible waveform distortions, which simplifies the gate calibration procedure. The total gate time is therefore
        \begin{align}
            t_{\mathrm{tot}}=t_{\mathrm{d}}+2t_{\mathrm{idle}}. \label{eq:total_gate_time}
        \end{align}

        During the flux drive, the qubit and coupler frequencies are dynamically shifted by the ac Zeeman effect, leading to systematic phase accumulation and exchange-axis misalignments during the gate. Since the iSWAP gate is phase-sensitive, these phase errors cause the acquired phase to deviate from that of the ideal iSWAP operation, which maps $\ket{01} \to -i\ket{10}$ and $\ket{10} \to -i\ket{01}$. We compensate these systematic phase shifts by applying frame updates to the qubit drives and the coupler flux drive. Here, the frame updates are phase adjustments applied to the digital oscillators that generate the drives so that subsequent pulses are referenced to the updated frame. For the qubit drives for Q1 and Q2, such frame updates are also known as virtual-$Z$ gates~\cite{mckay2017efficient}. See Appendix~\ref{sec:gate_characterization_details} for details of the tune-up procedures with error-amplification sequences.

    \subsection{Gate characterization} \label{sec:gate_characterization}
        \begin{figure}
            \centering
            \includegraphics[]{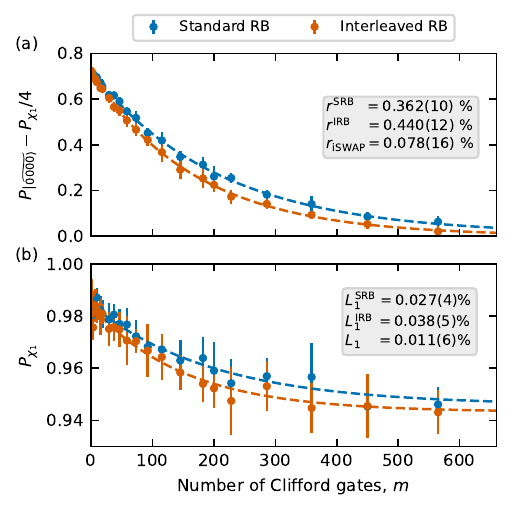}
            \caption{Randomized benchmarking of the iSWAP gate with a total gate duration $t_{\mathrm{tot}}$ of $112~\mathrm{ns}$. (a) Sequence fidelity $P_{\ket*{\widetilde{0000}}} - P_{\chi_1}/4$ as a function of the number of Clifford gates. The dashed lines are fits to Eq.~\eqref{eq:rb_fit}. (b) Leakage randomized benchmarking. The dashed lines are fits to Eq.~\eqref{eq:lrb_fit}.}
            \label{fig:rb_results}
        \end{figure}
        To assess the gate performance, we characterize the optimized iSWAP gate using standard and interleaved randomized benchmarking (SRB/IRB)~\cite{sung2021realization} and leakage randomized benchmarking (LRB)~\cite{li2024realization}. 
         
        We denote by $P_{\ket{\widetilde{0000}}}(m)$ the probability that all four modes of the qubits and coupler are found in the ground state after an $m$-Clifford sequence. We also denote by $P_{\chi_{1}}(m)$ the leakage-free probability that the system remains in the computational subspace of the two data qubits while both coupler modes remain in their ground state after an $m$-Clifford sequence. 
        
        To extract the iSWAP gate error taking leakage into account, we fit the sequence fidelity
        \begin{align}
            P_{\mathrm{seq}}(m) = P_{\ket{\widetilde{0000}}}(m) - \frac{P_{\chi_1}(m)}{d} 
        \end{align}
        defined in Ref.~\cite{li2024realization} to 
        \begin{equation}
            P_\mathrm{seq} = A_\mathrm{r} \lambda_\mathrm{r}^m + B_\mathrm{r}, \label{eq:rb_fit}
        \end{equation}
        where $d=4$ is the dimension of the computational subspace, and $A_{\mathrm{r}}, B_{\mathrm{r}}$, and $\lambda_{\mathrm{r}}$ are fitting parameters. Using the fitting results of $\lambda_{\mathrm{r}}$ in SRB and IRB, respectively denoted by $\lambda_{\mathrm{r}}^{\mathrm{SRB}}$ and $\lambda_{\mathrm{r}}^{\mathrm{IRB}}$, we obtain the iSWAP gate error $r_{\mathrm{iSWAP}}$ as 
        \begin{align}
            r_{\mathrm{iSWAP}} = 1 - \frac{1 - r^{\mathrm{IRB}}}{1 - r^{\mathrm{SRB}}},
        \end{align}
        where $r^{\mathrm{SRB/IRB}} = (1 - \lambda_{\mathrm{r}}^{\mathrm{SRB/IRB}})  (1 - 1/d)$. The results are shown in Fig.~\ref{fig:rb_results}(a).

        Similarly, we extract the leakage error by fitting the leakage-free probability $P_{\chi_1}(m)$ as
        \begin{align}
            P_{\chi_1}(m) = A_{\mathrm{L}} \lambda_{\mathrm{L}}^m + B_{\mathrm{L}}, \label{eq:lrb_fit}
        \end{align}
        where $A_{\mathrm{L}}, B_{\mathrm{L}}$, and $\lambda_{\mathrm{L}}$ are fitting parameters. From the fitted $\lambda_{\mathrm{L}}$ in SRB and IRB, respectively denoted by $\lambda_{\mathrm{L}}^{\mathrm{SRB}}$ and $\lambda_{\mathrm{L}}^{\mathrm{IRB}}$, the leakage error is evaluated as
        \begin{align}
            L_{1} = 1 - \frac{1 - L_1^{\mathrm{IRB}}}{1 - L_1^{\mathrm{SRB}}},
        \end{align}
        where $L_{1}^{\mathrm{SRB/IRB}} = (1 - \lambda_{\mathrm{L}}^{\mathrm{SRB/IRB}})
        (1 - 1/d)$. The results are shown in Fig.~\ref{fig:rb_results}(b).

       We thus obtain the iSWAP gate error $r_{\mathrm{iSWAP}}$ including both the depolarization-induced error $r^{\mathrm{D}}_{\mathrm{iSWAP}}$ and the leakage contribution~\cite{li2024realization}:
        \begin{align}
            r_{\mathrm{iSWAP}} = r^{\mathrm{D}}_{\mathrm{iSWAP}} + \frac{(d-1)L_{1}}{d}.
        \end{align}
        The average gate infidelity is then given by
        \begin{align}
            \epsilon
            = 1 - F_{\mathrm{avg}}
            = r^{\mathrm{D}}_{\mathrm{iSWAP}} + L_{1}
            = r_{\mathrm{iSWAP}} + \frac{L_{1}}{d}. \label{eq:rdiswap}
        \end{align}

        We characterize the gate for total gate durations $t_{\mathrm{tot}}$ ranging from $96~\mathrm{ns}$ to $320~\mathrm{ns}$, optimizing the gate parameters at each duration. The best performance is obtained at a total gate time of $112~\mathrm{ns}$, including $12~\mathrm{ns}$ of idling, yielding $r_{\mathrm{iSWAP}} = 0.078(15)\%$ and $L_{1} = 0.011(7)\%$. The resulting total gate infidelity is
        \begin{align}
            \epsilon = 1 - F_{\mathrm{avg}} = r_{\mathrm{iSWAP}} + \frac{L_{1}}{4} = 0.08(2)\%,
        \end{align}
        corresponding to an average gate fidelity of $F_{\mathrm{avg}} = 99.92(2)\%$.

\section{Gate-error analysis} \label{sec:error_analysis}

\begin{figure}
        \centering
        \includegraphics[]{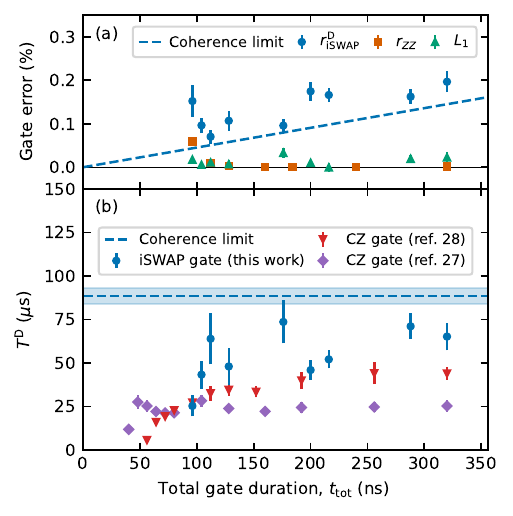}
        \caption{(a)~Experimentally measured iSWAP depolarization-induced error $r_\mathrm{iSWAP}^\mathrm{D}$, coherent $ZZ$ error $r_{ZZ}$, and leakage error $L_1$ as a function of the total iSWAP gate duration $t_\mathrm{tot}$. The blue dashed line denotes gate infidelities estimated from the qubit coherence times measured at zero flux bias. (b)~Effective coherence time $T^\mathrm{D}$~(blue dots) inferred from the depolarization-induced error for each total gate duration $t_{\mathrm{tot}}$. The blue dashed line shows the effective coherence time estimated from the same qubit coherence properties as in (a), and the shaded blue band indicates the $\pm 1\sigma$ uncertainty. The red and purple markers show the effective coherence times reported for CZ gates using a DTC~\cite{li2024realization} and CSDTC~\cite{li2025capacitively}, respectively.
        }
        \label{fig:coherence_time_fitting}
    \end{figure}

    We next analyze the error budget of the iSWAP gate by considering three contributions: coherent $ZZ$ error, leakage error, and incoherent error due to energy relaxation and dephasing. We derive analytic expressions for their contributions and compare the estimated total error with the experimentally measured gate errors at various total gate durations $t_{\mathrm{tot}}$ in Fig.~\ref{fig:coherence_time_fitting}(a). To estimate the gate fidelity analytically, we use the average gate fidelity~\cite{pedersen2007fidelity} given by
    \begin{align}
        F_{\mathrm{avg}} = \frac{\abs{\mathrm{Tr}\qty[\hat{U}_{\mathrm{iSWAP}}^\dagger \tilde{U}_{\mathrm{exp}}]}^2 + \mathrm{Tr}\qty[\tilde{U}_{\mathrm{exp}}^\dagger \tilde{U}_{\mathrm{exp}}]}{d(d+1)}, \label{eq:Fid_def}
    \end{align}
    where $d=4$ is the dimension of the Hilbert space of the two-qubit system, $\hat{U}_{\mathrm{iSWAP}}$ is the ideal iSWAP unitary matrix, and $\tilde{U}_{\mathrm{exp}}$ is the ${4\times 4}$ matrix representing the actual noisy gate operation.
    \subsection{Coherent $ZZ$ error}
        The residual $ZZ$ interaction during the iSWAP gate can be modeled as an accumulated phase error for $\ket{11}$. The resulting unitary evolution is
        \begin{align}
            \hat{U}(\phi_{ZZ}) =
            \begin{pmatrix}
                1 & 0 & 0 & 0 \\
                0 & 0 & -i & 0 \\
                0 & -i & 0 & 0 \\
                0 & 0 & 0 & e^{-i\phi_{ZZ}}
            \end{pmatrix}.
        \end{align}

        This coherent phase error can be partially compensated by applying equal-angle virtual-$Z$ gates on both data qubits,
        $\mathrm{VZ}_1(\gamma_{\mathrm{q}}) \otimes \mathrm{VZ}_2(\gamma_{\mathrm{q}})$, after the iSWAP gate. Because virtual-$Z$ gates are implemented by software-based frame updates, this correction does not introduce additional physical errors~\cite{lao2022software, google2025demonstrating}. The infidelity due to the $ZZ$ error is minimized when
        \begin{align}
            \gamma_{\mathrm{q}} = \frac{\phi_{ZZ}}{2},
        \end{align}
        which gives the following coherent $ZZ$ error:
        \begin{align}
            r_{ZZ}
            = 1 - \frac{4\abs{1 + e^{i\phi_{ZZ}/2}}^2 + 4}{20}
            \simeq \frac{\phi_{ZZ}^2}{20},
            \label{eq:rzz1}
        \end{align}
        where we have approximated the infidelity in the small-$\phi_{ZZ}$ limit. The coherent $ZZ$ error without the virtual-$Z$ compensation is ${3\phi_{ZZ}^2/20}$~\cite{ding2023high, fors2024comprehensive}, which is three times larger than the compensated value. In the experiment, this mitigation is included in the gate calibration: in Fig.~\ref{fig:tuneup}(h) of Appendix~\ref{sec:gate_characterization_details}, we optimize the frame update parameter $\gamma_{\mathrm{q}}$ using an ORBIT-based calibration~\cite{kelly2014optimal}.
        
        For each total gate duration $t_{\mathrm{tot}}$, we first calibrate the gate parameters using the tune-up protocol described above. We then measure the accumulated $ZZ$ phase $\phi_{ZZ}$ during the iSWAP gate using the characterization protocol shown in Fig.~\ref{fig:dynamic_zz}(a) below~\cite{ganzhorn2020benchmarking}. The coherent $ZZ$ error $r_{ZZ}$, obtained from the measured $\phi_{ZZ}$ using Eq.~\eqref{eq:rzz1}, is plotted in Fig.~\ref{fig:coherence_time_fitting}(a).

    \subsection{Leakage error}
        We perform the leakage randomized benchmarking~(LRB) described in the previous section for a range of gate durations~[Fig.~\ref{fig:coherence_time_fitting}(a)]. For the gate durations for which the fitting uncertainty is small, the extracted leakage error remains small at the level of $0.01\%$, which is in agreement with the numerical simulation using the full Hamiltonian in Eqs.~\eqref{eq-H}--\eqref{eq-Hm}. 

        For several data points, however, especially above a total gate duration of $t_{\mathrm{tot}} = 128~\mathrm{ns}$, the interleaved and standard LRB curves largely overlap within the measurement uncertainty. We therefore omit such points from the plot. To complement this experimental limitation, we also estimate the leakage error by numerical simulations using the full system Hamiltonian in Eqs.~\eqref{eq-H}--\eqref{eq-Hm}. The simulation results confirm that the leakage error remains small at the 0.01\% level for the gate durations studied here.
    \subsection{Idling error}
        The iSWAP gate waveform includes $12~\mathrm{ns}$ of zero-amplitude idling in the waveform defined in Eq.~\eqref{eq:flux_waveform_def}. 
        During this idling time, the data qubits are subject to incoherent errors at the zero-flux bias point. We characterize the idling decoherence of the $i$th qubit by the decoherence rate
        \begin{align}
        \Gamma_{\mathrm{idle}, i} = \Gamma_{1, i} + \Gamma_{\phi,i}^{*},
        \end{align}
        where $\Gamma_{1, i}$ and $\Gamma_{\phi,i}^{*}$ are the energy-relaxation and Ramsey pure-dephasing rates of the $i$th qubit at the zero-flux bias point, respectively. Using the qubit coherence properties measured at zero flux bias (see Table~\ref{tab:device_parameters} in Appendix~\ref{sec:coherence}), the effective coherence time during idling can be estimated as
        \begin{align}
            T_{\mathrm{idle}, 1}^{\mathrm{eff}} &= \frac{1}{\Gamma^{\mathrm{eff}}_{\mathrm{idle}, 1}} = 163(19)~\mathrm{\mu s}, \\
            T_{\mathrm{idle}, 2}^{\mathrm{eff}} &= \frac{1}{\Gamma^{\mathrm{eff}}_{\mathrm{idle}, 2}} = 133(16)~\mathrm{\mu s}.
        \end{align}
        Using these decoherence rates and the total idling time $2t_{\mathrm{idle}} = 12~\mathrm{ns}$, we estimate the infidelity contribution from the idling process as
        \begin{align}
            r_{\mathrm{idle}}^{\mathrm{eff}} = \frac{2\cdot 2 t_{\mathrm{idle}}}{5} \qty(\Gamma_{\mathrm{idle}, 1}^{\mathrm{eff}} + \Gamma_{\mathrm{idle}, 2}^{\mathrm{eff}}) = 0.0065(5)\%.
        \end{align}

    \subsection{Incoherent error as a function of gate time}
        \begin{figure}
            \centering
            \includegraphics[]{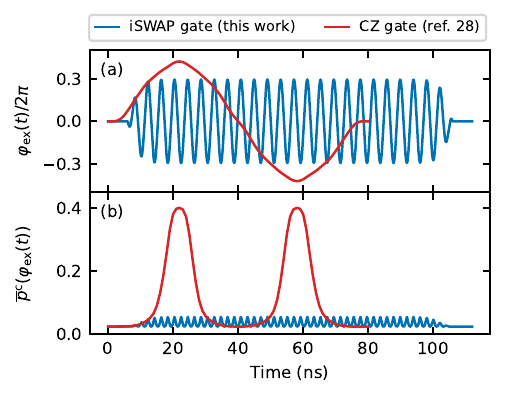}
            \caption{(a)~Optimized flux-drive waveform for the iSWAP gate ($t_\mathrm{tot}=112$~ns), together with the symmetric part of the BNZ waveform used to implement a CZ gate ($t_{\mathrm{tot}}=80$~ns) on the same device in Ref.~\citenum{li2025capacitively}. (b)~Numerically simulated coupler hybridization fraction as a function of time. The solid blue and red lines represent the hybridization fractions for the iSWAP- and CZ-gate waveforms in (a), respectively.}
            \label{fig:coupler_hybridization}
        \end{figure}
        The experimentally measured depolarization-induced iSWAP gate error $r^{\mathrm{D}}_{\mathrm{iSWAP}}$ is shown in Fig.~\ref{fig:coherence_time_fitting}(a) as a function of total gate duration $t_{\mathrm{tot}}$. Except for the shortest gate duration, where coherent $ZZ$ error becomes large because of the large qubit--coupler hybridization due to the large flux-drive amplitude, the gate error increases approximately linearly with the gate duration, indicating that the incoherent error is dominant. To quantify the incoherent error, we use the following effective formula~\cite{ding2023high,li2024realization}:
        \begin{align}
            r_{\mathrm{incoh}} = \frac{2 t_{\mathrm{tot}}}{5} \sum_{i=1,2} \left(\frac{1}{T^{\mathrm{eff}}_{1,i}} + \frac{1}{T^{\mathrm{eff}}_{\phi, i}}\right) \equiv \frac{2}{5} \frac{t_{\mathrm{tot}}}{T^{\mathrm{eff}}}, \label{eq:rincoh1}
        \end{align} 
        where $T_{1,i}^\mathrm{eff}$ and $T_{\phi, i}^\mathrm{eff}$ are the effective energy-relaxation and pure-dephasing times of the $i$th qubit under the flux drive, and we denote the overall effective coherence time as $T^{\mathrm{eff}}$. The effective coherence time depends on the flux-drive amplitude, which can be estimated by integrating over the flux-drive trajectory~\cite{marxer2023long, marxer2025above} or directly measuring the driven coherence~\cite{hong2020demonstration, fried2019assessing, didier2019ac}.
        
        The effective coherence time, which is hard to estimate accurately, is lower bounded by the coherence time defined with the depolarization-induced error $r^{\mathrm{D}}$~\cite{ding2023high,li2024realization} as
        \begin{align}
            T^{\mathrm{eff}} = \frac{2}{5}\frac{t_{\mathrm{tot}}}{r_{\mathrm{incoh}}} \geq \frac{2}{5}\frac{t_{\mathrm{tot}}}{r^{\mathrm{D}}} \equiv T^{\mathrm{D}}, \label{eq:teff_def}
        \end{align}
        even if we do not have access to purity data~\cite{wallman2015estimating, arute2019quantum}. Thus in the following, we use $T^{\mathrm{D}}$ as an effective coherence time instead of $T^{\mathrm{eff}}$, assuming that the contribution of coherent errors is small.

        The experimentally estimated effective coherence times $T^{\mathrm{D}}$ are shown in Fig.~\ref{fig:coherence_time_fitting}(b). For the iSWAP gate studied in this work, we obtain $T^{\mathrm{D}}_{\mathrm{iSWAP}} = 57(11)~\mathrm{\mu s}$ at the optimal gate duration, which is shorter than $88(4)~\mathrm{\mu s}$ estimated from Eq.~\eqref{eq:rincoh1} with the zero-flux coherence times in Table~\ref{tab:device_parameters}. This difference is expected because, during the flux drive, the qubits are exposed to additional decoherence channels that are absent or weaker at the idling point, including enhanced qubit--coupler hybridization, increased sensitivity to flux noise, and experimental noise originating from classical control electronics~\cite{fried2019assessing, didier2019ac}.
        
        We next compare our results with previous CZ-gate demonstrations using a DTC~\cite{li2024realization} and CSDTC~\cite{li2025capacitively}~[Fig.~\ref{fig:coherence_time_fitting}(b)]. Although the uncertainty in $T^{\mathrm{D}}_{\mathrm{iSWAP}}$ in our iSWAP gate is larger, the present effective coherence times are systematically longer than the previous two schemes across the range of gate durations studied here. The longer effective coherence times can be explained as follows: The parametric iSWAP gate is implemented with a smaller flux excursion from the zero-bias point than that for the CZ gate~\cite{li2025capacitively}. The reduced drive amplitude leads to smaller qubit--coupler hybridization, which then makes the qubits less affected by the short-lived coupler modes.
        
        To evaluate the qubit--coupler hybridization, we define the hybridization fraction of the coupler modes in a computational state $|ij\rangle$ as $p_{ij}^{\mathrm{c}}(\varphi_{\mathrm{ex}}(t))$, following the approach in Appendix~\ref{sec:hybridization_factor}. This fraction quantifies how much coupler excited states participate in the computational state at each flux bias $\varphi_{\mathrm{ex}}$. The averaged hybridization fraction for all the computational states is then defined as
        \begin{align}
            \bar{p}^{\mathrm{c}}(\varphi_{\mathrm{ex}}(t)) = \frac{1}{4} \sum_{i,j \in \qty{0, 1}} p_{ij}^{\mathrm{c}}(\varphi_{\mathrm{ex}}(t)). \label{eq:avg_hybridization}
        \end{align}

        Figures~\ref{fig:coupler_hybridization}(a) and (b) compare the optimized flux-drive waveforms for the iSWAP and CZ gates and the corresponding trajectories of $\bar{p}^{\mathrm{c}}(\varphi_{\mathrm{ex}}(t))$, respectively. Because the qubit--coupler hybridization sharply increases when the coupler M~mode frequency approaches the qubit frequencies, the hybridization fraction during the CZ gate is much larger than that for the iSWAP gate. The CZ gate in Ref.~\citenum{li2025capacitively} was implemented with a total gate time of $80~\mathrm{ns}$, and $\bar{p}^{\mathrm{c}} = 0.40$ at the peaks. In contrast, the present iSWAP gate is realized with a total gate time of $112~\mathrm{ns}$ and $\bar{p}^{\mathrm{c}} = 0.05$ at most. This order-of-magnitude reduction in the participation of the short-lived coupler modes is consistent with the longer effective coherence times and higher gate fidelity of the iSWAP gate, despite its longer duration.
       
\section{Suppression of $ZZ$ interaction during \NoCaseChange{iSWAP} gate} \label{sec:zz_suppression}
    In this section, we characterize the effective $ZZ$ interaction across the range of flux-drive amplitudes used for the iSWAP gate by measuring the accumulated phase on $\ket{11}$ using the pulse sequence shown in Fig.~\ref{fig:dynamic_zz}(a). The extracted effective $ZZ$ interaction $\zeta^{\mathrm{eff}}$ is shown in Fig.~\ref{fig:dynamic_zz}(b) as a function of the flux-drive amplitude $\varphi_{\mathrm{d}}$ together with the corresponding full-Hamiltonian simulation.

    \subsection{$ZZ$ interaction under flux drive}
        As a baseline, we first consider the averaged static $ZZ$ interaction in the iSWAP gate defined as
        \begin{align}
            \zeta^{\mathrm{avg}} = \frac{1}{t_{\mathrm{tot}}} \int_0^{t_{\mathrm{tot}}} \zeta(\varphi_{\mathrm{ex}}(t)) \dd{t}, \label{eq:avgzz}
        \end{align}
        where $t_{\mathrm{tot}}$ is the total gate duration including the idling time [Eq.~\eqref{eq:total_gate_time}]. 
        
        The blue solid line in Fig.~\ref{fig:dynamic_zz}(b) shows the averaged static $ZZ$ interaction as a function of the flux-drive amplitude $\varphi_{\mathrm{d}}$ numerically computed from the simulated energy-level structure. An advantage of the present CSDTC is that it exhibits two minima in the absolute value of its static $ZZ$ interaction as a function of the external flux bias $\varphi_{\mathrm{ex}}$~[Fig.~\ref{fig:device_and_spectrum}(d)]. As a result, the averaged static $ZZ$-interaction magnitude becomes smaller than that of the static $ZZ$ interaction at the zero-flux idle point. Therefore, the CSDTC can maintain a small $ZZ$ interaction even at a relatively large drive amplitude of $\varphi_{\mathrm{d}}/2\pi \simeq \pm 0.28$.
    \subsection{Dynamical suppression of $ZZ$ interaction}
        \begin{figure}
            \centering
            \includegraphics[]{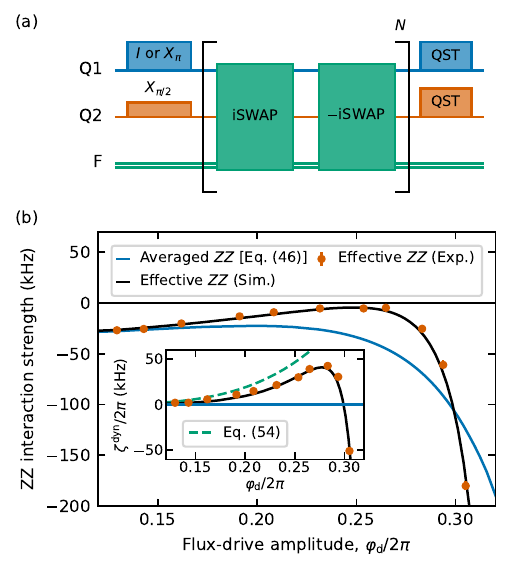}
            \caption{(a)~Pulse sequence used to evaluate the effective $ZZ$ interaction strength. F stands for the flux drive inducing the iSWAP and $-$iSWAP gates. (b)~Experimentally measured effective $ZZ$ interaction $\zeta^\mathrm{eff}$~(orange dots) and numerically calculated averaged static $ZZ$ interaction $\zeta^\mathrm{avg}$~[Eq.~\eqref{eq:avgzz}; blue curve] as a function of the flux-drive amplitude. The solid black curve is the full-Hamiltonian simulation of the effective $ZZ$ interaction, which shows quantitative agreement with the experimental data. The inset shows the dynamical contributions corresponding to the dots and curves in~(b). The green dashed curve is the analytic estimate of the dynamical $ZZ$ interaction from Eq.~\eqref{eq:sidebandzz}.}
            \label{fig:dynamic_zz}
        \end{figure}
        To examine the suppression of the effective $ZZ$ interaction, we define the dynamical $ZZ$ interaction as the difference between the effective $ZZ$ interaction and the averaged static $ZZ$ interaction:
        \begin{align}
            \zeta^{\mathrm{dyn}} = \zeta^{\mathrm{eff}} - \zeta^{\mathrm{avg}}.
        \end{align}

        The inset of Fig.~\ref{fig:dynamic_zz}(b) shows the dynamical $ZZ$ interaction $\zeta^{\mathrm{dyn}}$ as a function of the flux-drive amplitude. Over the operating range of our iSWAP gate \mbox{($\varphi_{\mathrm{d}}/2\pi \lesssim 0.28$)}, the dynamical $ZZ$ interaction is consistently positive, suppressing the effective $ZZ$ interaction by partially canceling the negative averaged static $ZZ$ interaction. Remarkably, the cancellation is almost perfect at $\varphi_{\mathrm{d}}/2\pi = 0.26$, where the effective $ZZ$ interaction reaches its smallest magnitude, $\zeta^{\mathrm{eff}}/2\pi = -4.7~\mathrm{kHz}$. The maximum dynamical $ZZ$ interaction is $\zeta^{\mathrm{dyn}}/2\pi = 42.4~\mathrm{kHz}$ at $\varphi_{\mathrm{d}}/2\pi = 0.28$, which is about $60\%$ of the magnitude of the averaged static $ZZ$ interaction at the same drive amplitude. 

        This positive dynamical $ZZ$ interaction can be explained qualitatively by the level repulsion from the sideband transitions near the flux-drive frequency. Since the drive frequency is comparable to the detuning between $\ket{11}$ and $\ket{20}$ and that between $\ket{11}$ and $\ket{02}$, we focus on the following interaction-Hamiltonian terms from Eq.~\eqref{eq_Heff} for the sideband-induced $ZZ$ interaction:
        \begin{align}
            \hat{H}_{ZZ}^{\mathrm{dyn}}(t)
            &\simeq 
            e^{i (\omega_{20} - \omega_{11}) t} \hbar g_{20,11}(t) |20 \rangle \langle 11 | + \mathrm{h.c.} \nonumber
            \\
            &+
            e^{i (\omega_{02} - \omega_{11}) t} \hbar g_{02,11}(t) |02 \rangle \langle 11 | + \mathrm{h.c.},
            \\
            g_{20,11}(t)
            &\simeq 
            \frac{E_{\mathrm{J}5} \varphi_{\mathrm{ex}}(t)^2}{2\hbar } \langle 20| \cos (2\hat{\varphi}_{\mathrm{m}}) |11\rangle, \label{eq:49}
            \\
            g_{02,11}(t)
            &\simeq
            \frac{E_{\mathrm{J}5} \varphi_{\mathrm{ex}}(t)^2}{2\hbar } \langle 02| \cos (2\hat{\varphi}_{\mathrm{m}}) |11\rangle. \label{eq:50}
        \end{align}

        Substituting $\varphi_{\mathrm{ex}}(t)$ in Eq.~\eqref{eq:flux_drive} into Eqs.~\eqref{eq:49} and \eqref{eq:50} and keeping the slowest oscillating terms, 
        we obtain 
        \begin{align}
            \hat{H}_{ZZ}^{\mathrm{dyn}}(t)
            &\simeq
            e^{i \alpha_1 t} \hbar \bar{g}_{20,11} |20 \rangle \langle 11 | + \mathrm{h.c.} \nonumber
            \\
            &+
            e^{i \alpha_2 t} \hbar \bar{g}_{02,11} |02 \rangle \langle 11 | + \mathrm{h.c.}, \label{eq-Hzz}
            \\
            \bar{g}_{20,11}
            &\simeq
            \frac{E_{\mathrm{J}5} \varphi_{\mathrm{d}}^2}{8\hbar } e^{2i \phi_{\mathrm{d}}} \langle 20| \cos (2\hat{\varphi}_{\mathrm{m}}) |11\rangle, 
            \\
            \bar{g}_{02,11}
            &\simeq
            \frac{E_{\mathrm{J}5} \varphi_{\mathrm{d}}^2}{8\hbar } e^{-2i \phi_{\mathrm{d}}} \langle 02| \cos (2\hat{\varphi}_{\mathrm{m}}) |11\rangle, 
        \end{align}
        where $\alpha_1=\omega_{20}-2\omega_{10}$ and $\alpha_2=\omega_{02}-2\omega_{01}$ are 
        the anharmonicities of Q1 and Q2 at zero flux bias, respectively. Here we have used the approximations $\omega_{11} \simeq \omega_{10} + \omega_{01}$ and $2\omega_{\mathrm{d}} \simeq \omega_{01} - \omega_{10}$.
        These off-resonant drives between $\ket{11}$ and $\qty{\ket{20}, \ket{02}}$ induce a frequency shift of $\ket{11}$, namely, the dynamical $ZZ$ interaction:
        \begin{align}
            &\zeta^{\mathrm{dyn}} 
            \simeq  \qty(\frac{E_{\mathrm{J}5}\varphi_{\mathrm{d}}^2}{8\hbar})^2
            \times 
            \Big[\frac{\left| \langle 20| \cos(2\hat{\varphi}_{\mathrm{m}}) |11\rangle \right|^2}{-\alpha_1} \nonumber \\
            &\qquad \qquad +
            \frac{\left|\langle 02| \cos(2\hat{\varphi}_{\mathrm{m}}) |11\rangle \right|^2}{-\alpha_2}\Big]. \label{eq:sidebandzz}
        \end{align}

        As $\alpha_{1,2}<0$ for the transmons, $\zeta^{\mathrm{dyn}}$ in Eq.~\eqref{eq:sidebandzz} is positive within this approximation, which explains why the absolute value of the effective $ZZ$ interaction is smaller than that of the averaged static $ZZ$ interaction. At the largest flux-drive amplitude in our experiment ($\varphi_{\mathrm{d}}/2\pi = 0.31$), where $\zeta^{\mathrm{dyn}}$ becomes negative in Fig.~\ref{fig:dynamic_zz}(b), the three-level approximation in Eq.~\eqref{eq-Hzz} apparently breaks down due to additional level repulsions involving coupler states. A discussion on the dynamical $ZZ$ interaction using a toy model is given in Appendix~\ref{sec:two_photon_iswap_toy_model}.

\section{Discussion and Conclusion}
    In this work, we realized a high-fidelity parametrically driven iSWAP gate with a CSDTC at the zero-flux sweet spot. By systematically characterizing the gate error as a function of the gate duration, we found an optimal condition where the gate fidelity reaches $99.92(2)\%$ with a total gate time of $112~\mathrm{ns}$ including $12~\mathrm{ns}$ of idling. The measured error was dominated by incoherent errors, while both leakage and coherent $ZZ$ errors were strongly suppressed. Notably, this performance was obtained using a simple analytically defined waveform without predistortion, substantially reducing the control complexity compared with baseband flux-activated two-qubit gates.

    We attribute the high-fidelity performance of the iSWAP gate to two key features of the proposed scheme. First, the CSDTC architecture benefits from a small static $ZZ$ interaction over a wide flux-bias range. The interaction is further reduced by the dynamical $ZZ$ interaction under the flux drive during the iSWAP gate. Second, the iSWAP gate uses a relatively small flux-pulse amplitude, thereby avoiding strong qubit--coupler hybridization that enhances incoherent errors. In addition to these features, the present iSWAP gate is activated via flux drive at the zero-flux sweet spot, reducing its sensitivity to low-frequency flux noise. Thus, the present scheme can achieve a high-fidelity iSWAP gate despite its relatively long gate duration.

    One notable feature of this work is the quantitative agreement between the experimental results and full-circuit numerical simulations using the Cooper-pair-number basis in a multi-transmon system. We successfully obtained significant agreement not only for the static energy spectrum and $ZZ$ interaction (Fig.~\ref{fig:device_and_spectrum}) but also for the time-dependent properties such as iSWAP interaction strength and frequency (Fig.~\ref{fig:iswapscaling}) and the dynamical $ZZ$ interaction (Fig.~\ref{fig:dynamic_zz}). These results demonstrate that the Cooper-pair-basis modeling can provide a reliable framework for designing and optimizing high-fidelity gates in superconducting quantum processors.

    Further increasing the gate speed and fidelity will require reducing both the effective $ZZ$ interaction and the increased dephasing under flux drive. In this work, we employed an analytically defined simple flux pulse envelope. An interesting next step is to use a more sophisticated pulse shape, such as a multi-tone drive to cancel the unwanted $ZZ$ interaction~\cite{jin2025superconducting}. In particular, a weak auxiliary tone near the $\ket{11}\!\Leftrightarrow\!\ket{02}$ or $\ket{11}\!\Leftrightarrow\!\ket{20}$ transition provides an additional knob for reducing the driven $ZZ$ interaction even at stronger drive amplitudes. On the other hand, however, suppressing the drive-induced dephasing in the present architecture remains an open question. One possible route is to engineer the energy-level structure under flux drive so as to realize an ac sweet spot near the operating amplitude~\cite{hong2020demonstration}.

\begin{acknowledgments}
    We thank Tarush Tiwari, Leonardo Ranzani, and Archana Kamal for sharing the results of their independent work prior to publication. We also acknowledge K. Kusuyama for Ta film deposition and NICT for wafer supply. We thank Z. Yan, S. Wang, H. Akimoto, and H. Mukai for helpful advice on troubleshooting the control electronics and the dilution refrigerator. S.I. acknowledges support from Forefront Physics and Mathematics Program to Drive Transformation (FoPM), a World-leading Innovative Graduate Study (WINGS) Program, the University of Tokyo, and RIKEN Junior Research Associate Program. This research was partly supported by the Ministry of Education, Culture, Sports, Science and Technology~(MEXT) Quantum Leap Flagship Program~(Q-LEAP) (Grant No.~JPMXS0118068682), and the Japan Science and Technology Agency (JST) as part of Adopting Sustainable Partnerships for Innovative Research Ecosystem (ASPIRE) (Grant Number JPMJAP2513).
\end{acknowledgments}

\appendix 

\section{Measurement setup}\label{sec:measurement_setup}
\begin{figure}[t]
        \centering
        \includegraphics[]{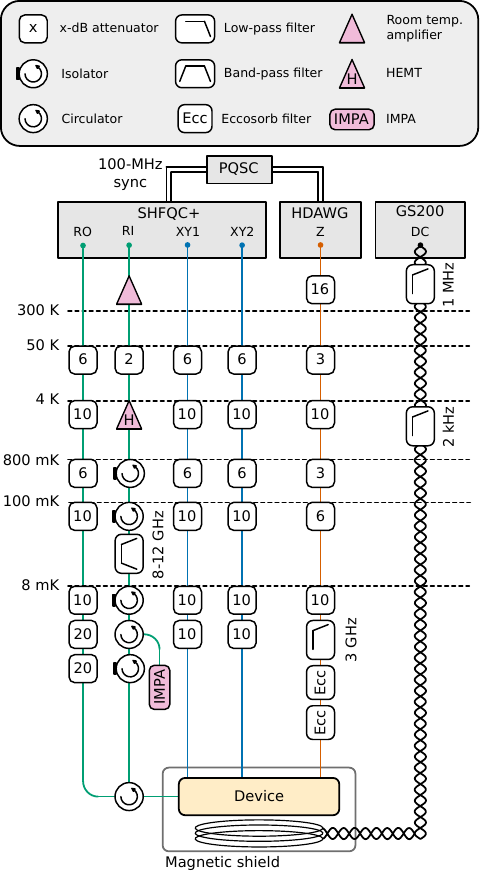}
        \caption{Room-temperature control electronics and cryogenic wiring. The SHFQC+ provides qubit drive tones ($\mathrm{XY}1$, $\mathrm{XY}2$) and generates/detects a readout tone ($\mathrm{RO}/\mathrm{RI}$), while the HDAWG delivers the parametric flux-drive waveform ($\mathrm{Z}$) to the coupler's flux loop.}
        \label{fig:fridge_wiring}
    \end{figure}

    All measurements are performed in a Bluefors LD400 dilution refrigerator with a base temperature of $\simeq 8~\mathrm{mK}$. The sample is mounted at the mixing-chamber stage inside a $\mu$-metal magnetic shield consisting of two layers of $\mu$-metal and one layer of Al and oxygen-free copper each from outside to inside~(Fig.~\ref{fig:fridge_wiring}). We use the same capacitively shunted double-transmon coupler (CSDTC) device as in our previous CZ-gate demonstration~\cite{li2025capacitively}.

    The room-temperature control electronics and the cryogenic wiring are shown in Fig.~\ref{fig:fridge_wiring}. Single-qubit control and qubit readout are performed using a Zurich Instruments SHFQC+. A flux drive applied to the coupler for the iSWAP interaction is generated by a Zurich Instruments HDAWG operated in the amplified mode. The SHFQC+ and HDAWG are phase-locked and synchronized by distributing a common $100$-$\mathrm{MHz}$ reference signal from a Zurich Instruments PQSC.
   
    The single-qubit control pulses are generated by the SHFQC+ and sent to the control lines, which are attenuated and thermalized at each temperature stage before reaching the individual qubit drive ports. Because the coupler modes do not have dedicated microwave control lines, we drive the coupler modes through the qubit drive lines via classical crosstalk for characterization purposes.

    A fast flux drive for activating the iSWAP interaction is delivered through a dedicated coupler flux line (Z), driven by the HDAWG in the amplified mode with the peak-to-peak output voltage $V_{\mathrm{pp}} = 4~\mathrm{V}$. The signal is first attenuated at room temperature and then further attenuated and thermalized at successive temperature stages. At the mixing chamber stage, the line is additionally filtered by a $3$-$\mathrm{GHz}$ low-pass filter~(RLC Electronics F-30-3000) and two Eccosorb filters to further suppress high-frequency noise. No digital predistortion is applied to the flux-drive waveform in this work. A global dc flux bias is provided by a Yokogawa GS200 operated in the constant-current mode and coupled to the sample by a superconducting global bias coil. The dc bias line is filtered at room temperature using a $\pi$-filter and further low-pass filtered at the $4$-$\mathrm{K}$ stage to suppress high-frequency noise.
    
    The readout pulse is generated at the $\mathrm{RO}$ port of the SHFQC+ and attenuated/thermalized before reaching the sample. The reflected readout signal is routed through cryogenic circulators and isolators and then amplified first by an impedance-matched Josephson parametric amplifier (IMPA) at the base-temperature stage, followed by a high-electron-mobility transistor (HEMT) amplifier~(Low Noise Factory LNF-LNC4\_16C) at $4~\mathrm{K}$ and an additional low-noise microwave amplifier at room temperature. The amplified signal is then demodulated and digitized at the SHFQC+ $\mathrm{RI}$ port for state discrimination.

\section{Device characterization} \label{sec:coherence}
    \begin{figure}[t]
        \centering
        \includegraphics[]{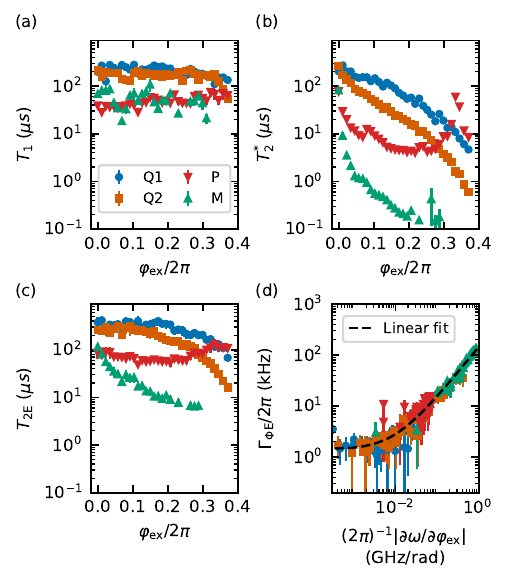}
        \caption{Flux dependence of the coherence properties of the two data qubits and the two coupler modes.
        (a)~Energy-relaxation time $T_{1}$,
        (b)~Ramsey dephasing time $T_{2}^{*}$, and
        (c)~Hahn-echo dephasing time $T_{2\mathrm{E}}$ measured as functions of the external flux bias $\varphi_{\mathrm{ex}}$ for Q1, Q2, the P mode, and the M mode.
        (d)~$1/f$-noise-induced pure-dephasing rate $\Gamma_{\Phi\mathrm{E}}$ extracted from the Hahn-echo decay curves as a function of the flux sensitivity of the corresponding energy level. The dashed line shows a linear fit with a positive intercept to the combined data.}
        \label{fig:coherence}
    \end{figure}
   
    \begin{table}
        \centering
        \caption{Device parameters of the full Hamiltonian extracted from the fit to the energy levels in Fig.~\ref{fig:device_and_spectrum}(c). (a) Capacitances and (b) Josephson-junction critical currents. For the geometrical definitions of the capacitances and critical currents of the Josephson junctions, see Ref.~\citenum{li2025capacitively}.}
        \label{tab:csdtc_circuit_parameters}
        \begin{minipage}{\columnwidth}
            \centering
            \text{(a) Capacitances (fF)} \\ [0.3em]
            {
            \setlength{\tabcolsep}{7pt} 
            \begin{tabular}{cccccc}
            \hline\hline
            & $C_{11}$ & $C_{22}$ & $C_{33}$ & $C_{44}$ & \\
            \hline
            & 108 & 80 & 90 & 90 & \\
            \hline \hline
            $C_{12}$ & $C_{14}$ & $C_{23}$ & $C_{13}$ &$C_{24}$ & $C_{34}$ \\
            \hline
            0.005 & 0.06 & 0.06 & 12.6 & 12.6 & 30.04 \\
            \hline\hline
            \end{tabular}
            }
        \end{minipage}
        \par\vspace{0.5em}
        \begin{minipage}{\columnwidth}
            \centering
            \text{(b) Critical currents (nA)}\\[0.3em]
            {
            \setlength{\tabcolsep}{7pt}
            \begin{tabular}{ccccc}
                \hline \hline
                $I_{\mathrm{c1}}$ & $I_{\mathrm{c2}}$ & $I_{\mathrm{c3}}$ & $I_{\mathrm{c4}}$ & $I_{\mathrm{c5}}$ \\
                \hline
                26.78 & 26.62 & 55.18 & 55.18 & 11.9 \\
                \hline \hline
            \end{tabular}
            }
        \end{minipage}
    \end{table}

    \begin{table}
        \centering
        \caption{Characterized parameters of the two data qubits, Q1 and Q2, and the coupler modes P and M. The coherence times are measured at zero flux bias. The numbers in parentheses indicate $\pm 1\sigma$ uncertainties measured over 15~hours. Due to multiple thermal cycles, we observe slight parameter shifts from the values in our previous work~\cite{li2025capacitively}.}
        \label{tab:device_parameters}
        \begin{tabular}{l r r r r}
            \hline \hline
            Parameter & Q1 & Q2 & P & M \\
            \hline
            $\omega / 2\pi$ (GHz) & $3.950$ & $4.448$ & $6.358$ & $5.987$ \\
            $\alpha / 2\pi$ (MHz) & $-176$ & $-213$ &  &  \\
            $T_{1}$ ($\mu$s) & $226(26)$ & $200(21)$ & $38(2)$ & $100(16)$ \\
            $T_{2}^{*}$ ($\mu$s) & $256(43)$ & $200(35)$ & $64(9)$ & $128(31)$ \\
            $T_{2\mathrm{E}}$ ($\mu$s) & $322(33)$ & $286(25)$ & $73(11)$ & $175(29)$ \\
            \hline \hline
        \end{tabular}
    \end{table}

    \subsection{Device parameters}
        For the spectroscopy measurements in Fig.~\ref{fig:device_and_spectrum}(c) and residual $ZZ$ interaction measurements in Fig.~\ref{fig:device_and_spectrum}(d), we fit the data with the model Hamiltonian in Eqs.~\eqref{eq-H}--\eqref{eq-Hm} to extract the device parameters. The fitting parameters are summarized in Table~\ref{tab:csdtc_circuit_parameters}.

    \subsection{Qubit coherence away from zero flux bias}
        We characterize the coherence of the qubits by measuring the energy-relaxation time $T_{1}$, the Ramsey dephasing time $T_{2}^{*}$, and the Hahn-echo dephasing time $T_{2\mathrm{E}}$ both at zero flux bias (Table~\ref{tab:device_parameters}) and as a function of the external flux bias $\varphi_{\mathrm{ex}}$~[Figs.~\ref{fig:coherence}(a)--\ref{fig:coherence}(c)].

        As the flux bias is increased away from zero, $T_{1}$ of the qubits decreases gradually. We attribute this trend to the enhanced participation of the coupler M~mode, whose frequency decreases with increasing flux bias and whose coherence is shorter than that of the data qubits. In contrast, $T_{2}^{*}$ and $T_{2\mathrm{E}}$ degrade more rapidly. This behavior is consistent with the increased first-order flux sensitivity of the qubit frequencies and dephasing due to $1/f$ flux noise. These measurements imply that, under a finite-amplitude flux drive, the effective coherence time becomes shorter than the optimal value at zero flux bias.

        Away from the zero-flux bias point, dephasing due to $1/f$ flux noise is expected to dominate over near-Markovian dephasing observed at zero flux bias. To separate the Markovian and $1/f$ contributions, we fit the Hahn-echo decay signal to
        \begin{align}
            P_{\ket{1}}(t) = A \exp[-\Gamma_{\mathrm{exp}} t - \qty(\Gamma_{\Phi\mathrm{E}} t)^2] + B,
        \end{align}
        where $t$ is the total free-evolution time in the echo sequence and $\Gamma_{\mathrm{exp}}$ contains contributions from energy relaxation and Markovian pure dephasing. From this fit, we extract the dephasing rate induced by $1/f$ flux noise, shown in Fig.~\ref{fig:coherence}(d),
        \begin{align}
            \Gamma_{\Phi\mathrm{E}} = \qty(\frac{2\pi}{\Phi_0}) \sqrt{A_{\Phi}\ln 2}\, \abs{\frac{\partial \omega}{\partial \varphi_{\mathrm{ex}}}},
        \end{align}
        where we obtain $\sqrt{A_{\Phi}} = 4.37(5)~\mu\Phi_0$, which is consistent with typical values reported for superconducting qubits~\cite{li2024realization, braumuller2020characterizing}. We note that we perform a linear fit with an offset to the combined data of all four modes, which share the same flux loop.

\section{Toy model for iSWAP gate}\label{sec:two_photon_iswap_toy_model}
    To facilitate an intuitive understanding of the parametrically driven iSWAP gate at zero flux bias, we provide an effective Hamiltonian description based on a toy model developed in Ref.~\citenum{li2024realization}. This toy model is not intended to reproduce the experimental results quantitatively, but rather to provide qualitative insight into the underlying iSWAP gate mechanism.
    
    In the following, we first derive the toy-model Hamiltonian in Eq.~\eqref{eq:Hsys}. Then, using a Jacobi--Anger expansion~\cite{blais2021circuit,huber2025parametric,ma2025parametric} and a time-dependent Schrieffer--Wolff transformation~\cite{malekakhlagh2020first, roth2017analysis, petrescu2023accurate}, we derive the effective Hamiltonian for the iSWAP gate. The resulting effective Hamiltonian contains multiple harmonics of the modulation frequency $\omega_{\mathrm{d}}$, including the iSWAP interaction term and the dynamical $ZZ$ interaction term due to the ac Zeeman shift caused by off-resonant driving of the sideband transitions $|11\rangle\!\Leftrightarrow\!|20\rangle$ and $|11\rangle\!\Leftrightarrow\!|02\rangle$.

    \subsection{Toy-model Hamiltonian}
        By expanding the cosine terms in Eqs.~\eqref{eq-H} and~\eqref{eq-H0} to fourth order in $\hat{\varphi}_1$ and $\hat{\varphi}_2$, and to second order in $\hat{\varphi}_{\mathrm{p}}$ and $\hat{\varphi}_{\mathrm{m}}$, $\hat{H}_0$ becomes
        \begin{align}
            \hat{H}_0
            &\simeq 4\hbar \hat{\mathrm{\mathbf{n}}}^T W \hat{\mathrm{\mathbf{n}}}  \nonumber \\
            &+ 
            \sum_{j=1,2} E_{\mathrm{J}j} \left( \frac{\hat{\varphi}_j^2}{2} -  \frac{\hat{\varphi}_j^4}{24} \right) 
            +\sum_{j=\mathrm{p},\mathrm{m}} E_{\mathrm{J}j} \frac{\hat{\varphi}_j^2}{2},
        \end{align}
        where we have dropped $(E_{\mathrm{J}3} - E_{\mathrm{J}4}) \hat{\varphi}_{\mathrm{p}}\hat{\varphi}_{\mathrm{m}}$ assuming $E_{\mathrm{J}3} = E_{\mathrm{J}4}$ and defined
        \begin{align}
            E_{\mathrm{Jp}} &= E_{\mathrm{J}3} + E_{\mathrm{J}4}, \\
            E_{\mathrm{Jm}} &= E_{\mathrm{J}3} + E_{\mathrm{J}4} + 4E_{\mathrm{J}5}.
        \end{align}
        We rewrite the charging-energy term $\hat{\mathrm{\mathbf{n}}}^T W \hat{\mathrm{\mathbf{n}}}$ using $\hat{n}_{\mathrm{p}}=\hat{n}_3+\hat{n}_{4}$ and $\hat{n}_{\mathrm{m}}=\hat{n}_3-\hat{n}_{4}$, and quantize the harmonic-oscillator terms~\cite{goto2022double}. In the rotating-wave approximation, we thus obtain
        \begin{align}
            \hat{H}_0 &\simeq \sum_{j=1,2,\mathrm{p}, \mathrm{m}} \hbar \omega_j \hat{a}_j^{\dagger} \hat{a}_j + \sum_{j=1,2} \frac{\hbar \alpha_j}{2} \hat{a}_j^{\dagger} \hat{a}_j^{\dagger} \hat{a}_j \hat{a}_j   \nonumber \\
            &+ \sum_{j=1,2} \sum_{k= \mathrm{p}, \mathrm{m}} \hbar g_{jk} \qty(\hat{a}_j^{\dagger} \hat{a}_k + \hat{a}_j \hat{a}_k^{\dagger}), \label{eq:Hboson}
        \end{align}
        where $\hat{a}_j$ and $\hat{a}_j^{\dagger}$ ($j=1,2,\mathrm{p},\mathrm{m}$) are the annihilation and creation operators of the two data qubits and the two hybridized coupler modes at zero flux bias. We have dropped the coupling terms $8\hbar W_{12} \hat{n}_1 \hat{n}_2$ and $2\hbar (W_{33}-W_{44}) \hat{n}_{\mathrm{p}} \hat{n}_{\mathrm{m}}$ and defined
        \begin{align}
                \omega_1& = \sqrt{8 W_{11} (E_{\mathrm{J}1}/\hbar)} - W_{11}, \label{eq-toy-o1} \\
                \omega_2& = \sqrt{8 W_{22} (E_{\mathrm{J}2}/\hbar)} - W_{22}, \\
                \omega_{\mathrm{p}}& = \sqrt{8 W_{\mathrm{pp}} (E_{\mathrm{Jp}}/\hbar)}, \\
                \omega_{\mathrm{m}}& = \sqrt{8 W_{\mathrm{mm}} (E_{\mathrm{Jm}}/\hbar)}, \\
                \alpha_j&= -W_{jj}, \\
                g_{jk}&= \frac{W_{jk}}{2} \sqrt{\frac{(\omega_j + W_{jj})\omega_k}{W_{jj} W_{kk}}}, \label{eq-gjk} \\
                W_{\mathrm{pp}} &= (W_{33}+W_{44}+2W_{34})/4, \\
                W_{\mathrm{mm}} &= (W_{33}+W_{44}-2W_{34})/4, \\
                W_{\mathrm{1p}} &= (W_{13}+W_{14})/2, \label{eq-c13}\\
                W_{\mathrm{1m}} &= (W_{13}-W_{14})/2, \\
                W_{\mathrm{2p}} &= (W_{23}+W_{24})/2, \\
                W_{\mathrm{2m}} &= (W_{23}-W_{24})/2. \label{eq-toy-w2m}
        \end{align}
        Here, $\omega_j$ is the transition frequency of the $j$th qubit or the coupler mode $j$, $\alpha_j$ is the anharmonicity of the $j$th data qubit, and $g_{jk}$ is the coupling rate between the $j$th qubit and the coupler mode $k$. The coupling signs satisfy $g_{1\mathrm{m}}, g_{1\mathrm{p}}, g_{2\mathrm{p}} > 0$ and $g_{2\mathrm{m}} < 0$~\cite{li2024realization} from Eqs.\eqref{eq-c13}--\eqref{eq-toy-w2m},  $W_{13} > W_{14}$, and $W_{24} > W_{23}$.

        Phenomenologically introducing the time-dependent flux drive into the Hamiltonian in Eq.~\eqref{eq:Hboson},
        we obtain the following toy-model Hamiltonian for the iSWAP gate:
        \begin{align}
            \hat{H}(\varphi_{\mathrm{ex}}) &= \sum_{j=1,2} \left( \hbar \omega_j \hat{a}_j^{\dagger} \hat{a}_j + \frac{\hbar \alpha_j}{2} \hat{a}_j^{\dagger} \hat{a}_j^{\dagger} \hat{a}_j \hat{a}_j \right) \nonumber \\
            &+ \hbar \omega_{\mathrm{p}} \hat{a}^\dagger_{\mathrm{p}}\hat{a}_{\mathrm{p}} + \hbar \omega_{\mathrm{m}}\qty(\varphi_{\mathrm{ex}}) \hat{a}_{\mathrm{m}}^{\dagger} \hat{a}_{\mathrm{m}}  \nonumber \\
            &+ \sum_{j=1,2} \sum_{k= \mathrm{p}, \mathrm{m}} \hbar g_{jk} \qty(\hat{a}_j^{\dagger} \hat{a}_k + \hat{a}_j \hat{a}_k^{\dagger}), \label{eq:Hsys}
        \end{align}
        where $\omega_j$ ($j=1,2,\mathrm{p},\mathrm{m}$) and $\alpha_j$ ($j=1,2$) are determined by the experimental values at zero flux bias, the flux-tunable M-mode frequency $\omega_{\mathrm{m}}(\varphi_{\mathrm{ex}})$ is determined such that it is in agreement with the experimental values in the reduced-flux range of $\varphi_{\mathrm{ex}}/2\pi \in [-0.3, 0.3]$, and $g_{jk}$~($j=1,2$ and $k=\mathrm{p},\mathrm{m}$) are given by Eq.~\eqref{eq-gjk}.
        
        To implement the iSWAP gate, we apply the flux drive in Eq.~\eqref{eq:flux_drive}. We first define the time-averaged coupler frequency under the iSWAP gate as
        \begin{align}
            \bar{\omega}_{\mathrm{m}}(\varphi_{\mathrm{d}}) 
            = \frac{1}{t_{\mathrm{tot}}}\int_0^{t_{\mathrm{tot}}} 
            \omega_{\mathrm{m}}\!\left(\varphi_{\mathrm{d}}\cos(\omega_{\mathrm{d}} t)\right) \dd{t},
            \label{eq:qscoupler}
        \end{align}
        where we set $\varphi_{\mathrm{d}} = 0$ and $t_{\mathrm{tot}}$ is the total duration of the iSWAP gate [Eq.~\eqref{eq:total_gate_time}]. Because $\omega_{\mathrm{m}}(\varphi_{\mathrm{ex}})$ is an even function of $\varphi_{\mathrm{ex}}$, the time-dependent coupler frequency modulation contains only even harmonics of $\omega_{\mathrm{d}}$:
        \begin{align}
            \omega_{\mathrm{m}}(\varphi_{\mathrm{ex}}(t)) &= 
            \bar{\omega}_{\mathrm{m}} + \sum_{k \geq 1} \varepsilon_{2k} \cos\!\big[2k (\omega_{\mathrm{d}} t + \phi_{\mathrm{d}})\big] \\
            &\simeq \bar{\omega}_{\mathrm{m}} + \varepsilon_2 \cos\!\big[2 (\omega_{\mathrm{d}} t + \phi_{\mathrm{d}})\big], \label{eq:omega_mod}
        \end{align}
        where $\varepsilon_{2k}$ is the amplitude of the $2k$th harmonic component. The second-harmonic component $\varepsilon_2$ at $2\omega_{\mathrm{d}} \simeq \Delta_{21}$ resonantly drives the $\ket{01}\! \Leftrightarrow\! \ket{10}$ exchange interaction.
        
        In the weak-drive limit, the leading second-harmonic frequency-modulation amplitude $\varepsilon_2$ is expressed as
        \begin{align}
            \varepsilon_2 \simeq \frac{1}{4}\,\frac{\partial^2 \omega_{\mathrm{m}}}{\partial \varphi_{\mathrm{ex}}^2}\Big|_{\varphi_{\mathrm{ex}}=0}\,\varphi_{\mathrm{d}}^2.
            \label{eq:f2_weakdrive}
        \end{align}

    \subsection{Quasi-static frame transformation}
        From Eq.~\eqref{eq:Hsys}, the coupler M-mode frequency modulation $\omega_{\mathrm{m}}(\varphi_{\mathrm{ex}}(t))$ under the parametric flux drive can be transformed into phase modulation of the exchange couplings via a unitary transformation
        \begin{align}
            \hat{U}_{\vartheta}(t)
            = \exp\!\Big[-i\vartheta(t)\hat n_{\mathrm{m}}\Big],
        \end{align}
        where $\hat{n}_{\mathrm{m}} = \hat{a}^\dagger_{\mathrm{m}} \hat{a}_{\mathrm{m}}$ is the number operator for the M~mode and the accumulated phase is expressed as 
        \begin{align}
            \vartheta(t)=\int_{0}^t \qty{\omega_{\mathrm{m}}(\varphi_{\mathrm{ex}}(s))-\overline{\omega}_{\mathrm{m}}(\varphi_{\mathrm{d}})} \dd{s}.
        \end{align}

        We refer to the frame defined by this transformation as the quasi-static frame. In this frame, the Hamiltonian becomes
        \begin{align}
            \hat{H}_{\vartheta}(t) &= \hat{U}_{\vartheta}^{\dagger}(t) (\hat{H} - i\hbar\partial_t) \hat{U}_{\vartheta}(t) \nonumber \\
            &= \sum_{i=1,2} \left[ \hbar\omega_i \hat{n}_i + \frac{\hbar\alpha_i}{2} \hat{n}_i (\hat{n}_i - 1) \right] + \hbar\omega_{\mathrm{p}} \hat{n}_{\mathrm{p}} + \hbar\bar{\omega}_{\mathrm{m}} \hat{n}_{\mathrm{m}} \nonumber \\
            &+ \hbar g_{1\mathrm{p}} \qty(\hat{a}_1^{\dagger} \hat{a}_{\mathrm{p}} + \hat{a}_1 \hat{a}_{\mathrm{p}}^{\dagger})
            + \hbar g_{2\mathrm{p}} \qty(\hat{a}_2^{\dagger} \hat{a}_{\mathrm{p}} + \hat{a}_2 \hat{a}_{\mathrm{p}}^{\dagger}) \nonumber \\
            &+\hbar  g_{1\mathrm{m}} \qty(\hat{a}_1^{\dagger} \hat{a}_{\mathrm{m}} e^{-i\vartheta(t)} + \hat{a}_1 \hat{a}_{\mathrm{m}}^{\dagger} e^{i\vartheta(t)}) \nonumber \\
            &+ \hbar g_{2\mathrm{m}} \qty(\hat{a}_2^{\dagger} \hat{a}_{\mathrm{m}} e^{-i\vartheta(t)} + \hat{a}_2 \hat{a}_{\mathrm{m}}^{\dagger} e^{i\vartheta(t)}),
        \end{align}
        where $\hat{n}_{i} = \hat{a}^\dagger_i \hat{a}_i$ is the number operator of the $i$th mode~($i=1,2,\mathrm{p}, \mathrm{m}$). In this quasi-static frame, the coupling terms with the M~mode are modulated by the phase factor $e^{\pm i \vartheta(t)}$, which contains multiple harmonics of the modulation frequency $\omega_{\mathrm{d}}$. As the second-harmonic component is dominant~[Eq.~\eqref{eq:omega_mod}], the phase modulation can be expanded using the Jacobi--Anger expansion,
        \begin{align}
            e^{\pm i \vartheta(t)} &= \sum_{n\in\mathbb{Z}} J_n(\eta) e^{\pm i n (2\omega_{\mathrm{d}} t + 2\phi_{\mathrm{d}})}, \label{eq-JA}
        \end{align}
        where $J_n(\eta)$ is the $n$th Bessel function of the first kind and $\eta = \varepsilon_2 / 2\omega_{\mathrm{d}}$ is the modulation index.
    
\subsection{Time-dependent Schrieffer--Wolff transformation} \label{subsec:tdswt}
    As the P~mode is not flux-driven in the toy model, we consider a subsystem Hamiltonian $\hat{H}_{\mathrm{m}}$ consisting of the two data qubits and the M~mode for simplicity.
    
    The subsystem Hamiltonian is given by
    \begin{align}
        \hat{H}_{\mathrm{m}}(t) &= \hat{H}_{\mathrm{m}, 0} + \hat{V}_{\mathrm{m}}(t),
    \end{align}
    whose static and time-dependent parts are defined, respectively, as 
    \begin{align}
        \hat{H}_{\mathrm{m}, 0} &= \sum_{i=1,2} \left[ \hbar \omega_i \hat{n}_i + \frac{\hbar \alpha_i}{2} \hat{n}_i (\hat{n}_i - 1)  \right] + \hbar \bar{\omega}_{\mathrm{m}} \hat{n}_{\mathrm{m}}, \\
        \hat{V}_{\mathrm{m}}(t) &= \sum_{i=1,2} \sum_{n \in \mathbb{Z}} \hbar g_{i\mathrm{m}} \xi_{n} \qty(\hat{a}_i^{\dagger} \hat{a}_{\mathrm{m}} e^{-2in(\omega_{\mathrm{d}} t + \phi_{\mathrm{d}})} + \text{h.c.}),
    \end{align}
    where we have defined $\xi_n = J_n(\eta)$ for brevity.

    Treating $\hat V_{\mathrm{m}}(t)$ as a perturbation, we apply a time-dependent Schrieffer--Wolff transformation~\cite{malekakhlagh2020first, roth2017analysis, petrescu2023accurate}:
    \begin{align}
        \hat U(t)=e^{-\lambda \hat S(t)}, \label{eq-SW-S}
    \end{align}
    where $\hat S(t)$ is an anti-Hermitian generator and $\lambda$ is a dummy parameter that tracks the perturbation order. The effective Hamiltonian is then defined as
    \begin{align}
        &\hat H_{\mathrm{m}}^{\mathrm{eff}}(t) \nonumber \\
        & = e^{\lambda \hat{S}(t)} \qty[\hat H_{\mathrm{m}}(t) - i\hbar \partial_t] e^{-\lambda \hat{S}(t)} \\ \label{eq:Heff_def}
        & = \hat{H}_{\mathrm{m}, 0} + \lambda \qty{\big[\hat{S}(t), \hat{H}_{\mathrm{m}, 0}\big] + i \hbar \dot{\hat{S}}(t)+ \hat{V}_{\mathrm{m}}(t)} \nonumber \\
        & + \lambda^2 \Big\{\frac{i\hbar}{2} \big[\hat{S}(t), \dot{\hat{S}}(t)] + \frac{1}{2}\qty[\hat{S}(t), \big[\hat{S}(t), \hat{H}_{\mathrm{m}, 0}]] \nonumber \\
        &+ \big[\hat{S}(t), \hat{V}_{\mathrm{m}}(t)] \Big\} + \mathcal{O}(\lambda^3).
    \end{align}

    We choose $\hat S(t)$ such that these first-order terms cancel, which yields the operator equation,
    \begin{align}
       \big[\hat{S}(t), \hat{H}_{\mathrm{m}, 0}\big] + i\hbar\dot{\hat{S}}(t)+ \hat{V}_{\mathrm{m}}(t) = 0.
        \label{eq:S_condition}
    \end{align}
    
    Then, to second order in $\lambda$, the effective Hamiltonian is expressed as
    \begin{align}
        \hat H_{\mathrm{m}}^{\mathrm{eff}}(t)
        =
        \hat H_{\mathrm{m},0}+\frac{\lambda^2}{2}[\hat S(t),\hat V_{\mathrm{m}}(t)] +\mathcal{O}(\lambda^3).
        \label{eq:Heff_second_order}
    \end{align}

    To find the generator $\hat{S}(t)$, we introduce the following operators
    \begin{align}
        \hat{\sigma}^{\uparrow}_{i,k}= \ket{k+1}_i\!\bra{k}_i,\quad
        \hat{\sigma}^{\downarrow}_{i,k}= \ket{k}_i\!\bra{k+1}_i,
        \label{eq:sigma_def}
    \end{align}
    where $\hat{\sigma}^{\uparrow}_{i,k}$ ($\hat{\sigma}^{\downarrow}_{i,k}$) is the raising (lowering) operator between the $k$th and $(k+1)$th levels of the $i$th qubit. The creation and annihilation operators are then expressed as 
    \begin{align}
        \hat{a}_{i} = \sum_{k\ge0} \sqrt{k+1}\,\hat{\sigma}^{\downarrow}_{i,k},\quad \hat{a}_{i}^\dagger = \sum_{k\ge0} \sqrt{k+1}\,\hat{\sigma}^{\uparrow}_{i,k}.
    \end{align}

    We assume the following ansatz for the generator $\hat S(t)$:
    \begin{align}
        \hat S(t)
        &= \sum_{i=1,2} \sum_{n \in \mathbb{Z}} \sum_{k \geq 0}
        \mu_{i,n,k}\sqrt{k+1} \nonumber \\
        &\qquad \times \qty(\hat{\sigma}^{\uparrow}_{i,k} \hat{a}_{\mathrm{m}} e^{-2in(\omega_{\mathrm{d}} t+\phi_{\mathrm{d}})} - \text{h.c.}). \label{eq:S_ansatz}
    \end{align}

    Solving for $\mu_{i,n,k}$ to satisfy Eq.~\eqref{eq:S_condition}, we obtain
    \begin{align}
        \mu_{i,n,k}
        &=\frac{g_{i\mathrm{m}}\xi_n}{\overline{\Delta}^{(n,k)}_{i\mathrm{m}}}, \nonumber \\
        \overline{\Delta}^{(n,k)}_{i\mathrm{m}}
        &= \omega_i+k\alpha_i-\bar\omega_{\mathrm{m}}-2n\omega_{\mathrm{d}} \nonumber \\
        &\simeq \omega_i + k \alpha_i - \bar{\omega}_{\mathrm{m}} - n \Delta_{21},
        \label{eq:lambda_solution}
    \end{align}
    where we have approximated $2\omega_{\mathrm{d}}$ by $\Delta_{21}$.
    
    Therefore, the second-order contribution is expressed as
    \begin{widetext}
        \begin{align}
            \frac{1}{2}\qty[\hat S(t),\hat V_{\mathrm{m}}(t)] 
            &= \sum_{i, i^\prime} \sum_{n, n^\prime} \sum_{k, k^\prime} \frac{ \hbar g_{i \mathrm{m}} g_{i^\prime \mathrm{m}} \xi_{n} \xi_{n^\prime} \sqrt{(k+1)(k^\prime+1)}}{2 \overline{\Delta}_{i\mathrm{m}}^{(n, k)}} \qty[\hat{\sigma}^{\uparrow}_{i,k} \hat{a}_{\mathrm{m}} e^{-2in(\omega_{\mathrm{d}} t+\phi_{\mathrm{d}})} - \text{h.c.}, \hat{\sigma}^{\uparrow}_{i^\prime,k^\prime} \hat{a}_{\mathrm{m}} e^{-2in^\prime(\omega_{\mathrm{d}} t+\phi_{\mathrm{d}})} + \text{h.c.}] \\
            &\simeq \sum_{i, i^\prime} \sum_{n, n^\prime} \sum_{k, k^\prime} \frac{\hbar g_{i \mathrm{m}} g_{i^\prime \mathrm{m}} \xi_{n} \xi_{n^\prime} \sqrt{(k+1)(k^\prime+1)}}{2 \overline{\Delta}_{i\mathrm{m}}^{(n, k)}} \qty(e^{-2i(n-n^\prime)(\omega_{\mathrm{d}} t+\phi_{\mathrm{d}})} \hat{\sigma}_{i,k}^\uparrow \hat{\sigma}_{i^\prime,k^\prime}^\downarrow + \text{h.c.}) \quad \text{for $\ev*{\hat{n}_{\mathrm{m}}} \simeq 0$},
            \label{eq:Heff_second_order2}
        \end{align}
    \end{widetext}
    where we have projected the Hamiltonian onto the coupler ground-state subspace, assuming $\ev*{\hat{n}_{\mathrm{m}}} \simeq 0$. Equation~\eqref{eq:Heff_second_order2} contains various harmonics of the second-harmonic modulation frequency $2\omega_{\mathrm{d}}$. In the weak-drive limit, the fast-oscillating terms are averaged out under the rotating-wave approximation. Therefore, to leading order, we only keep terms with $n-n^\prime=-1$ and $0$. 
    \subsubsection{iSWAP interaction}
        \begin{figure}
            \centering
            \includegraphics[]{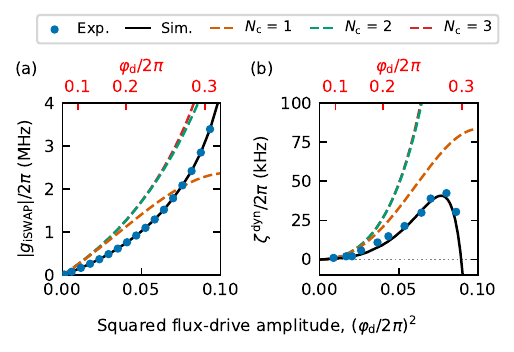}
            \caption{Comparison with analytic results obtained from the toy-model Hamiltonian. The blue dots and black solid curves are experimental data and full-Hamiltonian numerical simulation, respectively. $N_{\mathrm{c}}$ is the cutoff number, meaning that the analytic expressions include the summation over $n=-N_{\mathrm{c}},\ldots,N_{\mathrm{c}}-1$. The results for $N_{\mathrm{c}}=2$ and $3$ nearly overlap within the flux-bias range shown here, indicating that $N_{\mathrm{c}}=2$ is already sufficient to capture the essential physics. (a)~iSWAP interaction strength $g_{\mathrm{iSWAP}}$~[Eq.~\eqref{eq-iswap}] as a function of the squared flux-drive amplitude $\varphi_{\mathrm{d}}^2$. (b)~Dynamical $ZZ$ interaction strength $\zeta^{\mathrm{dyn}}$~[Eq.~\eqref{eq:dynamic_zz1}] as a function of the flux-drive amplitude $\varphi_{\mathrm{d}}$.}
            \label{fig:bosonic_exprs}
        \end{figure}
        The exchange interaction between the two data qubits in the computational subspace is obtained from Eq.~\eqref{eq:Heff_second_order2} by setting $k = k^\prime = 0$:
        \begin{align}
            \hat{H}_{\mathrm{iSWAP}}^{(2)} &=\sum_{n,n'\in\mathbb Z}
            \frac{\hbar g_{1\mathrm{m}}g_{2\mathrm{m}}\,\xi_n\xi_{n'}}{2}
            \left(
            \frac{1}{\overline{\Delta}^{(n,0)}_{1\mathrm{m}}}+\frac{1}{\overline{\Delta}^{(n',0)}_{2\mathrm{m}}}
            \right) \nonumber \\
            &\times \Big[
            e^{-2i(n-n')(\omega_{\mathrm{d}} t+\phi_{\mathrm{d}})}\hat{\sigma}^{\uparrow}_{1,0}\hat{\sigma}^{\downarrow}_{2,0}
            + \text{h.c.}
            \Big].
            \label{eq:H_ex_general}
        \end{align}

        To obtain the resonant iSWAP interaction, we examine the time dependence of Eq.~\eqref{eq:H_ex_general} in the rotating frame defined by the data qubits. The operator $\hat{\sigma}^\uparrow_{1,0} \hat{\sigma}^\downarrow_{2,0} = \ket{10}\bra{01}$ acquires a phase factor $e^{+i(\omega_1-\omega_2)t} = e^{-i\Delta_{21}t}$ under the free qubit Hamiltonian $\hat{H}_{\mathrm{m}, 0}$, so that the relevant term in the rotating frame oscillates as
        \begin{align}
            &e^{-2i(n-n^\prime)(\omega_{\mathrm{d}}t + \phi_{\mathrm{d}})} e^{-i\Delta_{21}t} \nonumber \\
            &\quad = e^{-i\left[ 2(n-n^\prime)\omega_{\mathrm{d}} + \Delta_{21} \right]t} e^{-2i(n-n^\prime)\phi_{\mathrm{d}}}. \label{eq:b28}
        \end{align}

        The exchange process becomes nearly static (and therefore survives the rotating-wave approximation) when the total oscillation frequency in Eq.~\eqref{eq:b28} is close to zero. In the present second-harmonic driving scheme, the modulation frequency satisfies the near-resonance condition $2\omega_{\mathrm{d}} \simeq \Delta_{21}$, so the dominant contribution arises from the terms obeying 
        \begin{align}
            2(n-n^\prime)\omega_{\mathrm{d}} + \Delta_{21} \simeq 0 \quad \Rightarrow \quad n - n^\prime = -1.
        \end{align}

        Substituting this condition into Eq.~\eqref{eq:H_ex_general}, we obtain a time-independent (resonant) exchange Hamiltonian,
        \begin{align}
            \hat{H}_{\mathrm{iSWAP}}^{(2)} &= \hbar g_{\mathrm{iSWAP}} \nonumber \\
            &\times \left(e^{+2i\phi_{\mathrm{d}}} |10 \rangle \!\langle 01| + e^{-2i\phi_{\mathrm{d}}} |01 \rangle \!\langle 10|\right),
            \label{eq:Heff_iswap}
        \end{align}
        where
        \begin{align}
            g_{\mathrm{iSWAP}}
            & = \sum_{n} \frac{g_{1\mathrm{m}}g_{2\mathrm{m}} \xi_{n} \xi_{n+1}}{2} \qty(\frac{1}{\overline{\Delta}^{(n,0)}_{1\mathrm{m}}}+\frac{1}{\overline{\Delta}^{(n+1,0)}_{2\mathrm{m}}}) \\
            &\simeq g_{1\mathrm{m}} g_{2\mathrm{m}} \sum_{n} \frac{\xi_{n} \xi_{n+1}}{\overline{\Delta}_{1\mathrm{m}} - n\Delta_{21}}. \label{eq-iswap}
        \end{align}
        In the weak-drive limit, the dominant contribution from $n= -1$ and $0$ is approximated as
        \begin{align}
            g_{\mathrm{iSWAP}} &= \frac{g_{1\mathrm{m}} g_{2\mathrm{m}} \eta}{2} \qty(\frac{1}{\overline{\Delta}_{1\mathrm{m}}} - \frac{1}{\overline{\Delta}_{2\mathrm{m}}}) + \mathcal{O}(\eta^3). \label{eq:iswap_rate_weak_drive1}
        \end{align}

        Furthermore, if we compute the next-order correction to this iSWAP rate by extending the sum to $n=-2,-1,0$, and $1$, we obtain
        \begin{align}
            g_{\mathrm{iSWAP}} &= \frac{g_{1\mathrm{m}} g_{2\mathrm{m}} \eta}{2} \qty(\frac{1}{\overline{\Delta}_{1\mathrm{m}}} - \frac{1}{\overline{\Delta}_{2\mathrm{m}}}) \nonumber \\
            &\times \qty[1 + \frac{3\eta^2 \Delta_{21}^{2}}{4(\overline{\Delta}_{1\mathrm{m}} - \Delta_{21})(\overline{\Delta}_{2\mathrm{m}} + \Delta_{21})}] + \mathcal{O}(\eta^5). \label{eq:iswap5}
        \end{align}

        Figure~\ref{fig:bosonic_exprs}(a) compares the analytic expression for $g_{\mathrm{iSWAP}}$~[Eq.~\eqref{eq-iswap}] with the experimental and full-Hamiltonian numerical results presented in the main text. The analytic expression overestimates the interaction strength across the entire range of flux-drive amplitudes. Nevertheless, it captures the qualitative dependence of the iSWAP interaction strength on the flux-drive amplitude $\varphi_{\mathrm{d}}$. For this comparison, the detunings in Eq.~\eqref{eq-iswap} are taken from the experimentally measured frequencies shown in Fig.~\ref{fig:device_and_spectrum}(c), the second-harmonic flux-modulation amplitude~$\varepsilon_{2}$~[Eq.~\eqref{eq:omega_mod}] is determined by numerically simulating the M-mode response under continuous-wave flux drive, and the coupling strengths are obtained from Eq.~\eqref{eq-gjk} with parameters in Table~\ref{tab:csdtc_circuit_parameters}.

    \subsubsection{Dynamical $ZZ$ interaction}
        The flux drive that activates the iSWAP gate also induces frequency shifts of the energy levels in the two-excitation manifold that lead to an additional $ZZ$ interaction~\cite{petrescu2023accurate,jin2025superconducting,ganzhorn2020benchmarking,han2020error}, which we refer to as a drive-induced dynamical $ZZ$ interaction $\zeta^{\mathrm{dyn}}$. The dynamical $ZZ$ interaction is caused by the ac Zeeman shift due to the off-resonant second-harmonic drive of the sideband transitions $\ket{11}\! \Leftrightarrow\! \ket{02}$ and $\ket{11}\!\Leftrightarrow\!\ket{20}$, with frequency detunings of $\alpha_{2}$ and $\alpha_{1}$, respectively. The corresponding correction terms arise from the contributions with $n-n^\prime=-1$ and $(k,k')=(1,0)$ or $(0,1)$ in Eq.~\eqref{eq:Heff_second_order2}. 
        \begin{align}
            &\hat{H}_{\mathrm{dyn}}^{(2)} =\sum_{n} \sum_{(k,k^\prime)=(1,0), (0,1)} \frac{ \hbar g_{1\mathrm{m}} g_{2\mathrm{m}} \xi_n \xi_{n+1} \sqrt{2}}{2} \nonumber \\
            &\quad \times \qty(\frac{1}{\overline{\Delta}^{(n,k)}_{1\mathrm{m}}} + \frac{1}{\overline{\Delta}^{(n+1,k^\prime)}_{2\mathrm{m}}}) \Big[
            e^{2i(\omega_{\mathrm{d}} t+\phi_{\mathrm{d}})}\hat{\sigma}^{\uparrow}_{1,k}\hat{\sigma}^{\downarrow}_{2,k'}
            + \text{h.c.}
            \Big].
        \end{align}
        
        Projecting onto the subspace involving $\ket{11}$ and the second-excited states $\ket{20}, \ket{02}$, the second-order off-resonant coupling is expressed as
        \begin{align}
            \hat{H}_{\mathrm{dyn}}^{(2)} &= \hbar g_{11,20}^{\mathrm{dyn}} \qty(e^{+2i(\omega_{\mathrm{d}}t + \phi_{\mathrm{d}})}\hat{\sigma}_{1,1}^\uparrow \hat{\sigma}_{2,0}^\downarrow + \text{h.c.}) \nonumber \\
            &+\hbar g_{11,02}^{\mathrm{dyn}} \qty(e^{+2i(\omega_{\mathrm{d}}t + \phi_{\mathrm{d}})}\hat{\sigma}_{1,0}^\uparrow \hat{\sigma}_{2,1}^\downarrow + \text{h.c.}),
        \end{align}
        where the coupling rates are given by
        \begin{align}
            g_{11,20}^{\mathrm{dyn}} &= \frac{g_{1\mathrm{m}} g_{2\mathrm{m}} \sqrt{2}}{2} \sum_{n} \xi_n \xi_{n+1} \qty(\frac{1}{\overline{\Delta}^{(n,1)}_{1\mathrm{m}}} + \frac{1}{\overline{\Delta}^{(n+1,0)}_{2\mathrm{m}}}), \\
            g_{11,02}^{\mathrm{dyn}} &= \frac{g_{1\mathrm{m}} g_{2\mathrm{m}} \sqrt{2}}{2} \sum_{n} \xi_n \xi_{n+1} \qty(\frac{1}{\overline{\Delta}^{(n,0)}_{1\mathrm{m}}} + \frac{1}{\overline{\Delta}^{(n+1,1)}_{2\mathrm{m}}}).
        \end{align}

        In the weak-drive limit, the dominant contribution originates from $n=-1$ and $0$, yielding
        \begin{align}
            &g_{11,20}^{\mathrm{dyn}} \nonumber \\
            &= \frac{g_{1\mathrm{m}} g_{2\mathrm{m}} \sqrt{2}\,\xi_0 \xi_1}{2} \nonumber \\
            &\times \qty(\frac{1}{\overline{\Delta}^{(1,0)}_{2\mathrm{m}}} + \frac{1}{\overline{\Delta}_{1\mathrm{m}}^{(0, 1)}} - \frac{1}{\overline{\Delta}^{(-1,1)}_{1\mathrm{m}}} - \frac{1}{\overline{\Delta}^{(0,0)}_{2\mathrm{m}}}) + \mathcal{O}(\eta^3), \nonumber \\
            &= \sqrt{2} g_{1\mathrm{m}} g_{2\mathrm{m}} \xi_0 \xi_1 \qty(\frac{1}{\overline{\Delta}_{1\mathrm{m}}} - \frac{1}{\overline{\Delta}_{2\mathrm{m}}}) \nonumber \\
            &\times \qty[1 - \frac{\alpha_1}{2}\qty(\frac{1}{\overline{\Delta}_{1\mathrm{m}}} + \frac{1}{\overline{\Delta}_{2\mathrm{m}}})] + \mathcal{O}(\alpha_1^2, \eta^3),
        \end{align}
        where we expand the expression assuming $\abs*{\alpha_i}\ll\abs*{\overline{\Delta}_{i\mathrm{m}}}$. Using the iSWAP interaction strength in Eq.~\eqref{eq:iswap_rate_weak_drive1}, we can simplify the above expression to
        \begin{align}
            g_{11,20}^{\mathrm{dyn}} &= \sqrt{2} \, g_{\mathrm{iSWAP}} \qty[1 - \frac{\alpha_1}{2}\qty(\frac{1}{\overline{\Delta}_{1\mathrm{m}}} + \frac{1}{\overline{\Delta}_{2\mathrm{m}}})] \nonumber \\
            & + \mathcal{O}(\alpha_1^2, \eta^3).
        \end{align}
        We similarly obtain the simplified expression for $g_{11,02}^{\mathrm{dyn}}$.
        
        In the rotating frame defined by the transition frequencies of Q1 and Q2, these coupling rates create sideband couplings that oscillate at frequencies of order $\alpha_{1,2}$. These couplings therefore generate an ac Zeeman energy shift on $\ket{11}$ given by
        \begin{align}
        \zeta^{\mathrm{dyn}}
        &= \frac{|g_{11,20}^{\mathrm{dyn}}|^2}{-\alpha_1}
            + \frac{|g_{11,02}^{\mathrm{dyn}}|^2}{-\alpha_2} \label{eq:dynamic_zz1} \\
        &\simeq 2g_{\mathrm{iSWAP}}^2
        \left(
        \frac{1}{-\alpha_1}+\frac{1}{-\alpha_2}
        +\frac{2}{\overline{\Delta}_{1\mathrm{m}}}+\frac{2}{\overline{\Delta}_{2\mathrm{m}}}
        \right),
        \end{align}
        where we note that this term provides a positive correction to the total $ZZ$ interaction because $\alpha_i, \overline{\Delta}_{i\mathrm{m}} <0$ and can partially cancel the effective $ZZ$ interaction during the iSWAP gate. 

        Figure~\ref{fig:bosonic_exprs}(b) compares the analytic expression for $\zeta^{\mathrm{dyn}}$~[Eq.~\eqref{eq:dynamic_zz1}] with the experimental and full-Hamiltonian numerical results presented in the main text. The analytic expression overestimates the dynamical $ZZ$ interaction. However, it captures the qualitative behavior of the drive-induced dynamical $ZZ$ interaction strength as a function of the flux-drive amplitude $\varphi_{\mathrm{d}}$.

\section{Details of Numerical Simulations} \label{sec:cooper_pair_numerical}
    In this section, we describe the numerical simulations performed in this work, which are based on the Cooper-pair number-basis model used in Refs.~\citenum{goto2022double, kubo2023fast, kubo2024high}. This model captures the full multi-level dynamics of the system and shows quantitative agreement with the experimental data not only on the spectroscopy but also on the time-domain dynamics of the iSWAP gate (see the main text for detailed comparisons between the simulation and experimental results).
    \subsection{Cooper-pair number-operator description}
        In the main text, we have omitted a term proportional to the time derivative of the modulated external flux $\dot{\varphi}_{\mathrm{ex}}$ in the Hamiltonian~[Eq.~\eqref{eq-Hm}] for simplicity. The full Hamiltonian including this term is given by
        \begin{align}
            \hat{H}(\varphi_{\mathrm{ex}}(t))
            &= 4\hbar \hat{\mathrm{\mathbf{n}}}^T W \hat{\mathrm{\mathbf{n}}} 
            +\hbar \frac{\hbar \dot{\varphi}_{\mathrm{ex}}(t)}{E_{\mathrm{C}_{34}}} (0~0~-1~1) W \hat{\mathrm{\mathbf{n}}}
            \nonumber\\
            &- \sum_{j=1}^4 E_{\mathrm{J}j} \cos \hat{\varphi}_j - E_{\mathrm{J}5} \cos\qty[\hat{\varphi}_4 - \hat{\varphi}_3- \varphi_{\mathrm{ex}}(t)],
            \label{eq:H_sim}
        \end{align}
        where $E_{\mathrm{C}_{34}}=e^2/(2C_{34})$ is the charging energy for the shunt capacitance between the two coupler transmons C3 and C4~\cite{goto2022double}. 

    \subsection{Matrix representations of operators}
    \label{sec-operator-matrix}
        The phase-difference and Cooper-pair number operators are quantized by the commutation relation 
        ${[\hat{\varphi}_i, \hat{n}_j ]=i \delta_{ij}}$ as follows.
        The Cooper-pair number operator $\hat{n}_i$ is represented by ${-i \frac{\partial}{\partial \varphi_i}}$ 
        and the eigenfunction of $\hat{n}_i$ is proportional to 
        $e^{i n_i \varphi_i}$. 
        In the basis of these eigenfunctions, 
        the operators are represented by the following matrices:
        \begin{align}
        \hat{n}_i &= 
        \begin{pmatrix}
        -N_{\mathrm{C}} & & \\
        & \ddots & \\
        & & N_{\mathrm{C}}
        \end{pmatrix},
        \label{eq-n}
        \\
        \cos \hat{\varphi}_i &= 
        \frac{1}{2}
        \begin{pmatrix}
        & 1 & & \\
        1 & & \ddots & \\
        & \ddots & & 1 \\
        & & 1 &
        \end{pmatrix},
        \label{eq-cos}
        \\
        \sin \hat{\varphi}_i &= 
        \frac{1}{2i} 
        \begin{pmatrix}
        & -1 & & \\
        1 & & \ddots & \\
        & \ddots & & -1 \\
        & & 1 & 
        \end{pmatrix},
        \label{eq-sin}
        \end{align}
        where we have truncated the Cooper-pair number at $\pm N_{\mathrm{C}}$. 
        This results in a $(2N_{\mathrm{C}}+1)^4 \times (2N_{\mathrm{C}}+1)^4$ matrix representation of the Hamiltonian in Eq.~\eqref{eq:H_sim}. In this work, we choose $N_{\mathrm{C}}=10$ for sufficient convergence of the energy spectrum. The present simulations can be accelerated by the dimension-reduction technique in Ref.~\citenum{kubo2024high}.
    \subsection{Energy spectrum and static $ZZ$ interaction}
        By numerically diagonalizing the Hamiltonian matrix $\hat{H}(\varphi_{\mathrm{ex}})$ with $\dot{\varphi}_{\mathrm{ex}}=0$, 
        we obtain the eigenstates $\ket*{\widetilde{ijkl} (\varphi_{\mathrm{ex}})}$, following the conventions in Sec.~\ref{sec:device}. The resulting energy spectra, $\omega_{\widetilde{ijkl}}(\varphi_\mathrm{ex})$, and the static $ZZ$ interaction,
        \begin{align}
            \zeta (\varphi_{\mathrm{ex}})
            &= \omega_{\widetilde{1100}}(\varphi_{\mathrm{ex}})
            - \omega_{\widetilde{1000}}(\varphi_{\mathrm{ex}}) \nonumber \\
            &\qquad - \omega_{\widetilde{0100}}(\varphi_{\mathrm{ex}})
            + \omega_{\widetilde{0000}}(\varphi_{\mathrm{ex}}),
        \end{align}
        are shown in Figs.~\ref{fig:device_and_spectrum}(c) and (d) in the main text.  
        Here, we use the circuit parameters summarized in Table~\ref{tab:csdtc_circuit_parameters}, 
        which are used throughout all numerical simulations in this study.

	 \subsection{iSWAP interaction rate} \label{sec:giSWAP}
        To evaluate the iSWAP interaction rate $g_{\mathrm{iSWAP}}$ shown in Fig.~\ref{fig:iswapscaling}(b), we compute the time evolution of the initial state $|01\rangle$ under a continuous-wave flux drive given by Eq.~\eqref{eq:flux_drive}, with the drive phase set as $\phi_{\mathrm{d}} = -\pi/2$.

        Starting from $|01\rangle$, we denote the quantum state at time $t$ by $|01(t)\rangle$. This state is obtained by solving the Schr\"odinger equation with the Hamiltonian defined in Eq.~\eqref{eq:H_sim}, using the \texttt{sesolve} function of QuTiP~\cite{johansson2012qutip}. 
        We then evaluate the population of $|10\rangle$ and $|01\rangle$ as 
        \begin{align}
            P_{10}(t) = \left|\langle 10|01(t)\rangle\right|^2, \quad P_{01}(t) = \left|\langle 01|01(t)\rangle \right|^2.
        \end{align}

        Fixing the amplitude $\varphi_{\mathrm{d}}$ and sweeping the flux-drive frequency $\omega_{\mathrm{d}}$, 
        we numerically generate chevron patterns by plotting $P_{10}(t)$ and $P_{01}(t)$ as functions of $t$ and $\omega_{\mathrm{d}}$, analogous to the experimental results shown in Fig.~\ref{fig:iswapscaling}(a). 
        From these patterns, we extract $g_{\mathrm{iSWAP}}$ in a similar manner to the experiment (see Sec.~\ref{subsec:gate_principle}). This procedure is repeated for different values of $\varphi_{\mathrm{d}}$, 
        yielding the amplitude dependence of $g_{\mathrm{iSWAP}}$ shown by the black curve in Fig.~\ref{fig:iswapscaling}(b).

     \subsection{Effective and dynamical $ZZ$ interactions}
        To numerically extract the effective and dynamical $ZZ$ interactions during the iSWAP gate, we compute the time evolution from four initial states $|ij\rangle$~($i,j\in\{0,1\}$) under the flux-drive pulse defined in Eqs.~\eqref{eq:flux_waveform_def} and~\eqref{eq:25}, using the same method as in Sec.~\ref{sec:giSWAP}. We denote the quantum state at the end of the gate time $t_{\mathrm{tot}}$ as $|ij(t_{\mathrm{tot}})\rangle$.  
        From these initial and final states, we construct the process unitary matrix $U$ in the computational subspace as
        \begin{align}
            U_{2i+j,\,2i'+j'} =
            \frac{|\langle 00|00(t_{\mathrm{tot}})\rangle|}
                {\langle 00|00(t_{\mathrm{tot}})\rangle}
            \langle ij|i'j'(t_{\mathrm{tot}})\rangle,
        \end{align}
        where we have chosen the overall phase factor such that $U_{0,0}$ is equal to $|U_{0,0}|$. To match the phases of $U_{1,2}$ and $U_{2,1}$ to those of the ideal iSWAP gate, while preserving $\arg(U_{0,0}) = 0$, 
        we apply appropriate single-qubit $Z$ rotations to $U$. 
        The resulting phase-corrected process matrix is denoted by $U'$. 
        We obtain the iSWAP gate fidelity $F_{\mathrm{avg}}$ for the numerical simulation by substituting $U'$ into $\tilde{U}_{\mathrm{exp}}$ in Eq.~\eqref{eq:Fid_def}.

        Using the waveform defined in Eqs.~\eqref{eq:flux_waveform_def} and~\eqref{eq:25} with $\beta/2\pi = 290/2\pi~\mathrm{MHz}$ and $\phi_{\mathrm{d}}=-\pi/2$, we optimize the flux-drive amplitude $\varphi_{\mathrm{d}}$ and frequency $\omega_{\mathrm{d}}$ by minimizing the infidelity $1-F_{\mathrm{avg}}$ using the L\texttt{-}BFGS\texttt{-}B optimizer implemented in the \texttt{scipy.optimize} package~\cite{virtanen2020scipy}. 
        This optimization yields the optimal pulse that implements the iSWAP gate for a given gate duration $t_{\mathrm{tot}}$.
        
        For each optimized pulse, we extract the $ZZ$ rotation angle as 
        $\phi_{ZZ} = \arg(-U'_{3,3})$, 
        and obtain the effective $ZZ$ interaction as
        \begin{align}
            \zeta^{\mathrm{eff}} = \frac{\phi_{ZZ}}{t_{\mathrm{tot}}}.
        \end{align}

        From the effective $ZZ$ interaction $\zeta^{\mathrm{eff}}$, we further obtain the dynamical $ZZ$ interaction
        $\zeta^{\mathrm{dyn}} = \zeta^{\mathrm{eff}} - \zeta^{\mathrm{avg}}$, where $\zeta^{\mathrm{avg}}$ is the averaged static $ZZ$ interaction defined by Eq.~\eqref{eq:avgzz}.
        The resulting $\zeta^{\mathrm{eff}}$ and $\zeta^{\mathrm{dyn}}$ as functions of $\varphi_{\mathrm{d}}$ are shown by the solid black curves in Fig.~\ref{fig:dynamic_zz}(b) and the inset.

    \subsection{Hybridization fractions} \label{sec:hybridization_factor}
        The hybridization fractions of the coupler modes in the two-qubit states, the average of which is shown in Fig.~\ref{fig:coupler_hybridization}(b) in the main text, are evaluated as follows. 

        We first rewrite the Hamiltonian in Eq.~\eqref{eq-H} as 
        \begin{widetext}
            \begin{align}
            H(\varphi_{\mathrm{ex}}) &= \sum_{j=1,2} H_j + H_{\mathrm{PM}}(\varphi_{\mathrm{ex}}) + H_{\mathrm{I}}, \\
            H_j &= 4\hbar W_{jj} \hat{n}_j^2 -E_{\mathrm{J}j} \cos \hat{\varphi}_j \quad (j=1,2), \\
            H_{\mathrm{PM}}(\varphi_{\mathrm{ex}}) &= \sum_{j=3,4} \! \! \left (4\hbar W_{jj} \hat{n}_j^2 
            -E_{\mathrm{J}j} \cos \hat{\varphi}_j \right) +8\hbar W_{34} \hat{n}_3 \hat{n}_4 
            -E_{\mathrm{J}5} \cos \qty(\hat{\varphi}_4 - \hat{\varphi}_3 - \varphi_{\mathrm{ex}})
            \nonumber \\
            &=\sum_{j=\mathrm{p,m}} \! \! 4\hbar W_{jj} \hat{n}_j^2 +4\hbar (W_{33}-W_{44}) \hat{n}_{\mathrm{p}} \hat{n}_{\mathrm{m}}
            -(E_{\mathrm{J}3}+E_{\mathrm{J}4}) \cos \hat{\varphi}_{\mathrm{p}} \cos \hat{\varphi}_{\mathrm{m}} \nonumber \\
            &+(E_{\mathrm{J}3}-E_{\mathrm{J}4}) \sin \hat{\varphi}_{\mathrm{p}} \sin \hat{\varphi}_{\mathrm{m}}
            -E_{\mathrm{J}5} \cos(2\hat{\varphi}_{\mathrm{m}} + \varphi_{\mathrm{ex}}), \label{eq-Hpm}
            \\
            H_{\mathrm{I}} &= 8\hbar \left( W_{12} \hat{n}_1 \hat{n}_2 + W_{13} \hat{n}_1 \hat{n}_3
            + W_{14} \hat{n}_1 \hat{n}_4
            + W_{23} \hat{n}_2 \hat{n}_3 + W_{24} \hat{n}_2 \hat{n}_4 \right)
            \nonumber \\
            &=8\hbar W_{12} \hat{n}_1 \hat{n}_2 + 8\hbar \sum_{j=1,2} \sum_{k=\mathrm{p,m}} W_{jk} \hat{n}_j \hat{n}_k,
            \end{align}
        \end{widetext}
        where $H_{1}, H_{2}$, and $H_{\mathrm{PM}}$ are the Hamiltonians of the data qubits and the coupler modes, respectively, and $H_{\mathrm{I}}$ contains the interaction terms between the data qubits and the coupler modes.

        Using the tensor products of the eigenvectors of $H_1$, $H_2$, and $H_{\mathrm{PM}}(\varphi_{\mathrm{ex}})$, which are denoted by $|i\rangle |j\rangle |\widetilde{kl}_{\mathrm{PM}}(\varphi_{\mathrm{ex}})\rangle $, 
        the hybridization fractions for $\ket*{ij(\varphi_{\mathrm{ex}})}$ are defined as follows:
        \begin{align}
            p_{i' j' k' l'|i k}(\varphi_{\mathrm{ex}})
            =
            |\langle i'| \langle j'| \langle \widetilde{k' l'}_{\mathrm{PM}}(\varphi_{\mathrm{ex}})|
            i j (\varphi_{\mathrm{ex}}) \rangle |^2.
        \end{align}

        The participation of the coupler modes in the computational state $\ket*{ij (\varphi_{\mathrm{ex}})}$ is quantified by summing the hybridization fractions over all states with at least one coupler excitation:
        \begin{align}
            p_{ij}^{\mathrm{c}}(\varphi_{\mathrm{ex}}) &= \sum_{i^\prime,\, j^\prime} \sum_{k^\prime + l^\prime \geq 1} p_{i'j' k' l'|ij}(\varphi_{\mathrm{ex}}). \label{eq:hybridization_fraction}
        \end{align}

        We numerically confirm that the contribution of the coupler P~mode is negligible over the entire flux-bias range. Thus Eq.~\eqref{eq:hybridization_fraction} effectively quantifies the hybridization between the data qubits and the M~mode. This is natural because the M~mode frequency, compared with the P-mode frequency, is closer to the data-qubit frequencies, in particular, at a flux bias far from zero. 
        
        Using the coupler fraction $p_{ij}^{\mathrm{c}}(\varphi_{\mathrm{ex}})$ for each computational state, we define the averaged hybridization fraction of the coupler modes for all the computational states, $\overline{p}^{\mathrm{c}}(\varphi_{\mathrm{ex}})$, by Eq.~\eqref{eq:avg_hybridization} in the main text.

\begin{figure}
            \centering
            \includegraphics[]{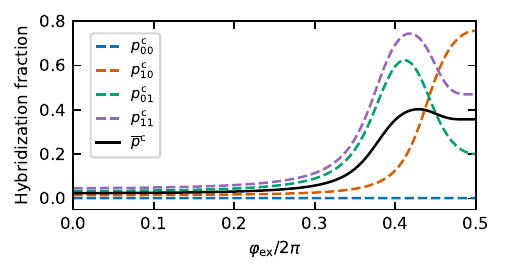}
            \caption{Hybridization fractions $p^{\mathrm{c}}_{ij}$ of the coupler modes in the computational states $|ij\rangle (\varphi_{\mathrm{ex}})$~[Eq.~\eqref{eq:hybridization_fraction}] as functions of the external flux $\varphi_{\mathrm{ex}}$. The solid black line shows the average $\bar{p}^{\mathrm{c}}$ defined in Eq.~\eqref{eq:avg_hybridization}.}
            \label{fig:hybridization_fraction}
        \end{figure}

        Figure~\ref{fig:hybridization_fraction} shows the numerically simulated coupler hybridization fractions for the computational states $\ket*{00(\varphi_{\mathrm{ex}})}$, $\ket*{01(\varphi_{\mathrm{ex}})}$, $\ket*{10(\varphi_{\mathrm{ex}})}$, and $\ket*{11(\varphi_{\mathrm{ex}})}$ as functions of $\varphi_{\mathrm{ex}}$. The averaged hybridization fraction $\overline{p}^{\mathrm{c}}$ exhibits a sharp increase after $\varphi_{\mathrm{ex}}/2\pi \gtrsim 0.3$, which corresponds to the onset of the strong hybridization regime, where the qubit and coupler frequencies are near-degenerate.

        The reason for dropping $\langle 01| \sin (2\hat{\varphi}_{\mathrm{m}}) |10\rangle$ in Eq.~(\ref{eq-g1001}) is explained using the hybridization as follows. The computational state $|10\rangle$ ($|01\rangle$) is the hybridized state of dominantly $|1\rangle |0\rangle |\widetilde{00}_{\mathrm{PM}}\rangle$ ($|0\rangle |1\rangle |\widetilde{00}_{\mathrm{PM}}\rangle$) and $|0\rangle |0\rangle |\widetilde{01}_{\mathrm{PM}}\rangle$. Therefore, $\langle 01| \sin (2\hat{\varphi}_{\mathrm{m}}) |10\rangle$ is approximately proportional to $\langle \widetilde{01}_{\mathrm{PM}}| \sin (2\hat{\varphi}_{\mathrm{m}}) |\widetilde{01}_{\mathrm{PM}}\rangle$. Here, the coupler Hamiltonian, $ H_{\mathrm{PM}}$, in Eq.~(\ref{eq-Hpm}) is an even function of  $\hat{\varphi}_{\mathrm{m}}$ and $\hat{n}_{\mathrm{m}}$ at the idle point where $\varphi_{\mathrm{ex}}=0$, under the approximations ${W_{33}-W_{44}=0}$ and ${E_{\mathrm{J}3}-E_{\mathrm{J}4}=0}$ by symmetry. Consequently, $|\widetilde{01}_{\mathrm{PM}}\rangle$, which is an eigenvector of $H_{\mathrm{PM}}$, has a specific parity from the parity symmetry of $H_{\mathrm{PM}}$. On the other hand, $\sin (2\hat{\varphi}_{\mathrm{m}})$ is an odd function of $\hat{\varphi}_{\mathrm{m}}$. Hence, the parity of $|\widetilde{01}_{\mathrm{PM}}\rangle$ is the opposite of that of $\sin (2\hat{\varphi}_{\mathrm{m}}) |\widetilde{01}_{\mathrm{PM}}\rangle$, and consequently $\langle \widetilde{01}_{\mathrm{PM}}| \sin (2\hat{\varphi}_{\mathrm{m}}) |\widetilde{01}_{\mathrm{PM}}\rangle$ [$\langle 01| \sin (2\hat{\varphi}_{\mathrm{m}}) |10\rangle$] vanishes.
        \vspace{1pt}

\section{Derivation of the second-order correction in Eq.~\eqref{eq:semi1}} \label{sec-giSWAPindirect}
    Under the sinusoidal flux drive given by Eq.~\eqref{eq:flux_drive}, the Hamiltonian in Eq.~\eqref{eq:H_virt} is expressed as
    \begin{widetext}
    \begin{align}
        \hat{H}_{\mathrm{I}}^{\mathrm{indir}}(t)
        &\simeq E_{\mathrm{J5}} \varphi_{\mathrm{ex}}(t) \sum_{j \neq 1,2} \sum_{k=1,2} e^{i\Delta_{jk} t} s_{jk} \hat{\sigma}_{jk} + \mathrm{h.c.} =
        E_{\mathrm{J}5} \varphi_{\mathrm{d}} \sum_{j \neq 1,2} \sum_{k=1,2} 
        \frac{e^{i [(\Delta_{jk}+\omega_{\mathrm{d}}) t + \phi_{\mathrm{d}}]} + e^{i [(\Delta_{jk}-\omega_{\mathrm{d}}) t - \phi_{\mathrm{d}}]}}{2}  
        s_{jk} \hat{\sigma}_{jk} + \mathrm{h.c.},
        \label{eq_Hindir2} 
    \end{align}
    where $s_{jk} = \bra*{\tilde{\Psi}_{j}} \sin(2\hat{\varphi}_{\mathrm{m}}) \ket*{\tilde{\Psi}_{k}}$ and $\hat{\sigma}_{jk} = |\tilde{\Psi}_{j}\rangle \langle \tilde{\Psi}_{k}|$.

    Note that the following resonance condition holds from ${\omega_{\mathrm{d}} \simeq (\widetilde{\omega}_2-\widetilde{\omega}_1)/2}$:
    \begin{align}
        \Delta_{j2}+\omega_{\mathrm{d}} \simeq \Delta_{j1}-\omega_{\mathrm{d}} \simeq \tilde{\omega}_j - \frac{\widetilde{\omega}_2+\widetilde{\omega}_1}{2}.
        \label{eq_resonance_condition} 
    \end{align}
    Hence, applying the time-dependent Schrieffer-Wolff (van Vleck) transformation with the following anti-Hermitian generator $\hat{S}(t)$ in Eq.~(\ref{eq-SW-S}):
    \begin{align}
        \hat{S}(t)=\frac{E_{\mathrm{J}5} \varphi_{\mathrm{d}}}{2\hbar} \sum_{j \neq 1,2} \sum_{k=1,2} 
        \left(
        \frac{e^{i [(\Delta_{jk}+\omega_{\mathrm{d}}) t + \phi_{\mathrm{d}}]}}{\Delta_{jk}+\omega_{\mathrm{d}}} +
        \frac{e^{i [(\Delta_{jk}-\omega_{\mathrm{d}}) t - \phi_{\mathrm{d}}]}}{\Delta_{jk}-\omega_{\mathrm{d}}}  \right)
        s_{jk} \hat{\sigma}_{jk} - \mathrm{h.c.},
        \label{eq_van_Vleck_S} 
    \end{align}
    and keeping only time-independent terms resulting from the resonance condition in Eq.~\eqref{eq_resonance_condition}, $\hat{H}_{\mathrm{I}}^{\mathrm{indir}}(t)$ is converted into
    \begin{align}
        \hat{H}_{\mathrm{I}}^{\mathrm{indir}}(t)
        \simeq
        \frac{E_{\mathrm{J}5}^2 \varphi_{\mathrm{d}}^2}{4\hbar} \sum_{j\neq1,2} 
        \frac{\left[ e^{i \phi_{\mathrm{d}}} s_{j2} \hat{\sigma}_{j2} + e^{-i\phi_{\mathrm{d}}} s_{j1} \hat{\sigma}_{j1}, e^{-i \phi_{\mathrm{d}}} s_{2j} \hat{\sigma}_{2j} + e^{i\phi_{\mathrm{d}}} s_{1j} \hat{\sigma}_{1j} \right] }{\widetilde{\omega}_j - (\widetilde{\omega}_2+\widetilde{\omega}_1)/2}.
        \label{eq_Hindir3} 
    \end{align}
    By retaining the terms related to the iSWAP transition, Eq.~(\ref{eq_Hindir3}) leads to the second-order correction in Eq.~\eqref{eq:semi1}.
    \end{widetext}

\section{Gate characterization details} \label{sec:gate_characterization_details}
    \subsection{Gate tune-up procedure}
        In this Appendix, we outline the gate tune-up protocol used in our experiments. 

\begin{figure*}
            \centering
            \includegraphics[]{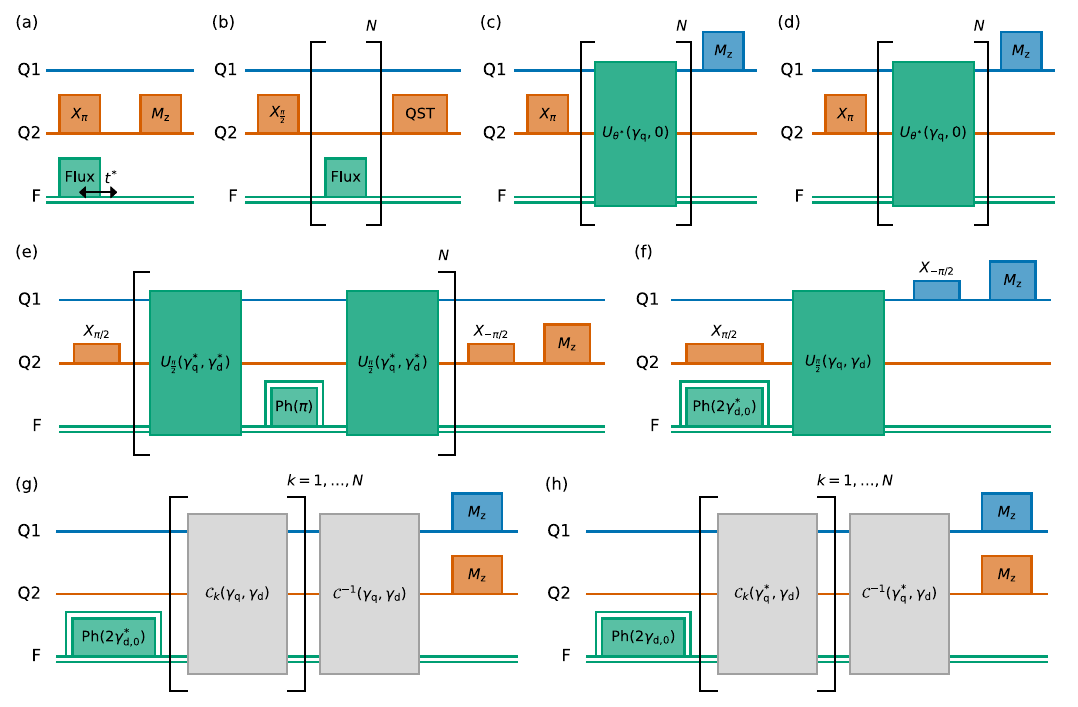}
            \caption{Pulse sequences used to calibrate the parametrically driven iSWAP gate. In each step, the parameter being optimized is marked by an asterisk~($^*$).
            (a)~Relative-delay~($t_\mathrm{delay}$) calibration between qubit drive lines and the flux-drive line.
            (b)~Zero-flux bias fine-tuning.
            (c,d)~iSWAP flux-drive amplitude and frequency calibrations, respectively, to tune up the population swap angle $\theta$. Starting from $\ket{01}$, we apply $N = 2n+1$ iSWAP pulses to amplify amplitude (frequency) errors.
            (e)~Per-gate frame-update calibration. Using an echoed cross-Ramsey-like sequence with repeated iSWAP blocks, we calibrate the per-gate frame updates, $\gamma_{\mathrm{q}}$ for the single-qubit frame and $\gamma_{\mathrm{d}}$ for the two-qubit frame.
            (f)~Initial frame-update calibration. Using a cross-Ramsey sequence, we calibrate the initial frame update on the two-qubit frame, $\gamma_{\mathrm{d},0}$.
            (g,h)~ORBIT-like fine-tuning, sweeping the initial frame update (g) and the per-gate frame update (h) around the calibrated operating point. The Clifford gates in the sequences, constructed using the calibrated iSWAP gate parameters $\gamma_{\mathrm{q}}$ and $\gamma_{\mathrm{d}}$, are denoted by $\mathcal{C}_k(\gamma_{\mathrm{q}}, \gamma_{\mathrm{d}})$ and generated according to the procedure described in Refs.~\citenum{corcoles2013process} and~\citenum{sung2021realization}.}
            \label{fig:tuneup}
        \end{figure*}

\begin{figure*}
            \centering
            \includegraphics[]{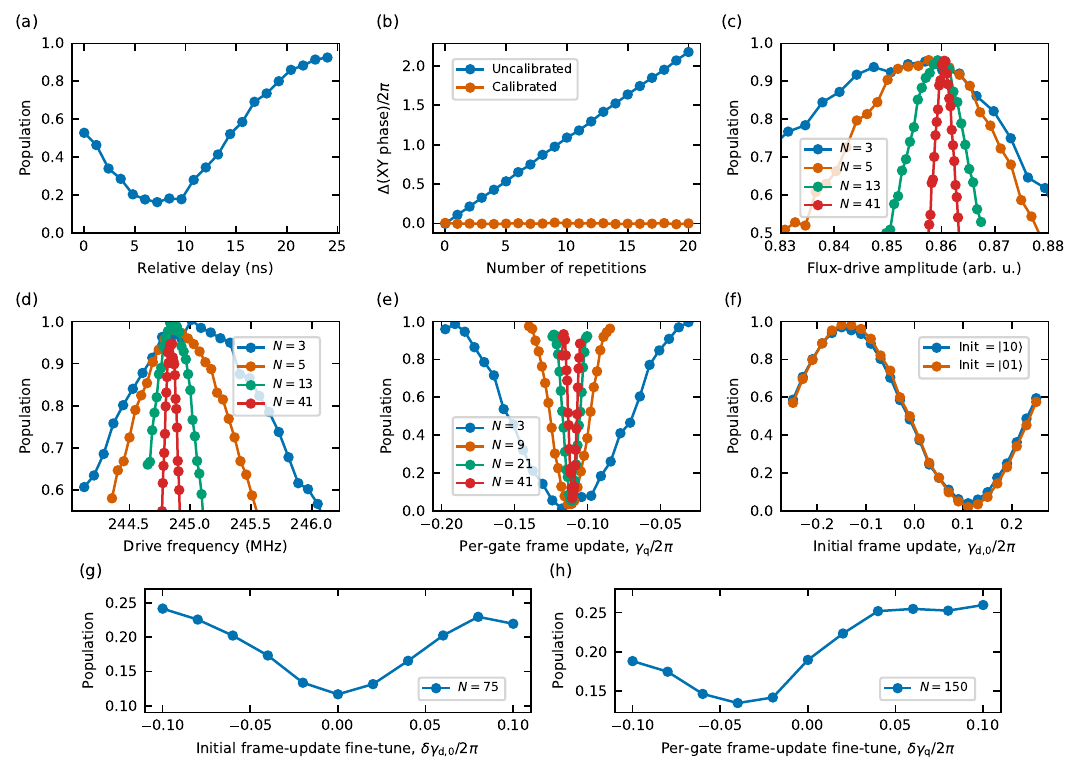}
            \caption{Representative calibration data corresponding to the tune-up sequences in Fig.~\ref{fig:tuneup}. (a)~Relative-delay calibration. (b)~Zero flux-bias fine-tuning. (c,d)~iSWAP flux-drive amplitude and frequency calibrations, respectively. Note that the oscillation patterns are averaged over interleaved per-gate frame updates to mitigate unwanted oscillations in the pattern. (e)~Per-gate frame-update calibration. Note that we only plot the results after calibrating the per-gate frame update of the flux-drive oscillator $\gamma_{\mathrm{d}}$. (f)~Initial frame-update calibration. (g)~Initial frame-update fine-tuning. (h)~Per-gate frame-update fine-tuning.}
            \label{fig:tuneup_exp}
        \end{figure*}

        First, we calibrate the relative timing between the qubit drive line and the coupler flux-drive line. We apply a $X_{\pi}$ pulse to a data qubit while almost simultaneously applying a baseband flux pulse whose envelope matches that of the qubit drive, and we sweep the relative delay between the two pulses. When the timing is aligned, the flux-induced transient frequency shift occurs during the microwave drive and suppresses the qubit excitation. We therefore choose the delay that minimizes the measured excitation~[Figs.~\ref{fig:tuneup}(a) and~\ref{fig:tuneup_exp}(a)]. After that, we precisely tune the dc flux bias to zero. We apply a train of equal-amplitude baseband flux pulses with opposite signs and compare the phase accumulated by the qubit in the positive (negative) amplitude case. As the Q2 frequency $\omega_{2}(\varphi_{\mathrm{ex}})$ is symmetric about the zero flux bias, the two accumulated phases are equal at the zero flux bias. To increase sensitivity, we repeat the pulse train $N$ times and fit the accumulated phase against $N$, then adjust the global coil current until the difference is minimized~[Figs.~\ref{fig:tuneup}(b) and~\ref{fig:tuneup_exp}(b)].

        Second, we calibrate the iSWAP flux-drive amplitude and frequency [Figs.~\ref{fig:tuneup}(c--d) and~\ref{fig:tuneup_exp}(c--d)]. Starting from an $X_{\pi}$ pulse on Q2, we apply $N=2n+1$ iSWAP pulses and measure the population of Q1 while sweeping either flux-drive amplitude or frequency. For a fixed $N$, we repeat amplitude calibration followed by frequency calibration twice for stability. We then increase $N$ up to $361$ to find the optimal amplitude and frequency with greater sensitivity. 
        
        Third, we calibrate phase misalignments of the iSWAP gate, which arise from (i) an initial phase offset among the Q1, Q2, and flux-drive control lines, and (ii) per-gate phase offsets caused by drive-induced frequency shifts. We compensate these phase misalignments by applying frame updates on each control line. Using the sequence in Fig.~\ref{fig:tuneup}(e)~[experiment in Fig.~\ref{fig:tuneup_exp}(e)], we calibrate the per-gate compensations by repeating an echoed $\pm \mathrm{iSWAP}$ pulse train while sweeping the frame updates. By preparing the initial state in either $\ket{10}$ or $\ket{01}$, we tune the calibration parameters so that the two fringes obtained from the two initial states reach their minima at the same frame-update values. Using the sequence in Fig.~\ref{fig:tuneup}(f)~[experiment in Fig.~\ref{fig:tuneup_exp}(f)], we then calibrate the initial phase offset by applying a single iSWAP pulse and sweeping the frame update on the flux-drive line. We extract the frame-update value that gives the minimum population of Q1 and Q2 at the same time. We discuss more detailed calibration protocols in Sec.~\ref{subsec:phase_correction}.

        Fourth, we fine-tune the initial phase offset and the per-gate phase compensation on the qubit control lines using an ORBIT sequence~\cite{kelly2014optimal}, which ideally projects the final state to $\ket{00}$. In the sequence in Fig.~\ref{fig:tuneup}(g), we sweep the initial phase offset on the flux line and measure the mean excitation probabilities of Q1 and Q2, and choose the value that minimizes them. In the sequence in Fig.~\ref{fig:tuneup}(h), we similarly sweep the per-gate phase compensation on the qubit drive lines and find the value that minimizes the mean excitation probability.

        Finally, as the last tune-up step, we perform ORBIT calibration by jointly optimizing four parameters: the iSWAP flux-drive amplitude, the initial phase offset on the flux line, and the per-gate phase compensation on the qubit drive lines and the flux-drive line. We use the same pulse sequence as in Fig.~\ref{fig:tuneup}(h) but measure the $\ket*{\widetilde{0000}}$ state probability including coupler modes via single-shot readout, and maximize this sequence fidelity using the Powell optimizer~\cite{powell1964efficient} implemented in the \texttt{scipy.optimize} package~\cite{virtanen2020scipy}.
        \begin{figure}
            \centering
            \includegraphics[]{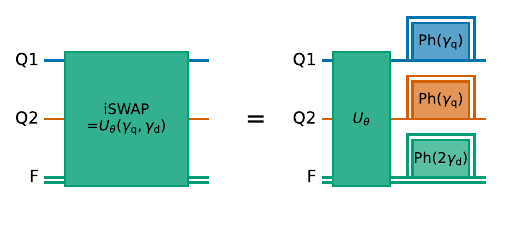}
            \caption{Decomposition of the iSWAP gate into primitive operations. F stands for the flux drive. The block $U_\theta$ represents the flux-driven unitary evolution acting on Q1 and Q2 with population swap angle $\theta$, and the subsequent double-line boxes denote software frame updates of the single-qubit frames $\mathrm{Ph}(\gamma_{\mathrm{q}})$ and the two-qubit frame relative to the flux-drive phase $\mathrm{Ph}(2\gamma_{\mathrm{d}})$ implemented by the classical phase advance in the corresponding oscillator. The frame updates applied to the single-qubit frames are also known as virtual-$Z$ gates~\cite{mckay2017efficient}.}
            \label{fig:iswap_equivalent}
        \end{figure}
    
    \subsection{Phase correction} \label{subsec:phase_correction}
        The parametrically driven iSWAP gate is phase-sensitive. The exchange axis is set by the relative phases of the control oscillators that generate the two qubit-drive waveforms and the flux-drive waveform.

        To implement the iSWAP gate with the correct exchange axis that yields a phase of $-i$ after population exchange, we must explicitly keep track of phase references throughout the experimental sequence, a procedure often referred to as frame tracking~\cite{mckay2016universal, arute2020observation, wei2024native, abrams2020implementation, bao2022fluxonium, krizan2025quantum}. Following Ref.~\citenum{abrams2020implementation}, we refer to the rotating frames defined by the Q1 and Q2 microwave drive oscillators as single-qubit frames. We also introduce a reference frame for the flux-drive waveform and refer to it as the two-qubit frame. The absolute phase of each frame can be chosen arbitrarily, but the relative phase between these frames becomes physically meaningful and needs careful calibration whenever the system undergoes an interaction that does not commute with single-qubit Z rotations, which is the case for our exchange-type gates. In practice, residual phase jitter or drift among the control oscillators therefore appears as stochastic or systematic gate errors and can limit the overall gate fidelity~\cite{ganzhorn2020benchmarking}. In our setup, the microwave and flux-drive sources are phase-locked to a common $100$-$\mathrm{MHz}$ reference clock, as described in Appendix~\ref{sec:measurement_setup}.
        
        We compensate these phase errors by applying frame updates. A frame update does not directly act on the quantum state; instead, it is implemented as a software update of the reference frame that affects the subsequent pulses generated by the corresponding control oscillator. For the qubit drive oscillators, such a frame update is also known as a virtual-$Z$ gate, which we denote by $\mathrm{VZ}_i(\gamma_i)$ for qubit $i=1,2$~\cite{mckay2017efficient}. For the flux-drive oscillator, a frame update simply advances the phase reference of the oscillator. While this frame update on the two-qubit frame does not correspond to any single- or two-qubit unitary by itself, it determines the phase of the subsequent exchange interaction and therefore affects the exchange axis of subsequent iSWAP pulses.
        
        Therefore, even after calibrating the iSWAP gate amplitude and frequency in the above-mentioned calibration steps, the realized two-qubit gate can still deviate from the ideal iSWAP gate due to exchange-axis misalignments. In our implementation, this misalignment originates from the initial phase offsets among the control oscillators and drive-induced shifts of the qubit transition frequencies during the gate, which generate dynamical physical $Z$ rotations on the qubits.
 
        \subsubsection{Phase conventions and notation}
            We denote the reference phase of the oscillator $l$ immediately before the iSWAP gate by $\phi_{l}$, where $l=1,2,\mathrm{d}$ labels the Q1 drive, the Q2 drive, and the flux-drive oscillator, respectively. During the flux drive, the qubit frequencies are shifted and the qubits acquire additional dynamical phases. We denote the corresponding per-gate phase on each qubit by $\delta \phi_{l}$ ($l = 1, 2$).
            
            In the rotating frame of the data qubits defined by the single-qubit frames, the iSWAP interaction implemented with phase parameters $(\phi_{l}, \, \delta \phi_{l})$ $(l=1,2,\mathrm{d})$ can be expressed as
            \begin{align}
                &\hat{U}_{\frac{\pi}{2}}(\phi_{\mathrm{mis}}, \delta \phi_{1}, \delta \phi_{2}, \phi_{ZZ}) = \nonumber \\
                &\begingroup
                \setlength{\arraycolsep}{0.5pt}
                \begin{pmatrix}
                1 & 0 & 0 & 0 \\
                0 & 0 & -i e^{-i(\delta \phi_2 + \phi_{\mathrm{mis}})} & 0 \\
                0 & -i e^{-i(\delta \phi_1 - \phi_{\mathrm{mis}})} & 0 & 0 \\
                0 & 0 & 0 & e^{-i(\delta \phi_{1} + \delta \phi_{2} + \phi_{ZZ})}
                \end{pmatrix}
                \endgroup ,
                \label{eq:iswap_with_phase}
            \end{align}
            where we denote the exchange-axis misalignment by
            \begin{align}
                 \phi_{\mathrm{mis}} = 2\phi_{\mathrm{d}} + \phi_{1} - \phi_{2},
            \end{align}
            which is determined by the phase references of the Q1-drive~($\phi_1$), Q2-drive~($\phi_2$), and flux-drive~($\phi_{\mathrm{d}}$) oscillators, respectively. The parameter $\phi_{ZZ}$ is the parasitic $ZZ$ phase accumulated during the gate. While $\delta \phi_{\mathrm{d}}$ does not explicitly appear in Eq.~\eqref{eq:iswap_with_phase}, it contributes to the exchange-axis misalignment of the next iSWAP pulse as $\phi_{\mathrm{mis}} \rightarrow \phi_{\mathrm{mis}} + 2\delta \phi_{\mathrm{d}}$, and must therefore be tracked together with the single-qubit frames.
            
            We use frame updates to compensate these phase errors (double-line boxes in Fig.~\ref{fig:iswap_equivalent}). We denote the per-gate single-qubit frame updates by $\gamma_1$ and $\gamma_2$, and the per-gate two-qubit frame update by $\gamma_{\mathrm{d}}$. In addition, we apply an initial frame update of the two-qubit frame at the beginning of a measurement sequence, which we denote by $\gamma_{\mathrm{d}, 0}$.
        \subsubsection{Commutation condition for frame updates}
            A frame update is a virtual operation, indicating that it does not directly change the underlying quantum state. Its physical effect only appears through the phase of subsequent control pulses generated by the same oscillator. As a result, a sequence of accumulated frame updates can be regarded as being applied immediately prior to the next gate operation. 
            
            For our phase-tracking convention to be consistent throughout experimental sequences, the single-qubit frame updates applied after an iSWAP gate must commute with the iSWAP gate. We therefore impose
            \begin{align}
                \qty[\hat{U}_{\mathrm{iSWAP}}, \mathrm{VZ}_1(\gamma_1) \otimes \mathrm{VZ}_2(\gamma_2)] = 0,
            \end{align}
            where $\hat{U}_{\mathrm{iSWAP}}$ denotes the ideal iSWAP operation, and $\mathrm{VZ}_i(\gamma_i)$ is the frame update on the $i$th qubit.

            The condition is satisfied when~\cite{abrams2020implementation}
            \begin{align}
                \gamma_1 - \gamma_2 \equiv 0~(\mathrm{mod}~2\pi),
            \end{align}
            and we therefore restrict the per-gate single-qubit frame updates applied to each qubit to a common value
            \begin{align}
                \gamma_{\mathrm{q}} = \gamma_1 = \gamma_2.
            \end{align}

            As shown in Fig.~\ref{fig:iswap_equivalent}, the iSWAP gate is implemented by the raw unitary evolution generated by the flux drive, followed by the frame updates that compensate the phase misalignments. The overall unitary evolution is therefore expressed as
            \begin{align}
                &\hat{U}_{\frac{\pi}{2}}(\gamma_{\mathrm{q}}, \gamma_{\mathrm{d}}) \nonumber \\
                &\quad = \qty[\mathrm{VZ}_1(\gamma_{\mathrm{q}}) \otimes \mathrm{VZ}_2(\gamma_{\mathrm{q}})] \circ \hat{U}_{\frac{\pi}{2}}(\phi_{\mathrm{mis}}, \delta \phi_{1}, \delta \phi_{2}, \phi_{ZZ}),
            \end{align}
            where $\gamma_{\mathrm{d}}$ does not explicitly appear on the right-hand side. However, it contributes to the exchange-axis misalignment of the subsequent iSWAP pulses.

        \subsubsection{Per-gate phase correction}
            We calibrate the per-gate frame updates $(\gamma_{\mathrm{q}}, \gamma_{\mathrm{d}})$ using an echoed cross-Ramsey-like sequence in Fig.~\ref{fig:tuneup}(e)~\cite{ganzhorn2020benchmarking, krizan2025quantum}. The sequence prepares a superposition state in either Q1 or Q2, applies a train of iSWAP pulses with a $\pi$ phase flip that reverses the sign of the subsequent pulses, and finally measures the accumulated phase by reading out the initially excited qubit. We repeat the echo block $N$ times to amplify the small residual phase errors.
            
            Let us denote the exchange-axis misalignment before the $n$th iSWAP repetition block as 
            \begin{align}
                \phi_{\mathrm{mis}}^{(n)} = 2 \phi_{\mathrm{d}}^{(n)} + \phi_{1}^{(n)} - \phi_{2}^{(n)}.
            \end{align}
            Then the unitary evolution of the $n$th repetition block is expressed as
            \begin{align}
                &\hat{U}_{\mathrm{block}}^{(n)} \nonumber \\
                &=\begingroup
                \setlength{\arraycolsep}{0.5pt}
                \begin{pmatrix}
                1 & 0 & 0 & 0 \\
                0 & 0 & +i e^{-i(\delta \phi_2 + \gamma_\mathrm{q} + \phi_{\mathrm{mis}}^{\prime(n)})} & 0 \\
                0 & +i e^{-i(\delta \phi_1 + \gamma_\mathrm{q}- \phi_{\mathrm{mis}}^{\prime(n)})} & 0 & 0 \\
                0 & 0 & 0 & *
                \end{pmatrix}
                \endgroup \nonumber \\
                &\quad \times \begingroup
                \setlength{\arraycolsep}{0.5pt}
                \begin{pmatrix}
                1 & 0 & 0 & 0 \\
                0 & 0 & -i e^{-i(\delta \phi_2 + \gamma_\mathrm{q} + \phi_{\mathrm{mis}}^{(n)})} & 0 \\
                0 & -i e^{-i(\delta \phi_1 + \gamma_\mathrm{q}- \phi_{\mathrm{mis}}^{(n)})} & 0 & 0 \\
                0 & 0 & 0 & *
                \end{pmatrix}
                \endgroup , \label{eq:iswap_zeeman_unitary}
            \end{align}
            where we omit the $\ket{11}$ term, because it does not affect the populations in the single-excitation subspace for this calibration. Here we express the exchange-axis misalignment in the second iSWAP pulse in the block as 
            \begin{align}
                \phi_{\mathrm{mis}}^{\prime (n)} &= \phi_{\mathrm{mis}}^{(n)} + 2\gamma_{\mathrm{d}} + (2 \delta \phi_{\mathrm{d}} + \delta \phi_1 - \delta \phi_2) \\
                &= \phi_{\mathrm{mis}}^{(n)} + \Delta \phi_{\mathrm{mis}},
            \end{align}
            where we denote the per-gate drift of the exchange-axis misalignment by $\Delta \phi_{\mathrm{mis}}$.
            
            The echoed block $\hat{U}_{\mathrm{block}}^{(n)}$ effectively applies a phase shift on $\ket{10}$ and $\ket{01}$ depending on the parameters $\gamma_{\mathrm{q}}$ and $\gamma_{\mathrm{d}}$. For two different initial states prepared by $X_{\pi/2}$ on Q1 and Q2, we obtain Ramsey-like fringes after repeating the echoed block $N$ times as 
            \begin{align}
                P_{\mathrm{Q1}}(N) = \sin^2\qty{N \qty[(\delta \phi_1 + \delta \phi_2) + 2\gamma_{\mathrm{q}} - \Delta \phi_{\mathrm{mis}}]}, \\
                P_{\mathrm{Q2}}(N) = \sin^2\qty{N \qty[(\delta \phi_1 + \delta \phi_2) + 2\gamma_{\mathrm{q}} + \Delta \phi_{\mathrm{mis}}]},
            \end{align}
            where $P_{\mathrm{Q1}(\mathrm{Q2})}(N)$ denotes the probability of finding Q1~(Q2) in the excited state at the end of the sequence when the initial superposition is prepared in Q1~(Q2). 

            In the experiment, we sweep $\gamma_{\mathrm{q}}$ and fit the phases that give minima in the two fringes, which we denote by $\gamma_{\mathrm{q}, 1}$ and $\gamma_{\mathrm{q}, 2}$ for the initial states prepared in Q1 and Q2, respectively. The fitted minima satisfy
            \begin{align}
                \gamma_{\mathrm{q}, 1}
                &\equiv
                \frac{-(\delta\phi_1+\delta\phi_2)+\Delta \phi_{\mathrm{mis}}}{2}
                \quad (\mathrm{mod}\ \pi),
                \\
                \gamma_{\mathrm{q}, 2}
                &\equiv
                \frac{-(\delta\phi_1+\delta\phi_2)-\Delta \phi_{\mathrm{mis}}}{2}
                \quad (\mathrm{mod}\ \pi).
            \end{align}
            Therefore, the difference between the two minima is the residual exchange-axis misalignment given by
            \begin{align}
                \Delta \phi_{\mathrm{mis}}
                \equiv
                \gamma_{\mathrm{q}, 1} - \gamma_{\mathrm{q}, 2}
                \quad (\mathrm{mod}\ \pi),
                \label{eq:Delta_phi_from_minima}
            \end{align}

            We repeat the calibration step a few times so that $\gamma_{\mathrm{d}}$ is well-calibrated and the two minima are close enough, which corresponds to $\Delta \phi_{\mathrm{mis}} = 0$. We then take the average of the two minima to determine the common per-gate single-qubit frame update as
            \begin{align}
                \gamma_{\mathrm{q}} = \frac{\gamma_{\mathrm{q}, 1}+\gamma_{\mathrm{q}, 2}}{2}.
            \end{align}
            
            We repeat this calibration procedure twice for a fixed $N$ to reliably determine the optimal $(\gamma_{\mathrm{q}}, \gamma_{\mathrm{d}})$ and then increase $N$ to increase the sensitivity. In this work, we use $N$ up to $361$.

        \subsubsection{Initial phase correction}
            After calibrating the per-gate frame updates, we calibrate the initial two-qubit frame update $\gamma_{\mathrm{d},0}$ applied at the beginning of the sequence. Using the cross-Ramsey sequence in Fig.~\ref{fig:tuneup}(f), we prepare a superposition state in Q1 or Q2, apply an iSWAP pulse using the calibrated parameters $(\gamma_{\mathrm{q}}, \gamma_{\mathrm{d}})$, apply an $X_{-\pi/2}$ pulse on the qubit that is not initially excited, and finally measure the qubit.

            Since the per-gate frame updates are already calibrated, the remaining exchange-axis misalignment is the initial misalignment set by the experimental setup. Applying the initial frame update $\gamma_{\mathrm{d},0}$ advances the two-qubit frame as
            \begin{align}
                \phi_{\mathrm{mis}}^{(0)} \rightarrow \phi_{\mathrm{mis}}^{(0)} + 2\gamma_{\mathrm{d},0}.
            \end{align}

            We sweep $\gamma_{\mathrm{d}, 0}$ and fit the resulting fringes to determine the value that minimizes the measured excitation probability. We accept the calibration when the fringes obtained from the two different initial states are consistent within the experimental uncertainty. Because $\gamma_{\mathrm{d}, 0}$ is a constant offset, it is not amplified by repeating the iSWAP block. We therefore run this calibration with one iSWAP gate and subsequently fine-tune $\gamma_{\mathrm{d},0}$ together with other parameters in the ORBIT optimization.

    \subsection{Quantum process tomography} 
        \begin{figure}
            \centering
            \includegraphics[]{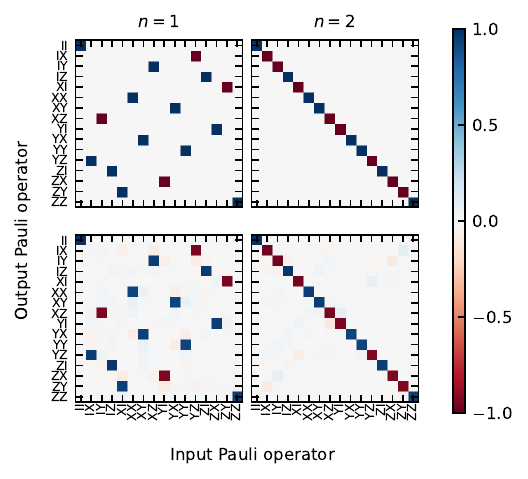}
            \caption{Pauli transfer matrices (PTMs) of the calibrated $\mathrm{iSWAP}^{n}$ gate from quantum process tomography. The two columns correspond to the number of successive iSWAP gates with $n=1, 2$. The top row shows the ideal PTMs, and the bottom row shows the experimentally reconstructed PTMs.}
            \label{fig:ptm2pt}
        \end{figure}
        After completing all calibrations, we verify that the iSWAP gate is correctly tuned by using quantum process tomography (QPT). From the QPT data, we reconstruct the Pauli transfer matrix (PTM) and compare it with the ideal PTM of the iSWAP gate. The experimentally reconstructed PTM, together with the theoretical PTMs for $\mathrm{iSWAP}^{n}\,(n=1,2)$, is shown in Fig.~\ref{fig:ptm2pt}.
        
        The process fidelity of a single iSWAP gate is $F_{\mathrm{process}} = 95.5\%$, which is mainly limited by the state-preparation and measurement (SPAM) error. Therefore, for more accurate estimation of the gate fidelity, we perform the randomized benchmarking discussed in the main text.

\section{Qubit state readout}
    \begin{figure*}
        \centering
        \includegraphics[]{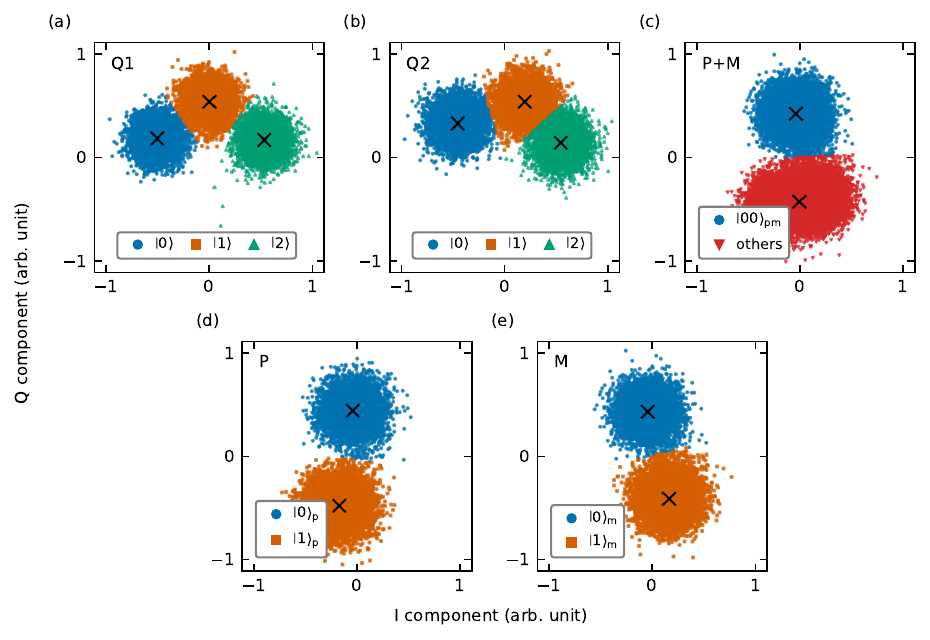}
        \caption{Single-shot IQ readout characterization. The black crosses mark the centers of the corresponding IQ clusters. Single-shot IQ distributions for data qubits (a) Q1 and (b) Q2. The state assignment for data qubits uses $k$-means clustering with $k=3$ to account for leakage into $\ket{2}$. (c) Single-shot IQ distributions for the coupler modes. Because the P and M modes share the same readout resonator, their signals are not well separated by mode. We therefore use a shared $k=2$ discriminator to separate the state in which both coupler modes are in the ground state from all other cases. The latter are treated as leakage states. (d,e) Single-shot IQ distributions for the P~mode and M~mode, respectively.}
        \label{fig:singleshot_readout}
    \end{figure*}
    \begin{table}
        \caption{Assignment probability matrix for the qubits and couplers used in this work. $P(m|n)$ is the probability to find the state in $\ket{m}$ starting from an initial state $\ket{n}$. Because the signals of the excited P-mode and M-mode states largely overlap in the IQ plane, we only discriminate $|00\rangle_{\mathrm{pm}}$ from other states.}
        \begin{tabular}{lllll}
        \hline \hline & & \multicolumn{3}{c}{Initial state $n$} \\
        \hline Q1 $P(m|n)$ & & $\ket{0}$ & $\ket{1}$ & $\ket{2}$ \\
        Readout state $m$ & $\ket{0}$ & 0.9931 & 0.0065 & 0.0004 \\
        $F_{\mathrm{assign}} = 0.987$ & $\ket{1}$ & 0.0107 & 0.9847 & 0.0046 \\
        & $\ket{2}$ & 0.0017 & 0.0151 & 0.9832 \\
        \hline Q2 $P(m|n)$ & & $\ket{0}$ & $\ket{1}$ & $\ket{2}$ \\
        Readout state $m$ & $\ket{0}$ & 0.9942 & 0.0058 & 0.0000 \\
        $F_{\mathrm{assign}} = 0.973$ & $\ket{1}$ & 0.0110 & 0.9656 & 0.0233 \\
        & $\ket{2}$ & 0.0023 & 0.0378 & 0.9599 \\
        \hline Couplers $P(m|n)$ & & $\ket{00}_{\mathrm{pm}}$ & others \\
        Readout state $m$ & $\ket{00}_{\mathrm{pm}}$ & 0.9977 & 0.0023 \\
        $F_{\mathrm{assign}} = 0.970$ & others & 0.0568 & 0.9432 \\
        \hline \hline
        \end{tabular}
        \label{tab:qubit_readout_fidelity}
    \end{table}
        
    Qubit state readout is performed using a $1000$-$\mathrm{ns}$ raised-cosine flattop pulse with $25$-$\mathrm{ns}$ rise and fall times. We independently perform single-shot qubit readout after preparing each qubit in either $\ket{0}$ or $\ket{1}$ state. The resulting data points in the IQ plane are shown in Fig.~\ref{fig:singleshot_readout}, and the corresponding readout assignment fidelities are summarized in Table~\ref{tab:qubit_readout_fidelity}. We use a $k$-means clustering method to discriminate qubit states with $k=3$ for the data qubits to account for leakage into $\ket{2}$, and $k=2$ for the coupler modes. For the state discrimination of the coupler modes, we employ a shared discriminator obtained from M-mode single-shot results. This is because the P~mode and M~mode share the same readout resonator and their signals in the IQ plane overlap significantly, as seen in Figs.~\ref{fig:singleshot_readout}(d) and (e).

\cleardoublepage
\bibliographystyle{apsrev4-2}
\bibliography{mybib}

\begin{thebibliography}{85}%
\makeatletter
\providecommand \@ifxundefined [1]{%
 \@ifx{#1\undefined}
}%
\providecommand \@ifnum [1]{%
 \ifnum #1\expandafter \@firstoftwo
 \else \expandafter \@secondoftwo
 \fi
}%
\providecommand \@ifx [1]{%
 \ifx #1\expandafter \@firstoftwo
 \else \expandafter \@secondoftwo
 \fi
}%
\providecommand \natexlab [1]{#1}%
\providecommand \enquote  [1]{``#1''}%
\providecommand \bibnamefont  [1]{#1}%
\providecommand \bibfnamefont [1]{#1}%
\providecommand \citenamefont [1]{#1}%
\providecommand \href@noop [0]{\@secondoftwo}%
\providecommand \href [0]{\begingroup \@sanitize@url \@href}%
\providecommand \@href[1]{\@@startlink{#1}\@@href}%
\providecommand \@@href[1]{\endgroup#1\@@endlink}%
\providecommand \@sanitize@url [0]{\catcode `\\12\catcode `\$12\catcode
  `\&12\catcode `\#12\catcode `\^12\catcode `\_12\catcode `\%12\relax}%
\providecommand \@@startlink[1]{}%
\providecommand \@@endlink[0]{}%
\providecommand \url  [0]{\begingroup\@sanitize@url \@url }%
\providecommand \@url [1]{\endgroup\@href {#1}{\urlprefix }}%
\providecommand \urlprefix  [0]{URL }%
\providecommand \Eprint [0]{\href }%
\providecommand \doibase [0]{https://doi.org/}%
\providecommand \selectlanguage [0]{\@gobble}%
\providecommand \bibinfo  [0]{\@secondoftwo}%
\providecommand \bibfield  [0]{\@secondoftwo}%
\providecommand \translation [1]{[#1]}%
\providecommand \BibitemOpen [0]{}%
\providecommand \bibitemStop [0]{}%
\providecommand \bibitemNoStop [0]{.\EOS\space}%
\providecommand \EOS [0]{\spacefactor3000\relax}%
\providecommand \BibitemShut  [1]{\csname bibitem#1\endcsname}%
\let\auto@bib@innerbib\@empty
\bibitem [{\citenamefont {Kim}\ \emph {et~al.}(2023)\citenamefont {Kim},
  \citenamefont {Eddins}, \citenamefont {Anand}, \citenamefont {Wei},
  \citenamefont {Van Den~Berg}, \citenamefont {Rosenblatt}, \citenamefont
  {Nayfeh}, \citenamefont {Wu}, \citenamefont {Zaletel}, \citenamefont
  {Temme},\ and\ \citenamefont {Kandala}}]{kim2023evidence}%
  \BibitemOpen
  \bibfield  {author} {\bibinfo {author} {\bibfnamefont {Y.}~\bibnamefont
  {Kim}}, \bibinfo {author} {\bibfnamefont {A.}~\bibnamefont {Eddins}},
  \bibinfo {author} {\bibfnamefont {S.}~\bibnamefont {Anand}}, \bibinfo
  {author} {\bibfnamefont {K.~X.}\ \bibnamefont {Wei}}, \bibinfo {author}
  {\bibfnamefont {E.}~\bibnamefont {Van Den~Berg}}, \bibinfo {author}
  {\bibfnamefont {S.}~\bibnamefont {Rosenblatt}}, \bibinfo {author}
  {\bibfnamefont {H.}~\bibnamefont {Nayfeh}}, \bibinfo {author} {\bibfnamefont
  {Y.}~\bibnamefont {Wu}}, \bibinfo {author} {\bibfnamefont {M.}~\bibnamefont
  {Zaletel}}, \bibinfo {author} {\bibfnamefont {K.}~\bibnamefont {Temme}},\
  and\ \bibinfo {author} {\bibfnamefont {A.}~\bibnamefont {Kandala}},\ }\href
  {https://doi.org/10.1038/s41586-023-06096-3} {\bibfield  {journal} {\bibinfo
  {journal} {Nature}\ }\textbf {\bibinfo {volume} {618}},\ \bibinfo {pages}
  {500} (\bibinfo {year} {2023})}\BibitemShut {NoStop}%
\bibitem [{\citenamefont {{Google Quantum AI and
  Collaborators}}(2025)}]{google2025observation}%
  \BibitemOpen
  \bibfield  {author} {\bibinfo {author} {\bibnamefont {{Google Quantum AI and
  Collaborators}}},\ }\href {https://doi.org/10.1038/s41586-025-09526-6}
  {\bibfield  {journal} {\bibinfo  {journal} {Nature}\ }\textbf {\bibinfo
  {volume} {646}},\ \bibinfo {pages} {825} (\bibinfo {year}
  {2025})}\BibitemShut {NoStop}%
\bibitem [{\citenamefont {Zhao}\ \emph {et~al.}(2022)\citenamefont {Zhao},
  \citenamefont {Ye}, \citenamefont {Huang}, \citenamefont {Zhang},
  \citenamefont {Wu}, \citenamefont {Guan}, \citenamefont {Zhu}, \citenamefont
  {Wei}, \citenamefont {He}, \citenamefont {Cao}, \citenamefont {Chen},
  \citenamefont {Chung}, \citenamefont {Deng}, \citenamefont {Fan},
  \citenamefont {Gong}, \citenamefont {Guo}, \citenamefont {Guo}, \citenamefont
  {Han}, \citenamefont {Li}, \citenamefont {Li}, \citenamefont {Li},
  \citenamefont {Liang}, \citenamefont {Lin}, \citenamefont {Qian},
  \citenamefont {Rong}, \citenamefont {Su}, \citenamefont {Sun}, \citenamefont
  {Wang}, \citenamefont {Wu}, \citenamefont {Xu}, \citenamefont {Ying},
  \citenamefont {Yu}, \citenamefont {Zha}, \citenamefont {Zhang}, \citenamefont
  {Huo}, \citenamefont {Lu}, \citenamefont {Peng}, \citenamefont {Zhu},\ and\
  \citenamefont {Pan}}]{zhao2022realization}%
  \BibitemOpen
  \bibfield  {author} {\bibinfo {author} {\bibfnamefont {Y.}~\bibnamefont
  {Zhao}}, \bibinfo {author} {\bibfnamefont {Y.}~\bibnamefont {Ye}}, \bibinfo
  {author} {\bibfnamefont {H.-L.}\ \bibnamefont {Huang}}, \bibinfo {author}
  {\bibfnamefont {Y.}~\bibnamefont {Zhang}}, \bibinfo {author} {\bibfnamefont
  {D.}~\bibnamefont {Wu}}, \bibinfo {author} {\bibfnamefont {H.}~\bibnamefont
  {Guan}}, \bibinfo {author} {\bibfnamefont {Q.}~\bibnamefont {Zhu}}, \bibinfo
  {author} {\bibfnamefont {Z.}~\bibnamefont {Wei}}, \bibinfo {author}
  {\bibfnamefont {T.}~\bibnamefont {He}}, \bibinfo {author} {\bibfnamefont
  {S.}~\bibnamefont {Cao}}, \bibinfo {author} {\bibfnamefont {F.}~\bibnamefont
  {Chen}}, \bibinfo {author} {\bibfnamefont {T.-H.}\ \bibnamefont {Chung}},
  \bibinfo {author} {\bibfnamefont {H.}~\bibnamefont {Deng}}, \bibinfo {author}
  {\bibfnamefont {D.}~\bibnamefont {Fan}}, \bibinfo {author} {\bibfnamefont
  {M.}~\bibnamefont {Gong}}, \bibinfo {author} {\bibfnamefont {C.}~\bibnamefont
  {Guo}}, \bibinfo {author} {\bibfnamefont {S.}~\bibnamefont {Guo}}, \bibinfo
  {author} {\bibfnamefont {L.}~\bibnamefont {Han}}, \bibinfo {author}
  {\bibfnamefont {N.}~\bibnamefont {Li}}, \bibinfo {author} {\bibfnamefont
  {S.}~\bibnamefont {Li}}, \bibinfo {author} {\bibfnamefont {Y.}~\bibnamefont
  {Li}}, \bibinfo {author} {\bibfnamefont {F.}~\bibnamefont {Liang}}, \bibinfo
  {author} {\bibfnamefont {J.}~\bibnamefont {Lin}}, \bibinfo {author}
  {\bibfnamefont {H.}~\bibnamefont {Qian}}, \bibinfo {author} {\bibfnamefont
  {H.}~\bibnamefont {Rong}}, \bibinfo {author} {\bibfnamefont {H.}~\bibnamefont
  {Su}}, \bibinfo {author} {\bibfnamefont {L.}~\bibnamefont {Sun}}, \bibinfo
  {author} {\bibfnamefont {S.}~\bibnamefont {Wang}}, \bibinfo {author}
  {\bibfnamefont {Y.}~\bibnamefont {Wu}}, \bibinfo {author} {\bibfnamefont
  {Y.}~\bibnamefont {Xu}}, \bibinfo {author} {\bibfnamefont {C.}~\bibnamefont
  {Ying}}, \bibinfo {author} {\bibfnamefont {J.}~\bibnamefont {Yu}}, \bibinfo
  {author} {\bibfnamefont {C.}~\bibnamefont {Zha}}, \bibinfo {author}
  {\bibfnamefont {K.}~\bibnamefont {Zhang}}, \bibinfo {author} {\bibfnamefont
  {Y.-H.}\ \bibnamefont {Huo}}, \bibinfo {author} {\bibfnamefont {C.-Y.}\
  \bibnamefont {Lu}}, \bibinfo {author} {\bibfnamefont {C.-Z.}\ \bibnamefont
  {Peng}}, \bibinfo {author} {\bibfnamefont {X.}~\bibnamefont {Zhu}},\ and\
  \bibinfo {author} {\bibfnamefont {J.-W.}\ \bibnamefont {Pan}},\ }\href
  {https://doi.org/10.1103/PhysRevLett.129.030501} {\bibfield  {journal}
  {\bibinfo  {journal} {Phys. Rev. Lett.}\ }\textbf {\bibinfo {volume} {129}},\
  \bibinfo {pages} {030501} (\bibinfo {year} {2022})}\BibitemShut {NoStop}%
\bibitem [{\citenamefont {Andersen}\ \emph {et~al.}(2020)\citenamefont
  {Andersen}, \citenamefont {Remm}, \citenamefont {Lazar}, \citenamefont
  {Krinner}, \citenamefont {Lacroix}, \citenamefont {Norris}, \citenamefont
  {Gabureac}, \citenamefont {Eichler},\ and\ \citenamefont
  {Wallraff}}]{andersen2020repeated}%
  \BibitemOpen
  \bibfield  {author} {\bibinfo {author} {\bibfnamefont {C.~K.}\ \bibnamefont
  {Andersen}}, \bibinfo {author} {\bibfnamefont {A.}~\bibnamefont {Remm}},
  \bibinfo {author} {\bibfnamefont {S.}~\bibnamefont {Lazar}}, \bibinfo
  {author} {\bibfnamefont {S.}~\bibnamefont {Krinner}}, \bibinfo {author}
  {\bibfnamefont {N.}~\bibnamefont {Lacroix}}, \bibinfo {author} {\bibfnamefont
  {G.~J.}\ \bibnamefont {Norris}}, \bibinfo {author} {\bibfnamefont
  {M.}~\bibnamefont {Gabureac}}, \bibinfo {author} {\bibfnamefont
  {C.}~\bibnamefont {Eichler}},\ and\ \bibinfo {author} {\bibfnamefont
  {A.}~\bibnamefont {Wallraff}},\ }\href
  {https://www.nature.com/articles/s41567-020-0920-y} {\bibfield  {journal}
  {\bibinfo  {journal} {Nature Physics}\ }\textbf {\bibinfo {volume} {16}},\
  \bibinfo {pages} {875} (\bibinfo {year} {2020})}\BibitemShut {NoStop}%
\bibitem [{\citenamefont {Krinner}\ \emph {et~al.}(2022)\citenamefont
  {Krinner}, \citenamefont {Lacroix}, \citenamefont {Remm}, \citenamefont
  {Di~Paolo}, \citenamefont {Genois}, \citenamefont {Leroux}, \citenamefont
  {Hellings}, \citenamefont {Lazar}, \citenamefont {Swiadek}, \citenamefont
  {Herrmann}, \citenamefont {Norris}, \citenamefont {Andersen}, \citenamefont
  {Müller}, \citenamefont {Blais}, \citenamefont {Eichler},\ and\
  \citenamefont {Wallraff}}]{krinner2022realizing}%
  \BibitemOpen
  \bibfield  {author} {\bibinfo {author} {\bibfnamefont {S.}~\bibnamefont
  {Krinner}}, \bibinfo {author} {\bibfnamefont {N.}~\bibnamefont {Lacroix}},
  \bibinfo {author} {\bibfnamefont {A.}~\bibnamefont {Remm}}, \bibinfo {author}
  {\bibfnamefont {A.}~\bibnamefont {Di~Paolo}}, \bibinfo {author}
  {\bibfnamefont {E.}~\bibnamefont {Genois}}, \bibinfo {author} {\bibfnamefont
  {C.}~\bibnamefont {Leroux}}, \bibinfo {author} {\bibfnamefont
  {C.}~\bibnamefont {Hellings}}, \bibinfo {author} {\bibfnamefont
  {S.}~\bibnamefont {Lazar}}, \bibinfo {author} {\bibfnamefont
  {F.}~\bibnamefont {Swiadek}}, \bibinfo {author} {\bibfnamefont
  {J.}~\bibnamefont {Herrmann}}, \bibinfo {author} {\bibfnamefont {G.~J.}\
  \bibnamefont {Norris}}, \bibinfo {author} {\bibfnamefont {C.~K.}\
  \bibnamefont {Andersen}}, \bibinfo {author} {\bibfnamefont {M.}~\bibnamefont
  {Müller}}, \bibinfo {author} {\bibfnamefont {A.}~\bibnamefont {Blais}},
  \bibinfo {author} {\bibfnamefont {C.}~\bibnamefont {Eichler}},\ and\ \bibinfo
  {author} {\bibfnamefont {A.}~\bibnamefont {Wallraff}},\ }\href
  {https://doi.org/10.1038/s41586-022-04566-8} {\bibfield  {journal} {\bibinfo
  {journal} {Nature}\ }\textbf {\bibinfo {volume} {605}},\ \bibinfo {pages}
  {669–674} (\bibinfo {year} {2022})}\BibitemShut {NoStop}%
\bibitem [{\citenamefont {{Google Quantum AI}.}(2023)}]{google2023suppressing}%
  \BibitemOpen
  \bibfield  {author} {\bibinfo {author} {\bibnamefont {{Google Quantum
  AI}.}},\ }\href {https://doi.org/10.1038/s41586-022-05434-1} {\bibfield
  {journal} {\bibinfo  {journal} {Nature}\ }\textbf {\bibinfo {volume} {614}},\
  \bibinfo {pages} {676} (\bibinfo {year} {2023})}\BibitemShut {NoStop}%
\bibitem [{\citenamefont {{Google Quantum AI and
  Collaborators}.}(2024)}]{google2024quantum}%
  \BibitemOpen
  \bibfield  {author} {\bibinfo {author} {\bibnamefont {{Google Quantum AI and
  Collaborators}.}},\ }\href {https://doi.org/10.1038/s41586-024-08449-y}
  {\bibfield  {journal} {\bibinfo  {journal} {Nature}\ }\textbf {\bibinfo
  {volume} {638}},\ \bibinfo {pages} {920} (\bibinfo {year}
  {2024})}\BibitemShut {NoStop}%
\bibitem [{\citenamefont {He}\ \emph {et~al.}(2025)\citenamefont {He},
  \citenamefont {Lin}, \citenamefont {Wang}, \citenamefont {Li}, \citenamefont
  {Bei}, \citenamefont {Cai}, \citenamefont {Cao}, \citenamefont {Chen},
  \citenamefont {Chen}, \citenamefont {Chen}, \citenamefont {Chen},
  \citenamefont {Chen}, \citenamefont {Chen}, \citenamefont {Chu},
  \citenamefont {Deng}, \citenamefont {Ding}, \citenamefont {Ding},
  \citenamefont {Fan}, \citenamefont {Fan}, \citenamefont {Fu}, \citenamefont
  {Gao}, \citenamefont {Gong}, \citenamefont {Gui}, \citenamefont {Guo},
  \citenamefont {Guo}, \citenamefont {Han}, \citenamefont {Hong}, \citenamefont
  {Hu}, \citenamefont {Huang}, \citenamefont {Huo}, \citenamefont {Jiang},
  \citenamefont {Jiang}, \citenamefont {Jiang}, \citenamefont {Jiang},
  \citenamefont {Jin}, \citenamefont {Li}, \citenamefont {Li}, \citenamefont
  {Li}, \citenamefont {Li}, \citenamefont {Li}, \citenamefont {Li},
  \citenamefont {Li}, \citenamefont {Li}, \citenamefont {Li}, \citenamefont
  {Liang}, \citenamefont {Liao}, \citenamefont {Lin}, \citenamefont {Liu},
  \citenamefont {Liu}, \citenamefont {Liu}, \citenamefont {Lou}, \citenamefont
  {Ma}, \citenamefont {Nan}, \citenamefont {Nie}, \citenamefont {Niu},
  \citenamefont {Peng}, \citenamefont {Qian}, \citenamefont {Rong},
  \citenamefont {Rong}, \citenamefont {Shen}, \citenamefont {Shen},
  \citenamefont {Su}, \citenamefont {Su}, \citenamefont {Sun}, \citenamefont
  {Sun}, \citenamefont {Sun}, \citenamefont {Sun}, \citenamefont {Tan},
  \citenamefont {Tan}, \citenamefont {Tu}, \citenamefont {Wang}, \citenamefont
  {Wang}, \citenamefont {Wang}, \citenamefont {Wang}, \citenamefont {Wang},
  \citenamefont {Wang}, \citenamefont {Wang}, \citenamefont {Wang},
  \citenamefont {Wei}, \citenamefont {Wu}, \citenamefont {Wu}, \citenamefont
  {Wu}, \citenamefont {Xu}, \citenamefont {Xue}, \citenamefont {Yan},
  \citenamefont {Yan}, \citenamefont {Yang}, \citenamefont {Yang},
  \citenamefont {Yang}, \citenamefont {Ye}, \citenamefont {Ye}, \citenamefont
  {Yi}, \citenamefont {Ying}, \citenamefont {Yu}, \citenamefont {Yu},
  \citenamefont {Zeng}, \citenamefont {Zha}, \citenamefont {Zhan},
  \citenamefont {Zhang}, \citenamefont {Zhang}, \citenamefont {Zhang},
  \citenamefont {Zhang}, \citenamefont {Zhang}, \citenamefont {Zhang},
  \citenamefont {Zhang}, \citenamefont {Zhao}, \citenamefont {Zhao},
  \citenamefont {Zhao}, \citenamefont {Zhao}, \citenamefont {Zheng},
  \citenamefont {Zhou}, \citenamefont {Zhou}, \citenamefont {Zhou},
  \citenamefont {Zhou}, \citenamefont {Zhu}, \citenamefont {Zhu}, \citenamefont
  {Zou}, \citenamefont {Zou}, \citenamefont {Zhang}, \citenamefont {Lu},
  \citenamefont {Peng}, \citenamefont {Chen}, \citenamefont {Zhu},\ and\
  \citenamefont {Pan}}]{he2025experimental}%
  \BibitemOpen
  \bibfield  {author} {\bibinfo {author} {\bibfnamefont {T.}~\bibnamefont
  {He}}, \bibinfo {author} {\bibfnamefont {W.}~\bibnamefont {Lin}}, \bibinfo
  {author} {\bibfnamefont {R.}~\bibnamefont {Wang}}, \bibinfo {author}
  {\bibfnamefont {Y.}~\bibnamefont {Li}}, \bibinfo {author} {\bibfnamefont
  {J.}~\bibnamefont {Bei}}, \bibinfo {author} {\bibfnamefont {J.}~\bibnamefont
  {Cai}}, \bibinfo {author} {\bibfnamefont {S.}~\bibnamefont {Cao}}, \bibinfo
  {author} {\bibfnamefont {D.}~\bibnamefont {Chen}}, \bibinfo {author}
  {\bibfnamefont {K.}~\bibnamefont {Chen}}, \bibinfo {author} {\bibfnamefont
  {X.}~\bibnamefont {Chen}}, \bibinfo {author} {\bibfnamefont {Z.}~\bibnamefont
  {Chen}}, \bibinfo {author} {\bibfnamefont {Z.}~\bibnamefont {Chen}}, \bibinfo
  {author} {\bibfnamefont {Z.}~\bibnamefont {Chen}}, \bibinfo {author}
  {\bibfnamefont {W.}~\bibnamefont {Chu}}, \bibinfo {author} {\bibfnamefont
  {H.}~\bibnamefont {Deng}}, \bibinfo {author} {\bibfnamefont {X.}~\bibnamefont
  {Ding}}, \bibinfo {author} {\bibfnamefont {Z.}~\bibnamefont {Ding}}, \bibinfo
  {author} {\bibfnamefont {B.}~\bibnamefont {Fan}}, \bibinfo {author}
  {\bibfnamefont {D.}~\bibnamefont {Fan}}, \bibinfo {author} {\bibfnamefont
  {Y.}~\bibnamefont {Fu}}, \bibinfo {author} {\bibfnamefont {D.}~\bibnamefont
  {Gao}}, \bibinfo {author} {\bibfnamefont {M.}~\bibnamefont {Gong}}, \bibinfo
  {author} {\bibfnamefont {J.}~\bibnamefont {Gui}}, \bibinfo {author}
  {\bibfnamefont {C.}~\bibnamefont {Guo}}, \bibinfo {author} {\bibfnamefont
  {S.}~\bibnamefont {Guo}}, \bibinfo {author} {\bibfnamefont {L.}~\bibnamefont
  {Han}}, \bibinfo {author} {\bibfnamefont {L.}~\bibnamefont {Hong}}, \bibinfo
  {author} {\bibfnamefont {Y.}~\bibnamefont {Hu}}, \bibinfo {author}
  {\bibfnamefont {H.-L.}\ \bibnamefont {Huang}}, \bibinfo {author}
  {\bibfnamefont {Y.-H.}\ \bibnamefont {Huo}}, \bibinfo {author} {\bibfnamefont
  {C.}~\bibnamefont {Jiang}}, \bibinfo {author} {\bibfnamefont
  {L.}~\bibnamefont {Jiang}}, \bibinfo {author} {\bibfnamefont
  {T.}~\bibnamefont {Jiang}}, \bibinfo {author} {\bibfnamefont
  {Z.}~\bibnamefont {Jiang}}, \bibinfo {author} {\bibfnamefont
  {H.}~\bibnamefont {Jin}}, \bibinfo {author} {\bibfnamefont {D.}~\bibnamefont
  {Li}}, \bibinfo {author} {\bibfnamefont {D.}~\bibnamefont {Li}}, \bibinfo
  {author} {\bibfnamefont {J.}~\bibnamefont {Li}}, \bibinfo {author}
  {\bibfnamefont {J.}~\bibnamefont {Li}}, \bibinfo {author} {\bibfnamefont
  {J.}~\bibnamefont {Li}}, \bibinfo {author} {\bibfnamefont {J.}~\bibnamefont
  {Li}}, \bibinfo {author} {\bibfnamefont {N.}~\bibnamefont {Li}}, \bibinfo
  {author} {\bibfnamefont {S.}~\bibnamefont {Li}}, \bibinfo {author}
  {\bibfnamefont {Y.}~\bibnamefont {Li}}, \bibinfo {author} {\bibfnamefont
  {F.}~\bibnamefont {Liang}}, \bibinfo {author} {\bibfnamefont
  {N.}~\bibnamefont {Liao}}, \bibinfo {author} {\bibfnamefont {J.}~\bibnamefont
  {Lin}}, \bibinfo {author} {\bibfnamefont {K.}~\bibnamefont {Liu}}, \bibinfo
  {author} {\bibfnamefont {M.}~\bibnamefont {Liu}}, \bibinfo {author}
  {\bibfnamefont {Y.}~\bibnamefont {Liu}}, \bibinfo {author} {\bibfnamefont
  {H.}~\bibnamefont {Lou}}, \bibinfo {author} {\bibfnamefont {Y.}~\bibnamefont
  {Ma}}, \bibinfo {author} {\bibfnamefont {K.}~\bibnamefont {Nan}}, \bibinfo
  {author} {\bibfnamefont {M.}~\bibnamefont {Nie}}, \bibinfo {author}
  {\bibfnamefont {L.}~\bibnamefont {Niu}}, \bibinfo {author} {\bibfnamefont
  {W.}~\bibnamefont {Peng}}, \bibinfo {author} {\bibfnamefont {H.}~\bibnamefont
  {Qian}}, \bibinfo {author} {\bibfnamefont {H.}~\bibnamefont {Rong}}, \bibinfo
  {author} {\bibfnamefont {T.}~\bibnamefont {Rong}}, \bibinfo {author}
  {\bibfnamefont {H.}~\bibnamefont {Shen}}, \bibinfo {author} {\bibfnamefont
  {Q.}~\bibnamefont {Shen}}, \bibinfo {author} {\bibfnamefont {H.}~\bibnamefont
  {Su}}, \bibinfo {author} {\bibfnamefont {F.}~\bibnamefont {Su}}, \bibinfo
  {author} {\bibfnamefont {C.}~\bibnamefont {Sun}}, \bibinfo {author}
  {\bibfnamefont {L.}~\bibnamefont {Sun}}, \bibinfo {author} {\bibfnamefont
  {T.}~\bibnamefont {Sun}}, \bibinfo {author} {\bibfnamefont {Y.}~\bibnamefont
  {Sun}}, \bibinfo {author} {\bibfnamefont {Y.}~\bibnamefont {Tan}}, \bibinfo
  {author} {\bibfnamefont {J.}~\bibnamefont {Tan}}, \bibinfo {author}
  {\bibfnamefont {W.}~\bibnamefont {Tu}}, \bibinfo {author} {\bibfnamefont
  {J.}~\bibnamefont {Wang}}, \bibinfo {author} {\bibfnamefont {B.}~\bibnamefont
  {Wang}}, \bibinfo {author} {\bibfnamefont {C.}~\bibnamefont {Wang}}, \bibinfo
  {author} {\bibfnamefont {C.}~\bibnamefont {Wang}}, \bibinfo {author}
  {\bibfnamefont {C.}~\bibnamefont {Wang}}, \bibinfo {author} {\bibfnamefont
  {J.}~\bibnamefont {Wang}}, \bibinfo {author} {\bibfnamefont {S.}~\bibnamefont
  {Wang}}, \bibinfo {author} {\bibfnamefont {X.}~\bibnamefont {Wang}}, \bibinfo
  {author} {\bibfnamefont {Z.}~\bibnamefont {Wei}}, \bibinfo {author}
  {\bibfnamefont {D.}~\bibnamefont {Wu}}, \bibinfo {author} {\bibfnamefont
  {G.}~\bibnamefont {Wu}}, \bibinfo {author} {\bibfnamefont {Y.}~\bibnamefont
  {Wu}}, \bibinfo {author} {\bibfnamefont {Y.}~\bibnamefont {Xu}}, \bibinfo
  {author} {\bibfnamefont {C.}~\bibnamefont {Xue}}, \bibinfo {author}
  {\bibfnamefont {K.}~\bibnamefont {Yan}}, \bibinfo {author} {\bibfnamefont
  {X.}~\bibnamefont {Yan}}, \bibinfo {author} {\bibfnamefont {W.}~\bibnamefont
  {Yang}}, \bibinfo {author} {\bibfnamefont {X.}~\bibnamefont {Yang}}, \bibinfo
  {author} {\bibfnamefont {Y.}~\bibnamefont {Yang}}, \bibinfo {author}
  {\bibfnamefont {Y.}~\bibnamefont {Ye}}, \bibinfo {author} {\bibfnamefont
  {Z.}~\bibnamefont {Ye}}, \bibinfo {author} {\bibfnamefont {Z.}~\bibnamefont
  {Yi}}, \bibinfo {author} {\bibfnamefont {C.}~\bibnamefont {Ying}}, \bibinfo
  {author} {\bibfnamefont {J.}~\bibnamefont {Yu}}, \bibinfo {author}
  {\bibfnamefont {Q.}~\bibnamefont {Yu}}, \bibinfo {author} {\bibfnamefont
  {X.}~\bibnamefont {Zeng}}, \bibinfo {author} {\bibfnamefont {C.}~\bibnamefont
  {Zha}}, \bibinfo {author} {\bibfnamefont {S.}~\bibnamefont {Zhan}}, \bibinfo
  {author} {\bibfnamefont {H.}~\bibnamefont {Zhang}}, \bibinfo {author}
  {\bibfnamefont {H.}~\bibnamefont {Zhang}}, \bibinfo {author} {\bibfnamefont
  {K.}~\bibnamefont {Zhang}}, \bibinfo {author} {\bibfnamefont
  {W.}~\bibnamefont {Zhang}}, \bibinfo {author} {\bibfnamefont
  {Y.}~\bibnamefont {Zhang}}, \bibinfo {author} {\bibfnamefont
  {Y.}~\bibnamefont {Zhang}}, \bibinfo {author} {\bibfnamefont
  {Z.}~\bibnamefont {Zhang}}, \bibinfo {author} {\bibfnamefont
  {G.}~\bibnamefont {Zhao}}, \bibinfo {author} {\bibfnamefont {X.}~\bibnamefont
  {Zhao}}, \bibinfo {author} {\bibfnamefont {Y.}~\bibnamefont {Zhao}}, \bibinfo
  {author} {\bibfnamefont {Z.}~\bibnamefont {Zhao}}, \bibinfo {author}
  {\bibfnamefont {L.}~\bibnamefont {Zheng}}, \bibinfo {author} {\bibfnamefont
  {F.}~\bibnamefont {Zhou}}, \bibinfo {author} {\bibfnamefont {L.}~\bibnamefont
  {Zhou}}, \bibinfo {author} {\bibfnamefont {N.}~\bibnamefont {Zhou}}, \bibinfo
  {author} {\bibfnamefont {N.}~\bibnamefont {Zhou}}, \bibinfo {author}
  {\bibfnamefont {C.}~\bibnamefont {Zhu}}, \bibinfo {author} {\bibfnamefont
  {Q.}~\bibnamefont {Zhu}}, \bibinfo {author} {\bibfnamefont {G.}~\bibnamefont
  {Zou}}, \bibinfo {author} {\bibfnamefont {H.}~\bibnamefont {Zou}}, \bibinfo
  {author} {\bibfnamefont {Q.}~\bibnamefont {Zhang}}, \bibinfo {author}
  {\bibfnamefont {C.-Y.}\ \bibnamefont {Lu}}, \bibinfo {author} {\bibfnamefont
  {C.-Z.}\ \bibnamefont {Peng}}, \bibinfo {author} {\bibfnamefont
  {F.}~\bibnamefont {Chen}}, \bibinfo {author} {\bibfnamefont {X.}~\bibnamefont
  {Zhu}},\ and\ \bibinfo {author} {\bibfnamefont {J.-W.}\ \bibnamefont {Pan}},\
  }\href {https://doi.org/10.1103/rqkg-dw31} {\bibfield  {journal} {\bibinfo
  {journal} {Phys. Rev. Lett.}\ }\textbf {\bibinfo {volume} {135}},\ \bibinfo
  {pages} {260601} (\bibinfo {year} {2025})}\BibitemShut {NoStop}%
\bibitem [{\citenamefont {{Google Quantum AI and
  Collaborators}.}(2025)}]{google2025demonstrating}%
  \BibitemOpen
  \bibfield  {author} {\bibinfo {author} {\bibnamefont {{Google Quantum AI and
  Collaborators}.}},\ }\href {https://doi.org/10.1038/s41567-025-03070-w}
  {\bibfield  {journal} {\bibinfo  {journal} {Nature Physics}\ ,\ \bibinfo
  {pages} {1}} (\bibinfo {year} {2025})}\BibitemShut {NoStop}%
\bibitem [{\citenamefont {Lacroix}\ \emph {et~al.}(2020)\citenamefont
  {Lacroix}, \citenamefont {Bourassa}, \citenamefont {Heras}, \citenamefont
  {Zhang}, \citenamefont {Bausch}, \citenamefont {Senior}, \citenamefont
  {Edlich}, \citenamefont {Shutty}, \citenamefont {Sivak}, \citenamefont
  {Bengtsson}, \citenamefont {{McEwen}}, \citenamefont {Higgott}, \citenamefont
  {Kafri}, \citenamefont {Claes}, \citenamefont {Morvan}, \citenamefont {Chen},
  \citenamefont {Zalcman}, \citenamefont {Madhuk}, \citenamefont {Acharya},
  \citenamefont {Aghababaie~Beni}, \citenamefont {Aigeldinger}, \citenamefont
  {Alcaraz}, \citenamefont {Andersen}, \citenamefont {Ansmann}, \citenamefont
  {Arute}, \citenamefont {Arya}, \citenamefont {Asfaw}, \citenamefont
  {Atalaya}, \citenamefont {Babbush}, \citenamefont {Ballard}, \citenamefont
  {Bardin}, \citenamefont {Bilmes}, \citenamefont {Blackwell}, \citenamefont
  {Bovaird}, \citenamefont {Bowers}, \citenamefont {Brill}, \citenamefont
  {Broughton}, \citenamefont {Browne}, \citenamefont {Buchea}, \citenamefont
  {Buckley}, \citenamefont {Burger}, \citenamefont {Burkett}, \citenamefont
  {Bushnell}, \citenamefont {Cabrera}, \citenamefont {Campero}, \citenamefont
  {Chang}, \citenamefont {Chiaro}, \citenamefont {Chih}, \citenamefont
  {Cleland}, \citenamefont {Cogan}, \citenamefont {Collins}, \citenamefont
  {Conner}, \citenamefont {Courtney}, \citenamefont {Crook}, \citenamefont
  {Curtin}, \citenamefont {Das}, \citenamefont {Demura}, \citenamefont
  {De~Lorenzo}, \citenamefont {Di~Paolo}, \citenamefont {Donohoe},
  \citenamefont {Drozdov}, \citenamefont {Dunsworth}, \citenamefont
  {Eickbusch}, \citenamefont {Elbag}, \citenamefont {Elzouka}, \citenamefont
  {Erickson}, \citenamefont {Ferreira}, \citenamefont {Flores~Burgos},
  \citenamefont {Forati}, \citenamefont {Fowler}, \citenamefont {Foxen},
  \citenamefont {Ganjam}, \citenamefont {Garcia}, \citenamefont {Gasca},
  \citenamefont {Genois}, \citenamefont {Giang}, \citenamefont {Gilboa},
  \citenamefont {Gosula}, \citenamefont {Grajales~Dau}, \citenamefont
  {Graumann}, \citenamefont {Greene}, \citenamefont {Gross}, \citenamefont
  {Ha}, \citenamefont {Habegger}, \citenamefont {Hansen}, \citenamefont
  {Harrigan}, \citenamefont {Harrington}, \citenamefont {Heslin}, \citenamefont
  {Heu}, \citenamefont {Hiltermann}, \citenamefont {Hilton}, \citenamefont
  {Hong}, \citenamefont {Huang}, \citenamefont {Huff}, \citenamefont {Huggins},
  \citenamefont {Jeffrey}, \citenamefont {Jiang}, \citenamefont {Jin},
  \citenamefont {Joshi}, \citenamefont {Juhas}, \citenamefont {Kabel},
  \citenamefont {Kang}, \citenamefont {Karamlou}, \citenamefont {Kechedzhi},
  \citenamefont {Khaire}, \citenamefont {Khattar}, \citenamefont {Khezri},
  \citenamefont {Kim}, \citenamefont {Klimov}, \citenamefont {Kobrin},
  \citenamefont {Korotkov}, \citenamefont {Kostritsa}, \citenamefont
  {Kreikebaum}, \citenamefont {Kurilovich}, \citenamefont {Landhuis},
  \citenamefont {Lange-Dei}, \citenamefont {Langley}, \citenamefont {Laptev},
  \citenamefont {Lau}, \citenamefont {Ledford}, \citenamefont {Lee},
  \citenamefont {Lester}, \citenamefont {Le~Guevel}, \citenamefont {Li},
  \citenamefont {Li}, \citenamefont {Lill}, \citenamefont {Livingston},
  \citenamefont {Locharla}, \citenamefont {Lucero}, \citenamefont {Lundahl},
  \citenamefont {Lunt}, \citenamefont {Maloney}, \citenamefont {Mandr{\`a}},
  \citenamefont {Martin}, \citenamefont {Martin}, \citenamefont {Maxfield},
  \citenamefont {{McClean}}, \citenamefont {Meeks}, \citenamefont {Megrant},
  \citenamefont {Miao}, \citenamefont {Molavi}, \citenamefont {Molina},
  \citenamefont {Montazeri}, \citenamefont {Movassagh}, \citenamefont {Neill},
  \citenamefont {Newman}, \citenamefont {Nguyen}, \citenamefont {Nguyen},
  \citenamefont {Ni}, \citenamefont {Niu}, \citenamefont {Oas}, \citenamefont
  {Oliver}, \citenamefont {Orosco}, \citenamefont {Ottosson}, \citenamefont
  {Pizzuto}, \citenamefont {Potter}, \citenamefont {Pritchard}, \citenamefont
  {Quintana}, \citenamefont {Ramachandran}, \citenamefont {Reagor},
  \citenamefont {Resnick}, \citenamefont {Rhodes}, \citenamefont {Roberts},
  \citenamefont {Rosenberg}, \citenamefont {Rosenfeld}, \citenamefont {Rossi},
  \citenamefont {Roushan}, \citenamefont {Sankaragomathi}, \citenamefont
  {Schurkus}, \citenamefont {Shearn}, \citenamefont {Shorter}, \citenamefont
  {Shvarts}, \citenamefont {Small}, \citenamefont {Smith}, \citenamefont
  {Springer}, \citenamefont {Sterling}, \citenamefont {Suchard}, \citenamefont
  {Szasz}, \citenamefont {Sztein}, \citenamefont {Thor}, \citenamefont
  {Tomita}, \citenamefont {Torres}, \citenamefont {Torunbalci}, \citenamefont
  {Vaishnav}, \citenamefont {Vargas}, \citenamefont {Vdovichev}, \citenamefont
  {Vidal}, \citenamefont {Vollgraff~Heidweiller}, \citenamefont {Waltman},
  \citenamefont {Waltz}, \citenamefont {Wang}, \citenamefont {Ware},
  \citenamefont {Weidel}, \citenamefont {White}, \citenamefont {Wong},
  \citenamefont {Woo}, \citenamefont {Woodson}, \citenamefont {Xing},
  \citenamefont {Yao}, \citenamefont {Yeh}, \citenamefont {Ying}, \citenamefont
  {Yoo}, \citenamefont {Yosri}, \citenamefont {Young}, \citenamefont {Zhang},
  \citenamefont {Zhu}, \citenamefont {Zobrist}, \citenamefont {Neven},
  \citenamefont {Kohli}, \citenamefont {Davies}, \citenamefont {Boixo},
  \citenamefont {Kelly}, \citenamefont {Jones}, \citenamefont {Gidney},\ and\
  \citenamefont {Satzinger}}]{lacroix2025scaling}%
  \BibitemOpen
  \bibfield  {author} {\bibinfo {author} {\bibfnamefont {N.}~\bibnamefont
  {Lacroix}}, \bibinfo {author} {\bibfnamefont {A.}~\bibnamefont {Bourassa}},
  \bibinfo {author} {\bibfnamefont {F.~J.~H.}\ \bibnamefont {Heras}}, \bibinfo
  {author} {\bibfnamefont {L.~M.}\ \bibnamefont {Zhang}}, \bibinfo {author}
  {\bibfnamefont {J.}~\bibnamefont {Bausch}}, \bibinfo {author} {\bibfnamefont
  {A.~W.}\ \bibnamefont {Senior}}, \bibinfo {author} {\bibfnamefont
  {T.}~\bibnamefont {Edlich}}, \bibinfo {author} {\bibfnamefont
  {N.}~\bibnamefont {Shutty}}, \bibinfo {author} {\bibfnamefont
  {V.}~\bibnamefont {Sivak}}, \bibinfo {author} {\bibfnamefont
  {A.}~\bibnamefont {Bengtsson}}, \bibinfo {author} {\bibfnamefont
  {M.}~\bibnamefont {{McEwen}}}, \bibinfo {author} {\bibfnamefont
  {O.}~\bibnamefont {Higgott}}, \bibinfo {author} {\bibfnamefont
  {D.}~\bibnamefont {Kafri}}, \bibinfo {author} {\bibfnamefont
  {J.}~\bibnamefont {Claes}}, \bibinfo {author} {\bibfnamefont
  {A.}~\bibnamefont {Morvan}}, \bibinfo {author} {\bibfnamefont
  {Z.}~\bibnamefont {Chen}}, \bibinfo {author} {\bibfnamefont {A.}~\bibnamefont
  {Zalcman}}, \bibinfo {author} {\bibfnamefont {S.}~\bibnamefont {Madhuk}},
  \bibinfo {author} {\bibfnamefont {R.}~\bibnamefont {Acharya}}, \bibinfo
  {author} {\bibfnamefont {L.}~\bibnamefont {Aghababaie~Beni}}, \bibinfo
  {author} {\bibfnamefont {G.}~\bibnamefont {Aigeldinger}}, \bibinfo {author}
  {\bibfnamefont {R.}~\bibnamefont {Alcaraz}}, \bibinfo {author} {\bibfnamefont
  {T.~I.}\ \bibnamefont {Andersen}}, \bibinfo {author} {\bibfnamefont
  {M.}~\bibnamefont {Ansmann}}, \bibinfo {author} {\bibfnamefont
  {F.}~\bibnamefont {Arute}}, \bibinfo {author} {\bibfnamefont
  {K.}~\bibnamefont {Arya}}, \bibinfo {author} {\bibfnamefont {A.}~\bibnamefont
  {Asfaw}}, \bibinfo {author} {\bibfnamefont {J.}~\bibnamefont {Atalaya}},
  \bibinfo {author} {\bibfnamefont {R.}~\bibnamefont {Babbush}}, \bibinfo
  {author} {\bibfnamefont {B.}~\bibnamefont {Ballard}}, \bibinfo {author}
  {\bibfnamefont {J.~C.}\ \bibnamefont {Bardin}}, \bibinfo {author}
  {\bibfnamefont {A.}~\bibnamefont {Bilmes}}, \bibinfo {author} {\bibfnamefont
  {S.}~\bibnamefont {Blackwell}}, \bibinfo {author} {\bibfnamefont
  {J.}~\bibnamefont {Bovaird}}, \bibinfo {author} {\bibfnamefont
  {D.}~\bibnamefont {Bowers}}, \bibinfo {author} {\bibfnamefont
  {L.}~\bibnamefont {Brill}}, \bibinfo {author} {\bibfnamefont
  {M.}~\bibnamefont {Broughton}}, \bibinfo {author} {\bibfnamefont {D.~A.}\
  \bibnamefont {Browne}}, \bibinfo {author} {\bibfnamefont {B.}~\bibnamefont
  {Buchea}}, \bibinfo {author} {\bibfnamefont {B.~B.}\ \bibnamefont {Buckley}},
  \bibinfo {author} {\bibfnamefont {T.}~\bibnamefont {Burger}}, \bibinfo
  {author} {\bibfnamefont {B.}~\bibnamefont {Burkett}}, \bibinfo {author}
  {\bibfnamefont {N.}~\bibnamefont {Bushnell}}, \bibinfo {author}
  {\bibfnamefont {A.}~\bibnamefont {Cabrera}}, \bibinfo {author} {\bibfnamefont
  {J.}~\bibnamefont {Campero}}, \bibinfo {author} {\bibfnamefont {H.-S.}\
  \bibnamefont {Chang}}, \bibinfo {author} {\bibfnamefont {B.}~\bibnamefont
  {Chiaro}}, \bibinfo {author} {\bibfnamefont {L.-Y.}\ \bibnamefont {Chih}},
  \bibinfo {author} {\bibfnamefont {A.~Y.}\ \bibnamefont {Cleland}}, \bibinfo
  {author} {\bibfnamefont {J.}~\bibnamefont {Cogan}}, \bibinfo {author}
  {\bibfnamefont {R.}~\bibnamefont {Collins}}, \bibinfo {author} {\bibfnamefont
  {P.}~\bibnamefont {Conner}}, \bibinfo {author} {\bibfnamefont
  {W.}~\bibnamefont {Courtney}}, \bibinfo {author} {\bibfnamefont {A.~L.}\
  \bibnamefont {Crook}}, \bibinfo {author} {\bibfnamefont {B.}~\bibnamefont
  {Curtin}}, \bibinfo {author} {\bibfnamefont {S.}~\bibnamefont {Das}},
  \bibinfo {author} {\bibfnamefont {S.}~\bibnamefont {Demura}}, \bibinfo
  {author} {\bibfnamefont {L.}~\bibnamefont {De~Lorenzo}}, \bibinfo {author}
  {\bibfnamefont {A.}~\bibnamefont {Di~Paolo}}, \bibinfo {author}
  {\bibfnamefont {P.}~\bibnamefont {Donohoe}}, \bibinfo {author} {\bibfnamefont
  {I.}~\bibnamefont {Drozdov}}, \bibinfo {author} {\bibfnamefont
  {A.}~\bibnamefont {Dunsworth}}, \bibinfo {author} {\bibfnamefont
  {A.}~\bibnamefont {Eickbusch}}, \bibinfo {author} {\bibfnamefont {A.~M.}\
  \bibnamefont {Elbag}}, \bibinfo {author} {\bibfnamefont {M.}~\bibnamefont
  {Elzouka}}, \bibinfo {author} {\bibfnamefont {C.}~\bibnamefont {Erickson}},
  \bibinfo {author} {\bibfnamefont {V.~S.}\ \bibnamefont {Ferreira}}, \bibinfo
  {author} {\bibfnamefont {L.}~\bibnamefont {Flores~Burgos}}, \bibinfo {author}
  {\bibfnamefont {E.}~\bibnamefont {Forati}}, \bibinfo {author} {\bibfnamefont
  {A.~G.}\ \bibnamefont {Fowler}}, \bibinfo {author} {\bibfnamefont
  {B.}~\bibnamefont {Foxen}}, \bibinfo {author} {\bibfnamefont
  {S.}~\bibnamefont {Ganjam}}, \bibinfo {author} {\bibfnamefont
  {G.}~\bibnamefont {Garcia}}, \bibinfo {author} {\bibfnamefont
  {R.}~\bibnamefont {Gasca}}, \bibinfo {author} {\bibfnamefont
  {{\'E}.}~\bibnamefont {Genois}}, \bibinfo {author} {\bibfnamefont
  {W.}~\bibnamefont {Giang}}, \bibinfo {author} {\bibfnamefont
  {D.}~\bibnamefont {Gilboa}}, \bibinfo {author} {\bibfnamefont
  {R.}~\bibnamefont {Gosula}}, \bibinfo {author} {\bibfnamefont
  {A.}~\bibnamefont {Grajales~Dau}}, \bibinfo {author} {\bibfnamefont
  {D.}~\bibnamefont {Graumann}}, \bibinfo {author} {\bibfnamefont
  {A.}~\bibnamefont {Greene}}, \bibinfo {author} {\bibfnamefont {J.~A.}\
  \bibnamefont {Gross}}, \bibinfo {author} {\bibfnamefont {T.}~\bibnamefont
  {Ha}}, \bibinfo {author} {\bibfnamefont {S.}~\bibnamefont {Habegger}},
  \bibinfo {author} {\bibfnamefont {M.}~\bibnamefont {Hansen}}, \bibinfo
  {author} {\bibfnamefont {M.~P.}\ \bibnamefont {Harrigan}}, \bibinfo {author}
  {\bibfnamefont {S.~D.}\ \bibnamefont {Harrington}}, \bibinfo {author}
  {\bibfnamefont {S.}~\bibnamefont {Heslin}}, \bibinfo {author} {\bibfnamefont
  {P.}~\bibnamefont {Heu}}, \bibinfo {author} {\bibfnamefont {R.}~\bibnamefont
  {Hiltermann}}, \bibinfo {author} {\bibfnamefont {J.}~\bibnamefont {Hilton}},
  \bibinfo {author} {\bibfnamefont {S.}~\bibnamefont {Hong}}, \bibinfo {author}
  {\bibfnamefont {H.-Y.}\ \bibnamefont {Huang}}, \bibinfo {author}
  {\bibfnamefont {A.}~\bibnamefont {Huff}}, \bibinfo {author} {\bibfnamefont
  {W.~J.}\ \bibnamefont {Huggins}}, \bibinfo {author} {\bibfnamefont
  {E.}~\bibnamefont {Jeffrey}}, \bibinfo {author} {\bibfnamefont
  {Z.}~\bibnamefont {Jiang}}, \bibinfo {author} {\bibfnamefont
  {X.}~\bibnamefont {Jin}}, \bibinfo {author} {\bibfnamefont {C.}~\bibnamefont
  {Joshi}}, \bibinfo {author} {\bibfnamefont {P.}~\bibnamefont {Juhas}},
  \bibinfo {author} {\bibfnamefont {A.}~\bibnamefont {Kabel}}, \bibinfo
  {author} {\bibfnamefont {H.}~\bibnamefont {Kang}}, \bibinfo {author}
  {\bibfnamefont {A.~H.}\ \bibnamefont {Karamlou}}, \bibinfo {author}
  {\bibfnamefont {K.}~\bibnamefont {Kechedzhi}}, \bibinfo {author}
  {\bibfnamefont {T.}~\bibnamefont {Khaire}}, \bibinfo {author} {\bibfnamefont
  {T.}~\bibnamefont {Khattar}}, \bibinfo {author} {\bibfnamefont
  {M.}~\bibnamefont {Khezri}}, \bibinfo {author} {\bibfnamefont
  {S.}~\bibnamefont {Kim}}, \bibinfo {author} {\bibfnamefont {P.~V.}\
  \bibnamefont {Klimov}}, \bibinfo {author} {\bibfnamefont {B.}~\bibnamefont
  {Kobrin}}, \bibinfo {author} {\bibfnamefont {A.~N.}\ \bibnamefont
  {Korotkov}}, \bibinfo {author} {\bibfnamefont {F.}~\bibnamefont {Kostritsa}},
  \bibinfo {author} {\bibfnamefont {J.~M.}\ \bibnamefont {Kreikebaum}},
  \bibinfo {author} {\bibfnamefont {V.~D.}\ \bibnamefont {Kurilovich}},
  \bibinfo {author} {\bibfnamefont {D.}~\bibnamefont {Landhuis}}, \bibinfo
  {author} {\bibfnamefont {T.}~\bibnamefont {Lange-Dei}}, \bibinfo {author}
  {\bibfnamefont {B.~W.}\ \bibnamefont {Langley}}, \bibinfo {author}
  {\bibfnamefont {P.}~\bibnamefont {Laptev}}, \bibinfo {author} {\bibfnamefont
  {K.-M.}\ \bibnamefont {Lau}}, \bibinfo {author} {\bibfnamefont
  {J.}~\bibnamefont {Ledford}}, \bibinfo {author} {\bibfnamefont
  {K.}~\bibnamefont {Lee}}, \bibinfo {author} {\bibfnamefont {B.~J.}\
  \bibnamefont {Lester}}, \bibinfo {author} {\bibfnamefont {L.}~\bibnamefont
  {Le~Guevel}}, \bibinfo {author} {\bibfnamefont {W.~Y.}\ \bibnamefont {Li}},
  \bibinfo {author} {\bibfnamefont {Y.}~\bibnamefont {Li}}, \bibinfo {author}
  {\bibfnamefont {A.~T.}\ \bibnamefont {Lill}}, \bibinfo {author}
  {\bibfnamefont {W.~P.}\ \bibnamefont {Livingston}}, \bibinfo {author}
  {\bibfnamefont {A.}~\bibnamefont {Locharla}}, \bibinfo {author}
  {\bibfnamefont {E.}~\bibnamefont {Lucero}}, \bibinfo {author} {\bibfnamefont
  {D.}~\bibnamefont {Lundahl}}, \bibinfo {author} {\bibfnamefont
  {A.}~\bibnamefont {Lunt}}, \bibinfo {author} {\bibfnamefont {A.}~\bibnamefont
  {Maloney}}, \bibinfo {author} {\bibfnamefont {S.}~\bibnamefont {Mandr{\`a}}},
  \bibinfo {author} {\bibfnamefont {L.~S.}\ \bibnamefont {Martin}}, \bibinfo
  {author} {\bibfnamefont {O.}~\bibnamefont {Martin}}, \bibinfo {author}
  {\bibfnamefont {C.}~\bibnamefont {Maxfield}}, \bibinfo {author}
  {\bibfnamefont {J.~R.}\ \bibnamefont {{McClean}}}, \bibinfo {author}
  {\bibfnamefont {S.}~\bibnamefont {Meeks}}, \bibinfo {author} {\bibfnamefont
  {A.}~\bibnamefont {Megrant}}, \bibinfo {author} {\bibfnamefont {K.~C.}\
  \bibnamefont {Miao}}, \bibinfo {author} {\bibfnamefont {R.}~\bibnamefont
  {Molavi}}, \bibinfo {author} {\bibfnamefont {S.}~\bibnamefont {Molina}},
  \bibinfo {author} {\bibfnamefont {S.}~\bibnamefont {Montazeri}}, \bibinfo
  {author} {\bibfnamefont {R.}~\bibnamefont {Movassagh}}, \bibinfo {author}
  {\bibfnamefont {C.}~\bibnamefont {Neill}}, \bibinfo {author} {\bibfnamefont
  {M.}~\bibnamefont {Newman}}, \bibinfo {author} {\bibfnamefont
  {A.}~\bibnamefont {Nguyen}}, \bibinfo {author} {\bibfnamefont
  {M.}~\bibnamefont {Nguyen}}, \bibinfo {author} {\bibfnamefont {C.-H.}\
  \bibnamefont {Ni}}, \bibinfo {author} {\bibfnamefont {M.~Y.}\ \bibnamefont
  {Niu}}, \bibinfo {author} {\bibfnamefont {L.}~\bibnamefont {Oas}}, \bibinfo
  {author} {\bibfnamefont {W.~D.}\ \bibnamefont {Oliver}}, \bibinfo {author}
  {\bibfnamefont {R.}~\bibnamefont {Orosco}}, \bibinfo {author} {\bibfnamefont
  {K.}~\bibnamefont {Ottosson}}, \bibinfo {author} {\bibfnamefont
  {A.}~\bibnamefont {Pizzuto}}, \bibinfo {author} {\bibfnamefont
  {R.}~\bibnamefont {Potter}}, \bibinfo {author} {\bibfnamefont
  {O.}~\bibnamefont {Pritchard}}, \bibinfo {author} {\bibfnamefont
  {C.}~\bibnamefont {Quintana}}, \bibinfo {author} {\bibfnamefont
  {G.}~\bibnamefont {Ramachandran}}, \bibinfo {author} {\bibfnamefont {M.~J.}\
  \bibnamefont {Reagor}}, \bibinfo {author} {\bibfnamefont {R.}~\bibnamefont
  {Resnick}}, \bibinfo {author} {\bibfnamefont {D.~M.}\ \bibnamefont {Rhodes}},
  \bibinfo {author} {\bibfnamefont {G.}~\bibnamefont {Roberts}}, \bibinfo
  {author} {\bibfnamefont {E.}~\bibnamefont {Rosenberg}}, \bibinfo {author}
  {\bibfnamefont {E.}~\bibnamefont {Rosenfeld}}, \bibinfo {author}
  {\bibfnamefont {E.}~\bibnamefont {Rossi}}, \bibinfo {author} {\bibfnamefont
  {P.}~\bibnamefont {Roushan}}, \bibinfo {author} {\bibfnamefont
  {K.}~\bibnamefont {Sankaragomathi}}, \bibinfo {author} {\bibfnamefont
  {H.~F.}\ \bibnamefont {Schurkus}}, \bibinfo {author} {\bibfnamefont {M.~J.}\
  \bibnamefont {Shearn}}, \bibinfo {author} {\bibfnamefont {A.}~\bibnamefont
  {Shorter}}, \bibinfo {author} {\bibfnamefont {V.}~\bibnamefont {Shvarts}},
  \bibinfo {author} {\bibfnamefont {S.}~\bibnamefont {Small}}, \bibinfo
  {author} {\bibfnamefont {W.~C.}\ \bibnamefont {Smith}}, \bibinfo {author}
  {\bibfnamefont {S.}~\bibnamefont {Springer}}, \bibinfo {author}
  {\bibfnamefont {G.}~\bibnamefont {Sterling}}, \bibinfo {author}
  {\bibfnamefont {J.}~\bibnamefont {Suchard}}, \bibinfo {author} {\bibfnamefont
  {A.}~\bibnamefont {Szasz}}, \bibinfo {author} {\bibfnamefont
  {A.}~\bibnamefont {Sztein}}, \bibinfo {author} {\bibfnamefont
  {D.}~\bibnamefont {Thor}}, \bibinfo {author} {\bibfnamefont {E.}~\bibnamefont
  {Tomita}}, \bibinfo {author} {\bibfnamefont {A.}~\bibnamefont {Torres}},
  \bibinfo {author} {\bibfnamefont {M.~M.}\ \bibnamefont {Torunbalci}},
  \bibinfo {author} {\bibfnamefont {A.}~\bibnamefont {Vaishnav}}, \bibinfo
  {author} {\bibfnamefont {J.}~\bibnamefont {Vargas}}, \bibinfo {author}
  {\bibfnamefont {S.}~\bibnamefont {Vdovichev}}, \bibinfo {author}
  {\bibfnamefont {G.}~\bibnamefont {Vidal}}, \bibinfo {author} {\bibfnamefont
  {C.}~\bibnamefont {Vollgraff~Heidweiller}}, \bibinfo {author} {\bibfnamefont
  {S.}~\bibnamefont {Waltman}}, \bibinfo {author} {\bibfnamefont
  {J.}~\bibnamefont {Waltz}}, \bibinfo {author} {\bibfnamefont {S.~X.}\
  \bibnamefont {Wang}}, \bibinfo {author} {\bibfnamefont {B.}~\bibnamefont
  {Ware}}, \bibinfo {author} {\bibfnamefont {T.}~\bibnamefont {Weidel}},
  \bibinfo {author} {\bibfnamefont {T.}~\bibnamefont {White}}, \bibinfo
  {author} {\bibfnamefont {K.}~\bibnamefont {Wong}}, \bibinfo {author}
  {\bibfnamefont {B.~W.~K.}\ \bibnamefont {Woo}}, \bibinfo {author}
  {\bibfnamefont {M.}~\bibnamefont {Woodson}}, \bibinfo {author} {\bibfnamefont
  {C.}~\bibnamefont {Xing}}, \bibinfo {author} {\bibfnamefont {Z.~J.}\
  \bibnamefont {Yao}}, \bibinfo {author} {\bibfnamefont {P.}~\bibnamefont
  {Yeh}}, \bibinfo {author} {\bibfnamefont {B.}~\bibnamefont {Ying}}, \bibinfo
  {author} {\bibfnamefont {J.}~\bibnamefont {Yoo}}, \bibinfo {author}
  {\bibfnamefont {N.}~\bibnamefont {Yosri}}, \bibinfo {author} {\bibfnamefont
  {G.}~\bibnamefont {Young}}, \bibinfo {author} {\bibfnamefont
  {Y.}~\bibnamefont {Zhang}}, \bibinfo {author} {\bibfnamefont
  {N.}~\bibnamefont {Zhu}}, \bibinfo {author} {\bibfnamefont {N.}~\bibnamefont
  {Zobrist}}, \bibinfo {author} {\bibfnamefont {H.}~\bibnamefont {Neven}},
  \bibinfo {author} {\bibfnamefont {P.}~\bibnamefont {Kohli}}, \bibinfo
  {author} {\bibfnamefont {A.}~\bibnamefont {Davies}}, \bibinfo {author}
  {\bibfnamefont {S.}~\bibnamefont {Boixo}}, \bibinfo {author} {\bibfnamefont
  {J.}~\bibnamefont {Kelly}}, \bibinfo {author} {\bibfnamefont
  {C.}~\bibnamefont {Jones}}, \bibinfo {author} {\bibfnamefont
  {C.}~\bibnamefont {Gidney}},\ and\ \bibinfo {author} {\bibfnamefont {K.~J.}\
  \bibnamefont {Satzinger}},\ }\href
  {https://www.nature.com/articles/s41586-025-09061-4} {\bibfield  {journal}
  {\bibinfo  {journal} {Nature Physics}\ }\textbf {\bibinfo {volume} {16}},\
  \bibinfo {pages} {875} (\bibinfo {year} {2020})}\BibitemShut {NoStop}%
\bibitem [{\citenamefont {Sheldon}\ \emph {et~al.}(2016)\citenamefont
  {Sheldon}, \citenamefont {Magesan}, \citenamefont {Chow},\ and\ \citenamefont
  {Gambetta}}]{sheldon2016procedure}%
  \BibitemOpen
  \bibfield  {author} {\bibinfo {author} {\bibfnamefont {S.}~\bibnamefont
  {Sheldon}}, \bibinfo {author} {\bibfnamefont {E.}~\bibnamefont {Magesan}},
  \bibinfo {author} {\bibfnamefont {J.~M.}\ \bibnamefont {Chow}},\ and\
  \bibinfo {author} {\bibfnamefont {J.~M.}\ \bibnamefont {Gambetta}},\ }\href
  {https://doi.org/10.1103/PhysRevA.93.060302} {\bibfield  {journal} {\bibinfo
  {journal} {Phys. Rev. A}\ }\textbf {\bibinfo {volume} {93}},\ \bibinfo
  {pages} {060302} (\bibinfo {year} {2016})}\BibitemShut {NoStop}%
\bibitem [{\citenamefont {Wei}\ \emph {et~al.}(2022)\citenamefont {Wei},
  \citenamefont {Magesan}, \citenamefont {Lauer}, \citenamefont {Srinivasan},
  \citenamefont {Bogorin}, \citenamefont {Carnevale}, \citenamefont {Keefe},
  \citenamefont {Kim}, \citenamefont {Klaus}, \citenamefont {Landers},
  \citenamefont {Sundaresan}, \citenamefont {Wang}, \citenamefont {Zhang},
  \citenamefont {Steffen}, \citenamefont {Dial}, \citenamefont {McKay},\ and\
  \citenamefont {Kandala}}]{wei2022hamiltonian}%
  \BibitemOpen
  \bibfield  {author} {\bibinfo {author} {\bibfnamefont {K.~X.}\ \bibnamefont
  {Wei}}, \bibinfo {author} {\bibfnamefont {E.}~\bibnamefont {Magesan}},
  \bibinfo {author} {\bibfnamefont {I.}~\bibnamefont {Lauer}}, \bibinfo
  {author} {\bibfnamefont {S.}~\bibnamefont {Srinivasan}}, \bibinfo {author}
  {\bibfnamefont {D.~F.}\ \bibnamefont {Bogorin}}, \bibinfo {author}
  {\bibfnamefont {S.}~\bibnamefont {Carnevale}}, \bibinfo {author}
  {\bibfnamefont {G.~A.}\ \bibnamefont {Keefe}}, \bibinfo {author}
  {\bibfnamefont {Y.}~\bibnamefont {Kim}}, \bibinfo {author} {\bibfnamefont
  {D.}~\bibnamefont {Klaus}}, \bibinfo {author} {\bibfnamefont
  {W.}~\bibnamefont {Landers}}, \bibinfo {author} {\bibfnamefont
  {N.}~\bibnamefont {Sundaresan}}, \bibinfo {author} {\bibfnamefont
  {C.}~\bibnamefont {Wang}}, \bibinfo {author} {\bibfnamefont {E.~J.}\
  \bibnamefont {Zhang}}, \bibinfo {author} {\bibfnamefont {M.}~\bibnamefont
  {Steffen}}, \bibinfo {author} {\bibfnamefont {O.~E.}\ \bibnamefont {Dial}},
  \bibinfo {author} {\bibfnamefont {D.~C.}\ \bibnamefont {McKay}},\ and\
  \bibinfo {author} {\bibfnamefont {A.}~\bibnamefont {Kandala}},\ }\href
  {https://doi.org/10.1103/PhysRevLett.129.060501} {\bibfield  {journal}
  {\bibinfo  {journal} {Phys. Rev. Lett.}\ }\textbf {\bibinfo {volume} {129}},\
  \bibinfo {pages} {060501} (\bibinfo {year} {2022})}\BibitemShut {NoStop}%
\bibitem [{\citenamefont {Tripathi}\ \emph {et~al.}(2019)\citenamefont
  {Tripathi}, \citenamefont {Khezri},\ and\ \citenamefont
  {Korotkov}}]{tripathi2019operation}%
  \BibitemOpen
  \bibfield  {author} {\bibinfo {author} {\bibfnamefont {V.}~\bibnamefont
  {Tripathi}}, \bibinfo {author} {\bibfnamefont {M.}~\bibnamefont {Khezri}},\
  and\ \bibinfo {author} {\bibfnamefont {A.~N.}\ \bibnamefont {Korotkov}},\
  }\href {https://doi.org/10.1103/PhysRevA.100.012301} {\bibfield  {journal}
  {\bibinfo  {journal} {Phys. Rev. A}\ }\textbf {\bibinfo {volume} {100}},\
  \bibinfo {pages} {012301} (\bibinfo {year} {2019})}\BibitemShut {NoStop}%
\bibitem [{\citenamefont {Malekakhlagh}\ \emph {et~al.}(2020)\citenamefont
  {Malekakhlagh}, \citenamefont {Magesan},\ and\ \citenamefont
  {McKay}}]{malekakhlagh2020first}%
  \BibitemOpen
  \bibfield  {author} {\bibinfo {author} {\bibfnamefont {M.}~\bibnamefont
  {Malekakhlagh}}, \bibinfo {author} {\bibfnamefont {E.}~\bibnamefont
  {Magesan}},\ and\ \bibinfo {author} {\bibfnamefont {D.~C.}\ \bibnamefont
  {McKay}},\ }\href {https://doi.org/10.1103/PhysRevA.102.042605} {\bibfield
  {journal} {\bibinfo  {journal} {Phys. Rev. A}\ }\textbf {\bibinfo {volume}
  {102}},\ \bibinfo {pages} {042605} (\bibinfo {year} {2020})}\BibitemShut
  {NoStop}%
\bibitem [{\citenamefont {Hertzberg}\ \emph {et~al.}(2021)\citenamefont
  {Hertzberg}, \citenamefont {Zhang}, \citenamefont {Rosenblatt}, \citenamefont
  {Magesan}, \citenamefont {Smolin}, \citenamefont {Yau}, \citenamefont
  {Adiga}, \citenamefont {Sandberg}, \citenamefont {Brink}, \citenamefont
  {Chow},\ and\ \citenamefont {Orcutt}}]{hertzberg2021laser}%
  \BibitemOpen
  \bibfield  {author} {\bibinfo {author} {\bibfnamefont {J.~B.}\ \bibnamefont
  {Hertzberg}}, \bibinfo {author} {\bibfnamefont {E.~J.}\ \bibnamefont
  {Zhang}}, \bibinfo {author} {\bibfnamefont {S.}~\bibnamefont {Rosenblatt}},
  \bibinfo {author} {\bibfnamefont {E.}~\bibnamefont {Magesan}}, \bibinfo
  {author} {\bibfnamefont {J.~A.}\ \bibnamefont {Smolin}}, \bibinfo {author}
  {\bibfnamefont {J.-B.}\ \bibnamefont {Yau}}, \bibinfo {author} {\bibfnamefont
  {V.~P.}\ \bibnamefont {Adiga}}, \bibinfo {author} {\bibfnamefont
  {M.}~\bibnamefont {Sandberg}}, \bibinfo {author} {\bibfnamefont
  {M.}~\bibnamefont {Brink}}, \bibinfo {author} {\bibfnamefont {J.~M.}\
  \bibnamefont {Chow}},\ and\ \bibinfo {author} {\bibfnamefont {J.~S.}\
  \bibnamefont {Orcutt}},\ }\href {https://doi.org/10.1038/s41534-021-00464-5}
  {\bibfield  {journal} {\bibinfo  {journal} {npj Quantum Information}\
  }\textbf {\bibinfo {volume} {7}},\ \bibinfo {pages} {129} (\bibinfo {year}
  {2021})}\BibitemShut {NoStop}%
\bibitem [{\citenamefont {Morvan}\ \emph {et~al.}(2022)\citenamefont {Morvan},
  \citenamefont {Chen}, \citenamefont {Larson}, \citenamefont {Santiago},\ and\
  \citenamefont {Siddiqi}}]{morvan2022optimizing}%
  \BibitemOpen
  \bibfield  {author} {\bibinfo {author} {\bibfnamefont {A.}~\bibnamefont
  {Morvan}}, \bibinfo {author} {\bibfnamefont {L.}~\bibnamefont {Chen}},
  \bibinfo {author} {\bibfnamefont {J.~M.}\ \bibnamefont {Larson}}, \bibinfo
  {author} {\bibfnamefont {D.~I.}\ \bibnamefont {Santiago}},\ and\ \bibinfo
  {author} {\bibfnamefont {I.}~\bibnamefont {Siddiqi}},\ }\href
  {https://doi.org/10.1103/PhysRevResearch.4.023079} {\bibfield  {journal}
  {\bibinfo  {journal} {Phys. Rev. Res.}\ }\textbf {\bibinfo {volume} {4}},\
  \bibinfo {pages} {023079} (\bibinfo {year} {2022})}\BibitemShut {NoStop}%
\bibitem [{\citenamefont {Zhang}\ \emph {et~al.}(2022)\citenamefont {Zhang},
  \citenamefont {Srinivasan}, \citenamefont {Sundaresan}, \citenamefont
  {Bogorin}, \citenamefont {Martin}, \citenamefont {Hertzberg}, \citenamefont
  {Timmerwilke}, \citenamefont {Pritchett}, \citenamefont {Yau}, \citenamefont
  {Wang}, \citenamefont {Landers}, \citenamefont {Lewandowski}, \citenamefont
  {Narasgond}, \citenamefont {Rosenblatt}, \citenamefont {Keefe}, \citenamefont
  {Lauer}, \citenamefont {Rothwell}, \citenamefont {McClure}, \citenamefont
  {Dial}, \citenamefont {Orcutt}, \citenamefont {Brink},\ and\ \citenamefont
  {Chow}}]{zhang2022high}%
  \BibitemOpen
  \bibfield  {author} {\bibinfo {author} {\bibfnamefont {E.~J.}\ \bibnamefont
  {Zhang}}, \bibinfo {author} {\bibfnamefont {S.}~\bibnamefont {Srinivasan}},
  \bibinfo {author} {\bibfnamefont {N.}~\bibnamefont {Sundaresan}}, \bibinfo
  {author} {\bibfnamefont {D.~F.}\ \bibnamefont {Bogorin}}, \bibinfo {author}
  {\bibfnamefont {Y.}~\bibnamefont {Martin}}, \bibinfo {author} {\bibfnamefont
  {J.~B.}\ \bibnamefont {Hertzberg}}, \bibinfo {author} {\bibfnamefont
  {J.}~\bibnamefont {Timmerwilke}}, \bibinfo {author} {\bibfnamefont {E.~J.}\
  \bibnamefont {Pritchett}}, \bibinfo {author} {\bibfnamefont {J.-B.}\
  \bibnamefont {Yau}}, \bibinfo {author} {\bibfnamefont {C.}~\bibnamefont
  {Wang}}, \bibinfo {author} {\bibfnamefont {W.}~\bibnamefont {Landers}},
  \bibinfo {author} {\bibfnamefont {E.~P.}\ \bibnamefont {Lewandowski}},
  \bibinfo {author} {\bibfnamefont {A.}~\bibnamefont {Narasgond}}, \bibinfo
  {author} {\bibfnamefont {S.}~\bibnamefont {Rosenblatt}}, \bibinfo {author}
  {\bibfnamefont {G.~A.}\ \bibnamefont {Keefe}}, \bibinfo {author}
  {\bibfnamefont {I.}~\bibnamefont {Lauer}}, \bibinfo {author} {\bibfnamefont
  {M.~B.}\ \bibnamefont {Rothwell}}, \bibinfo {author} {\bibfnamefont {D.~T.}\
  \bibnamefont {McClure}}, \bibinfo {author} {\bibfnamefont {O.~E.}\
  \bibnamefont {Dial}}, \bibinfo {author} {\bibfnamefont {J.~S.}\ \bibnamefont
  {Orcutt}}, \bibinfo {author} {\bibfnamefont {M.}~\bibnamefont {Brink}},\ and\
  \bibinfo {author} {\bibfnamefont {J.~M.}\ \bibnamefont {Chow}},\ }\href
  {https://doi.org/10.1126/sciadv.abi6690} {\bibfield  {journal} {\bibinfo
  {journal} {Science Advances}\ }\textbf {\bibinfo {volume} {8}},\ \bibinfo
  {pages} {eabi6690} (\bibinfo {year} {2022})}\BibitemShut {NoStop}%
\bibitem [{\citenamefont {Zhang}\ \emph {et~al.}(2025)\citenamefont {Zhang},
  \citenamefont {Gokhale},\ and\ \citenamefont {Larson}}]{zhang2025efficient}%
  \BibitemOpen
  \bibfield  {author} {\bibinfo {author} {\bibfnamefont {Z.}~\bibnamefont
  {Zhang}}, \bibinfo {author} {\bibfnamefont {P.}~\bibnamefont {Gokhale}},\
  and\ \bibinfo {author} {\bibfnamefont {J.~M.}\ \bibnamefont {Larson}},\
  }\href {https://doi.org/10.1103/PhysRevA.111.012619} {\bibfield  {journal}
  {\bibinfo  {journal} {Phys. Rev. A}\ }\textbf {\bibinfo {volume} {111}},\
  \bibinfo {pages} {012619} (\bibinfo {year} {2025})}\BibitemShut {NoStop}%
\bibitem [{\citenamefont {Kandala}\ \emph {et~al.}(2021)\citenamefont
  {Kandala}, \citenamefont {Wei}, \citenamefont {Srinivasan}, \citenamefont
  {Magesan}, \citenamefont {Carnevale}, \citenamefont {Keefe}, \citenamefont
  {Klaus}, \citenamefont {Dial},\ and\ \citenamefont
  {McKay}}]{kandala2021demonstration}%
  \BibitemOpen
  \bibfield  {author} {\bibinfo {author} {\bibfnamefont {A.}~\bibnamefont
  {Kandala}}, \bibinfo {author} {\bibfnamefont {K.~X.}\ \bibnamefont {Wei}},
  \bibinfo {author} {\bibfnamefont {S.}~\bibnamefont {Srinivasan}}, \bibinfo
  {author} {\bibfnamefont {E.}~\bibnamefont {Magesan}}, \bibinfo {author}
  {\bibfnamefont {S.}~\bibnamefont {Carnevale}}, \bibinfo {author}
  {\bibfnamefont {G.~A.}\ \bibnamefont {Keefe}}, \bibinfo {author}
  {\bibfnamefont {D.}~\bibnamefont {Klaus}}, \bibinfo {author} {\bibfnamefont
  {O.}~\bibnamefont {Dial}},\ and\ \bibinfo {author} {\bibfnamefont {D.~C.}\
  \bibnamefont {McKay}},\ }\href
  {https://doi.org/10.1103/PhysRevLett.127.130501} {\bibfield  {journal}
  {\bibinfo  {journal} {Phys. Rev. Lett.}\ }\textbf {\bibinfo {volume} {127}},\
  \bibinfo {pages} {130501} (\bibinfo {year} {2021})}\BibitemShut {NoStop}%
\bibitem [{\citenamefont {Mundada}\ \emph {et~al.}(2019)\citenamefont
  {Mundada}, \citenamefont {Zhang}, \citenamefont {Hazard},\ and\ \citenamefont
  {Houck}}]{mundada2019suppression}%
  \BibitemOpen
  \bibfield  {author} {\bibinfo {author} {\bibfnamefont {P.}~\bibnamefont
  {Mundada}}, \bibinfo {author} {\bibfnamefont {G.}~\bibnamefont {Zhang}},
  \bibinfo {author} {\bibfnamefont {T.}~\bibnamefont {Hazard}},\ and\ \bibinfo
  {author} {\bibfnamefont {A.}~\bibnamefont {Houck}},\ }\href
  {https://doi.org/10.1103/PhysRevApplied.12.054023} {\bibfield  {journal}
  {\bibinfo  {journal} {Phys. Rev. Appl.}\ }\textbf {\bibinfo {volume} {12}},\
  \bibinfo {pages} {054023} (\bibinfo {year} {2019})}\BibitemShut {NoStop}%
\bibitem [{\citenamefont {Kumph}\ \emph {et~al.}(2024)\citenamefont {Kumph},
  \citenamefont {Raftery}, \citenamefont {Finck}, \citenamefont {Blair},
  \citenamefont {Carniol}, \citenamefont {Carnevale}, \citenamefont {Keefe},
  \citenamefont {Arena}, \citenamefont {Hall}, \citenamefont {McKay},\ and\
  \citenamefont {Stehlik}}]{kumph2024demonstration}%
  \BibitemOpen
  \bibfield  {author} {\bibinfo {author} {\bibfnamefont {M.}~\bibnamefont
  {Kumph}}, \bibinfo {author} {\bibfnamefont {J.}~\bibnamefont {Raftery}},
  \bibinfo {author} {\bibfnamefont {A.}~\bibnamefont {Finck}}, \bibinfo
  {author} {\bibfnamefont {J.}~\bibnamefont {Blair}}, \bibinfo {author}
  {\bibfnamefont {A.}~\bibnamefont {Carniol}}, \bibinfo {author} {\bibfnamefont
  {S.}~\bibnamefont {Carnevale}}, \bibinfo {author} {\bibfnamefont {G.~A.}\
  \bibnamefont {Keefe}}, \bibinfo {author} {\bibfnamefont {V.}~\bibnamefont
  {Arena}}, \bibinfo {author} {\bibfnamefont {S.}~\bibnamefont {Hall}},
  \bibinfo {author} {\bibfnamefont {D.}~\bibnamefont {McKay}},\ and\ \bibinfo
  {author} {\bibfnamefont {G.}~\bibnamefont {Stehlik}},\ }\href
  {https://arxiv.org/abs/2406.11770} {\bibfield  {journal} {\bibinfo  {journal}
  {arXiv preprint arXiv:2406.11770}\ } (\bibinfo {year} {2024})}\BibitemShut
  {NoStop}%
\bibitem [{\citenamefont {Shirai}\ \emph {et~al.}(2023)\citenamefont {Shirai},
  \citenamefont {Okubo}, \citenamefont {Matsuura}, \citenamefont {Osada},
  \citenamefont {Nakamura},\ and\ \citenamefont {Noguchi}}]{shirai2023all}%
  \BibitemOpen
  \bibfield  {author} {\bibinfo {author} {\bibfnamefont {S.}~\bibnamefont
  {Shirai}}, \bibinfo {author} {\bibfnamefont {Y.}~\bibnamefont {Okubo}},
  \bibinfo {author} {\bibfnamefont {K.}~\bibnamefont {Matsuura}}, \bibinfo
  {author} {\bibfnamefont {A.}~\bibnamefont {Osada}}, \bibinfo {author}
  {\bibfnamefont {Y.}~\bibnamefont {Nakamura}},\ and\ \bibinfo {author}
  {\bibfnamefont {A.}~\bibnamefont {Noguchi}},\ }\href
  {https://doi.org/10.1103/PhysRevLett.130.260601} {\bibfield  {journal}
  {\bibinfo  {journal} {Phys. Rev. Lett.}\ }\textbf {\bibinfo {volume} {130}},\
  \bibinfo {pages} {260601} (\bibinfo {year} {2023})}\BibitemShut {NoStop}%
\bibitem [{\citenamefont {Shirai}\ \emph {et~al.}(2025)\citenamefont {Shirai},
  \citenamefont {Inoue}, \citenamefont {Tamate}, \citenamefont {Li},
  \citenamefont {Nakamura},\ and\ \citenamefont {Noguchi}}]{shirai2025high}%
  \BibitemOpen
  \bibfield  {author} {\bibinfo {author} {\bibfnamefont {S.}~\bibnamefont
  {Shirai}}, \bibinfo {author} {\bibfnamefont {S.}~\bibnamefont {Inoue}},
  \bibinfo {author} {\bibfnamefont {S.}~\bibnamefont {Tamate}}, \bibinfo
  {author} {\bibfnamefont {R.}~\bibnamefont {Li}}, \bibinfo {author}
  {\bibfnamefont {Y.}~\bibnamefont {Nakamura}},\ and\ \bibinfo {author}
  {\bibfnamefont {A.}~\bibnamefont {Noguchi}},\ }\href
  {https://doi.org/10.48550/arXiv.2511.01260} {\bibfield  {journal} {\bibinfo
  {journal} {arXiv preprint arXiv:2511.01260}\ } (\bibinfo {year}
  {2025})}\BibitemShut {NoStop}%
\bibitem [{\citenamefont {Goto}(2022)}]{goto2022double}%
  \BibitemOpen
  \bibfield  {author} {\bibinfo {author} {\bibfnamefont {H.}~\bibnamefont
  {Goto}},\ }\href {https://doi.org/10.1103/PhysRevApplied.18.034038}
  {\bibfield  {journal} {\bibinfo  {journal} {Phys. Rev. Appl.}\ }\textbf
  {\bibinfo {volume} {18}},\ \bibinfo {pages} {034038} (\bibinfo {year}
  {2022})}\BibitemShut {NoStop}%
\bibitem [{\citenamefont {Kubo}\ and\ \citenamefont
  {Goto}(2023)}]{kubo2023fast}%
  \BibitemOpen
  \bibfield  {author} {\bibinfo {author} {\bibfnamefont {K.}~\bibnamefont
  {Kubo}}\ and\ \bibinfo {author} {\bibfnamefont {H.}~\bibnamefont {Goto}},\
  }\href {https://doi.org/10.1063/5.0138699} {\bibfield  {journal} {\bibinfo
  {journal} {Applied Physics Letters}\ }\textbf {\bibinfo {volume} {122}},\
  \bibinfo {pages} {064001} (\bibinfo {year} {2023})}\BibitemShut {NoStop}%
\bibitem [{\citenamefont {Kubo}\ \emph {et~al.}(2024)\citenamefont {Kubo},
  \citenamefont {Ho},\ and\ \citenamefont {Goto}}]{kubo2024high}%
  \BibitemOpen
  \bibfield  {author} {\bibinfo {author} {\bibfnamefont {K.}~\bibnamefont
  {Kubo}}, \bibinfo {author} {\bibfnamefont {Y.}~\bibnamefont {Ho}},\ and\
  \bibinfo {author} {\bibfnamefont {H.}~\bibnamefont {Goto}},\ }\href
  {https://doi.org/10.1103/PhysRevApplied.22.024057} {\bibfield  {journal}
  {\bibinfo  {journal} {Phys. Rev. Appl.}\ }\textbf {\bibinfo {volume} {22}},\
  \bibinfo {pages} {024057} (\bibinfo {year} {2024})}\BibitemShut {NoStop}%
\bibitem [{\citenamefont {Li}\ \emph {et~al.}(2024)\citenamefont {Li},
  \citenamefont {Kubo}, \citenamefont {Ho}, \citenamefont {Yan}, \citenamefont
  {Nakamura},\ and\ \citenamefont {Goto}}]{li2024realization}%
  \BibitemOpen
  \bibfield  {author} {\bibinfo {author} {\bibfnamefont {R.}~\bibnamefont
  {Li}}, \bibinfo {author} {\bibfnamefont {K.}~\bibnamefont {Kubo}}, \bibinfo
  {author} {\bibfnamefont {Y.}~\bibnamefont {Ho}}, \bibinfo {author}
  {\bibfnamefont {Z.}~\bibnamefont {Yan}}, \bibinfo {author} {\bibfnamefont
  {Y.}~\bibnamefont {Nakamura}},\ and\ \bibinfo {author} {\bibfnamefont
  {H.}~\bibnamefont {Goto}},\ }\href
  {https://doi.org/10.1103/PhysRevX.14.041050} {\bibfield  {journal} {\bibinfo
  {journal} {Phys. Rev. X}\ }\textbf {\bibinfo {volume} {14}},\ \bibinfo
  {pages} {041050} (\bibinfo {year} {2024})}\BibitemShut {NoStop}%
\bibitem [{\citenamefont {Li}\ \emph {et~al.}(2025)\citenamefont {Li},
  \citenamefont {Kubo}, \citenamefont {Ho}, \citenamefont {Yan}, \citenamefont
  {Inoue}, \citenamefont {Nakamura},\ and\ \citenamefont
  {Goto}}]{li2025capacitively}%
  \BibitemOpen
  \bibfield  {author} {\bibinfo {author} {\bibfnamefont {R.}~\bibnamefont
  {Li}}, \bibinfo {author} {\bibfnamefont {K.}~\bibnamefont {Kubo}}, \bibinfo
  {author} {\bibfnamefont {Y.}~\bibnamefont {Ho}}, \bibinfo {author}
  {\bibfnamefont {Z.}~\bibnamefont {Yan}}, \bibinfo {author} {\bibfnamefont
  {S.}~\bibnamefont {Inoue}}, \bibinfo {author} {\bibfnamefont
  {Y.}~\bibnamefont {Nakamura}},\ and\ \bibinfo {author} {\bibfnamefont
  {H.}~\bibnamefont {Goto}},\ }\href {https://doi.org/10.1103/l8tq-7sb3}
  {\bibfield  {journal} {\bibinfo  {journal} {Phys. Rev. Appl.}\ }\textbf
  {\bibinfo {volume} {23}},\ \bibinfo {pages} {064069} (\bibinfo {year}
  {2025})}\BibitemShut {NoStop}%
\bibitem [{\citenamefont {Campbell}\ \emph {et~al.}(2023)\citenamefont
  {Campbell}, \citenamefont {Kamal}, \citenamefont {Ranzani}, \citenamefont
  {Senatore},\ and\ \citenamefont {LaHaye}}]{campbell2023modular}%
  \BibitemOpen
  \bibfield  {author} {\bibinfo {author} {\bibfnamefont {D.~L.}\ \bibnamefont
  {Campbell}}, \bibinfo {author} {\bibfnamefont {A.}~\bibnamefont {Kamal}},
  \bibinfo {author} {\bibfnamefont {L.}~\bibnamefont {Ranzani}}, \bibinfo
  {author} {\bibfnamefont {M.}~\bibnamefont {Senatore}},\ and\ \bibinfo
  {author} {\bibfnamefont {M.~D.}\ \bibnamefont {LaHaye}},\ }\href
  {https://doi.org/10.1103/PhysRevApplied.19.064043} {\bibfield  {journal}
  {\bibinfo  {journal} {Phys. Rev. Appl.}\ }\textbf {\bibinfo {volume} {19}},\
  \bibinfo {pages} {064043} (\bibinfo {year} {2023})}\BibitemShut {NoStop}%
\bibitem [{\citenamefont {Cai}\ \emph {et~al.}(2025)\citenamefont {Cai},
  \citenamefont {Chen}, \citenamefont {Bu}, \citenamefont {Huai}, \citenamefont
  {Yang}, \citenamefont {Zong}, \citenamefont {Li}, \citenamefont {Zhang},
  \citenamefont {Zheng},\ and\ \citenamefont {Zhang}}]{cai2025multiplexed}%
  \BibitemOpen
  \bibfield  {author} {\bibinfo {author} {\bibfnamefont {T.}~\bibnamefont
  {Cai}}, \bibinfo {author} {\bibfnamefont {C.}~\bibnamefont {Chen}}, \bibinfo
  {author} {\bibfnamefont {K.}~\bibnamefont {Bu}}, \bibinfo {author}
  {\bibfnamefont {S.}~\bibnamefont {Huai}}, \bibinfo {author} {\bibfnamefont
  {X.}~\bibnamefont {Yang}}, \bibinfo {author} {\bibfnamefont {Z.}~\bibnamefont
  {Zong}}, \bibinfo {author} {\bibfnamefont {Y.}~\bibnamefont {Li}}, \bibinfo
  {author} {\bibfnamefont {Z.}~\bibnamefont {Zhang}}, \bibinfo {author}
  {\bibfnamefont {Y.-C.}\ \bibnamefont {Zheng}},\ and\ \bibinfo {author}
  {\bibfnamefont {S.}~\bibnamefont {Zhang}},\ }\href
  {https://arxiv.org/abs/2511.02249} {\bibfield  {journal} {\bibinfo  {journal}
  {arXiv preprint arXiv:2511.02249}\ } (\bibinfo {year} {2025})}\BibitemShut
  {NoStop}%
\bibitem [{\citenamefont {Chen}\ \emph {et~al.}(2014)\citenamefont {Chen},
  \citenamefont {Neill}, \citenamefont {Roushan}, \citenamefont {Leung},
  \citenamefont {Fang}, \citenamefont {Barends}, \citenamefont {Kelly},
  \citenamefont {Campbell}, \citenamefont {Chen}, \citenamefont {Chiaro},
  \citenamefont {Dunsworth}, \citenamefont {Jeffrey}, \citenamefont {Megrant},
  \citenamefont {Mutus}, \citenamefont {O'Malley}, \citenamefont {Quintana},
  \citenamefont {Sank}, \citenamefont {Vainsencher}, \citenamefont {Wenner},
  \citenamefont {White}, \citenamefont {Geller}, \citenamefont {Cleland},\ and\
  \citenamefont {Martinis}}]{chen2014qubit}%
  \BibitemOpen
  \bibfield  {author} {\bibinfo {author} {\bibfnamefont {Y.}~\bibnamefont
  {Chen}}, \bibinfo {author} {\bibfnamefont {C.}~\bibnamefont {Neill}},
  \bibinfo {author} {\bibfnamefont {P.}~\bibnamefont {Roushan}}, \bibinfo
  {author} {\bibfnamefont {N.}~\bibnamefont {Leung}}, \bibinfo {author}
  {\bibfnamefont {M.}~\bibnamefont {Fang}}, \bibinfo {author} {\bibfnamefont
  {R.}~\bibnamefont {Barends}}, \bibinfo {author} {\bibfnamefont
  {J.}~\bibnamefont {Kelly}}, \bibinfo {author} {\bibfnamefont
  {B.}~\bibnamefont {Campbell}}, \bibinfo {author} {\bibfnamefont
  {Z.}~\bibnamefont {Chen}}, \bibinfo {author} {\bibfnamefont {B.}~\bibnamefont
  {Chiaro}}, \bibinfo {author} {\bibfnamefont {A.}~\bibnamefont {Dunsworth}},
  \bibinfo {author} {\bibfnamefont {E.}~\bibnamefont {Jeffrey}}, \bibinfo
  {author} {\bibfnamefont {A.}~\bibnamefont {Megrant}}, \bibinfo {author}
  {\bibfnamefont {J.~Y.}\ \bibnamefont {Mutus}}, \bibinfo {author}
  {\bibfnamefont {P.~J.~J.}\ \bibnamefont {O'Malley}}, \bibinfo {author}
  {\bibfnamefont {C.~M.}\ \bibnamefont {Quintana}}, \bibinfo {author}
  {\bibfnamefont {D.}~\bibnamefont {Sank}}, \bibinfo {author} {\bibfnamefont
  {A.}~\bibnamefont {Vainsencher}}, \bibinfo {author} {\bibfnamefont
  {J.}~\bibnamefont {Wenner}}, \bibinfo {author} {\bibfnamefont {T.~C.}\
  \bibnamefont {White}}, \bibinfo {author} {\bibfnamefont {M.~R.}\ \bibnamefont
  {Geller}}, \bibinfo {author} {\bibfnamefont {A.~N.}\ \bibnamefont
  {Cleland}},\ and\ \bibinfo {author} {\bibfnamefont {J.~M.}\ \bibnamefont
  {Martinis}},\ }\href {https://doi.org/10.1103/PhysRevLett.113.220502}
  {\bibfield  {journal} {\bibinfo  {journal} {Phys. Rev. Lett.}\ }\textbf
  {\bibinfo {volume} {113}},\ \bibinfo {pages} {220502} (\bibinfo {year}
  {2014})}\BibitemShut {NoStop}%
\bibitem [{\citenamefont {Yan}\ \emph {et~al.}(2018)\citenamefont {Yan},
  \citenamefont {Krantz}, \citenamefont {Sung}, \citenamefont {Kjaergaard},
  \citenamefont {Campbell}, \citenamefont {Orlando}, \citenamefont
  {Gustavsson},\ and\ \citenamefont {Oliver}}]{yan2018tunable}%
  \BibitemOpen
  \bibfield  {author} {\bibinfo {author} {\bibfnamefont {F.}~\bibnamefont
  {Yan}}, \bibinfo {author} {\bibfnamefont {P.}~\bibnamefont {Krantz}},
  \bibinfo {author} {\bibfnamefont {Y.}~\bibnamefont {Sung}}, \bibinfo {author}
  {\bibfnamefont {M.}~\bibnamefont {Kjaergaard}}, \bibinfo {author}
  {\bibfnamefont {D.~L.}\ \bibnamefont {Campbell}}, \bibinfo {author}
  {\bibfnamefont {T.~P.}\ \bibnamefont {Orlando}}, \bibinfo {author}
  {\bibfnamefont {S.}~\bibnamefont {Gustavsson}},\ and\ \bibinfo {author}
  {\bibfnamefont {W.~D.}\ \bibnamefont {Oliver}},\ }\href
  {https://doi.org/10.1103/PhysRevApplied.10.054062} {\bibfield  {journal}
  {\bibinfo  {journal} {Phys. Rev. Appl.}\ }\textbf {\bibinfo {volume} {10}},\
  \bibinfo {pages} {054062} (\bibinfo {year} {2018})}\BibitemShut {NoStop}%
\bibitem [{\citenamefont {Li}\ \emph {et~al.}(2019)\citenamefont {Li},
  \citenamefont {Clark}, \citenamefont {Wang}, \citenamefont {Wu},
  \citenamefont {Gong}, \citenamefont {Yan}, \citenamefont {Rong},
  \citenamefont {Deng}, \citenamefont {Zha}, \citenamefont {Guo}, \citenamefont
  {Sun}, \citenamefont {Peng}, \citenamefont {Zhu},\ and\ \citenamefont
  {Pan}}]{li2019realisation}%
  \BibitemOpen
  \bibfield  {author} {\bibinfo {author} {\bibfnamefont {S.}~\bibnamefont
  {Li}}, \bibinfo {author} {\bibfnamefont {J.}~\bibnamefont {Clark}}, \bibinfo
  {author} {\bibfnamefont {S.}~\bibnamefont {Wang}}, \bibinfo {author}
  {\bibfnamefont {Y.}~\bibnamefont {Wu}}, \bibinfo {author} {\bibfnamefont
  {M.}~\bibnamefont {Gong}}, \bibinfo {author} {\bibfnamefont {Z.}~\bibnamefont
  {Yan}}, \bibinfo {author} {\bibfnamefont {H.}~\bibnamefont {Rong}}, \bibinfo
  {author} {\bibfnamefont {H.}~\bibnamefont {Deng}}, \bibinfo {author}
  {\bibfnamefont {C.}~\bibnamefont {Zha}}, \bibinfo {author} {\bibfnamefont
  {C.}~\bibnamefont {Guo}}, \bibinfo {author} {\bibfnamefont {L.}~\bibnamefont
  {Sun}}, \bibinfo {author} {\bibfnamefont {C.}~\bibnamefont {Peng}}, \bibinfo
  {author} {\bibfnamefont {X.}~\bibnamefont {Zhu}},\ and\ \bibinfo {author}
  {\bibfnamefont {J.-W.}\ \bibnamefont {Pan}},\ }\href
  {https://doi.org/10.1038/s41534-019-0202-7} {\bibfield  {journal} {\bibinfo
  {journal} {npj Quantum Information}\ }\textbf {\bibinfo {volume} {5}},\
  \bibinfo {pages} {84} (\bibinfo {year} {2019})}\BibitemShut {NoStop}%
\bibitem [{\citenamefont {Marxer}\ \emph {et~al.}(2023)\citenamefont {Marxer},
  \citenamefont {Veps\"al\"ainen}, \citenamefont {Jolin}, \citenamefont
  {Tuorila}, \citenamefont {Landra}, \citenamefont {Ockeloen-Korppi},
  \citenamefont {Liu}, \citenamefont {Ahonen}, \citenamefont {Auer},
  \citenamefont {Belzane}, \citenamefont {Bergholm}, \citenamefont {Chan},
  \citenamefont {Chan}, \citenamefont {Hiltunen}, \citenamefont {Hotari},
  \citenamefont {Hyypp\"a}, \citenamefont {Ikonen}, \citenamefont {Janzso},
  \citenamefont {Koistinen}, \citenamefont {Kotilahti}, \citenamefont {Li},
  \citenamefont {Luus}, \citenamefont {Papic}, \citenamefont {Partanen},
  \citenamefont {R\"abin\"a}, \citenamefont {Rosti}, \citenamefont {Savytskyi},
  \citenamefont {Sepp\"al\"a}, \citenamefont {Sevriuk}, \citenamefont {Takala},
  \citenamefont {Tarasinski}, \citenamefont {Thapa}, \citenamefont {Tosto},
  \citenamefont {Vorobeva}, \citenamefont {Yu}, \citenamefont {Tan},
  \citenamefont {Hassel}, \citenamefont {M\"ott\"onen},\ and\ \citenamefont
  {Heinsoo}}]{marxer2023long}%
  \BibitemOpen
  \bibfield  {author} {\bibinfo {author} {\bibfnamefont {F.}~\bibnamefont
  {Marxer}}, \bibinfo {author} {\bibfnamefont {A.}~\bibnamefont
  {Veps\"al\"ainen}}, \bibinfo {author} {\bibfnamefont {S.~W.}\ \bibnamefont
  {Jolin}}, \bibinfo {author} {\bibfnamefont {J.}~\bibnamefont {Tuorila}},
  \bibinfo {author} {\bibfnamefont {A.}~\bibnamefont {Landra}}, \bibinfo
  {author} {\bibfnamefont {C.}~\bibnamefont {Ockeloen-Korppi}}, \bibinfo
  {author} {\bibfnamefont {W.}~\bibnamefont {Liu}}, \bibinfo {author}
  {\bibfnamefont {O.}~\bibnamefont {Ahonen}}, \bibinfo {author} {\bibfnamefont
  {A.}~\bibnamefont {Auer}}, \bibinfo {author} {\bibfnamefont {L.}~\bibnamefont
  {Belzane}}, \bibinfo {author} {\bibfnamefont {V.}~\bibnamefont {Bergholm}},
  \bibinfo {author} {\bibfnamefont {C.~F.}\ \bibnamefont {Chan}}, \bibinfo
  {author} {\bibfnamefont {K.~W.}\ \bibnamefont {Chan}}, \bibinfo {author}
  {\bibfnamefont {T.}~\bibnamefont {Hiltunen}}, \bibinfo {author}
  {\bibfnamefont {J.}~\bibnamefont {Hotari}}, \bibinfo {author} {\bibfnamefont
  {E.}~\bibnamefont {Hyypp\"a}}, \bibinfo {author} {\bibfnamefont
  {J.}~\bibnamefont {Ikonen}}, \bibinfo {author} {\bibfnamefont
  {D.}~\bibnamefont {Janzso}}, \bibinfo {author} {\bibfnamefont
  {M.}~\bibnamefont {Koistinen}}, \bibinfo {author} {\bibfnamefont
  {J.}~\bibnamefont {Kotilahti}}, \bibinfo {author} {\bibfnamefont
  {T.}~\bibnamefont {Li}}, \bibinfo {author} {\bibfnamefont {J.}~\bibnamefont
  {Luus}}, \bibinfo {author} {\bibfnamefont {M.}~\bibnamefont {Papic}},
  \bibinfo {author} {\bibfnamefont {M.}~\bibnamefont {Partanen}}, \bibinfo
  {author} {\bibfnamefont {J.}~\bibnamefont {R\"abin\"a}}, \bibinfo {author}
  {\bibfnamefont {J.}~\bibnamefont {Rosti}}, \bibinfo {author} {\bibfnamefont
  {M.}~\bibnamefont {Savytskyi}}, \bibinfo {author} {\bibfnamefont
  {M.}~\bibnamefont {Sepp\"al\"a}}, \bibinfo {author} {\bibfnamefont
  {V.}~\bibnamefont {Sevriuk}}, \bibinfo {author} {\bibfnamefont
  {E.}~\bibnamefont {Takala}}, \bibinfo {author} {\bibfnamefont
  {B.}~\bibnamefont {Tarasinski}}, \bibinfo {author} {\bibfnamefont {M.~J.}\
  \bibnamefont {Thapa}}, \bibinfo {author} {\bibfnamefont {F.}~\bibnamefont
  {Tosto}}, \bibinfo {author} {\bibfnamefont {N.}~\bibnamefont {Vorobeva}},
  \bibinfo {author} {\bibfnamefont {L.}~\bibnamefont {Yu}}, \bibinfo {author}
  {\bibfnamefont {K.~Y.}\ \bibnamefont {Tan}}, \bibinfo {author} {\bibfnamefont
  {J.}~\bibnamefont {Hassel}}, \bibinfo {author} {\bibfnamefont
  {M.}~\bibnamefont {M\"ott\"onen}},\ and\ \bibinfo {author} {\bibfnamefont
  {J.}~\bibnamefont {Heinsoo}},\ }\href
  {https://doi.org/10.1103/PRXQuantum.4.010314} {\bibfield  {journal} {\bibinfo
   {journal} {PRX Quantum}\ }\textbf {\bibinfo {volume} {4}},\ \bibinfo {pages}
  {010314} (\bibinfo {year} {2023})}\BibitemShut {NoStop}%
\bibitem [{\citenamefont {Glaser}\ \emph {et~al.}(2025)\citenamefont {Glaser},
  \citenamefont {Roy}, \citenamefont {Tsitsilin}, \citenamefont {Koch},
  \citenamefont {Bruckmoser}, \citenamefont {Schirk}, \citenamefont {Romeiro},
  \citenamefont {Huber}, \citenamefont {Wallner}, \citenamefont {Singh},
  \citenamefont {Krylov}, \citenamefont {Marx}, \citenamefont {S\"odergren},
  \citenamefont {Schneider}, \citenamefont {Werninghaus},\ and\ \citenamefont
  {Filipp}}]{glaser2024sensitivity}%
  \BibitemOpen
  \bibfield  {author} {\bibinfo {author} {\bibfnamefont {N.}~\bibnamefont
  {Glaser}}, \bibinfo {author} {\bibfnamefont {F.}~\bibnamefont {Roy}},
  \bibinfo {author} {\bibfnamefont {I.}~\bibnamefont {Tsitsilin}}, \bibinfo
  {author} {\bibfnamefont {L.}~\bibnamefont {Koch}}, \bibinfo {author}
  {\bibfnamefont {N.}~\bibnamefont {Bruckmoser}}, \bibinfo {author}
  {\bibfnamefont {J.}~\bibnamefont {Schirk}}, \bibinfo {author} {\bibfnamefont
  {J.}~\bibnamefont {Romeiro}}, \bibinfo {author} {\bibfnamefont
  {G.}~\bibnamefont {Huber}}, \bibinfo {author} {\bibfnamefont
  {F.}~\bibnamefont {Wallner}}, \bibinfo {author} {\bibfnamefont
  {M.}~\bibnamefont {Singh}}, \bibinfo {author} {\bibfnamefont
  {G.}~\bibnamefont {Krylov}}, \bibinfo {author} {\bibfnamefont
  {A.}~\bibnamefont {Marx}}, \bibinfo {author} {\bibfnamefont {L.}~\bibnamefont
  {S\"odergren}}, \bibinfo {author} {\bibfnamefont {C.}~\bibnamefont
  {Schneider}}, \bibinfo {author} {\bibfnamefont {M.}~\bibnamefont
  {Werninghaus}},\ and\ \bibinfo {author} {\bibfnamefont {S.}~\bibnamefont
  {Filipp}},\ }\href {https://doi.org/10.1103/pckq-2csc} {\bibfield  {journal}
  {\bibinfo  {journal} {Phys. Rev. Appl.}\ }\textbf {\bibinfo {volume} {24}},\
  \bibinfo {pages} {024048} (\bibinfo {year} {2025})}\BibitemShut {NoStop}%
\bibitem [{\citenamefont {Sung}\ \emph {et~al.}(2021)\citenamefont {Sung},
  \citenamefont {Ding}, \citenamefont {Braum\"uller}, \citenamefont
  {Veps\"al\"ainen}, \citenamefont {Kannan}, \citenamefont {Kjaergaard},
  \citenamefont {Greene}, \citenamefont {Samach}, \citenamefont {McNally},
  \citenamefont {Kim}, \citenamefont {Melville}, \citenamefont {Niedzielski},
  \citenamefont {Schwartz}, \citenamefont {Yoder}, \citenamefont {Orlando},
  \citenamefont {Gustavsson},\ and\ \citenamefont
  {Oliver}}]{sung2021realization}%
  \BibitemOpen
  \bibfield  {author} {\bibinfo {author} {\bibfnamefont {Y.}~\bibnamefont
  {Sung}}, \bibinfo {author} {\bibfnamefont {L.}~\bibnamefont {Ding}}, \bibinfo
  {author} {\bibfnamefont {J.}~\bibnamefont {Braum\"uller}}, \bibinfo {author}
  {\bibfnamefont {A.}~\bibnamefont {Veps\"al\"ainen}}, \bibinfo {author}
  {\bibfnamefont {B.}~\bibnamefont {Kannan}}, \bibinfo {author} {\bibfnamefont
  {M.}~\bibnamefont {Kjaergaard}}, \bibinfo {author} {\bibfnamefont
  {A.}~\bibnamefont {Greene}}, \bibinfo {author} {\bibfnamefont {G.~O.}\
  \bibnamefont {Samach}}, \bibinfo {author} {\bibfnamefont {C.}~\bibnamefont
  {McNally}}, \bibinfo {author} {\bibfnamefont {D.}~\bibnamefont {Kim}},
  \bibinfo {author} {\bibfnamefont {A.}~\bibnamefont {Melville}}, \bibinfo
  {author} {\bibfnamefont {B.~M.}\ \bibnamefont {Niedzielski}}, \bibinfo
  {author} {\bibfnamefont {M.~E.}\ \bibnamefont {Schwartz}}, \bibinfo {author}
  {\bibfnamefont {J.~L.}\ \bibnamefont {Yoder}}, \bibinfo {author}
  {\bibfnamefont {T.~P.}\ \bibnamefont {Orlando}}, \bibinfo {author}
  {\bibfnamefont {S.}~\bibnamefont {Gustavsson}},\ and\ \bibinfo {author}
  {\bibfnamefont {W.~D.}\ \bibnamefont {Oliver}},\ }\href
  {https://doi.org/10.1103/PhysRevX.11.021058} {\bibfield  {journal} {\bibinfo
  {journal} {Phys. Rev. X}\ }\textbf {\bibinfo {volume} {11}},\ \bibinfo
  {pages} {021058} (\bibinfo {year} {2021})}\BibitemShut {NoStop}%
\bibitem [{\citenamefont {Xu}\ \emph {et~al.}(2020)\citenamefont {Xu},
  \citenamefont {Chu}, \citenamefont {Yuan}, \citenamefont {Qiu}, \citenamefont
  {Zhou}, \citenamefont {Zhang}, \citenamefont {Tan}, \citenamefont {Yu},
  \citenamefont {Liu}, \citenamefont {Li}, \citenamefont {Yan},\ and\
  \citenamefont {Yu}}]{xu2020high}%
  \BibitemOpen
  \bibfield  {author} {\bibinfo {author} {\bibfnamefont {Y.}~\bibnamefont
  {Xu}}, \bibinfo {author} {\bibfnamefont {J.}~\bibnamefont {Chu}}, \bibinfo
  {author} {\bibfnamefont {J.}~\bibnamefont {Yuan}}, \bibinfo {author}
  {\bibfnamefont {J.}~\bibnamefont {Qiu}}, \bibinfo {author} {\bibfnamefont
  {Y.}~\bibnamefont {Zhou}}, \bibinfo {author} {\bibfnamefont {L.}~\bibnamefont
  {Zhang}}, \bibinfo {author} {\bibfnamefont {X.}~\bibnamefont {Tan}}, \bibinfo
  {author} {\bibfnamefont {Y.}~\bibnamefont {Yu}}, \bibinfo {author}
  {\bibfnamefont {S.}~\bibnamefont {Liu}}, \bibinfo {author} {\bibfnamefont
  {J.}~\bibnamefont {Li}}, \bibinfo {author} {\bibfnamefont {F.}~\bibnamefont
  {Yan}},\ and\ \bibinfo {author} {\bibfnamefont {D.}~\bibnamefont {Yu}},\
  }\href {https://doi.org/10.1103/PhysRevLett.125.240503} {\bibfield  {journal}
  {\bibinfo  {journal} {Phys. Rev. Lett.}\ }\textbf {\bibinfo {volume} {125}},\
  \bibinfo {pages} {240503} (\bibinfo {year} {2020})}\BibitemShut {NoStop}%
\bibitem [{\citenamefont {Xu}\ \emph {et~al.}(2021)\citenamefont {Xu},
  \citenamefont {Liu}, \citenamefont {Li}, \citenamefont {Han}, \citenamefont
  {Zhang}, \citenamefont {Linghu}, \citenamefont {Li}, \citenamefont {Chen},
  \citenamefont {Yang}, \citenamefont {Wang}, \citenamefont {Ma}, \citenamefont
  {Xue}, \citenamefont {Jin},\ and\ \citenamefont {Yu}}]{xu2021realization}%
  \BibitemOpen
  \bibfield  {author} {\bibinfo {author} {\bibfnamefont {H.}~\bibnamefont
  {Xu}}, \bibinfo {author} {\bibfnamefont {W.}~\bibnamefont {Liu}}, \bibinfo
  {author} {\bibfnamefont {Z.}~\bibnamefont {Li}}, \bibinfo {author}
  {\bibfnamefont {J.}~\bibnamefont {Han}}, \bibinfo {author} {\bibfnamefont
  {J.}~\bibnamefont {Zhang}}, \bibinfo {author} {\bibfnamefont
  {K.}~\bibnamefont {Linghu}}, \bibinfo {author} {\bibfnamefont
  {Y.}~\bibnamefont {Li}}, \bibinfo {author} {\bibfnamefont {M.}~\bibnamefont
  {Chen}}, \bibinfo {author} {\bibfnamefont {Z.}~\bibnamefont {Yang}}, \bibinfo
  {author} {\bibfnamefont {J.}~\bibnamefont {Wang}}, \bibinfo {author}
  {\bibfnamefont {T.}~\bibnamefont {Ma}}, \bibinfo {author} {\bibfnamefont
  {G.}~\bibnamefont {Xue}}, \bibinfo {author} {\bibfnamefont {Y.}~\bibnamefont
  {Jin}},\ and\ \bibinfo {author} {\bibfnamefont {H.}~\bibnamefont {Yu}},\
  }\href {https://doi.org/10.1088/1674-1056/abf03a} {\bibfield  {journal}
  {\bibinfo  {journal} {Chinese Physics B}\ }\textbf {\bibinfo {volume} {30}},\
  \bibinfo {pages} {044212} (\bibinfo {year} {2021})}\BibitemShut {NoStop}%
\bibitem [{\citenamefont {Rol}\ \emph {et~al.}(2019)\citenamefont {Rol},
  \citenamefont {Battistel}, \citenamefont {Malinowski}, \citenamefont
  {Bultink}, \citenamefont {Tarasinski}, \citenamefont {Vollmer}, \citenamefont
  {Haider}, \citenamefont {Muthusubramanian}, \citenamefont {Bruno},
  \citenamefont {Terhal},\ and\ \citenamefont {DiCarlo}}]{rol2019fast}%
  \BibitemOpen
  \bibfield  {author} {\bibinfo {author} {\bibfnamefont {M.~A.}\ \bibnamefont
  {Rol}}, \bibinfo {author} {\bibfnamefont {F.}~\bibnamefont {Battistel}},
  \bibinfo {author} {\bibfnamefont {F.~K.}\ \bibnamefont {Malinowski}},
  \bibinfo {author} {\bibfnamefont {C.~C.}\ \bibnamefont {Bultink}}, \bibinfo
  {author} {\bibfnamefont {B.~M.}\ \bibnamefont {Tarasinski}}, \bibinfo
  {author} {\bibfnamefont {R.}~\bibnamefont {Vollmer}}, \bibinfo {author}
  {\bibfnamefont {N.}~\bibnamefont {Haider}}, \bibinfo {author} {\bibfnamefont
  {N.}~\bibnamefont {Muthusubramanian}}, \bibinfo {author} {\bibfnamefont
  {A.}~\bibnamefont {Bruno}}, \bibinfo {author} {\bibfnamefont {B.~M.}\
  \bibnamefont {Terhal}},\ and\ \bibinfo {author} {\bibfnamefont
  {L.}~\bibnamefont {DiCarlo}},\ }\href
  {https://doi.org/10.1103/PhysRevLett.123.120502} {\bibfield  {journal}
  {\bibinfo  {journal} {Phys. Rev. Lett.}\ }\textbf {\bibinfo {volume} {123}},\
  \bibinfo {pages} {120502} (\bibinfo {year} {2019})}\BibitemShut {NoStop}%
\bibitem [{\citenamefont {Collodo}\ \emph {et~al.}(2020)\citenamefont
  {Collodo}, \citenamefont {Herrmann}, \citenamefont {Lacroix}, \citenamefont
  {Andersen}, \citenamefont {Remm}, \citenamefont {Lazar}, \citenamefont
  {Besse}, \citenamefont {Walter}, \citenamefont {Wallraff},\ and\
  \citenamefont {Eichler}}]{collodo2020implementation}%
  \BibitemOpen
  \bibfield  {author} {\bibinfo {author} {\bibfnamefont {M.~C.}\ \bibnamefont
  {Collodo}}, \bibinfo {author} {\bibfnamefont {J.}~\bibnamefont {Herrmann}},
  \bibinfo {author} {\bibfnamefont {N.}~\bibnamefont {Lacroix}}, \bibinfo
  {author} {\bibfnamefont {C.~K.}\ \bibnamefont {Andersen}}, \bibinfo {author}
  {\bibfnamefont {A.}~\bibnamefont {Remm}}, \bibinfo {author} {\bibfnamefont
  {S.}~\bibnamefont {Lazar}}, \bibinfo {author} {\bibfnamefont {J.-C.}\
  \bibnamefont {Besse}}, \bibinfo {author} {\bibfnamefont {T.}~\bibnamefont
  {Walter}}, \bibinfo {author} {\bibfnamefont {A.}~\bibnamefont {Wallraff}},\
  and\ \bibinfo {author} {\bibfnamefont {C.}~\bibnamefont {Eichler}},\ }\href
  {https://doi.org/10.1103/PhysRevLett.125.240502} {\bibfield  {journal}
  {\bibinfo  {journal} {Phys. Rev. Lett.}\ }\textbf {\bibinfo {volume} {125}},\
  \bibinfo {pages} {240502} (\bibinfo {year} {2020})}\BibitemShut {NoStop}%
\bibitem [{\citenamefont {Stehlik}\ \emph {et~al.}(2021)\citenamefont
  {Stehlik}, \citenamefont {Zajac}, \citenamefont {Underwood}, \citenamefont
  {Phung}, \citenamefont {Blair}, \citenamefont {Carnevale}, \citenamefont
  {Klaus}, \citenamefont {Keefe}, \citenamefont {Carniol}, \citenamefont
  {Kumph}, \citenamefont {Steffen},\ and\ \citenamefont
  {Dial}}]{stehlik2021tunable}%
  \BibitemOpen
  \bibfield  {author} {\bibinfo {author} {\bibfnamefont {J.}~\bibnamefont
  {Stehlik}}, \bibinfo {author} {\bibfnamefont {D.~M.}\ \bibnamefont {Zajac}},
  \bibinfo {author} {\bibfnamefont {D.~L.}\ \bibnamefont {Underwood}}, \bibinfo
  {author} {\bibfnamefont {T.}~\bibnamefont {Phung}}, \bibinfo {author}
  {\bibfnamefont {J.}~\bibnamefont {Blair}}, \bibinfo {author} {\bibfnamefont
  {S.}~\bibnamefont {Carnevale}}, \bibinfo {author} {\bibfnamefont
  {D.}~\bibnamefont {Klaus}}, \bibinfo {author} {\bibfnamefont {G.~A.}\
  \bibnamefont {Keefe}}, \bibinfo {author} {\bibfnamefont {A.}~\bibnamefont
  {Carniol}}, \bibinfo {author} {\bibfnamefont {M.}~\bibnamefont {Kumph}},
  \bibinfo {author} {\bibfnamefont {M.}~\bibnamefont {Steffen}},\ and\ \bibinfo
  {author} {\bibfnamefont {O.~E.}\ \bibnamefont {Dial}},\ }\href
  {https://doi.org/10.1103/PhysRevLett.127.080505} {\bibfield  {journal}
  {\bibinfo  {journal} {Phys. Rev. Lett.}\ }\textbf {\bibinfo {volume} {127}},\
  \bibinfo {pages} {080505} (\bibinfo {year} {2021})}\BibitemShut {NoStop}%
\bibitem [{\citenamefont {Neg\^{\i}rneac}\ \emph {et~al.}(2021)\citenamefont
  {Neg\^{\i}rneac}, \citenamefont {Ali}, \citenamefont {Muthusubramanian},
  \citenamefont {Battistel}, \citenamefont {Sagastizabal}, \citenamefont
  {Moreira}, \citenamefont {Marques}, \citenamefont {Vlothuizen}, \citenamefont
  {Beekman}, \citenamefont {Zachariadis}, \citenamefont {Haider}, \citenamefont
  {Bruno},\ and\ \citenamefont {DiCarlo}}]{negirneac2021high}%
  \BibitemOpen
  \bibfield  {author} {\bibinfo {author} {\bibfnamefont {V.}~\bibnamefont
  {Neg\^{\i}rneac}}, \bibinfo {author} {\bibfnamefont {H.}~\bibnamefont {Ali}},
  \bibinfo {author} {\bibfnamefont {N.}~\bibnamefont {Muthusubramanian}},
  \bibinfo {author} {\bibfnamefont {F.}~\bibnamefont {Battistel}}, \bibinfo
  {author} {\bibfnamefont {R.}~\bibnamefont {Sagastizabal}}, \bibinfo {author}
  {\bibfnamefont {M.~S.}\ \bibnamefont {Moreira}}, \bibinfo {author}
  {\bibfnamefont {J.~F.}\ \bibnamefont {Marques}}, \bibinfo {author}
  {\bibfnamefont {W.~J.}\ \bibnamefont {Vlothuizen}}, \bibinfo {author}
  {\bibfnamefont {M.}~\bibnamefont {Beekman}}, \bibinfo {author} {\bibfnamefont
  {C.}~\bibnamefont {Zachariadis}}, \bibinfo {author} {\bibfnamefont
  {N.}~\bibnamefont {Haider}}, \bibinfo {author} {\bibfnamefont
  {A.}~\bibnamefont {Bruno}},\ and\ \bibinfo {author} {\bibfnamefont
  {L.}~\bibnamefont {DiCarlo}},\ }\href
  {https://doi.org/10.1103/PhysRevLett.126.220502} {\bibfield  {journal}
  {\bibinfo  {journal} {Phys. Rev. Lett.}\ }\textbf {\bibinfo {volume} {126}},\
  \bibinfo {pages} {220502} (\bibinfo {year} {2021})}\BibitemShut {NoStop}%
\bibitem [{\citenamefont {Barends}\ \emph {et~al.}(2019)\citenamefont
  {Barends}, \citenamefont {Quintana}, \citenamefont {Petukhov}, \citenamefont
  {Chen}, \citenamefont {Kafri}, \citenamefont {Kechedzhi}, \citenamefont
  {Collins}, \citenamefont {Naaman}, \citenamefont {Boixo}, \citenamefont
  {Arute}, \citenamefont {Arya}, \citenamefont {Buell}, \citenamefont
  {Burkett}, \citenamefont {Chen}, \citenamefont {Chiaro}, \citenamefont
  {Dunsworth}, \citenamefont {Foxen}, \citenamefont {Fowler}, \citenamefont
  {Gidney}, \citenamefont {Giustina}, \citenamefont {Graff}, \citenamefont
  {Huang}, \citenamefont {Jeffrey}, \citenamefont {Kelly}, \citenamefont
  {Klimov}, \citenamefont {Kostritsa}, \citenamefont {Landhuis}, \citenamefont
  {Lucero}, \citenamefont {McEwen}, \citenamefont {Megrant}, \citenamefont
  {Mi}, \citenamefont {Mutus}, \citenamefont {Neeley}, \citenamefont {Neill},
  \citenamefont {Ostby}, \citenamefont {Roushan}, \citenamefont {Sank},
  \citenamefont {Satzinger}, \citenamefont {Vainsencher}, \citenamefont
  {White}, \citenamefont {Yao}, \citenamefont {Yeh}, \citenamefont {Zalcman},
  \citenamefont {Neven}, \citenamefont {Smelyanskiy},\ and\ \citenamefont
  {Martinis}}]{barends2019diabatic}%
  \BibitemOpen
  \bibfield  {author} {\bibinfo {author} {\bibfnamefont {R.}~\bibnamefont
  {Barends}}, \bibinfo {author} {\bibfnamefont {C.~M.}\ \bibnamefont
  {Quintana}}, \bibinfo {author} {\bibfnamefont {A.~G.}\ \bibnamefont
  {Petukhov}}, \bibinfo {author} {\bibfnamefont {Y.}~\bibnamefont {Chen}},
  \bibinfo {author} {\bibfnamefont {D.}~\bibnamefont {Kafri}}, \bibinfo
  {author} {\bibfnamefont {K.}~\bibnamefont {Kechedzhi}}, \bibinfo {author}
  {\bibfnamefont {R.}~\bibnamefont {Collins}}, \bibinfo {author} {\bibfnamefont
  {O.}~\bibnamefont {Naaman}}, \bibinfo {author} {\bibfnamefont
  {S.}~\bibnamefont {Boixo}}, \bibinfo {author} {\bibfnamefont
  {F.}~\bibnamefont {Arute}}, \bibinfo {author} {\bibfnamefont
  {K.}~\bibnamefont {Arya}}, \bibinfo {author} {\bibfnamefont {D.}~\bibnamefont
  {Buell}}, \bibinfo {author} {\bibfnamefont {B.}~\bibnamefont {Burkett}},
  \bibinfo {author} {\bibfnamefont {Z.}~\bibnamefont {Chen}}, \bibinfo {author}
  {\bibfnamefont {B.}~\bibnamefont {Chiaro}}, \bibinfo {author} {\bibfnamefont
  {A.}~\bibnamefont {Dunsworth}}, \bibinfo {author} {\bibfnamefont
  {B.}~\bibnamefont {Foxen}}, \bibinfo {author} {\bibfnamefont
  {A.}~\bibnamefont {Fowler}}, \bibinfo {author} {\bibfnamefont
  {C.}~\bibnamefont {Gidney}}, \bibinfo {author} {\bibfnamefont
  {M.}~\bibnamefont {Giustina}}, \bibinfo {author} {\bibfnamefont
  {R.}~\bibnamefont {Graff}}, \bibinfo {author} {\bibfnamefont
  {T.}~\bibnamefont {Huang}}, \bibinfo {author} {\bibfnamefont
  {E.}~\bibnamefont {Jeffrey}}, \bibinfo {author} {\bibfnamefont
  {J.}~\bibnamefont {Kelly}}, \bibinfo {author} {\bibfnamefont {P.~V.}\
  \bibnamefont {Klimov}}, \bibinfo {author} {\bibfnamefont {F.}~\bibnamefont
  {Kostritsa}}, \bibinfo {author} {\bibfnamefont {D.}~\bibnamefont {Landhuis}},
  \bibinfo {author} {\bibfnamefont {E.}~\bibnamefont {Lucero}}, \bibinfo
  {author} {\bibfnamefont {M.}~\bibnamefont {McEwen}}, \bibinfo {author}
  {\bibfnamefont {A.}~\bibnamefont {Megrant}}, \bibinfo {author} {\bibfnamefont
  {X.}~\bibnamefont {Mi}}, \bibinfo {author} {\bibfnamefont {J.}~\bibnamefont
  {Mutus}}, \bibinfo {author} {\bibfnamefont {M.}~\bibnamefont {Neeley}},
  \bibinfo {author} {\bibfnamefont {C.}~\bibnamefont {Neill}}, \bibinfo
  {author} {\bibfnamefont {E.}~\bibnamefont {Ostby}}, \bibinfo {author}
  {\bibfnamefont {P.}~\bibnamefont {Roushan}}, \bibinfo {author} {\bibfnamefont
  {D.}~\bibnamefont {Sank}}, \bibinfo {author} {\bibfnamefont {K.~J.}\
  \bibnamefont {Satzinger}}, \bibinfo {author} {\bibfnamefont {A.}~\bibnamefont
  {Vainsencher}}, \bibinfo {author} {\bibfnamefont {T.}~\bibnamefont {White}},
  \bibinfo {author} {\bibfnamefont {J.}~\bibnamefont {Yao}}, \bibinfo {author}
  {\bibfnamefont {P.}~\bibnamefont {Yeh}}, \bibinfo {author} {\bibfnamefont
  {A.}~\bibnamefont {Zalcman}}, \bibinfo {author} {\bibfnamefont
  {H.}~\bibnamefont {Neven}}, \bibinfo {author} {\bibfnamefont {V.~N.}\
  \bibnamefont {Smelyanskiy}},\ and\ \bibinfo {author} {\bibfnamefont {J.~M.}\
  \bibnamefont {Martinis}},\ }\href
  {https://doi.org/10.1103/PhysRevLett.123.210501} {\bibfield  {journal}
  {\bibinfo  {journal} {Phys. Rev. Lett.}\ }\textbf {\bibinfo {volume} {123}},\
  \bibinfo {pages} {210501} (\bibinfo {year} {2019})}\BibitemShut {NoStop}%
\bibitem [{\citenamefont {Foxen}\ \emph {et~al.}(2020)\citenamefont {Foxen},
  \citenamefont {Neill}, \citenamefont {Dunsworth}, \citenamefont {Roushan},
  \citenamefont {Chiaro}, \citenamefont {Megrant}, \citenamefont {Kelly},
  \citenamefont {Chen}, \citenamefont {Satzinger}, \citenamefont {Barends},
  \citenamefont {Arute}, \citenamefont {Arya}, \citenamefont {Babbush},
  \citenamefont {Bacon}, \citenamefont {Bardin}, \citenamefont {Boixo},
  \citenamefont {Buell}, \citenamefont {Burkett}, \citenamefont {Chen},
  \citenamefont {Collins}, \citenamefont {Farhi}, \citenamefont {Fowler},
  \citenamefont {Gidney}, \citenamefont {Giustina}, \citenamefont {Graff},
  \citenamefont {Harrigan}, \citenamefont {Huang}, \citenamefont {Isakov},
  \citenamefont {Jeffrey}, \citenamefont {Jiang}, \citenamefont {Kafri},
  \citenamefont {Kechedzhi}, \citenamefont {Klimov}, \citenamefont {Korotkov},
  \citenamefont {Kostritsa}, \citenamefont {Landhuis}, \citenamefont {Lucero},
  \citenamefont {McClean}, \citenamefont {McEwen}, \citenamefont {Mi},
  \citenamefont {Mohseni}, \citenamefont {Mutus}, \citenamefont {Naaman},
  \citenamefont {Neeley}, \citenamefont {Niu}, \citenamefont {Petukhov},
  \citenamefont {Quintana}, \citenamefont {Rubin}, \citenamefont {Sank},
  \citenamefont {Smelyanskiy}, \citenamefont {Vainsencher}, \citenamefont
  {White}, \citenamefont {Yao}, \citenamefont {Yeh}, \citenamefont {Zalcman},
  \citenamefont {Neven},\ and\ \citenamefont
  {Martinis}}]{foxen2020demonstrating}%
  \BibitemOpen
  \bibfield  {author} {\bibinfo {author} {\bibfnamefont {B.}~\bibnamefont
  {Foxen}}, \bibinfo {author} {\bibfnamefont {C.}~\bibnamefont {Neill}},
  \bibinfo {author} {\bibfnamefont {A.}~\bibnamefont {Dunsworth}}, \bibinfo
  {author} {\bibfnamefont {P.}~\bibnamefont {Roushan}}, \bibinfo {author}
  {\bibfnamefont {B.}~\bibnamefont {Chiaro}}, \bibinfo {author} {\bibfnamefont
  {A.}~\bibnamefont {Megrant}}, \bibinfo {author} {\bibfnamefont
  {J.}~\bibnamefont {Kelly}}, \bibinfo {author} {\bibfnamefont
  {Z.}~\bibnamefont {Chen}}, \bibinfo {author} {\bibfnamefont {K.}~\bibnamefont
  {Satzinger}}, \bibinfo {author} {\bibfnamefont {R.}~\bibnamefont {Barends}},
  \bibinfo {author} {\bibfnamefont {F.}~\bibnamefont {Arute}}, \bibinfo
  {author} {\bibfnamefont {K.}~\bibnamefont {Arya}}, \bibinfo {author}
  {\bibfnamefont {R.}~\bibnamefont {Babbush}}, \bibinfo {author} {\bibfnamefont
  {D.}~\bibnamefont {Bacon}}, \bibinfo {author} {\bibfnamefont {J.~C.}\
  \bibnamefont {Bardin}}, \bibinfo {author} {\bibfnamefont {S.}~\bibnamefont
  {Boixo}}, \bibinfo {author} {\bibfnamefont {D.}~\bibnamefont {Buell}},
  \bibinfo {author} {\bibfnamefont {B.}~\bibnamefont {Burkett}}, \bibinfo
  {author} {\bibfnamefont {Y.}~\bibnamefont {Chen}}, \bibinfo {author}
  {\bibfnamefont {R.}~\bibnamefont {Collins}}, \bibinfo {author} {\bibfnamefont
  {E.}~\bibnamefont {Farhi}}, \bibinfo {author} {\bibfnamefont
  {A.}~\bibnamefont {Fowler}}, \bibinfo {author} {\bibfnamefont
  {C.}~\bibnamefont {Gidney}}, \bibinfo {author} {\bibfnamefont
  {M.}~\bibnamefont {Giustina}}, \bibinfo {author} {\bibfnamefont
  {R.}~\bibnamefont {Graff}}, \bibinfo {author} {\bibfnamefont
  {M.}~\bibnamefont {Harrigan}}, \bibinfo {author} {\bibfnamefont
  {T.}~\bibnamefont {Huang}}, \bibinfo {author} {\bibfnamefont {S.~V.}\
  \bibnamefont {Isakov}}, \bibinfo {author} {\bibfnamefont {E.}~\bibnamefont
  {Jeffrey}}, \bibinfo {author} {\bibfnamefont {Z.}~\bibnamefont {Jiang}},
  \bibinfo {author} {\bibfnamefont {D.}~\bibnamefont {Kafri}}, \bibinfo
  {author} {\bibfnamefont {K.}~\bibnamefont {Kechedzhi}}, \bibinfo {author}
  {\bibfnamefont {P.}~\bibnamefont {Klimov}}, \bibinfo {author} {\bibfnamefont
  {A.}~\bibnamefont {Korotkov}}, \bibinfo {author} {\bibfnamefont
  {F.}~\bibnamefont {Kostritsa}}, \bibinfo {author} {\bibfnamefont
  {D.}~\bibnamefont {Landhuis}}, \bibinfo {author} {\bibfnamefont
  {E.}~\bibnamefont {Lucero}}, \bibinfo {author} {\bibfnamefont
  {J.}~\bibnamefont {McClean}}, \bibinfo {author} {\bibfnamefont
  {M.}~\bibnamefont {McEwen}}, \bibinfo {author} {\bibfnamefont
  {X.}~\bibnamefont {Mi}}, \bibinfo {author} {\bibfnamefont {M.}~\bibnamefont
  {Mohseni}}, \bibinfo {author} {\bibfnamefont {J.~Y.}\ \bibnamefont {Mutus}},
  \bibinfo {author} {\bibfnamefont {O.}~\bibnamefont {Naaman}}, \bibinfo
  {author} {\bibfnamefont {M.}~\bibnamefont {Neeley}}, \bibinfo {author}
  {\bibfnamefont {M.}~\bibnamefont {Niu}}, \bibinfo {author} {\bibfnamefont
  {A.}~\bibnamefont {Petukhov}}, \bibinfo {author} {\bibfnamefont
  {C.}~\bibnamefont {Quintana}}, \bibinfo {author} {\bibfnamefont
  {N.}~\bibnamefont {Rubin}}, \bibinfo {author} {\bibfnamefont
  {D.}~\bibnamefont {Sank}}, \bibinfo {author} {\bibfnamefont {V.}~\bibnamefont
  {Smelyanskiy}}, \bibinfo {author} {\bibfnamefont {A.}~\bibnamefont
  {Vainsencher}}, \bibinfo {author} {\bibfnamefont {T.~C.}\ \bibnamefont
  {White}}, \bibinfo {author} {\bibfnamefont {Z.}~\bibnamefont {Yao}}, \bibinfo
  {author} {\bibfnamefont {P.}~\bibnamefont {Yeh}}, \bibinfo {author}
  {\bibfnamefont {A.}~\bibnamefont {Zalcman}}, \bibinfo {author} {\bibfnamefont
  {H.}~\bibnamefont {Neven}},\ and\ \bibinfo {author} {\bibfnamefont {J.~M.}\
  \bibnamefont {Martinis}} (\bibinfo {collaboration} {Google AI Quantum}),\
  }\href {https://doi.org/10.1103/PhysRevLett.125.120504} {\bibfield  {journal}
  {\bibinfo  {journal} {Phys. Rev. Lett.}\ }\textbf {\bibinfo {volume} {125}},\
  \bibinfo {pages} {120504} (\bibinfo {year} {2020})}\BibitemShut {NoStop}%
\bibitem [{\citenamefont {Scarato}\ \emph {et~al.}(2025)\citenamefont
  {Scarato}, \citenamefont {Hanke}, \citenamefont {Remm}, \citenamefont
  {Laz\ifmmode~\u{a}\else \u{a}\fi{}r}, \citenamefont {Lacroix}, \citenamefont
  {Colao~Zanuz}, \citenamefont {Flasby}, \citenamefont {Wallraff},\ and\
  \citenamefont {Hellings}}]{scarato2025realizing}%
  \BibitemOpen
  \bibfield  {author} {\bibinfo {author} {\bibfnamefont {C.}~\bibnamefont
  {Scarato}}, \bibinfo {author} {\bibfnamefont {K.}~\bibnamefont {Hanke}},
  \bibinfo {author} {\bibfnamefont {A.}~\bibnamefont {Remm}}, \bibinfo {author}
  {\bibfnamefont {S.}~\bibnamefont {Laz\ifmmode~\u{a}\else \u{a}\fi{}r}},
  \bibinfo {author} {\bibfnamefont {N.}~\bibnamefont {Lacroix}}, \bibinfo
  {author} {\bibfnamefont {D.}~\bibnamefont {Colao~Zanuz}}, \bibinfo {author}
  {\bibfnamefont {A.}~\bibnamefont {Flasby}}, \bibinfo {author} {\bibfnamefont
  {A.}~\bibnamefont {Wallraff}},\ and\ \bibinfo {author} {\bibfnamefont
  {C.}~\bibnamefont {Hellings}},\ }\href {https://doi.org/10.1103/h7cv-xgw2}
  {\bibfield  {journal} {\bibinfo  {journal} {PRX Quantum}\ }\textbf {\bibinfo
  {volume} {6}},\ \bibinfo {pages} {040317} (\bibinfo {year}
  {2025})}\BibitemShut {NoStop}%
\bibitem [{\citenamefont {Rol}\ \emph {et~al.}(2020)\citenamefont {Rol},
  \citenamefont {Ciorciaro}, \citenamefont {Malinowski}, \citenamefont
  {Tarasinski}, \citenamefont {Sagastizabal}, \citenamefont {Bultink},
  \citenamefont {Salathe}, \citenamefont {Haandbaek}, \citenamefont {Sedivy},\
  and\ \citenamefont {DiCarlo}}]{rol2020time}%
  \BibitemOpen
  \bibfield  {author} {\bibinfo {author} {\bibfnamefont {M.~A.}\ \bibnamefont
  {Rol}}, \bibinfo {author} {\bibfnamefont {L.}~\bibnamefont {Ciorciaro}},
  \bibinfo {author} {\bibfnamefont {F.~K.}\ \bibnamefont {Malinowski}},
  \bibinfo {author} {\bibfnamefont {B.~M.}\ \bibnamefont {Tarasinski}},
  \bibinfo {author} {\bibfnamefont {R.~E.}\ \bibnamefont {Sagastizabal}},
  \bibinfo {author} {\bibfnamefont {C.~C.}\ \bibnamefont {Bultink}}, \bibinfo
  {author} {\bibfnamefont {Y.}~\bibnamefont {Salathe}}, \bibinfo {author}
  {\bibfnamefont {N.}~\bibnamefont {Haandbaek}}, \bibinfo {author}
  {\bibfnamefont {J.}~\bibnamefont {Sedivy}},\ and\ \bibinfo {author}
  {\bibfnamefont {L.}~\bibnamefont {DiCarlo}},\ }\href
  {https://doi.org/10.1063/1.5133894} {\bibfield  {journal} {\bibinfo
  {journal} {Applied Physics Letters}\ }\textbf {\bibinfo {volume} {116}},\
  \bibinfo {pages} {054001} (\bibinfo {year} {2020})}\BibitemShut {NoStop}%
\bibitem [{\citenamefont {Barends}\ \emph {et~al.}(2014)\citenamefont
  {Barends}, \citenamefont {Kelly}, \citenamefont {Megrant}, \citenamefont
  {Veitia}, \citenamefont {Sank}, \citenamefont {Jeffrey}, \citenamefont
  {White}, \citenamefont {Mutus}, \citenamefont {Fowler}, \citenamefont
  {Campbell}, \citenamefont {Chen}, \citenamefont {Chen}, \citenamefont
  {Chiaro}, \citenamefont {Dunsworth}, \citenamefont {Neill}, \citenamefont
  {O’Malley}, \citenamefont {Roushan}, \citenamefont {Vainsencher},
  \citenamefont {Wenner}, \citenamefont {Korotkov}, \citenamefont {Cleland},\
  and\ \citenamefont {Martinis}}]{barends2014superconducting}%
  \BibitemOpen
  \bibfield  {author} {\bibinfo {author} {\bibfnamefont {R.}~\bibnamefont
  {Barends}}, \bibinfo {author} {\bibfnamefont {J.}~\bibnamefont {Kelly}},
  \bibinfo {author} {\bibfnamefont {A.}~\bibnamefont {Megrant}}, \bibinfo
  {author} {\bibfnamefont {A.}~\bibnamefont {Veitia}}, \bibinfo {author}
  {\bibfnamefont {D.}~\bibnamefont {Sank}}, \bibinfo {author} {\bibfnamefont
  {E.}~\bibnamefont {Jeffrey}}, \bibinfo {author} {\bibfnamefont {T.~C.}\
  \bibnamefont {White}}, \bibinfo {author} {\bibfnamefont {J.}~\bibnamefont
  {Mutus}}, \bibinfo {author} {\bibfnamefont {A.~G.}\ \bibnamefont {Fowler}},
  \bibinfo {author} {\bibfnamefont {B.}~\bibnamefont {Campbell}}, \bibinfo
  {author} {\bibfnamefont {Y.}~\bibnamefont {Chen}}, \bibinfo {author}
  {\bibfnamefont {Z.}~\bibnamefont {Chen}}, \bibinfo {author} {\bibfnamefont
  {B.}~\bibnamefont {Chiaro}}, \bibinfo {author} {\bibfnamefont
  {A.}~\bibnamefont {Dunsworth}}, \bibinfo {author} {\bibfnamefont
  {C.}~\bibnamefont {Neill}}, \bibinfo {author} {\bibfnamefont
  {P.}~\bibnamefont {O’Malley}}, \bibinfo {author} {\bibfnamefont
  {P.}~\bibnamefont {Roushan}}, \bibinfo {author} {\bibfnamefont
  {A.}~\bibnamefont {Vainsencher}}, \bibinfo {author} {\bibfnamefont
  {J.}~\bibnamefont {Wenner}}, \bibinfo {author} {\bibfnamefont {A.~N.}\
  \bibnamefont {Korotkov}}, \bibinfo {author} {\bibfnamefont {A.~N.}\
  \bibnamefont {Cleland}},\ and\ \bibinfo {author} {\bibfnamefont {J.~M.}\
  \bibnamefont {Martinis}},\ }\href {https://doi.org/10.1038/nature13171}
  {\bibfield  {journal} {\bibinfo  {journal} {Nature}\ }\textbf {\bibinfo
  {volume} {508}},\ \bibinfo {pages} {500} (\bibinfo {year}
  {2014})}\BibitemShut {NoStop}%
\bibitem [{\citenamefont {Hellings}\ \emph {et~al.}(2025)\citenamefont
  {Hellings}, \citenamefont {Lacroix}, \citenamefont {Remm}, \citenamefont
  {Boell}, \citenamefont {Herrmann}, \citenamefont {Laz\ifmmode~\u{a}\else
  \u{a}\fi{}r}, \citenamefont {Krinner}, \citenamefont {Swiadek}, \citenamefont
  {Andersen}, \citenamefont {Eichler},\ and\ \citenamefont
  {Wallraff}}]{hellings2025calibrating}%
  \BibitemOpen
  \bibfield  {author} {\bibinfo {author} {\bibfnamefont {C.}~\bibnamefont
  {Hellings}}, \bibinfo {author} {\bibfnamefont {N.}~\bibnamefont {Lacroix}},
  \bibinfo {author} {\bibfnamefont {A.}~\bibnamefont {Remm}}, \bibinfo {author}
  {\bibfnamefont {R.}~\bibnamefont {Boell}}, \bibinfo {author} {\bibfnamefont
  {J.}~\bibnamefont {Herrmann}}, \bibinfo {author} {\bibfnamefont
  {S.}~\bibnamefont {Laz\ifmmode~\u{a}\else \u{a}\fi{}r}}, \bibinfo {author}
  {\bibfnamefont {S.}~\bibnamefont {Krinner}}, \bibinfo {author} {\bibfnamefont
  {F.~m.~c.}\ \bibnamefont {Swiadek}}, \bibinfo {author} {\bibfnamefont
  {C.~K.}\ \bibnamefont {Andersen}}, \bibinfo {author} {\bibfnamefont
  {C.}~\bibnamefont {Eichler}},\ and\ \bibinfo {author} {\bibfnamefont
  {A.}~\bibnamefont {Wallraff}},\ }\href {https://doi.org/10.1103/1qhb-r4fb}
  {\bibfield  {journal} {\bibinfo  {journal} {Phys. Rev. Res.}\ }\textbf
  {\bibinfo {volume} {7}},\ \bibinfo {pages} {043142} (\bibinfo {year}
  {2025})}\BibitemShut {NoStop}%
\bibitem [{\citenamefont {McKay}\ \emph {et~al.}(2016)\citenamefont {McKay},
  \citenamefont {Filipp}, \citenamefont {Mezzacapo}, \citenamefont {Magesan},
  \citenamefont {Chow},\ and\ \citenamefont {Gambetta}}]{mckay2016universal}%
  \BibitemOpen
  \bibfield  {author} {\bibinfo {author} {\bibfnamefont {D.~C.}\ \bibnamefont
  {McKay}}, \bibinfo {author} {\bibfnamefont {S.}~\bibnamefont {Filipp}},
  \bibinfo {author} {\bibfnamefont {A.}~\bibnamefont {Mezzacapo}}, \bibinfo
  {author} {\bibfnamefont {E.}~\bibnamefont {Magesan}}, \bibinfo {author}
  {\bibfnamefont {J.~M.}\ \bibnamefont {Chow}},\ and\ \bibinfo {author}
  {\bibfnamefont {J.~M.}\ \bibnamefont {Gambetta}},\ }\href
  {https://doi.org/10.1103/PhysRevApplied.6.064007} {\bibfield  {journal}
  {\bibinfo  {journal} {Phys. Rev. Appl.}\ }\textbf {\bibinfo {volume} {6}},\
  \bibinfo {pages} {064007} (\bibinfo {year} {2016})}\BibitemShut {NoStop}%
\bibitem [{\citenamefont {Roth}\ \emph {et~al.}(2017)\citenamefont {Roth},
  \citenamefont {Ganzhorn}, \citenamefont {Moll}, \citenamefont {Filipp},
  \citenamefont {Salis},\ and\ \citenamefont {Schmidt}}]{roth2017analysis}%
  \BibitemOpen
  \bibfield  {author} {\bibinfo {author} {\bibfnamefont {M.}~\bibnamefont
  {Roth}}, \bibinfo {author} {\bibfnamefont {M.}~\bibnamefont {Ganzhorn}},
  \bibinfo {author} {\bibfnamefont {N.}~\bibnamefont {Moll}}, \bibinfo {author}
  {\bibfnamefont {S.}~\bibnamefont {Filipp}}, \bibinfo {author} {\bibfnamefont
  {G.}~\bibnamefont {Salis}},\ and\ \bibinfo {author} {\bibfnamefont
  {S.}~\bibnamefont {Schmidt}},\ }\href
  {https://doi.org/10.1103/PhysRevA.96.062323} {\bibfield  {journal} {\bibinfo
  {journal} {Phys. Rev. A}\ }\textbf {\bibinfo {volume} {96}},\ \bibinfo
  {pages} {062323} (\bibinfo {year} {2017})}\BibitemShut {NoStop}%
\bibitem [{\citenamefont {Ganzhorn}\ \emph {et~al.}(2020)\citenamefont
  {Ganzhorn}, \citenamefont {Salis}, \citenamefont {Egger}, \citenamefont
  {Fuhrer}, \citenamefont {Mergenthaler}, \citenamefont {M\"uller},
  \citenamefont {M\"uller}, \citenamefont {Paredes}, \citenamefont {Pechal},
  \citenamefont {Werninghaus},\ and\ \citenamefont
  {Filipp}}]{ganzhorn2020benchmarking}%
  \BibitemOpen
  \bibfield  {author} {\bibinfo {author} {\bibfnamefont {M.}~\bibnamefont
  {Ganzhorn}}, \bibinfo {author} {\bibfnamefont {G.}~\bibnamefont {Salis}},
  \bibinfo {author} {\bibfnamefont {D.~J.}\ \bibnamefont {Egger}}, \bibinfo
  {author} {\bibfnamefont {A.}~\bibnamefont {Fuhrer}}, \bibinfo {author}
  {\bibfnamefont {M.}~\bibnamefont {Mergenthaler}}, \bibinfo {author}
  {\bibfnamefont {C.}~\bibnamefont {M\"uller}}, \bibinfo {author}
  {\bibfnamefont {P.}~\bibnamefont {M\"uller}}, \bibinfo {author}
  {\bibfnamefont {S.}~\bibnamefont {Paredes}}, \bibinfo {author} {\bibfnamefont
  {M.}~\bibnamefont {Pechal}}, \bibinfo {author} {\bibfnamefont
  {M.}~\bibnamefont {Werninghaus}},\ and\ \bibinfo {author} {\bibfnamefont
  {S.}~\bibnamefont {Filipp}},\ }\href
  {https://doi.org/10.1103/PhysRevResearch.2.033447} {\bibfield  {journal}
  {\bibinfo  {journal} {Phys. Rev. Res.}\ }\textbf {\bibinfo {volume} {2}},\
  \bibinfo {pages} {033447} (\bibinfo {year} {2020})}\BibitemShut {NoStop}%
\bibitem [{\citenamefont {Han}\ \emph {et~al.}(2020)\citenamefont {Han},
  \citenamefont {Cai}, \citenamefont {Li}, \citenamefont {Wu}, \citenamefont
  {Ma}, \citenamefont {Ma}, \citenamefont {Wang}, \citenamefont {Zhang},
  \citenamefont {Song},\ and\ \citenamefont {Duan}}]{han2020error}%
  \BibitemOpen
  \bibfield  {author} {\bibinfo {author} {\bibfnamefont {X.~Y.}\ \bibnamefont
  {Han}}, \bibinfo {author} {\bibfnamefont {T.~Q.}\ \bibnamefont {Cai}},
  \bibinfo {author} {\bibfnamefont {X.~G.}\ \bibnamefont {Li}}, \bibinfo
  {author} {\bibfnamefont {Y.~K.}\ \bibnamefont {Wu}}, \bibinfo {author}
  {\bibfnamefont {Y.~W.}\ \bibnamefont {Ma}}, \bibinfo {author} {\bibfnamefont
  {Y.~L.}\ \bibnamefont {Ma}}, \bibinfo {author} {\bibfnamefont {J.~H.}\
  \bibnamefont {Wang}}, \bibinfo {author} {\bibfnamefont {H.~Y.}\ \bibnamefont
  {Zhang}}, \bibinfo {author} {\bibfnamefont {Y.~P.}\ \bibnamefont {Song}},\
  and\ \bibinfo {author} {\bibfnamefont {L.~M.}\ \bibnamefont {Duan}},\ }\href
  {https://doi.org/10.1103/PhysRevA.102.022619} {\bibfield  {journal} {\bibinfo
   {journal} {Phys. Rev. A}\ }\textbf {\bibinfo {volume} {102}},\ \bibinfo
  {pages} {022619} (\bibinfo {year} {2020})}\BibitemShut {NoStop}%
\bibitem [{\citenamefont {Caldwell}\ \emph {et~al.}(2018)\citenamefont
  {Caldwell}, \citenamefont {Didier}, \citenamefont {Ryan}, \citenamefont
  {Sete}, \citenamefont {Hudson}, \citenamefont {Karalekas}, \citenamefont
  {Manenti}, \citenamefont {da~Silva}, \citenamefont {Sinclair}, \citenamefont
  {Acala}, \citenamefont {Alidoust}, \citenamefont {Angeles}, \citenamefont
  {Bestwick}, \citenamefont {Block}, \citenamefont {Bloom}, \citenamefont
  {Bradley}, \citenamefont {Bui}, \citenamefont {Capelluto}, \citenamefont
  {Chilcott}, \citenamefont {Cordova}, \citenamefont {Crossman}, \citenamefont
  {Curtis}, \citenamefont {Deshpande}, \citenamefont {Bouayadi}, \citenamefont
  {Girshovich}, \citenamefont {Hong}, \citenamefont {Kuang}, \citenamefont
  {Lenihan}, \citenamefont {Manning}, \citenamefont {Marchenkov}, \citenamefont
  {Marshall}, \citenamefont {Maydra}, \citenamefont {Mohan}, \citenamefont
  {O'Brien}, \citenamefont {Osborn}, \citenamefont {Otterbach}, \citenamefont
  {Papageorge}, \citenamefont {Paquette}, \citenamefont {Pelstring},
  \citenamefont {Polloreno}, \citenamefont {Prawiroatmodjo}, \citenamefont
  {Rawat}, \citenamefont {Reagor}, \citenamefont {Renzas}, \citenamefont
  {Rubin}, \citenamefont {Russell}, \citenamefont {Rust}, \citenamefont
  {Scarabelli}, \citenamefont {Scheer}, \citenamefont {Selvanayagam},
  \citenamefont {Smith}, \citenamefont {Staley}, \citenamefont {Suska},
  \citenamefont {Tezak}, \citenamefont {Thompson}, \citenamefont {To},
  \citenamefont {Vahidpour}, \citenamefont {Vodrahalli}, \citenamefont
  {Whyland}, \citenamefont {Yadav}, \citenamefont {Zeng},\ and\ \citenamefont
  {Rigetti}}]{caldwell2018parametrically}%
  \BibitemOpen
  \bibfield  {author} {\bibinfo {author} {\bibfnamefont {S.~A.}\ \bibnamefont
  {Caldwell}}, \bibinfo {author} {\bibfnamefont {N.}~\bibnamefont {Didier}},
  \bibinfo {author} {\bibfnamefont {C.~A.}\ \bibnamefont {Ryan}}, \bibinfo
  {author} {\bibfnamefont {E.~A.}\ \bibnamefont {Sete}}, \bibinfo {author}
  {\bibfnamefont {A.}~\bibnamefont {Hudson}}, \bibinfo {author} {\bibfnamefont
  {P.}~\bibnamefont {Karalekas}}, \bibinfo {author} {\bibfnamefont
  {R.}~\bibnamefont {Manenti}}, \bibinfo {author} {\bibfnamefont {M.~P.}\
  \bibnamefont {da~Silva}}, \bibinfo {author} {\bibfnamefont {R.}~\bibnamefont
  {Sinclair}}, \bibinfo {author} {\bibfnamefont {E.}~\bibnamefont {Acala}},
  \bibinfo {author} {\bibfnamefont {N.}~\bibnamefont {Alidoust}}, \bibinfo
  {author} {\bibfnamefont {J.}~\bibnamefont {Angeles}}, \bibinfo {author}
  {\bibfnamefont {A.}~\bibnamefont {Bestwick}}, \bibinfo {author}
  {\bibfnamefont {M.}~\bibnamefont {Block}}, \bibinfo {author} {\bibfnamefont
  {B.}~\bibnamefont {Bloom}}, \bibinfo {author} {\bibfnamefont
  {A.}~\bibnamefont {Bradley}}, \bibinfo {author} {\bibfnamefont
  {C.}~\bibnamefont {Bui}}, \bibinfo {author} {\bibfnamefont {L.}~\bibnamefont
  {Capelluto}}, \bibinfo {author} {\bibfnamefont {R.}~\bibnamefont {Chilcott}},
  \bibinfo {author} {\bibfnamefont {J.}~\bibnamefont {Cordova}}, \bibinfo
  {author} {\bibfnamefont {G.}~\bibnamefont {Crossman}}, \bibinfo {author}
  {\bibfnamefont {M.}~\bibnamefont {Curtis}}, \bibinfo {author} {\bibfnamefont
  {S.}~\bibnamefont {Deshpande}}, \bibinfo {author} {\bibfnamefont {T.~E.}\
  \bibnamefont {Bouayadi}}, \bibinfo {author} {\bibfnamefont {D.}~\bibnamefont
  {Girshovich}}, \bibinfo {author} {\bibfnamefont {S.}~\bibnamefont {Hong}},
  \bibinfo {author} {\bibfnamefont {K.}~\bibnamefont {Kuang}}, \bibinfo
  {author} {\bibfnamefont {M.}~\bibnamefont {Lenihan}}, \bibinfo {author}
  {\bibfnamefont {T.}~\bibnamefont {Manning}}, \bibinfo {author} {\bibfnamefont
  {A.}~\bibnamefont {Marchenkov}}, \bibinfo {author} {\bibfnamefont
  {J.}~\bibnamefont {Marshall}}, \bibinfo {author} {\bibfnamefont
  {R.}~\bibnamefont {Maydra}}, \bibinfo {author} {\bibfnamefont
  {Y.}~\bibnamefont {Mohan}}, \bibinfo {author} {\bibfnamefont
  {W.}~\bibnamefont {O'Brien}}, \bibinfo {author} {\bibfnamefont
  {C.}~\bibnamefont {Osborn}}, \bibinfo {author} {\bibfnamefont
  {J.}~\bibnamefont {Otterbach}}, \bibinfo {author} {\bibfnamefont
  {A.}~\bibnamefont {Papageorge}}, \bibinfo {author} {\bibfnamefont {J.-P.}\
  \bibnamefont {Paquette}}, \bibinfo {author} {\bibfnamefont {M.}~\bibnamefont
  {Pelstring}}, \bibinfo {author} {\bibfnamefont {A.}~\bibnamefont
  {Polloreno}}, \bibinfo {author} {\bibfnamefont {G.}~\bibnamefont
  {Prawiroatmodjo}}, \bibinfo {author} {\bibfnamefont {V.}~\bibnamefont
  {Rawat}}, \bibinfo {author} {\bibfnamefont {M.}~\bibnamefont {Reagor}},
  \bibinfo {author} {\bibfnamefont {R.}~\bibnamefont {Renzas}}, \bibinfo
  {author} {\bibfnamefont {N.}~\bibnamefont {Rubin}}, \bibinfo {author}
  {\bibfnamefont {D.}~\bibnamefont {Russell}}, \bibinfo {author} {\bibfnamefont
  {M.}~\bibnamefont {Rust}}, \bibinfo {author} {\bibfnamefont {D.}~\bibnamefont
  {Scarabelli}}, \bibinfo {author} {\bibfnamefont {M.}~\bibnamefont {Scheer}},
  \bibinfo {author} {\bibfnamefont {M.}~\bibnamefont {Selvanayagam}}, \bibinfo
  {author} {\bibfnamefont {R.}~\bibnamefont {Smith}}, \bibinfo {author}
  {\bibfnamefont {A.}~\bibnamefont {Staley}}, \bibinfo {author} {\bibfnamefont
  {M.}~\bibnamefont {Suska}}, \bibinfo {author} {\bibfnamefont
  {N.}~\bibnamefont {Tezak}}, \bibinfo {author} {\bibfnamefont {D.~C.}\
  \bibnamefont {Thompson}}, \bibinfo {author} {\bibfnamefont {T.-W.}\
  \bibnamefont {To}}, \bibinfo {author} {\bibfnamefont {M.}~\bibnamefont
  {Vahidpour}}, \bibinfo {author} {\bibfnamefont {N.}~\bibnamefont
  {Vodrahalli}}, \bibinfo {author} {\bibfnamefont {T.}~\bibnamefont {Whyland}},
  \bibinfo {author} {\bibfnamefont {K.}~\bibnamefont {Yadav}}, \bibinfo
  {author} {\bibfnamefont {W.}~\bibnamefont {Zeng}},\ and\ \bibinfo {author}
  {\bibfnamefont {C.}~\bibnamefont {Rigetti}},\ }\href
  {https://doi.org/10.1103/PhysRevApplied.10.034050} {\bibfield  {journal}
  {\bibinfo  {journal} {Phys. Rev. Appl.}\ }\textbf {\bibinfo {volume} {10}},\
  \bibinfo {pages} {034050} (\bibinfo {year} {2018})}\BibitemShut {NoStop}%
\bibitem [{\citenamefont {Abrams}\ \emph {et~al.}(2020)\citenamefont {Abrams},
  \citenamefont {Didier}, \citenamefont {Johnson}, \citenamefont {Silva},\ and\
  \citenamefont {Ryan}}]{abrams2020implementation}%
  \BibitemOpen
  \bibfield  {author} {\bibinfo {author} {\bibfnamefont {D.~M.}\ \bibnamefont
  {Abrams}}, \bibinfo {author} {\bibfnamefont {N.}~\bibnamefont {Didier}},
  \bibinfo {author} {\bibfnamefont {B.~R.}\ \bibnamefont {Johnson}}, \bibinfo
  {author} {\bibfnamefont {M.~P.~d.}\ \bibnamefont {Silva}},\ and\ \bibinfo
  {author} {\bibfnamefont {C.~A.}\ \bibnamefont {Ryan}},\ }\href
  {https://doi.org/10.1038/s41928-020-00498-1} {\bibfield  {journal} {\bibinfo
  {journal} {Nature Electronics}\ }\textbf {\bibinfo {volume} {3}},\ \bibinfo
  {pages} {744} (\bibinfo {year} {2020})}\BibitemShut {NoStop}%
\bibitem [{\citenamefont {Reagor}\ \emph {et~al.}(2018)\citenamefont {Reagor},
  \citenamefont {Osborn}, \citenamefont {Tezak}, \citenamefont {Staley},
  \citenamefont {Prawiroatmodjo}, \citenamefont {Scheer}, \citenamefont
  {Alidoust}, \citenamefont {Sete}, \citenamefont {Didier}, \citenamefont
  {da~Silva}, \citenamefont {Acala}, \citenamefont {Angeles}, \citenamefont
  {Bestwick}, \citenamefont {Block}, \citenamefont {Bloom}, \citenamefont
  {Bradley}, \citenamefont {Bui}, \citenamefont {Caldwell}, \citenamefont
  {Capelluto}, \citenamefont {Chilcott}, \citenamefont {Cordova}, \citenamefont
  {Crossman}, \citenamefont {Curtis}, \citenamefont {Deshpande}, \citenamefont
  {Bouayadi}, \citenamefont {Girshovich}, \citenamefont {Hong}, \citenamefont
  {Hudson}, \citenamefont {Karalekas}, \citenamefont {Kuang}, \citenamefont
  {Lenihan}, \citenamefont {Manenti}, \citenamefont {Manning}, \citenamefont
  {Marshall}, \citenamefont {Mohan}, \citenamefont {O’Brien}, \citenamefont
  {Otterbach}, \citenamefont {Papageorge}, \citenamefont {Paquette},
  \citenamefont {Pelstring}, \citenamefont {Polloreno}, \citenamefont {Rawat},
  \citenamefont {Ryan}, \citenamefont {Renzas}, \citenamefont {Rubin},
  \citenamefont {Russel}, \citenamefont {Rust}, \citenamefont {Scarabelli},
  \citenamefont {Selvanayagam}, \citenamefont {Sinclair}, \citenamefont
  {Smith}, \citenamefont {Suska}, \citenamefont {To}, \citenamefont
  {Vahidpour}, \citenamefont {Vodrahalli}, \citenamefont {Whyland},
  \citenamefont {Yadav}, \citenamefont {Zeng},\ and\ \citenamefont
  {Rigetti}}]{reagor2018demonstration}%
  \BibitemOpen
  \bibfield  {author} {\bibinfo {author} {\bibfnamefont {M.}~\bibnamefont
  {Reagor}}, \bibinfo {author} {\bibfnamefont {C.~B.}\ \bibnamefont {Osborn}},
  \bibinfo {author} {\bibfnamefont {N.}~\bibnamefont {Tezak}}, \bibinfo
  {author} {\bibfnamefont {A.}~\bibnamefont {Staley}}, \bibinfo {author}
  {\bibfnamefont {G.}~\bibnamefont {Prawiroatmodjo}}, \bibinfo {author}
  {\bibfnamefont {M.}~\bibnamefont {Scheer}}, \bibinfo {author} {\bibfnamefont
  {N.}~\bibnamefont {Alidoust}}, \bibinfo {author} {\bibfnamefont {E.~A.}\
  \bibnamefont {Sete}}, \bibinfo {author} {\bibfnamefont {N.}~\bibnamefont
  {Didier}}, \bibinfo {author} {\bibfnamefont {M.~P.}\ \bibnamefont
  {da~Silva}}, \bibinfo {author} {\bibfnamefont {E.}~\bibnamefont {Acala}},
  \bibinfo {author} {\bibfnamefont {J.}~\bibnamefont {Angeles}}, \bibinfo
  {author} {\bibfnamefont {A.}~\bibnamefont {Bestwick}}, \bibinfo {author}
  {\bibfnamefont {M.}~\bibnamefont {Block}}, \bibinfo {author} {\bibfnamefont
  {B.}~\bibnamefont {Bloom}}, \bibinfo {author} {\bibfnamefont
  {A.}~\bibnamefont {Bradley}}, \bibinfo {author} {\bibfnamefont
  {C.}~\bibnamefont {Bui}}, \bibinfo {author} {\bibfnamefont {S.}~\bibnamefont
  {Caldwell}}, \bibinfo {author} {\bibfnamefont {L.}~\bibnamefont {Capelluto}},
  \bibinfo {author} {\bibfnamefont {R.}~\bibnamefont {Chilcott}}, \bibinfo
  {author} {\bibfnamefont {J.}~\bibnamefont {Cordova}}, \bibinfo {author}
  {\bibfnamefont {G.}~\bibnamefont {Crossman}}, \bibinfo {author}
  {\bibfnamefont {M.}~\bibnamefont {Curtis}}, \bibinfo {author} {\bibfnamefont
  {S.}~\bibnamefont {Deshpande}}, \bibinfo {author} {\bibfnamefont {T.~E.}\
  \bibnamefont {Bouayadi}}, \bibinfo {author} {\bibfnamefont {D.}~\bibnamefont
  {Girshovich}}, \bibinfo {author} {\bibfnamefont {S.}~\bibnamefont {Hong}},
  \bibinfo {author} {\bibfnamefont {A.}~\bibnamefont {Hudson}}, \bibinfo
  {author} {\bibfnamefont {P.}~\bibnamefont {Karalekas}}, \bibinfo {author}
  {\bibfnamefont {K.}~\bibnamefont {Kuang}}, \bibinfo {author} {\bibfnamefont
  {M.}~\bibnamefont {Lenihan}}, \bibinfo {author} {\bibfnamefont
  {R.}~\bibnamefont {Manenti}}, \bibinfo {author} {\bibfnamefont
  {T.}~\bibnamefont {Manning}}, \bibinfo {author} {\bibfnamefont
  {J.}~\bibnamefont {Marshall}}, \bibinfo {author} {\bibfnamefont
  {Y.}~\bibnamefont {Mohan}}, \bibinfo {author} {\bibfnamefont
  {W.}~\bibnamefont {O’Brien}}, \bibinfo {author} {\bibfnamefont
  {J.}~\bibnamefont {Otterbach}}, \bibinfo {author} {\bibfnamefont
  {A.}~\bibnamefont {Papageorge}}, \bibinfo {author} {\bibfnamefont {J.-P.}\
  \bibnamefont {Paquette}}, \bibinfo {author} {\bibfnamefont {M.}~\bibnamefont
  {Pelstring}}, \bibinfo {author} {\bibfnamefont {A.}~\bibnamefont
  {Polloreno}}, \bibinfo {author} {\bibfnamefont {V.}~\bibnamefont {Rawat}},
  \bibinfo {author} {\bibfnamefont {C.~A.}\ \bibnamefont {Ryan}}, \bibinfo
  {author} {\bibfnamefont {R.}~\bibnamefont {Renzas}}, \bibinfo {author}
  {\bibfnamefont {N.}~\bibnamefont {Rubin}}, \bibinfo {author} {\bibfnamefont
  {D.}~\bibnamefont {Russel}}, \bibinfo {author} {\bibfnamefont
  {M.}~\bibnamefont {Rust}}, \bibinfo {author} {\bibfnamefont {D.}~\bibnamefont
  {Scarabelli}}, \bibinfo {author} {\bibfnamefont {M.}~\bibnamefont
  {Selvanayagam}}, \bibinfo {author} {\bibfnamefont {R.}~\bibnamefont
  {Sinclair}}, \bibinfo {author} {\bibfnamefont {R.}~\bibnamefont {Smith}},
  \bibinfo {author} {\bibfnamefont {M.}~\bibnamefont {Suska}}, \bibinfo
  {author} {\bibfnamefont {T.-W.}\ \bibnamefont {To}}, \bibinfo {author}
  {\bibfnamefont {M.}~\bibnamefont {Vahidpour}}, \bibinfo {author}
  {\bibfnamefont {N.}~\bibnamefont {Vodrahalli}}, \bibinfo {author}
  {\bibfnamefont {T.}~\bibnamefont {Whyland}}, \bibinfo {author} {\bibfnamefont
  {K.}~\bibnamefont {Yadav}}, \bibinfo {author} {\bibfnamefont
  {W.}~\bibnamefont {Zeng}},\ and\ \bibinfo {author} {\bibfnamefont {C.~T.}\
  \bibnamefont {Rigetti}},\ }\href {https://doi.org/10.1126/sciadv.aao3603}
  {\bibfield  {journal} {\bibinfo  {journal} {Science Advances}\ }\textbf
  {\bibinfo {volume} {4}},\ \bibinfo {pages} {eaao3603} (\bibinfo {year}
  {2018})}\BibitemShut {NoStop}%
\bibitem [{\citenamefont {Hong}\ \emph {et~al.}(2020)\citenamefont {Hong},
  \citenamefont {Papageorge}, \citenamefont {Sivarajah}, \citenamefont
  {Crossman}, \citenamefont {Didier}, \citenamefont {Polloreno}, \citenamefont
  {Sete}, \citenamefont {Turkowski}, \citenamefont {da~Silva},\ and\
  \citenamefont {Johnson}}]{hong2020demonstration}%
  \BibitemOpen
  \bibfield  {author} {\bibinfo {author} {\bibfnamefont {S.~S.}\ \bibnamefont
  {Hong}}, \bibinfo {author} {\bibfnamefont {A.~T.}\ \bibnamefont
  {Papageorge}}, \bibinfo {author} {\bibfnamefont {P.}~\bibnamefont
  {Sivarajah}}, \bibinfo {author} {\bibfnamefont {G.}~\bibnamefont {Crossman}},
  \bibinfo {author} {\bibfnamefont {N.}~\bibnamefont {Didier}}, \bibinfo
  {author} {\bibfnamefont {A.~M.}\ \bibnamefont {Polloreno}}, \bibinfo {author}
  {\bibfnamefont {E.~A.}\ \bibnamefont {Sete}}, \bibinfo {author}
  {\bibfnamefont {S.~W.}\ \bibnamefont {Turkowski}}, \bibinfo {author}
  {\bibfnamefont {M.~P.}\ \bibnamefont {da~Silva}},\ and\ \bibinfo {author}
  {\bibfnamefont {B.~R.}\ \bibnamefont {Johnson}},\ }\href
  {https://doi.org/10.1103/PhysRevA.101.012302} {\bibfield  {journal} {\bibinfo
   {journal} {Phys. Rev. A}\ }\textbf {\bibinfo {volume} {101}},\ \bibinfo
  {pages} {012302} (\bibinfo {year} {2020})}\BibitemShut {NoStop}%
\bibitem [{\citenamefont {Sete}\ \emph {et~al.}(2024)\citenamefont {Sete},
  \citenamefont {Tripathi}, \citenamefont {Valery}, \citenamefont {Lidar},\
  and\ \citenamefont {Mutus}}]{sete2024error}%
  \BibitemOpen
  \bibfield  {author} {\bibinfo {author} {\bibfnamefont {E.~A.}\ \bibnamefont
  {Sete}}, \bibinfo {author} {\bibfnamefont {V.}~\bibnamefont {Tripathi}},
  \bibinfo {author} {\bibfnamefont {J.~A.}\ \bibnamefont {Valery}}, \bibinfo
  {author} {\bibfnamefont {D.}~\bibnamefont {Lidar}},\ and\ \bibinfo {author}
  {\bibfnamefont {J.~Y.}\ \bibnamefont {Mutus}},\ }\href
  {https://doi.org/10.1103/PhysRevApplied.22.014059} {\bibfield  {journal}
  {\bibinfo  {journal} {Phys. Rev. Appl.}\ }\textbf {\bibinfo {volume} {22}},\
  \bibinfo {pages} {014059} (\bibinfo {year} {2024})}\BibitemShut {NoStop}%
\bibitem [{\citenamefont {Sete}\ \emph {et~al.}(2021)\citenamefont {Sete},
  \citenamefont {Didier}, \citenamefont {Chen}, \citenamefont {Kulshreshtha},
  \citenamefont {Manenti},\ and\ \citenamefont {Poletto}}]{sete2021parametric}%
  \BibitemOpen
  \bibfield  {author} {\bibinfo {author} {\bibfnamefont {E.~A.}\ \bibnamefont
  {Sete}}, \bibinfo {author} {\bibfnamefont {N.}~\bibnamefont {Didier}},
  \bibinfo {author} {\bibfnamefont {A.~Q.}\ \bibnamefont {Chen}}, \bibinfo
  {author} {\bibfnamefont {S.}~\bibnamefont {Kulshreshtha}}, \bibinfo {author}
  {\bibfnamefont {R.}~\bibnamefont {Manenti}},\ and\ \bibinfo {author}
  {\bibfnamefont {S.}~\bibnamefont {Poletto}},\ }\href
  {https://doi.org/10.1103/PhysRevApplied.16.024050} {\bibfield  {journal}
  {\bibinfo  {journal} {Phys. Rev. Appl.}\ }\textbf {\bibinfo {volume} {16}},\
  \bibinfo {pages} {024050} (\bibinfo {year} {2021})}\BibitemShut {NoStop}%
\bibitem [{\citenamefont {Leskes}\ \emph {et~al.}(2010)\citenamefont {Leskes},
  \citenamefont {Madhu},\ and\ \citenamefont {Vega}}]{leskes2010floquet}%
  \BibitemOpen
  \bibfield  {author} {\bibinfo {author} {\bibfnamefont {M.}~\bibnamefont
  {Leskes}}, \bibinfo {author} {\bibfnamefont {P.}~\bibnamefont {Madhu}},\ and\
  \bibinfo {author} {\bibfnamefont {S.}~\bibnamefont {Vega}},\ }\href
  {https://doi.org/https://doi.org/10.1016/j.pnmrs.2010.06.002} {\bibfield
  {journal} {\bibinfo  {journal} {Progress in Nuclear Magnetic Resonance
  Spectroscopy}\ }\textbf {\bibinfo {volume} {57}},\ \bibinfo {pages} {345}
  (\bibinfo {year} {2010})}\BibitemShut {NoStop}%
\bibitem [{\citenamefont {Eckardt}\ and\ \citenamefont
  {Anisimovas}(2015)}]{eckardt2015high}%
  \BibitemOpen
  \bibfield  {author} {\bibinfo {author} {\bibfnamefont {A.}~\bibnamefont
  {Eckardt}}\ and\ \bibinfo {author} {\bibfnamefont {E.}~\bibnamefont
  {Anisimovas}},\ }\href {https://doi.org/10.1088/1367-2630/17/9/093039}
  {\bibfield  {journal} {\bibinfo  {journal} {New Journal of Physics}\ }\textbf
  {\bibinfo {volume} {17}},\ \bibinfo {pages} {093039} (\bibinfo {year}
  {2015})}\BibitemShut {NoStop}%
\bibitem [{\citenamefont {McKay}\ \emph {et~al.}(2017)\citenamefont {McKay},
  \citenamefont {Wood}, \citenamefont {Sheldon}, \citenamefont {Chow},\ and\
  \citenamefont {Gambetta}}]{mckay2017efficient}%
  \BibitemOpen
  \bibfield  {author} {\bibinfo {author} {\bibfnamefont {D.~C.}\ \bibnamefont
  {McKay}}, \bibinfo {author} {\bibfnamefont {C.~J.}\ \bibnamefont {Wood}},
  \bibinfo {author} {\bibfnamefont {S.}~\bibnamefont {Sheldon}}, \bibinfo
  {author} {\bibfnamefont {J.~M.}\ \bibnamefont {Chow}},\ and\ \bibinfo
  {author} {\bibfnamefont {J.~M.}\ \bibnamefont {Gambetta}},\ }\href
  {https://doi.org/10.1103/PhysRevA.96.022330} {\bibfield  {journal} {\bibinfo
  {journal} {Phys. Rev. A}\ }\textbf {\bibinfo {volume} {96}},\ \bibinfo
  {pages} {022330} (\bibinfo {year} {2017})}\BibitemShut {NoStop}%
\bibitem [{\citenamefont {Pedersen}\ \emph {et~al.}(2007)\citenamefont
  {Pedersen}, \citenamefont {Møller},\ and\ \citenamefont
  {Mølmer}}]{pedersen2007fidelity}%
  \BibitemOpen
  \bibfield  {author} {\bibinfo {author} {\bibfnamefont {L.~H.}\ \bibnamefont
  {Pedersen}}, \bibinfo {author} {\bibfnamefont {N.~M.}\ \bibnamefont
  {Møller}},\ and\ \bibinfo {author} {\bibfnamefont {K.}~\bibnamefont
  {Mølmer}},\ }\href
  {https://doi.org/https://doi.org/10.1016/j.physleta.2007.02.069} {\bibfield
  {journal} {\bibinfo  {journal} {Physics Letters A}\ }\textbf {\bibinfo
  {volume} {367}},\ \bibinfo {pages} {47} (\bibinfo {year} {2007})}\BibitemShut
  {NoStop}%
\bibitem [{\citenamefont {Lao}\ \emph {et~al.}(2022)\citenamefont {Lao},
  \citenamefont {Korotkov}, \citenamefont {Jiang}, \citenamefont
  {Mruczkiewicz}, \citenamefont {O'Brien},\ and\ \citenamefont
  {Browne}}]{lao2022software}%
  \BibitemOpen
  \bibfield  {author} {\bibinfo {author} {\bibfnamefont {L.}~\bibnamefont
  {Lao}}, \bibinfo {author} {\bibfnamefont {A.}~\bibnamefont {Korotkov}},
  \bibinfo {author} {\bibfnamefont {Z.}~\bibnamefont {Jiang}}, \bibinfo
  {author} {\bibfnamefont {W.}~\bibnamefont {Mruczkiewicz}}, \bibinfo {author}
  {\bibfnamefont {T.~E.}\ \bibnamefont {O'Brien}},\ and\ \bibinfo {author}
  {\bibfnamefont {D.~E.}\ \bibnamefont {Browne}},\ }\href
  {https://doi.org/10.1088/2058-9565/ac57f1} {\bibfield  {journal} {\bibinfo
  {journal} {Quantum Science and Technology}\ }\textbf {\bibinfo {volume}
  {7}},\ \bibinfo {pages} {025021} (\bibinfo {year} {2022})}\BibitemShut
  {NoStop}%
\bibitem [{\citenamefont {Ding}\ \emph {et~al.}(2023)\citenamefont {Ding},
  \citenamefont {Hays}, \citenamefont {Sung}, \citenamefont {Kannan},
  \citenamefont {An}, \citenamefont {Di~Paolo}, \citenamefont {Karamlou},
  \citenamefont {Hazard}, \citenamefont {Azar}, \citenamefont {Kim},
  \citenamefont {Niedzielski}, \citenamefont {Melville}, \citenamefont
  {Schwartz}, \citenamefont {Yoder}, \citenamefont {Orlando}, \citenamefont
  {Gustavsson}, \citenamefont {Grover}, \citenamefont {Serniak},\ and\
  \citenamefont {Oliver}}]{ding2023high}%
  \BibitemOpen
  \bibfield  {author} {\bibinfo {author} {\bibfnamefont {L.}~\bibnamefont
  {Ding}}, \bibinfo {author} {\bibfnamefont {M.}~\bibnamefont {Hays}}, \bibinfo
  {author} {\bibfnamefont {Y.}~\bibnamefont {Sung}}, \bibinfo {author}
  {\bibfnamefont {B.}~\bibnamefont {Kannan}}, \bibinfo {author} {\bibfnamefont
  {J.}~\bibnamefont {An}}, \bibinfo {author} {\bibfnamefont {A.}~\bibnamefont
  {Di~Paolo}}, \bibinfo {author} {\bibfnamefont {A.~H.}\ \bibnamefont
  {Karamlou}}, \bibinfo {author} {\bibfnamefont {T.~M.}\ \bibnamefont
  {Hazard}}, \bibinfo {author} {\bibfnamefont {K.}~\bibnamefont {Azar}},
  \bibinfo {author} {\bibfnamefont {D.~K.}\ \bibnamefont {Kim}}, \bibinfo
  {author} {\bibfnamefont {B.~M.}\ \bibnamefont {Niedzielski}}, \bibinfo
  {author} {\bibfnamefont {A.}~\bibnamefont {Melville}}, \bibinfo {author}
  {\bibfnamefont {M.~E.}\ \bibnamefont {Schwartz}}, \bibinfo {author}
  {\bibfnamefont {J.~L.}\ \bibnamefont {Yoder}}, \bibinfo {author}
  {\bibfnamefont {T.~P.}\ \bibnamefont {Orlando}}, \bibinfo {author}
  {\bibfnamefont {S.}~\bibnamefont {Gustavsson}}, \bibinfo {author}
  {\bibfnamefont {J.~A.}\ \bibnamefont {Grover}}, \bibinfo {author}
  {\bibfnamefont {K.}~\bibnamefont {Serniak}},\ and\ \bibinfo {author}
  {\bibfnamefont {W.~D.}\ \bibnamefont {Oliver}},\ }\href
  {https://doi.org/10.1103/PhysRevX.13.031035} {\bibfield  {journal} {\bibinfo
  {journal} {Phys. Rev. X}\ }\textbf {\bibinfo {volume} {13}},\ \bibinfo
  {pages} {031035} (\bibinfo {year} {2023})}\BibitemShut {NoStop}%
\bibitem [{\citenamefont {Fors}\ \emph {et~al.}(2024)\citenamefont {Fors},
  \citenamefont {Fern{\'a}ndez-Pend{\'a}s},\ and\ \citenamefont
  {Kockum}}]{fors2024comprehensive}%
  \BibitemOpen
  \bibfield  {author} {\bibinfo {author} {\bibfnamefont {S.~P.}\ \bibnamefont
  {Fors}}, \bibinfo {author} {\bibfnamefont {J.}~\bibnamefont
  {Fern{\'a}ndez-Pend{\'a}s}},\ and\ \bibinfo {author} {\bibfnamefont {A.~F.}\
  \bibnamefont {Kockum}},\ }\href {https://doi.org/10.48550/arXiv.2408.15402}
  {\bibfield  {journal} {\bibinfo  {journal} {arXiv preprint arXiv:2408.15402}\
  } (\bibinfo {year} {2024})}\BibitemShut {NoStop}%
\bibitem [{\citenamefont {Kelly}\ \emph {et~al.}(2014)\citenamefont {Kelly},
  \citenamefont {Barends}, \citenamefont {Campbell}, \citenamefont {Chen},
  \citenamefont {Chen}, \citenamefont {Chiaro}, \citenamefont {Dunsworth},
  \citenamefont {Fowler}, \citenamefont {Hoi}, \citenamefont {Jeffrey},
  \citenamefont {Megrant}, \citenamefont {Mutus}, \citenamefont {Neill},
  \citenamefont {O'Malley}, \citenamefont {Quintana}, \citenamefont {Roushan},
  \citenamefont {Sank}, \citenamefont {Vainsencher}, \citenamefont {Wenner},
  \citenamefont {White}, \citenamefont {Cleland},\ and\ \citenamefont
  {Martinis}}]{kelly2014optimal}%
  \BibitemOpen
  \bibfield  {author} {\bibinfo {author} {\bibfnamefont {J.}~\bibnamefont
  {Kelly}}, \bibinfo {author} {\bibfnamefont {R.}~\bibnamefont {Barends}},
  \bibinfo {author} {\bibfnamefont {B.}~\bibnamefont {Campbell}}, \bibinfo
  {author} {\bibfnamefont {Y.}~\bibnamefont {Chen}}, \bibinfo {author}
  {\bibfnamefont {Z.}~\bibnamefont {Chen}}, \bibinfo {author} {\bibfnamefont
  {B.}~\bibnamefont {Chiaro}}, \bibinfo {author} {\bibfnamefont
  {A.}~\bibnamefont {Dunsworth}}, \bibinfo {author} {\bibfnamefont {A.~G.}\
  \bibnamefont {Fowler}}, \bibinfo {author} {\bibfnamefont {I.-C.}\
  \bibnamefont {Hoi}}, \bibinfo {author} {\bibfnamefont {E.}~\bibnamefont
  {Jeffrey}}, \bibinfo {author} {\bibfnamefont {A.}~\bibnamefont {Megrant}},
  \bibinfo {author} {\bibfnamefont {J.}~\bibnamefont {Mutus}}, \bibinfo
  {author} {\bibfnamefont {C.}~\bibnamefont {Neill}}, \bibinfo {author}
  {\bibfnamefont {P.~J.~J.}\ \bibnamefont {O'Malley}}, \bibinfo {author}
  {\bibfnamefont {C.}~\bibnamefont {Quintana}}, \bibinfo {author}
  {\bibfnamefont {P.}~\bibnamefont {Roushan}}, \bibinfo {author} {\bibfnamefont
  {D.}~\bibnamefont {Sank}}, \bibinfo {author} {\bibfnamefont {A.}~\bibnamefont
  {Vainsencher}}, \bibinfo {author} {\bibfnamefont {J.}~\bibnamefont {Wenner}},
  \bibinfo {author} {\bibfnamefont {T.~C.}\ \bibnamefont {White}}, \bibinfo
  {author} {\bibfnamefont {A.~N.}\ \bibnamefont {Cleland}},\ and\ \bibinfo
  {author} {\bibfnamefont {J.~M.}\ \bibnamefont {Martinis}},\ }\href
  {https://doi.org/10.1103/PhysRevLett.112.240504} {\bibfield  {journal}
  {\bibinfo  {journal} {Phys. Rev. Lett.}\ }\textbf {\bibinfo {volume} {112}},\
  \bibinfo {pages} {240504} (\bibinfo {year} {2014})}\BibitemShut {NoStop}%
\bibitem [{\citenamefont {Marxer}\ \emph {et~al.}(2025)\citenamefont {Marxer},
  \citenamefont {Mrożek}, \citenamefont {Andersson}, \citenamefont
  {Abdurakhimov}, \citenamefont {Adam}, \citenamefont {Bergholm}, \citenamefont
  {Beriwal}, \citenamefont {Chan}, \citenamefont {Dahl}, \citenamefont {Das},
  \citenamefont {Deppe}, \citenamefont {Fedorets}, \citenamefont {Gao},
  \citenamefont {Frieiro}, \citenamefont {Gusenkova}, \citenamefont {Guthrie},
  \citenamefont {Hiltunen}, \citenamefont {Hsu}, \citenamefont {Hyyppä},
  \citenamefont {Ikonen}, \citenamefont {Inel}, \citenamefont {Jolin},
  \citenamefont {Karis}, \citenamefont {Kim}, \citenamefont {Kindel},
  \citenamefont {Komlev}, \citenamefont {Koistinen}, \citenamefont
  {Kokkoniemi}, \citenamefont {Kumar}, \citenamefont {Ku}, \citenamefont
  {Lamprich}, \citenamefont {Laine}, \citenamefont {Landra}, \citenamefont
  {Lee}, \citenamefont {Lethif}, \citenamefont {Liebermann}, \citenamefont
  {Liu}, \citenamefont {Mitra}, \citenamefont {Mylläri}, \citenamefont
  {Ockeloen-Korppi}, \citenamefont {Orell}, \citenamefont {Plyshch},
  \citenamefont {Räbinä}, \citenamefont {Rebello}, \citenamefont {Renger},
  \citenamefont {Reentilä}, \citenamefont {Ritvas}, \citenamefont {Saarinen},
  \citenamefont {Salmenkivi}, \citenamefont {Sarsby}, \citenamefont
  {Savytskyi}, \citenamefont {Selinmaa}, \citenamefont {Steggles},
  \citenamefont {Takala}, \citenamefont {Takmakov}, \citenamefont {Tarasinski},
  \citenamefont {Tuorila}, \citenamefont {Välimaa}, \citenamefont {Verjauw},
  \citenamefont {Wesdorp}, \citenamefont {Wurz}, \citenamefont {Qiu},
  \citenamefont {Zhu}, \citenamefont {Hassel}, \citenamefont {Heinsoo},
  \citenamefont {Geresdi},\ and\ \citenamefont
  {Vepsäläinen}}]{marxer2025above}%
  \BibitemOpen
  \bibfield  {author} {\bibinfo {author} {\bibfnamefont {F.}~\bibnamefont
  {Marxer}}, \bibinfo {author} {\bibfnamefont {J.}~\bibnamefont {Mrożek}},
  \bibinfo {author} {\bibfnamefont {J.}~\bibnamefont {Andersson}}, \bibinfo
  {author} {\bibfnamefont {L.}~\bibnamefont {Abdurakhimov}}, \bibinfo {author}
  {\bibfnamefont {J.}~\bibnamefont {Adam}}, \bibinfo {author} {\bibfnamefont
  {V.}~\bibnamefont {Bergholm}}, \bibinfo {author} {\bibfnamefont
  {R.}~\bibnamefont {Beriwal}}, \bibinfo {author} {\bibfnamefont {C.~F.}\
  \bibnamefont {Chan}}, \bibinfo {author} {\bibfnamefont {S.}~\bibnamefont
  {Dahl}}, \bibinfo {author} {\bibfnamefont {S.~R.}\ \bibnamefont {Das}},
  \bibinfo {author} {\bibfnamefont {F.}~\bibnamefont {Deppe}}, \bibinfo
  {author} {\bibfnamefont {O.}~\bibnamefont {Fedorets}}, \bibinfo {author}
  {\bibfnamefont {Z.}~\bibnamefont {Gao}}, \bibinfo {author} {\bibfnamefont
  {A.~G.}\ \bibnamefont {Frieiro}}, \bibinfo {author} {\bibfnamefont
  {D.}~\bibnamefont {Gusenkova}}, \bibinfo {author} {\bibfnamefont
  {A.}~\bibnamefont {Guthrie}}, \bibinfo {author} {\bibfnamefont
  {T.}~\bibnamefont {Hiltunen}}, \bibinfo {author} {\bibfnamefont
  {H.}~\bibnamefont {Hsu}}, \bibinfo {author} {\bibfnamefont {E.}~\bibnamefont
  {Hyyppä}}, \bibinfo {author} {\bibfnamefont {J.}~\bibnamefont {Ikonen}},
  \bibinfo {author} {\bibfnamefont {S.}~\bibnamefont {Inel}}, \bibinfo {author}
  {\bibfnamefont {S.~W.}\ \bibnamefont {Jolin}}, \bibinfo {author}
  {\bibfnamefont {A.}~\bibnamefont {Karis}}, \bibinfo {author} {\bibfnamefont
  {S.-G.}\ \bibnamefont {Kim}}, \bibinfo {author} {\bibfnamefont
  {W.}~\bibnamefont {Kindel}}, \bibinfo {author} {\bibfnamefont
  {A.}~\bibnamefont {Komlev}}, \bibinfo {author} {\bibfnamefont
  {M.}~\bibnamefont {Koistinen}}, \bibinfo {author} {\bibfnamefont
  {R.}~\bibnamefont {Kokkoniemi}}, \bibinfo {author} {\bibfnamefont
  {S.}~\bibnamefont {Kumar}}, \bibinfo {author} {\bibfnamefont {H.-S.}\
  \bibnamefont {Ku}}, \bibinfo {author} {\bibfnamefont {J.}~\bibnamefont
  {Lamprich}}, \bibinfo {author} {\bibfnamefont {S.}~\bibnamefont {Laine}},
  \bibinfo {author} {\bibfnamefont {A.}~\bibnamefont {Landra}}, \bibinfo
  {author} {\bibfnamefont {L.-H.}\ \bibnamefont {Lee}}, \bibinfo {author}
  {\bibfnamefont {N.}~\bibnamefont {Lethif}}, \bibinfo {author} {\bibfnamefont
  {P.}~\bibnamefont {Liebermann}}, \bibinfo {author} {\bibfnamefont
  {W.}~\bibnamefont {Liu}}, \bibinfo {author} {\bibfnamefont {K.}~\bibnamefont
  {Mitra}}, \bibinfo {author} {\bibfnamefont {T.}~\bibnamefont {Mylläri}},
  \bibinfo {author} {\bibfnamefont {C.}~\bibnamefont {Ockeloen-Korppi}},
  \bibinfo {author} {\bibfnamefont {T.}~\bibnamefont {Orell}}, \bibinfo
  {author} {\bibfnamefont {A.}~\bibnamefont {Plyshch}}, \bibinfo {author}
  {\bibfnamefont {J.}~\bibnamefont {Räbinä}}, \bibinfo {author}
  {\bibfnamefont {A.}~\bibnamefont {Rebello}}, \bibinfo {author} {\bibfnamefont
  {M.}~\bibnamefont {Renger}}, \bibinfo {author} {\bibfnamefont
  {O.}~\bibnamefont {Reentilä}}, \bibinfo {author} {\bibfnamefont
  {J.}~\bibnamefont {Ritvas}}, \bibinfo {author} {\bibfnamefont
  {S.}~\bibnamefont {Saarinen}}, \bibinfo {author} {\bibfnamefont
  {O.}~\bibnamefont {Salmenkivi}}, \bibinfo {author} {\bibfnamefont
  {M.}~\bibnamefont {Sarsby}}, \bibinfo {author} {\bibfnamefont
  {M.}~\bibnamefont {Savytskyi}}, \bibinfo {author} {\bibfnamefont
  {V.}~\bibnamefont {Selinmaa}}, \bibinfo {author} {\bibfnamefont
  {M.}~\bibnamefont {Steggles}}, \bibinfo {author} {\bibfnamefont
  {E.}~\bibnamefont {Takala}}, \bibinfo {author} {\bibfnamefont
  {I.}~\bibnamefont {Takmakov}}, \bibinfo {author} {\bibfnamefont
  {B.}~\bibnamefont {Tarasinski}}, \bibinfo {author} {\bibfnamefont
  {J.}~\bibnamefont {Tuorila}}, \bibinfo {author} {\bibfnamefont
  {A.}~\bibnamefont {Välimaa}}, \bibinfo {author} {\bibfnamefont
  {J.}~\bibnamefont {Verjauw}}, \bibinfo {author} {\bibfnamefont
  {J.}~\bibnamefont {Wesdorp}}, \bibinfo {author} {\bibfnamefont
  {N.}~\bibnamefont {Wurz}}, \bibinfo {author} {\bibfnamefont {W.}~\bibnamefont
  {Qiu}}, \bibinfo {author} {\bibfnamefont {L.}~\bibnamefont {Zhu}}, \bibinfo
  {author} {\bibfnamefont {J.}~\bibnamefont {Hassel}}, \bibinfo {author}
  {\bibfnamefont {J.}~\bibnamefont {Heinsoo}}, \bibinfo {author} {\bibfnamefont
  {A.}~\bibnamefont {Geresdi}},\ and\ \bibinfo {author} {\bibfnamefont
  {A.}~\bibnamefont {Vepsäläinen}},\ }\href
  {https://doi.org/10.48550/arXiv.2508.16437} {\bibfield  {journal} {\bibinfo
  {journal} {arXiv preprint arXiv:2508.16437}\ } (\bibinfo {year}
  {2025})}\BibitemShut {NoStop}%
\bibitem [{\citenamefont {Fried}\ \emph {et~al.}(2019)\citenamefont {Fried},
  \citenamefont {Sivarajah}, \citenamefont {Didier}, \citenamefont {Sete},
  \citenamefont {da~Silva}, \citenamefont {Johnson},\ and\ \citenamefont
  {Ryan}}]{fried2019assessing}%
  \BibitemOpen
  \bibfield  {author} {\bibinfo {author} {\bibfnamefont {E.~S.}\ \bibnamefont
  {Fried}}, \bibinfo {author} {\bibfnamefont {P.}~\bibnamefont {Sivarajah}},
  \bibinfo {author} {\bibfnamefont {N.}~\bibnamefont {Didier}}, \bibinfo
  {author} {\bibfnamefont {E.~A.}\ \bibnamefont {Sete}}, \bibinfo {author}
  {\bibfnamefont {M.~P.}\ \bibnamefont {da~Silva}}, \bibinfo {author}
  {\bibfnamefont {B.~R.}\ \bibnamefont {Johnson}},\ and\ \bibinfo {author}
  {\bibfnamefont {C.~A.}\ \bibnamefont {Ryan}},\ }\href
  {https://arxiv.org/abs/1908.11370} {\bibfield  {journal} {\bibinfo  {journal}
  {arXiv preprint arXiv:1908.11370}\ } (\bibinfo {year} {2019})}\BibitemShut
  {NoStop}%
\bibitem [{\citenamefont {Didier}\ \emph {et~al.}(2019)\citenamefont {Didier},
  \citenamefont {Sete}, \citenamefont {Combes},\ and\ \citenamefont
  {da~Silva}}]{didier2019ac}%
  \BibitemOpen
  \bibfield  {author} {\bibinfo {author} {\bibfnamefont {N.}~\bibnamefont
  {Didier}}, \bibinfo {author} {\bibfnamefont {E.~A.}\ \bibnamefont {Sete}},
  \bibinfo {author} {\bibfnamefont {J.}~\bibnamefont {Combes}},\ and\ \bibinfo
  {author} {\bibfnamefont {M.~P.}\ \bibnamefont {da~Silva}},\ }\href
  {https://doi.org/10.1103/PhysRevApplied.12.054015} {\bibfield  {journal}
  {\bibinfo  {journal} {Phys. Rev. Appl.}\ }\textbf {\bibinfo {volume} {12}},\
  \bibinfo {pages} {054015} (\bibinfo {year} {2019})}\BibitemShut {NoStop}%
\bibitem [{\citenamefont {Wallman}\ \emph {et~al.}(2015)\citenamefont
  {Wallman}, \citenamefont {Granade}, \citenamefont {Harper},\ and\
  \citenamefont {Flammia}}]{wallman2015estimating}%
  \BibitemOpen
  \bibfield  {author} {\bibinfo {author} {\bibfnamefont {J.}~\bibnamefont
  {Wallman}}, \bibinfo {author} {\bibfnamefont {C.}~\bibnamefont {Granade}},
  \bibinfo {author} {\bibfnamefont {R.}~\bibnamefont {Harper}},\ and\ \bibinfo
  {author} {\bibfnamefont {S.~T.}\ \bibnamefont {Flammia}},\ }\href
  {https://doi.org/10.1088/1367-2630/17/11/113020} {\bibfield  {journal}
  {\bibinfo  {journal} {New Journal of Physics}\ }\textbf {\bibinfo {volume}
  {17}},\ \bibinfo {pages} {113020} (\bibinfo {year} {2015})}\BibitemShut
  {NoStop}%
\bibitem [{\citenamefont {Arute}\ \emph {et~al.}(2019)\citenamefont {Arute},
  \citenamefont {Arya}, \citenamefont {Babbush}, \citenamefont {Bacon},
  \citenamefont {Bardin}, \citenamefont {Barends}, \citenamefont {Biswas},
  \citenamefont {Boixo}, \citenamefont {Brandao}, \citenamefont {Buell},
  \citenamefont {Burkett}, \citenamefont {Chen}, \citenamefont {Chen},
  \citenamefont {Chiaro}, \citenamefont {Collins}, \citenamefont {Courtney},
  \citenamefont {Dunsworth}, \citenamefont {Farhi}, \citenamefont {Foxen},
  \citenamefont {Fowler}, \citenamefont {Gidney}, \citenamefont {Giustina},
  \citenamefont {Graff}, \citenamefont {Guerin}, \citenamefont {Habegger},
  \citenamefont {Harrigan}, \citenamefont {Hartmann}, \citenamefont {Ho},
  \citenamefont {Hoffmann}, \citenamefont {Huang}, \citenamefont {Humble},
  \citenamefont {Isakov}, \citenamefont {Jeffrey}, \citenamefont {Jiang},
  \citenamefont {Kafri}, \citenamefont {Kechedzhi}, \citenamefont {Kelly},
  \citenamefont {Klimov}, \citenamefont {Knysh}, \citenamefont {Korotkov},
  \citenamefont {Kostritsa}, \citenamefont {Landhuis}, \citenamefont
  {Lindmark}, \citenamefont {Lucero}, \citenamefont {Lyakh}, \citenamefont
  {Mandr\`a}, \citenamefont {{McClean}}, \citenamefont {{McEwen}},
  \citenamefont {Megrant}, \citenamefont {Mi}, \citenamefont {Michielsen},
  \citenamefont {Mohseni}, \citenamefont {Mutus}, \citenamefont {Naaman},
  \citenamefont {Neeley}, \citenamefont {Neill}, \citenamefont {Niu},
  \citenamefont {Ostby}, \citenamefont {Petukhov}, \citenamefont {Platt},
  \citenamefont {Quintana}, \citenamefont {Rieffel}, \citenamefont {Roushan},
  \citenamefont {Rubin}, \citenamefont {Sank}, \citenamefont {Satzinger},
  \citenamefont {Smelyanskiy}, \citenamefont {Sung}, \citenamefont
  {Trevithick}, \citenamefont {Vainsencher}, \citenamefont {Villalonga},
  \citenamefont {White}, \citenamefont {Yao}, \citenamefont {Yeh},
  \citenamefont {Zalcman}, \citenamefont {Neven},\ and\ \citenamefont
  {Martinis}}]{arute2019quantum}%
  \BibitemOpen
  \bibfield  {author} {\bibinfo {author} {\bibfnamefont {F.}~\bibnamefont
  {Arute}}, \bibinfo {author} {\bibfnamefont {K.}~\bibnamefont {Arya}},
  \bibinfo {author} {\bibfnamefont {R.}~\bibnamefont {Babbush}}, \bibinfo
  {author} {\bibfnamefont {D.}~\bibnamefont {Bacon}}, \bibinfo {author}
  {\bibfnamefont {J.~C.}\ \bibnamefont {Bardin}}, \bibinfo {author}
  {\bibfnamefont {R.}~\bibnamefont {Barends}}, \bibinfo {author} {\bibfnamefont
  {R.}~\bibnamefont {Biswas}}, \bibinfo {author} {\bibfnamefont
  {S.}~\bibnamefont {Boixo}}, \bibinfo {author} {\bibfnamefont {F.~G. S.~L.}\
  \bibnamefont {Brandao}}, \bibinfo {author} {\bibfnamefont {D.~A.}\
  \bibnamefont {Buell}}, \bibinfo {author} {\bibfnamefont {B.}~\bibnamefont
  {Burkett}}, \bibinfo {author} {\bibfnamefont {Y.}~\bibnamefont {Chen}},
  \bibinfo {author} {\bibfnamefont {Z.}~\bibnamefont {Chen}}, \bibinfo {author}
  {\bibfnamefont {B.}~\bibnamefont {Chiaro}}, \bibinfo {author} {\bibfnamefont
  {R.}~\bibnamefont {Collins}}, \bibinfo {author} {\bibfnamefont
  {W.}~\bibnamefont {Courtney}}, \bibinfo {author} {\bibfnamefont
  {A.}~\bibnamefont {Dunsworth}}, \bibinfo {author} {\bibfnamefont
  {E.}~\bibnamefont {Farhi}}, \bibinfo {author} {\bibfnamefont
  {B.}~\bibnamefont {Foxen}}, \bibinfo {author} {\bibfnamefont
  {A.}~\bibnamefont {Fowler}}, \bibinfo {author} {\bibfnamefont
  {C.}~\bibnamefont {Gidney}}, \bibinfo {author} {\bibfnamefont
  {M.}~\bibnamefont {Giustina}}, \bibinfo {author} {\bibfnamefont
  {R.}~\bibnamefont {Graff}}, \bibinfo {author} {\bibfnamefont
  {K.}~\bibnamefont {Guerin}}, \bibinfo {author} {\bibfnamefont
  {S.}~\bibnamefont {Habegger}}, \bibinfo {author} {\bibfnamefont {M.~P.}\
  \bibnamefont {Harrigan}}, \bibinfo {author} {\bibfnamefont {M.~J.}\
  \bibnamefont {Hartmann}}, \bibinfo {author} {\bibfnamefont {A.}~\bibnamefont
  {Ho}}, \bibinfo {author} {\bibfnamefont {M.}~\bibnamefont {Hoffmann}},
  \bibinfo {author} {\bibfnamefont {T.}~\bibnamefont {Huang}}, \bibinfo
  {author} {\bibfnamefont {T.~S.}\ \bibnamefont {Humble}}, \bibinfo {author}
  {\bibfnamefont {S.~V.}\ \bibnamefont {Isakov}}, \bibinfo {author}
  {\bibfnamefont {E.}~\bibnamefont {Jeffrey}}, \bibinfo {author} {\bibfnamefont
  {Z.}~\bibnamefont {Jiang}}, \bibinfo {author} {\bibfnamefont
  {D.}~\bibnamefont {Kafri}}, \bibinfo {author} {\bibfnamefont
  {K.}~\bibnamefont {Kechedzhi}}, \bibinfo {author} {\bibfnamefont
  {J.}~\bibnamefont {Kelly}}, \bibinfo {author} {\bibfnamefont {P.~V.}\
  \bibnamefont {Klimov}}, \bibinfo {author} {\bibfnamefont {S.}~\bibnamefont
  {Knysh}}, \bibinfo {author} {\bibfnamefont {A.}~\bibnamefont {Korotkov}},
  \bibinfo {author} {\bibfnamefont {F.}~\bibnamefont {Kostritsa}}, \bibinfo
  {author} {\bibfnamefont {D.}~\bibnamefont {Landhuis}}, \bibinfo {author}
  {\bibfnamefont {M.}~\bibnamefont {Lindmark}}, \bibinfo {author}
  {\bibfnamefont {E.}~\bibnamefont {Lucero}}, \bibinfo {author} {\bibfnamefont
  {D.}~\bibnamefont {Lyakh}}, \bibinfo {author} {\bibfnamefont
  {S.}~\bibnamefont {Mandr\`a}}, \bibinfo {author} {\bibfnamefont {J.~R.}\
  \bibnamefont {{McClean}}}, \bibinfo {author} {\bibfnamefont {M.}~\bibnamefont
  {{McEwen}}}, \bibinfo {author} {\bibfnamefont {A.}~\bibnamefont {Megrant}},
  \bibinfo {author} {\bibfnamefont {X.}~\bibnamefont {Mi}}, \bibinfo {author}
  {\bibfnamefont {K.}~\bibnamefont {Michielsen}}, \bibinfo {author}
  {\bibfnamefont {M.}~\bibnamefont {Mohseni}}, \bibinfo {author} {\bibfnamefont
  {J.}~\bibnamefont {Mutus}}, \bibinfo {author} {\bibfnamefont
  {O.}~\bibnamefont {Naaman}}, \bibinfo {author} {\bibfnamefont
  {M.}~\bibnamefont {Neeley}}, \bibinfo {author} {\bibfnamefont
  {C.}~\bibnamefont {Neill}}, \bibinfo {author} {\bibfnamefont {M.~Y.}\
  \bibnamefont {Niu}}, \bibinfo {author} {\bibfnamefont {E.}~\bibnamefont
  {Ostby}}, \bibinfo {author} {\bibfnamefont {A.}~\bibnamefont {Petukhov}},
  \bibinfo {author} {\bibfnamefont {J.~C.}\ \bibnamefont {Platt}}, \bibinfo
  {author} {\bibfnamefont {C.}~\bibnamefont {Quintana}}, \bibinfo {author}
  {\bibfnamefont {E.~G.}\ \bibnamefont {Rieffel}}, \bibinfo {author}
  {\bibfnamefont {P.}~\bibnamefont {Roushan}}, \bibinfo {author} {\bibfnamefont
  {N.~C.}\ \bibnamefont {Rubin}}, \bibinfo {author} {\bibfnamefont
  {D.}~\bibnamefont {Sank}}, \bibinfo {author} {\bibfnamefont {K.~J.}\
  \bibnamefont {Satzinger}}, \bibinfo {author} {\bibfnamefont {V.}~\bibnamefont
  {Smelyanskiy}}, \bibinfo {author} {\bibfnamefont {K.~J.}\ \bibnamefont
  {Sung}}, \bibinfo {author} {\bibfnamefont {M.~D.}\ \bibnamefont
  {Trevithick}}, \bibinfo {author} {\bibfnamefont {A.}~\bibnamefont
  {Vainsencher}}, \bibinfo {author} {\bibfnamefont {B.}~\bibnamefont
  {Villalonga}}, \bibinfo {author} {\bibfnamefont {T.}~\bibnamefont {White}},
  \bibinfo {author} {\bibfnamefont {Z.~J.}\ \bibnamefont {Yao}}, \bibinfo
  {author} {\bibfnamefont {P.}~\bibnamefont {Yeh}}, \bibinfo {author}
  {\bibfnamefont {A.}~\bibnamefont {Zalcman}}, \bibinfo {author} {\bibfnamefont
  {H.}~\bibnamefont {Neven}},\ and\ \bibinfo {author} {\bibfnamefont {J.~M.}\
  \bibnamefont {Martinis}},\ }\href
  {https://www.nature.com/articles/s41586-019-1666-5} {\bibfield  {journal}
  {\bibinfo  {journal} {nature}\ }\textbf {\bibinfo {volume} {574}},\ \bibinfo
  {pages} {505} (\bibinfo {year} {2019})}\BibitemShut {NoStop}%
\bibitem [{\citenamefont {Jin}\ \emph {et~al.}(2025)\citenamefont {Jin},
  \citenamefont {Parrott}, \citenamefont {Cicak}, \citenamefont {Kotler},
  \citenamefont {Lecocq}, \citenamefont {Teufel}, \citenamefont {Aumentado},
  \citenamefont {Kapit},\ and\ \citenamefont
  {Simmonds}}]{jin2025superconducting}%
  \BibitemOpen
  \bibfield  {author} {\bibinfo {author} {\bibfnamefont {X.}~\bibnamefont
  {Jin}}, \bibinfo {author} {\bibfnamefont {Z.}~\bibnamefont {Parrott}},
  \bibinfo {author} {\bibfnamefont {K.}~\bibnamefont {Cicak}}, \bibinfo
  {author} {\bibfnamefont {S.}~\bibnamefont {Kotler}}, \bibinfo {author}
  {\bibfnamefont {F.}~\bibnamefont {Lecocq}}, \bibinfo {author} {\bibfnamefont
  {J.}~\bibnamefont {Teufel}}, \bibinfo {author} {\bibfnamefont
  {J.}~\bibnamefont {Aumentado}}, \bibinfo {author} {\bibfnamefont
  {E.}~\bibnamefont {Kapit}},\ and\ \bibinfo {author} {\bibfnamefont
  {R.}~\bibnamefont {Simmonds}},\ }\href {https://doi.org/10.1103/kmls-lgp5}
  {\bibfield  {journal} {\bibinfo  {journal} {Phys. Rev. Appl.}\ }\textbf
  {\bibinfo {volume} {24}},\ \bibinfo {pages} {064026} (\bibinfo {year}
  {2025})}\BibitemShut {NoStop}%
\bibitem [{\citenamefont {Braum\"uller}\ \emph {et~al.}(2020)\citenamefont
  {Braum\"uller}, \citenamefont {Ding}, \citenamefont {Veps\"al\"ainen},
  \citenamefont {Sung}, \citenamefont {Kjaergaard}, \citenamefont {Menke},
  \citenamefont {Winik}, \citenamefont {Kim}, \citenamefont {Niedzielski},
  \citenamefont {Melville}, \citenamefont {Yoder}, \citenamefont
  {Hirjibehedin}, \citenamefont {Orlando}, \citenamefont {Gustavsson},\ and\
  \citenamefont {Oliver}}]{braumuller2020characterizing}%
  \BibitemOpen
  \bibfield  {author} {\bibinfo {author} {\bibfnamefont {J.}~\bibnamefont
  {Braum\"uller}}, \bibinfo {author} {\bibfnamefont {L.}~\bibnamefont {Ding}},
  \bibinfo {author} {\bibfnamefont {A.~P.}\ \bibnamefont {Veps\"al\"ainen}},
  \bibinfo {author} {\bibfnamefont {Y.}~\bibnamefont {Sung}}, \bibinfo {author}
  {\bibfnamefont {M.}~\bibnamefont {Kjaergaard}}, \bibinfo {author}
  {\bibfnamefont {T.}~\bibnamefont {Menke}}, \bibinfo {author} {\bibfnamefont
  {R.}~\bibnamefont {Winik}}, \bibinfo {author} {\bibfnamefont
  {D.}~\bibnamefont {Kim}}, \bibinfo {author} {\bibfnamefont {B.~M.}\
  \bibnamefont {Niedzielski}}, \bibinfo {author} {\bibfnamefont
  {A.}~\bibnamefont {Melville}}, \bibinfo {author} {\bibfnamefont {J.~L.}\
  \bibnamefont {Yoder}}, \bibinfo {author} {\bibfnamefont {C.~F.}\ \bibnamefont
  {Hirjibehedin}}, \bibinfo {author} {\bibfnamefont {T.~P.}\ \bibnamefont
  {Orlando}}, \bibinfo {author} {\bibfnamefont {S.}~\bibnamefont
  {Gustavsson}},\ and\ \bibinfo {author} {\bibfnamefont {W.~D.}\ \bibnamefont
  {Oliver}},\ }\href {https://doi.org/10.1103/PhysRevApplied.13.054079}
  {\bibfield  {journal} {\bibinfo  {journal} {Phys. Rev. Appl.}\ }\textbf
  {\bibinfo {volume} {13}},\ \bibinfo {pages} {054079} (\bibinfo {year}
  {2020})}\BibitemShut {NoStop}%
\bibitem [{\citenamefont {Blais}\ \emph {et~al.}(2021)\citenamefont {Blais},
  \citenamefont {Grimsmo}, \citenamefont {Girvin},\ and\ \citenamefont
  {Wallraff}}]{blais2021circuit}%
  \BibitemOpen
  \bibfield  {author} {\bibinfo {author} {\bibfnamefont {A.}~\bibnamefont
  {Blais}}, \bibinfo {author} {\bibfnamefont {A.~L.}\ \bibnamefont {Grimsmo}},
  \bibinfo {author} {\bibfnamefont {S.~M.}\ \bibnamefont {Girvin}},\ and\
  \bibinfo {author} {\bibfnamefont {A.}~\bibnamefont {Wallraff}},\ }\href
  {https://doi.org/10.1103/RevModPhys.93.025005} {\bibfield  {journal}
  {\bibinfo  {journal} {Rev. Mod. Phys.}\ }\textbf {\bibinfo {volume} {93}},\
  \bibinfo {pages} {025005} (\bibinfo {year} {2021})}\BibitemShut {NoStop}%
\bibitem [{\citenamefont {Huber}\ \emph {et~al.}(2025)\citenamefont {Huber},
  \citenamefont {Roy}, \citenamefont {Koch}, \citenamefont {Tsitsilin},
  \citenamefont {Schirk}, \citenamefont {Glaser}, \citenamefont {Bruckmoser},
  \citenamefont {Schweizer}, \citenamefont {Romeiro}, \citenamefont {Krylov},
  \citenamefont {Singh}, \citenamefont {Haslbeck}, \citenamefont {Knudsen},
  \citenamefont {Marx}, \citenamefont {Pfeiffer}, \citenamefont {Schneider},
  \citenamefont {Wallner}, \citenamefont {Bunch}, \citenamefont {Richard},
  \citenamefont {S\"odergren}, \citenamefont {Liegener}, \citenamefont
  {Werninghaus},\ and\ \citenamefont {Filipp}}]{huber2025parametric}%
  \BibitemOpen
  \bibfield  {author} {\bibinfo {author} {\bibfnamefont {G.}~\bibnamefont
  {Huber}}, \bibinfo {author} {\bibfnamefont {F.}~\bibnamefont {Roy}}, \bibinfo
  {author} {\bibfnamefont {L.}~\bibnamefont {Koch}}, \bibinfo {author}
  {\bibfnamefont {I.}~\bibnamefont {Tsitsilin}}, \bibinfo {author}
  {\bibfnamefont {J.}~\bibnamefont {Schirk}}, \bibinfo {author} {\bibfnamefont
  {N.}~\bibnamefont {Glaser}}, \bibinfo {author} {\bibfnamefont
  {N.}~\bibnamefont {Bruckmoser}}, \bibinfo {author} {\bibfnamefont
  {C.}~\bibnamefont {Schweizer}}, \bibinfo {author} {\bibfnamefont
  {J.}~\bibnamefont {Romeiro}}, \bibinfo {author} {\bibfnamefont
  {G.}~\bibnamefont {Krylov}}, \bibinfo {author} {\bibfnamefont
  {M.}~\bibnamefont {Singh}}, \bibinfo {author} {\bibfnamefont
  {F.}~\bibnamefont {Haslbeck}}, \bibinfo {author} {\bibfnamefont
  {M.}~\bibnamefont {Knudsen}}, \bibinfo {author} {\bibfnamefont
  {A.}~\bibnamefont {Marx}}, \bibinfo {author} {\bibfnamefont {F.}~\bibnamefont
  {Pfeiffer}}, \bibinfo {author} {\bibfnamefont {C.}~\bibnamefont {Schneider}},
  \bibinfo {author} {\bibfnamefont {F.}~\bibnamefont {Wallner}}, \bibinfo
  {author} {\bibfnamefont {D.}~\bibnamefont {Bunch}}, \bibinfo {author}
  {\bibfnamefont {L.}~\bibnamefont {Richard}}, \bibinfo {author} {\bibfnamefont
  {L.}~\bibnamefont {S\"odergren}}, \bibinfo {author} {\bibfnamefont
  {K.}~\bibnamefont {Liegener}}, \bibinfo {author} {\bibfnamefont
  {M.}~\bibnamefont {Werninghaus}},\ and\ \bibinfo {author} {\bibfnamefont
  {S.}~\bibnamefont {Filipp}},\ }\href {https://doi.org/10.1103/9shv-l4cx}
  {\bibfield  {journal} {\bibinfo  {journal} {PRX Quantum}\ }\textbf {\bibinfo
  {volume} {6}},\ \bibinfo {pages} {030313} (\bibinfo {year}
  {2025})}\BibitemShut {NoStop}%
\bibitem [{\citenamefont {Ma}\ \emph {et~al.}(2025)\citenamefont {Ma},
  \citenamefont {Li}, \citenamefont {Shi}, \citenamefont {Guo}, \citenamefont
  {Xu}, \citenamefont {Tan},\ and\ \citenamefont {Yu}}]{ma2025parametric}%
  \BibitemOpen
  \bibfield  {author} {\bibinfo {author} {\bibfnamefont {Z.}~\bibnamefont
  {Ma}}, \bibinfo {author} {\bibfnamefont {X.}~\bibnamefont {Li}}, \bibinfo
  {author} {\bibfnamefont {H.}~\bibnamefont {Shi}}, \bibinfo {author}
  {\bibfnamefont {R.}~\bibnamefont {Guo}}, \bibinfo {author} {\bibfnamefont
  {J.}~\bibnamefont {Xu}}, \bibinfo {author} {\bibfnamefont {X.}~\bibnamefont
  {Tan}},\ and\ \bibinfo {author} {\bibfnamefont {Y.}~\bibnamefont {Yu}},\
  }\href {https://doi.org/10.1103/q9wv-wz9m} {\bibfield  {journal} {\bibinfo
  {journal} {Phys. Rev. Appl.}\ }\textbf {\bibinfo {volume} {24}},\ \bibinfo
  {pages} {054067} (\bibinfo {year} {2025})}\BibitemShut {NoStop}%
\bibitem [{\citenamefont {Petrescu}\ \emph {et~al.}(2023)\citenamefont
  {Petrescu}, \citenamefont {Le~Calonnec}, \citenamefont {Leroux},
  \citenamefont {Di~Paolo}, \citenamefont {Mundada}, \citenamefont {Sussman},
  \citenamefont {Vrajitoarea}, \citenamefont {Houck},\ and\ \citenamefont
  {Blais}}]{petrescu2023accurate}%
  \BibitemOpen
  \bibfield  {author} {\bibinfo {author} {\bibfnamefont {A.}~\bibnamefont
  {Petrescu}}, \bibinfo {author} {\bibfnamefont {C.}~\bibnamefont
  {Le~Calonnec}}, \bibinfo {author} {\bibfnamefont {C.}~\bibnamefont {Leroux}},
  \bibinfo {author} {\bibfnamefont {A.}~\bibnamefont {Di~Paolo}}, \bibinfo
  {author} {\bibfnamefont {P.}~\bibnamefont {Mundada}}, \bibinfo {author}
  {\bibfnamefont {S.}~\bibnamefont {Sussman}}, \bibinfo {author} {\bibfnamefont
  {A.}~\bibnamefont {Vrajitoarea}}, \bibinfo {author} {\bibfnamefont {A.~A.}\
  \bibnamefont {Houck}},\ and\ \bibinfo {author} {\bibfnamefont
  {A.}~\bibnamefont {Blais}},\ }\href
  {https://doi.org/10.1103/PhysRevApplied.19.044003} {\bibfield  {journal}
  {\bibinfo  {journal} {Phys. Rev. Appl.}\ }\textbf {\bibinfo {volume} {19}},\
  \bibinfo {pages} {044003} (\bibinfo {year} {2023})}\BibitemShut {NoStop}%
\bibitem [{\citenamefont {Johansson}\ \emph {et~al.}(2012)\citenamefont
  {Johansson}, \citenamefont {Nation},\ and\ \citenamefont
  {Nori}}]{johansson2012qutip}%
  \BibitemOpen
  \bibfield  {author} {\bibinfo {author} {\bibfnamefont {J.}~\bibnamefont
  {Johansson}}, \bibinfo {author} {\bibfnamefont {P.}~\bibnamefont {Nation}},\
  and\ \bibinfo {author} {\bibfnamefont {F.}~\bibnamefont {Nori}},\ }\href
  {https://doi.org/https://doi.org/10.1016/j.cpc.2012.02.021} {\bibfield
  {journal} {\bibinfo  {journal} {Computer Physics Communications}\ }\textbf
  {\bibinfo {volume} {183}},\ \bibinfo {pages} {1760} (\bibinfo {year}
  {2012})}\BibitemShut {NoStop}%
\bibitem [{\citenamefont {Virtanen}\ \emph {et~al.}(2020)\citenamefont
  {Virtanen}, \citenamefont {Gommers}, \citenamefont {Oliphant}, \citenamefont
  {Haberland}, \citenamefont {Reddy}, \citenamefont {Cournapeau}, \citenamefont
  {Burovski}, \citenamefont {Peterson}, \citenamefont {Weckesser},
  \citenamefont {Bright}, \citenamefont {{van der Walt}}, \citenamefont
  {Brett}, \citenamefont {Wilson}, \citenamefont {Millman}, \citenamefont
  {Mayorov}, \citenamefont {Nelson}, \citenamefont {Jones}, \citenamefont
  {Kern}, \citenamefont {Larson}, \citenamefont {Carey}, \citenamefont {Polat},
  \citenamefont {Feng}, \citenamefont {Moore}, \citenamefont {{VanderPlas}},
  \citenamefont {Laxalde}, \citenamefont {Perktold}, \citenamefont {Cimrman},
  \citenamefont {Henriksen}, \citenamefont {Quintero}, \citenamefont {Harris},
  \citenamefont {Archibald}, \citenamefont {Ribeiro}, \citenamefont
  {Pedregosa}, \citenamefont {{van Mulbregt}},\ and\ \citenamefont {{SciPy 1.0
  Contributors}}}]{virtanen2020scipy}%
  \BibitemOpen
  \bibfield  {author} {\bibinfo {author} {\bibfnamefont {P.}~\bibnamefont
  {Virtanen}}, \bibinfo {author} {\bibfnamefont {R.}~\bibnamefont {Gommers}},
  \bibinfo {author} {\bibfnamefont {T.~E.}\ \bibnamefont {Oliphant}}, \bibinfo
  {author} {\bibfnamefont {M.}~\bibnamefont {Haberland}}, \bibinfo {author}
  {\bibfnamefont {T.}~\bibnamefont {Reddy}}, \bibinfo {author} {\bibfnamefont
  {D.}~\bibnamefont {Cournapeau}}, \bibinfo {author} {\bibfnamefont
  {E.}~\bibnamefont {Burovski}}, \bibinfo {author} {\bibfnamefont
  {P.}~\bibnamefont {Peterson}}, \bibinfo {author} {\bibfnamefont
  {W.}~\bibnamefont {Weckesser}}, \bibinfo {author} {\bibfnamefont
  {J.}~\bibnamefont {Bright}}, \bibinfo {author} {\bibfnamefont {S.~J.}\
  \bibnamefont {{van der Walt}}}, \bibinfo {author} {\bibfnamefont
  {M.}~\bibnamefont {Brett}}, \bibinfo {author} {\bibfnamefont
  {J.}~\bibnamefont {Wilson}}, \bibinfo {author} {\bibfnamefont {K.~J.}\
  \bibnamefont {Millman}}, \bibinfo {author} {\bibfnamefont {N.}~\bibnamefont
  {Mayorov}}, \bibinfo {author} {\bibfnamefont {A.~R.~J.}\ \bibnamefont
  {Nelson}}, \bibinfo {author} {\bibfnamefont {E.}~\bibnamefont {Jones}},
  \bibinfo {author} {\bibfnamefont {R.}~\bibnamefont {Kern}}, \bibinfo {author}
  {\bibfnamefont {E.}~\bibnamefont {Larson}}, \bibinfo {author} {\bibfnamefont
  {C.~J.}\ \bibnamefont {Carey}}, \bibinfo {author} {\bibfnamefont
  {{\.I}.}~\bibnamefont {Polat}}, \bibinfo {author} {\bibfnamefont
  {Y.}~\bibnamefont {Feng}}, \bibinfo {author} {\bibfnamefont {E.~W.}\
  \bibnamefont {Moore}}, \bibinfo {author} {\bibfnamefont {J.}~\bibnamefont
  {{VanderPlas}}}, \bibinfo {author} {\bibfnamefont {D.}~\bibnamefont
  {Laxalde}}, \bibinfo {author} {\bibfnamefont {J.}~\bibnamefont {Perktold}},
  \bibinfo {author} {\bibfnamefont {R.}~\bibnamefont {Cimrman}}, \bibinfo
  {author} {\bibfnamefont {I.}~\bibnamefont {Henriksen}}, \bibinfo {author}
  {\bibfnamefont {E.~A.}\ \bibnamefont {Quintero}}, \bibinfo {author}
  {\bibfnamefont {C.~R.}\ \bibnamefont {Harris}}, \bibinfo {author}
  {\bibfnamefont {A.~M.}\ \bibnamefont {Archibald}}, \bibinfo {author}
  {\bibfnamefont {A.~H.}\ \bibnamefont {Ribeiro}}, \bibinfo {author}
  {\bibfnamefont {F.}~\bibnamefont {Pedregosa}}, \bibinfo {author}
  {\bibfnamefont {P.}~\bibnamefont {{van Mulbregt}}},\ and\ \bibinfo {author}
  {\bibnamefont {{SciPy 1.0 Contributors}}},\ }\href
  {https://doi.org/10.1038/s41592-019-0686-2} {\bibfield  {journal} {\bibinfo
  {journal} {Nature Methods}\ }\textbf {\bibinfo {volume} {17}},\ \bibinfo
  {pages} {261} (\bibinfo {year} {2020})}\BibitemShut {NoStop}%
\bibitem [{\citenamefont {Powell}(1964)}]{powell1964efficient}%
  \BibitemOpen
  \bibfield  {author} {\bibinfo {author} {\bibfnamefont {M.~J.~D.}\
  \bibnamefont {Powell}},\ }\href {https://doi.org/10.1093/comjnl/7.2.155}
  {\bibfield  {journal} {\bibinfo  {journal} {The Computer Journal}\ }\textbf
  {\bibinfo {volume} {7}},\ \bibinfo {pages} {155} (\bibinfo {year}
  {1964})}\BibitemShut {NoStop}%
\bibitem [{\citenamefont {C\'orcoles}\ \emph {et~al.}(2013)\citenamefont
  {C\'orcoles}, \citenamefont {Gambetta}, \citenamefont {Chow}, \citenamefont
  {Smolin}, \citenamefont {Ware}, \citenamefont {Strand}, \citenamefont
  {Plourde},\ and\ \citenamefont {Steffen}}]{corcoles2013process}%
  \BibitemOpen
  \bibfield  {author} {\bibinfo {author} {\bibfnamefont {A.~D.}\ \bibnamefont
  {C\'orcoles}}, \bibinfo {author} {\bibfnamefont {J.~M.}\ \bibnamefont
  {Gambetta}}, \bibinfo {author} {\bibfnamefont {J.~M.}\ \bibnamefont {Chow}},
  \bibinfo {author} {\bibfnamefont {J.~A.}\ \bibnamefont {Smolin}}, \bibinfo
  {author} {\bibfnamefont {M.}~\bibnamefont {Ware}}, \bibinfo {author}
  {\bibfnamefont {J.}~\bibnamefont {Strand}}, \bibinfo {author} {\bibfnamefont
  {B.~L.~T.}\ \bibnamefont {Plourde}},\ and\ \bibinfo {author} {\bibfnamefont
  {M.}~\bibnamefont {Steffen}},\ }\href
  {https://doi.org/10.1103/PhysRevA.87.030301} {\bibfield  {journal} {\bibinfo
  {journal} {Phys. Rev. A}\ }\textbf {\bibinfo {volume} {87}},\ \bibinfo
  {pages} {030301} (\bibinfo {year} {2013})}\BibitemShut {NoStop}%
\bibitem [{\citenamefont {Arute}\ \emph {et~al.}(2020)\citenamefont {Arute},
  \citenamefont {Arya}, \citenamefont {Babbush}, \citenamefont {Bacon},
  \citenamefont {Bardin}, \citenamefont {Barends}, \citenamefont {Bengtsson},
  \citenamefont {Boixo}, \citenamefont {Broughton}, \citenamefont {Buckley},
  \citenamefont {Buell}, \citenamefont {Burkett}, \citenamefont {Bushnell},
  \citenamefont {Chen}, \citenamefont {Chen}, \citenamefont {Chen},
  \citenamefont {Chiaro}, \citenamefont {Collins}, \citenamefont {Cotton},
  \citenamefont {Courtney}, \citenamefont {Demura}, \citenamefont {Derk},
  \citenamefont {Dunsworth}, \citenamefont {Eppens}, \citenamefont {Eckl},
  \citenamefont {Erickson}, \citenamefont {Farhi}, \citenamefont {Fowler},
  \citenamefont {Foxen}, \citenamefont {Gidney}, \citenamefont {Giustina},
  \citenamefont {Graff}, \citenamefont {Gross}, \citenamefont {Habegger},
  \citenamefont {Harrigan}, \citenamefont {Ho}, \citenamefont {Hong},
  \citenamefont {Huang}, \citenamefont {Huggins}, \citenamefont {Ioffe},
  \citenamefont {Isakov}, \citenamefont {Jeffrey}, \citenamefont {Jiang},
  \citenamefont {Jones}, \citenamefont {Kafri}, \citenamefont {Kechedzhi},
  \citenamefont {Kelly}, \citenamefont {Kim}, \citenamefont {Klimov},
  \citenamefont {Korotkov}, \citenamefont {Kostritsa}, \citenamefont
  {Landhuis}, \citenamefont {Laptev}, \citenamefont {Lindmark}, \citenamefont
  {Lucero}, \citenamefont {Marthaler}, \citenamefont {Martin}, \citenamefont
  {Martinis}, \citenamefont {Marusczyk}, \citenamefont {McArdle}, \citenamefont
  {McClean}, \citenamefont {McCourt}, \citenamefont {McEwen}, \citenamefont
  {Megrant}, \citenamefont {Mejuto-Zaera}, \citenamefont {Mi}, \citenamefont
  {Mohseni}, \citenamefont {Mruczkiewicz}, \citenamefont {Mutus}, \citenamefont
  {Naaman}, \citenamefont {Neeley}, \citenamefont {Neill}, \citenamefont
  {Neven}, \citenamefont {Newman}, \citenamefont {Niu}, \citenamefont
  {O'Brien}, \citenamefont {Ostby}, \citenamefont {Pató}, \citenamefont
  {Petukhov}, \citenamefont {Putterman}, \citenamefont {Quintana},
  \citenamefont {Reiner}, \citenamefont {Roushan}, \citenamefont {Rubin},
  \citenamefont {Sank}, \citenamefont {Satzinger}, \citenamefont {Smelyanskiy},
  \citenamefont {Strain}, \citenamefont {Sung}, \citenamefont {Schmitteckert},
  \citenamefont {Szalay}, \citenamefont {Tubman}, \citenamefont {Vainsencher},
  \citenamefont {White}, \citenamefont {Vogt}, \citenamefont {Yao},
  \citenamefont {Yeh}, \citenamefont {Zalcman},\ and\ \citenamefont
  {Zanker}}]{arute2020observation}%
  \BibitemOpen
  \bibfield  {author} {\bibinfo {author} {\bibfnamefont {F.}~\bibnamefont
  {Arute}}, \bibinfo {author} {\bibfnamefont {K.}~\bibnamefont {Arya}},
  \bibinfo {author} {\bibfnamefont {R.}~\bibnamefont {Babbush}}, \bibinfo
  {author} {\bibfnamefont {D.}~\bibnamefont {Bacon}}, \bibinfo {author}
  {\bibfnamefont {J.~C.}\ \bibnamefont {Bardin}}, \bibinfo {author}
  {\bibfnamefont {R.}~\bibnamefont {Barends}}, \bibinfo {author} {\bibfnamefont
  {A.}~\bibnamefont {Bengtsson}}, \bibinfo {author} {\bibfnamefont
  {S.}~\bibnamefont {Boixo}}, \bibinfo {author} {\bibfnamefont
  {M.}~\bibnamefont {Broughton}}, \bibinfo {author} {\bibfnamefont {B.~B.}\
  \bibnamefont {Buckley}}, \bibinfo {author} {\bibfnamefont {D.~A.}\
  \bibnamefont {Buell}}, \bibinfo {author} {\bibfnamefont {B.}~\bibnamefont
  {Burkett}}, \bibinfo {author} {\bibfnamefont {N.}~\bibnamefont {Bushnell}},
  \bibinfo {author} {\bibfnamefont {Y.}~\bibnamefont {Chen}}, \bibinfo {author}
  {\bibfnamefont {Z.}~\bibnamefont {Chen}}, \bibinfo {author} {\bibfnamefont
  {Y.-A.}\ \bibnamefont {Chen}}, \bibinfo {author} {\bibfnamefont
  {B.}~\bibnamefont {Chiaro}}, \bibinfo {author} {\bibfnamefont
  {R.}~\bibnamefont {Collins}}, \bibinfo {author} {\bibfnamefont {S.~J.}\
  \bibnamefont {Cotton}}, \bibinfo {author} {\bibfnamefont {W.}~\bibnamefont
  {Courtney}}, \bibinfo {author} {\bibfnamefont {S.}~\bibnamefont {Demura}},
  \bibinfo {author} {\bibfnamefont {A.}~\bibnamefont {Derk}}, \bibinfo {author}
  {\bibfnamefont {A.}~\bibnamefont {Dunsworth}}, \bibinfo {author}
  {\bibfnamefont {D.}~\bibnamefont {Eppens}}, \bibinfo {author} {\bibfnamefont
  {T.}~\bibnamefont {Eckl}}, \bibinfo {author} {\bibfnamefont {C.}~\bibnamefont
  {Erickson}}, \bibinfo {author} {\bibfnamefont {E.}~\bibnamefont {Farhi}},
  \bibinfo {author} {\bibfnamefont {A.}~\bibnamefont {Fowler}}, \bibinfo
  {author} {\bibfnamefont {B.}~\bibnamefont {Foxen}}, \bibinfo {author}
  {\bibfnamefont {C.}~\bibnamefont {Gidney}}, \bibinfo {author} {\bibfnamefont
  {M.}~\bibnamefont {Giustina}}, \bibinfo {author} {\bibfnamefont
  {R.}~\bibnamefont {Graff}}, \bibinfo {author} {\bibfnamefont {J.~A.}\
  \bibnamefont {Gross}}, \bibinfo {author} {\bibfnamefont {S.}~\bibnamefont
  {Habegger}}, \bibinfo {author} {\bibfnamefont {M.~P.}\ \bibnamefont
  {Harrigan}}, \bibinfo {author} {\bibfnamefont {A.}~\bibnamefont {Ho}},
  \bibinfo {author} {\bibfnamefont {S.}~\bibnamefont {Hong}}, \bibinfo {author}
  {\bibfnamefont {T.}~\bibnamefont {Huang}}, \bibinfo {author} {\bibfnamefont
  {W.}~\bibnamefont {Huggins}}, \bibinfo {author} {\bibfnamefont {L.~B.}\
  \bibnamefont {Ioffe}}, \bibinfo {author} {\bibfnamefont {S.~V.}\ \bibnamefont
  {Isakov}}, \bibinfo {author} {\bibfnamefont {E.}~\bibnamefont {Jeffrey}},
  \bibinfo {author} {\bibfnamefont {Z.}~\bibnamefont {Jiang}}, \bibinfo
  {author} {\bibfnamefont {C.}~\bibnamefont {Jones}}, \bibinfo {author}
  {\bibfnamefont {D.}~\bibnamefont {Kafri}}, \bibinfo {author} {\bibfnamefont
  {K.}~\bibnamefont {Kechedzhi}}, \bibinfo {author} {\bibfnamefont
  {J.}~\bibnamefont {Kelly}}, \bibinfo {author} {\bibfnamefont
  {S.}~\bibnamefont {Kim}}, \bibinfo {author} {\bibfnamefont {P.~V.}\
  \bibnamefont {Klimov}}, \bibinfo {author} {\bibfnamefont {A.~N.}\
  \bibnamefont {Korotkov}}, \bibinfo {author} {\bibfnamefont {F.}~\bibnamefont
  {Kostritsa}}, \bibinfo {author} {\bibfnamefont {D.}~\bibnamefont {Landhuis}},
  \bibinfo {author} {\bibfnamefont {P.}~\bibnamefont {Laptev}}, \bibinfo
  {author} {\bibfnamefont {M.}~\bibnamefont {Lindmark}}, \bibinfo {author}
  {\bibfnamefont {E.}~\bibnamefont {Lucero}}, \bibinfo {author} {\bibfnamefont
  {M.}~\bibnamefont {Marthaler}}, \bibinfo {author} {\bibfnamefont
  {O.}~\bibnamefont {Martin}}, \bibinfo {author} {\bibfnamefont {J.~M.}\
  \bibnamefont {Martinis}}, \bibinfo {author} {\bibfnamefont {A.}~\bibnamefont
  {Marusczyk}}, \bibinfo {author} {\bibfnamefont {S.}~\bibnamefont {McArdle}},
  \bibinfo {author} {\bibfnamefont {J.~R.}\ \bibnamefont {McClean}}, \bibinfo
  {author} {\bibfnamefont {T.}~\bibnamefont {McCourt}}, \bibinfo {author}
  {\bibfnamefont {M.}~\bibnamefont {McEwen}}, \bibinfo {author} {\bibfnamefont
  {A.}~\bibnamefont {Megrant}}, \bibinfo {author} {\bibfnamefont
  {C.}~\bibnamefont {Mejuto-Zaera}}, \bibinfo {author} {\bibfnamefont
  {X.}~\bibnamefont {Mi}}, \bibinfo {author} {\bibfnamefont {M.}~\bibnamefont
  {Mohseni}}, \bibinfo {author} {\bibfnamefont {W.}~\bibnamefont
  {Mruczkiewicz}}, \bibinfo {author} {\bibfnamefont {J.}~\bibnamefont {Mutus}},
  \bibinfo {author} {\bibfnamefont {O.}~\bibnamefont {Naaman}}, \bibinfo
  {author} {\bibfnamefont {M.}~\bibnamefont {Neeley}}, \bibinfo {author}
  {\bibfnamefont {C.}~\bibnamefont {Neill}}, \bibinfo {author} {\bibfnamefont
  {H.}~\bibnamefont {Neven}}, \bibinfo {author} {\bibfnamefont
  {M.}~\bibnamefont {Newman}}, \bibinfo {author} {\bibfnamefont {M.~Y.}\
  \bibnamefont {Niu}}, \bibinfo {author} {\bibfnamefont {T.~E.}\ \bibnamefont
  {O'Brien}}, \bibinfo {author} {\bibfnamefont {E.}~\bibnamefont {Ostby}},
  \bibinfo {author} {\bibfnamefont {B.}~\bibnamefont {Pató}}, \bibinfo
  {author} {\bibfnamefont {A.}~\bibnamefont {Petukhov}}, \bibinfo {author}
  {\bibfnamefont {H.}~\bibnamefont {Putterman}}, \bibinfo {author}
  {\bibfnamefont {C.}~\bibnamefont {Quintana}}, \bibinfo {author}
  {\bibfnamefont {J.-M.}\ \bibnamefont {Reiner}}, \bibinfo {author}
  {\bibfnamefont {P.}~\bibnamefont {Roushan}}, \bibinfo {author} {\bibfnamefont
  {N.~C.}\ \bibnamefont {Rubin}}, \bibinfo {author} {\bibfnamefont
  {D.}~\bibnamefont {Sank}}, \bibinfo {author} {\bibfnamefont {K.~J.}\
  \bibnamefont {Satzinger}}, \bibinfo {author} {\bibfnamefont {V.}~\bibnamefont
  {Smelyanskiy}}, \bibinfo {author} {\bibfnamefont {D.}~\bibnamefont {Strain}},
  \bibinfo {author} {\bibfnamefont {K.~J.}\ \bibnamefont {Sung}}, \bibinfo
  {author} {\bibfnamefont {P.}~\bibnamefont {Schmitteckert}}, \bibinfo {author}
  {\bibfnamefont {M.}~\bibnamefont {Szalay}}, \bibinfo {author} {\bibfnamefont
  {N.~M.}\ \bibnamefont {Tubman}}, \bibinfo {author} {\bibfnamefont
  {A.}~\bibnamefont {Vainsencher}}, \bibinfo {author} {\bibfnamefont
  {T.}~\bibnamefont {White}}, \bibinfo {author} {\bibfnamefont
  {N.}~\bibnamefont {Vogt}}, \bibinfo {author} {\bibfnamefont {Z.~J.}\
  \bibnamefont {Yao}}, \bibinfo {author} {\bibfnamefont {P.}~\bibnamefont
  {Yeh}}, \bibinfo {author} {\bibfnamefont {A.}~\bibnamefont {Zalcman}},\ and\
  \bibinfo {author} {\bibfnamefont {S.}~\bibnamefont {Zanker}},\ }\href
  {https://arxiv.org/abs/2010.07965} {\bibfield  {journal} {\bibinfo  {journal}
  {arXiv preprint arXiv:2010.07965}\ } (\bibinfo {year} {2020})}\BibitemShut
  {NoStop}%
\bibitem [{\citenamefont {Wei}\ \emph {et~al.}(2024)\citenamefont {Wei},
  \citenamefont {Lauer}, \citenamefont {Pritchett}, \citenamefont {Shanks},
  \citenamefont {McKay},\ and\ \citenamefont {Javadi-Abhari}}]{wei2024native}%
  \BibitemOpen
  \bibfield  {author} {\bibinfo {author} {\bibfnamefont {K.~X.}\ \bibnamefont
  {Wei}}, \bibinfo {author} {\bibfnamefont {I.}~\bibnamefont {Lauer}}, \bibinfo
  {author} {\bibfnamefont {E.}~\bibnamefont {Pritchett}}, \bibinfo {author}
  {\bibfnamefont {W.}~\bibnamefont {Shanks}}, \bibinfo {author} {\bibfnamefont
  {D.~C.}\ \bibnamefont {McKay}},\ and\ \bibinfo {author} {\bibfnamefont
  {A.}~\bibnamefont {Javadi-Abhari}},\ }\href
  {https://doi.org/10.1103/PRXQuantum.5.020338} {\bibfield  {journal} {\bibinfo
   {journal} {PRX Quantum}\ }\textbf {\bibinfo {volume} {5}},\ \bibinfo {pages}
  {020338} (\bibinfo {year} {2024})}\BibitemShut {NoStop}%
\bibitem [{\citenamefont {Bao}\ \emph {et~al.}(2022)\citenamefont {Bao},
  \citenamefont {Deng}, \citenamefont {Ding}, \citenamefont {Gao},
  \citenamefont {Gao}, \citenamefont {Huang}, \citenamefont {Jiang},
  \citenamefont {Ku}, \citenamefont {Li}, \citenamefont {Ma}, \citenamefont
  {Ni}, \citenamefont {Qin}, \citenamefont {Song}, \citenamefont {Sun},
  \citenamefont {Tang}, \citenamefont {Wang}, \citenamefont {Wu}, \citenamefont
  {Xia}, \citenamefont {Yu}, \citenamefont {Zhang}, \citenamefont {Zhang},
  \citenamefont {Zhang}, \citenamefont {Zhou}, \citenamefont {Zhu},
  \citenamefont {Shi}, \citenamefont {Chen}, \citenamefont {Zhao},\ and\
  \citenamefont {Deng}}]{bao2022fluxonium}%
  \BibitemOpen
  \bibfield  {author} {\bibinfo {author} {\bibfnamefont {F.}~\bibnamefont
  {Bao}}, \bibinfo {author} {\bibfnamefont {H.}~\bibnamefont {Deng}}, \bibinfo
  {author} {\bibfnamefont {D.}~\bibnamefont {Ding}}, \bibinfo {author}
  {\bibfnamefont {R.}~\bibnamefont {Gao}}, \bibinfo {author} {\bibfnamefont
  {X.}~\bibnamefont {Gao}}, \bibinfo {author} {\bibfnamefont {C.}~\bibnamefont
  {Huang}}, \bibinfo {author} {\bibfnamefont {X.}~\bibnamefont {Jiang}},
  \bibinfo {author} {\bibfnamefont {H.-S.}\ \bibnamefont {Ku}}, \bibinfo
  {author} {\bibfnamefont {Z.}~\bibnamefont {Li}}, \bibinfo {author}
  {\bibfnamefont {X.}~\bibnamefont {Ma}}, \bibinfo {author} {\bibfnamefont
  {X.}~\bibnamefont {Ni}}, \bibinfo {author} {\bibfnamefont {J.}~\bibnamefont
  {Qin}}, \bibinfo {author} {\bibfnamefont {Z.}~\bibnamefont {Song}}, \bibinfo
  {author} {\bibfnamefont {H.}~\bibnamefont {Sun}}, \bibinfo {author}
  {\bibfnamefont {C.}~\bibnamefont {Tang}}, \bibinfo {author} {\bibfnamefont
  {T.}~\bibnamefont {Wang}}, \bibinfo {author} {\bibfnamefont {F.}~\bibnamefont
  {Wu}}, \bibinfo {author} {\bibfnamefont {T.}~\bibnamefont {Xia}}, \bibinfo
  {author} {\bibfnamefont {W.}~\bibnamefont {Yu}}, \bibinfo {author}
  {\bibfnamefont {F.}~\bibnamefont {Zhang}}, \bibinfo {author} {\bibfnamefont
  {G.}~\bibnamefont {Zhang}}, \bibinfo {author} {\bibfnamefont
  {X.}~\bibnamefont {Zhang}}, \bibinfo {author} {\bibfnamefont
  {J.}~\bibnamefont {Zhou}}, \bibinfo {author} {\bibfnamefont {X.}~\bibnamefont
  {Zhu}}, \bibinfo {author} {\bibfnamefont {Y.}~\bibnamefont {Shi}}, \bibinfo
  {author} {\bibfnamefont {J.}~\bibnamefont {Chen}}, \bibinfo {author}
  {\bibfnamefont {H.-H.}\ \bibnamefont {Zhao}},\ and\ \bibinfo {author}
  {\bibfnamefont {C.}~\bibnamefont {Deng}},\ }\href
  {https://doi.org/10.1103/PhysRevLett.129.010502} {\bibfield  {journal}
  {\bibinfo  {journal} {Phys. Rev. Lett.}\ }\textbf {\bibinfo {volume} {129}},\
  \bibinfo {pages} {010502} (\bibinfo {year} {2022})}\BibitemShut {NoStop}%
\bibitem [{\citenamefont {Križan}\ \emph {et~al.}(2025)\citenamefont
  {Križan}, \citenamefont {Biznárová}, \citenamefont {Chen}, \citenamefont
  {Hogedal}, \citenamefont {Osman}, \citenamefont {Warren}, \citenamefont
  {Kosen}, \citenamefont {Li}, \citenamefont {Abad}, \citenamefont {Aggarwal},
  \citenamefont {Caputo}, \citenamefont {Fernández-Pendás}, \citenamefont
  {Gaikwad}, \citenamefont {Grönberg}, \citenamefont {Nylander}, \citenamefont
  {Rehammar}, \citenamefont {Rommel}, \citenamefont {Yuzephovich},
  \citenamefont {Frisk~Kockum}, \citenamefont {Govenius}, \citenamefont
  {Tancredi},\ and\ \citenamefont {Bylander}}]{krizan2025quantum}%
  \BibitemOpen
  \bibfield  {author} {\bibinfo {author} {\bibfnamefont {C.}~\bibnamefont
  {Križan}}, \bibinfo {author} {\bibfnamefont {J.}~\bibnamefont
  {Biznárová}}, \bibinfo {author} {\bibfnamefont {L.}~\bibnamefont {Chen}},
  \bibinfo {author} {\bibfnamefont {E.}~\bibnamefont {Hogedal}}, \bibinfo
  {author} {\bibfnamefont {A.}~\bibnamefont {Osman}}, \bibinfo {author}
  {\bibfnamefont {C.~W.}\ \bibnamefont {Warren}}, \bibinfo {author}
  {\bibfnamefont {S.}~\bibnamefont {Kosen}}, \bibinfo {author} {\bibfnamefont
  {H.-X.}\ \bibnamefont {Li}}, \bibinfo {author} {\bibfnamefont
  {T.}~\bibnamefont {Abad}}, \bibinfo {author} {\bibfnamefont {A.}~\bibnamefont
  {Aggarwal}}, \bibinfo {author} {\bibfnamefont {M.}~\bibnamefont {Caputo}},
  \bibinfo {author} {\bibfnamefont {J.}~\bibnamefont {Fernández-Pendás}},
  \bibinfo {author} {\bibfnamefont {A.}~\bibnamefont {Gaikwad}}, \bibinfo
  {author} {\bibfnamefont {L.}~\bibnamefont {Grönberg}}, \bibinfo {author}
  {\bibfnamefont {A.}~\bibnamefont {Nylander}}, \bibinfo {author}
  {\bibfnamefont {R.}~\bibnamefont {Rehammar}}, \bibinfo {author}
  {\bibfnamefont {M.}~\bibnamefont {Rommel}}, \bibinfo {author} {\bibfnamefont
  {O.~I.}\ \bibnamefont {Yuzephovich}}, \bibinfo {author} {\bibfnamefont
  {A.}~\bibnamefont {Frisk~Kockum}}, \bibinfo {author} {\bibfnamefont
  {J.}~\bibnamefont {Govenius}}, \bibinfo {author} {\bibfnamefont
  {G.}~\bibnamefont {Tancredi}},\ and\ \bibinfo {author} {\bibfnamefont
  {J.}~\bibnamefont {Bylander}},\ }\href
  {https://doi.org/10.1088/1367-2630/adeba7} {\bibfield  {journal} {\bibinfo
  {journal} {New Journal of Physics}\ }\textbf {\bibinfo {volume} {27}},\
  \bibinfo {pages} {074507} (\bibinfo {year} {2025})}\BibitemShut {NoStop}%
\end{thebibliography}%

\end{document}